\documentclass[preprint,3p,sort&compress]{elsarticle}
\usepackage{bbold}
\usepackage[T1]{fontenc}
\usepackage{lmodern}
\usepackage[english]{babel}
\usepackage{amsfonts,amsmath,amssymb,mathtools,graphicx,verbatim}
\usepackage{psfrag}
\usepackage{xcolor}
\usepackage{graphicx}
\usepackage{booktabs}
\usepackage{fontawesome5}

\usepackage{pythonhighlight}
\usepackage{physics}
\usepackage{qcircuit}
\usepackage[colorlinks=true,linkcolor=magenta,citecolor=magenta,urlcolor=magenta,linktocpage=True]{hyperref}
\definecolor{darkred}{rgb}{0.5,0,0}
\definecolor{darkgreen}{rgb}{0,0.5,0}
\definecolor{darkblue}{rgb}{0,0,0.5}
\hypersetup{%
    linkcolor=darkred,
    filecolor=darkgreen,
    urlcolor=darkblue,
    citecolor=darkgreen,
    pdfauthor={Neill Lambert, Eric Giguère, Paul Menczel, Boxi Li, Patrick Hopf, Gerardo Suárez, Marc Gali, Jake Lishman, Rushiraj Gadhvi, Rochisha Agarwal, Asier Galicia, Nathan Shammah, Paul Nation, J. R. Johansson, Shahnawaz Ahmed, Simon Cross, Alexander Pitchford, Franco Nori},
}

\usepackage{listings}

\lstnewenvironment{numbered_python}[1][]{\lstset{style=mypython,numbers=left}}{}

\definecolor{codegreen}{rgb}{0,0.6,0}
\definecolor{codegray}{rgb}{0.5,0.5,0.5}
\definecolor{codepurple}{rgb}{0.58,0,0.82}
\definecolor{backcolour}{rgb}{0.95,0.95,0.92}
\lstdefinestyle{mystyle}{
    backgroundcolor=\color{backcolour},   
    commentstyle=\color{codegreen},
    keywordstyle=\color{magenta},
    numberstyle=\tiny\color{codegray},
    stringstyle=\color{codepurple},
    basicstyle=\ttfamily\footnotesize,
    breakatwhitespace=false,         
    breaklines=true,                 
    captionpos=b,                    
    keepspaces=true,                 
    numbers=none,                    
    numbersep=5pt,                  
    showspaces=false,                
    showstringspaces=false,
    showtabs=false,                  
    tabsize=2,
    escapechar=?
}
\def\code#1{\texttt{#1}}

\newcommand{\nocontentsline}[3]{}
\newcommand{\tocless}[2]{\bgroup\let\addcontentsline=\nocontentsline#1{#2}\egroup}

\usepackage{makecell}

\newcommand{\fig}[1]{Fig.~\ref{#1}}

\DeclarePairedDelimiter{\ex}{\langle}{\rangle}

\begin{document}
\title{QuTiP 5: The Quantum Toolbox in Python}
\author[8b]{Neill Lambert\corref{cor1}}
\ead{nwlambert@gmail.com}
\author[3]{Eric Giguère}
\author[8b]{Paul Menczel\corref{cor1}}
\ead{paul@menczel.net}
\author[2]{Boxi Li}
\author[8b,9]{Patrick Hopf}
\author[8b,4]{Gerardo Su\'arez}
\author[6]{Marc Gali}  
\author[5]{Jake Lishman} 
\author[15]{Rushiraj Gadhvi} 
\author[8b]{Rochisha Agarwal} 
\author[2b]{Asier Galicia} 
\author[7]{Nathan Shammah} 
\author[13]{Paul Nation}
\author[12]{J. R. Johansson}
\author[11]{Shahnawaz Ahmed}
\author[16]{Simon Cross}
\author[1]{Alexander Pitchford}
\author[8b,14]{Franco Nori\corref{cor1}}
\ead{fnori@riken.jp}
\cortext[cor1]{Corresponding authors}
\date{\today}
\affiliation[8b]{organization={Quantum Information Physics Theory Research Team, \\RIKEN Center for Quantum Computing, RIKEN, Wakoshi, Saitama 351-0198, Japan}} 
\affiliation[3]{organization={Institut quantique, Université de Sherbrooke, Sherbrooke J1K 2R1 Quebec, Canada}}
\affiliation[2]{organization={Peter Gr\"unberg Institute -Quantum Control (PGI-8), Forschungszentrum J\"ulich GmbH, D-52425 J\"ulich, Germany}}
\affiliation[9]{organization={Technical University of Munich, Munich, Germany}}
\affiliation[4]{organization={International Centre for Theory of Quantum Technologies (ICTQT), University of Gdansk, 80-308 Gdansk, Poland}}
\affiliation[6]{organization={Global Research and Development Center for Business by Quantum-AI Technology (G-QuAT), National Institute of Advanced Industrial Science and Technology (AIST), Tsukuba, Ibaraki 305-8568, Japan}}
\affiliation[5]{organization={IBM Quantum, IBM Research Europe, Hursley, United Kingdom}}
\affiliation[15]{organization={Plaksha University, Mohali, India}}
\affiliation[2b]{organization={Peter Gr\"unberg Institute -Functional Quantum Systems (PGI-13), \\ Forschungszentrum J\"ulich GmbH, D-52425 J\"ulich, Germany}}
\affiliation[7]{organization={Unitary Foundation, Walnut, California 91789, USA}}
\affiliation[13]{organization={IBM Quantum, IBM T.\ J.\ Watson Research Center, Yorktown Heights, New York 10598, USA}}
\affiliation[12]{organization={Data and AI Division, Rakuten, Tokyo, Japan}}
\affiliation[11]{organization={Wallenberg Centre for Quantum Technology, Department of Microtechnology and Nanoscience, Chalmers University of Technology, 412 96 Gothenburg, Sweden}}
\affiliation[16]{organization={Zurich Instruments, Zurich, Switzerland}}
\affiliation[1]{organization={Department of Mathematics, Aberystwyth University, Penglais Campus, Aberystwyth, SY23 3BZ, Wales, United Kingdom}}
\affiliation[14]{organization={Quantum Research Insitute and Physics Department, University of Michigan, Ann Arbor, MI 48109-1040, USA}}

\begin{abstract}
QuTiP, the Quantum Toolbox in Python \cite{Johansson12,Johansson13}, has been at the forefront of open-source quantum software for the past 13 years. It is used as a research, teaching, and industrial tool, and has been downloaded millions of times by users around the world. Here we introduce the latest developments in QuTiP v5, which are set to have a large impact on the future of QuTiP and enable it to be a modern, continuously developed and popular tool for another decade and more. We summarize the code design and fundamental data layer changes as well as efficiency improvements, new solvers, applications to quantum circuits with QuTiP-QIP,  and new quantum control tools with QuTiP-QOC.  Additional flexibility in the data layer underlying all ``quantum objects'' in QuTiP allows us to harness the power of  state-of-the-art data formats and packages like JAX, CuPy, and more. We explain these new features with a series of both well-known and new examples. The code for these examples is available in a static form on GitHub \cite{codeexamplesstatic} and as continuously updated and documented notebooks in the qutip-tutorials package \cite{tutorials}.   
\end{abstract}
\maketitle

\newpage 
\tableofcontents
\section{Introduction}

Open-source software plays an important role across a range of scientific disciplines, and is important for reproducibility in scientific research \cite{harris2020array}, enabling scientific education, and the transfer of academic ideas into industrial applications. Examples include the KWANT library for condensed matter physics~\cite{kwant}, Quantum ESPRESSO for density functional theory~\cite{espresso}, MDTraj for molecular dynamics~\cite{mdtraj}, and the The Astropy Project for astrophysics~\cite{astropy}, to name just a few. With the increasing interest in quantum computing, the need for open-source tools for the study of quantum noise, quantum dynamics and quantum circuits has exploded \cite{Zeng_2017_Nature, quantuminsider, steiger2016projectq, mcclean2017openfermion,2024qiskit}.
Among these tools, QuTiP, the Quantum Toolbox in Python, has remained one of the most widely used, consistently maintained, and academically independent.

In its first release about 13 years ago \cite{Johansson12,Johansson13}, QuTiP originally aimed to reproduce, in Python, the functionality of the famous Quantum Toolbox for Matlab \cite{qotbmtl}. QuTiP's initial design approach was focused around a flexible ``quantum object'' class which represents quantum states, operators and superoperators, and which allows the user to quickly and easily solve many of the standard problems that occur in the fields of quantum optics and open quantum systems \cite{breuer2002theory, Lidar2019}. For this purpose, it relied -- and still relies -- on the extensive scientific computing infrastructure available in Python, including libraries such as SciPy~\cite{virtanen2020scipy}, NumPy~\cite{harris2020array},  Cython~\cite{behnel2010cython}, and MatPlotLib~\cite{hunter2007matplotlib}. Many of QuTiP's original examples focused on traditional models from cavity quantum electrodynamics, like Lindblad master equation simulations of the open Jaynes-Cummings, Rabi and Dicke models, and tools to quickly calculate standard observable quantities associated with such systems, such as the Wigner function or the photonic $g^{(2)}(t)$ function.

Continuous development over the last 13 years by a large team of international contributors expanded the scope of QuTiP beyond traditional quantum optics, adding important features like:
\begin{itemize}
    \item optimal control methods, 
    \item quantum circuit simulators \cite{li2021pulselevel}, 
    \item a solver for the hierarchical equation of motion that describe non-Markovian dynamics \cite{lambert2020bofinheom}, 
    \item a solver that takes advantage of permutational symmetries \cite{Shammah_2018}, 
    \item stochastic master equation solvers,
    \item Floquet methods
    \end{itemize}
    and more.  Many of these improvements were done by students and first-time developers at Franco Nori's group in RIKEN, including some who were supported by the Google Summer of Code program \cite{gsoc}. For a more detailed overview of this era of QuTiP's development history, we refer to Sec.~\ref{sec:history} below.

\subsection{QuTiP v5}

In the last several years, the development team has pushed towards a new milestone release for QuTiP, v5, which combines multiple deep changes to the internals of QuTiP. These changes were originally designed and spearheaded in a project by Jake Lishman under the supervision of Eric Giguère and Alex Pitchford. The changes primarily focus on a generalization of the QuTiP data layer, which previously only supported a single data format: the Compressed Sparse Row (CSR) format. This format is a good choice for many applications, but other formats had to be introduced in an ad-hoc way sometimes. An example is QuTiP's optimal control library, where the exponentiation and multiplication of many small matrices can be performed more efficiently with a dense data format.
\subsubsection{A new flexible data layer}
It was thus recognized that a flexible data format would make significant performance improvements possible, and that it would also allow for the use of packages like CuPy, JAX, and cuquantum, which make integration with GPU and XLA hardware easier and provide powerful features like automatic differentiation. For example, an earlier effort \cite{qgrad} to make use of JAX had required sweeping changes throughout the entire QuTiP repository, which implied having to copy and maintain the code in parallel. 

In QuTiP v5, a new flexible data layer takes center stage and facilitates new features to be included easily and quickly. At the time of the release of QuTiP v5, it already supports several new data formats in addition to the existing CSR format: the native ``Dia'' and ``Dense'' formats (for diagonal sparse matrices and for dense matrices, respectively), and two JAX-based formats for diagonal sparse and dense matrices provided through the QuTiP-JAX optional sub-package.  In this article, we will carefully demonstrate, using a variety of examples and benchmarks, the circumstances in which these different data formats can be used and taken advantage of.
\subsubsection{Additional substantial changes}
In addition, QuTiP v5 includes a solver class interface, new methods for the integration of differential equations, updated tutorials and examples (in the form of Jupyter notebooks), and a large amount of miscellaneous improvements and bug fixes identified by new unit tests.  
Furthermore, starting with QuTiP v4.7, there has been a concerted effort to reduce the complexity and weight of the core QuTiP package by moving feature-rich aspects into their own sub-packages. This is exemplified by QuTiP-QIP and QuTiP-QOC, which now contain the most recent versions of the circuit simulator and optimal control libraries, respectively. This strategy reduces the maintenance cost of the core package, and the chance that dependencies in these sub-packages will break or interrupt core features.

The aim of this article is to provide a detailed explanation, with examples, of existing and new features of QuTiP. We chose examples which either demonstrate unusual use cases for QuTiP not covered already in the documentation, or ones which allow us to compare the regime of validity of different solvers. We end with an outlook and strategy for the future development of QuTiP into the next decade.

\section{The QuTiP project} \label{sec:history}

QuTiP began more than ten years ago as a collaborative project between Robert Johansson and Paul Nation, then two postdoctoral researchers in the group of Franco Nori in RIKEN, Japan. At that time, well-maintained and easy-to-use open source software packages for implementing common numerical methods in the fields of open quantum systems, quantum optics and quantum information were limited. One of the most widely known packages was the Quantum Toolbox for Matlab, developed by Sze M.\ Tan at the University of Auckland.  However, it had not seen active development since 2002, and relied on the commercial closed-source Matlab environment.

The programming language Python has seen extensive adoption across academia, particularly in the data science community. Its easy-to-read philosophy and its quickly increasing support for scientific calculations through packages like SciPy and NumPy made it very appealing as a platform for the development of a modern re-implementation of the Quantum Toolbox. To make the new package appealing to the community, its developers adopted the philosophy of mirroring the feature set and some of the syntax of the Quantum Toolbox in Matlab (nowadays, new packages like the quantumtoolbox.jl library for the Julia programming language are, in turn, being built to mirror QuTiP). 
This culminated in the first release of QuTiP in 2012 with a feature-set comparable to Matlab's toolbox \cite{Johansson12}. 

Only one year later, QuTiP v2 was released \cite{Johansson13}. Alongside API and efficiency improvements, it included solvers supporting arbitrary time-dependent Hamiltonians and collapse operators, and new solvers for Bloch–Redfield and Floquet–Markov master equations.

QuTiP v3 was released in 2014, including stochastic master equation and stochastic Schrödinger equation solvers, a broader range of methods for finding steady states, and the first version of a circuit simulator called \code{qutip.qip}.

The road from QuTiP v3 to v4 took more time, with the release of version 4.0 occurring only in late 2016. Minor releases in between introduced important new features like a hierarchical equations of motion (HEOM) solver and a new module for optimal control.
The optimal control module, then named \code{qutip.control}, included support for the powerful gradient ascent pulse engineering (GRAPE) and chopped random basis (CRAB) algorithms.

After the release of version 4.0, the development of QuTiP underwent a transition to a series of many minor releases.   In addition, the development team grew to a larger international team including full-time developers, volunteers contributing specialized functionality (like improvements of the optimal control and HEOM modules, or the permutational invariant quantum solver) and students who joined the team through internships at RIKEN or through Google Summer of Code projects.

Significant releases during this time include: 
\begin{itemize}
  \item  a time-dependent Bloch-Redfield equation solver in v4.2 (2017), 
    \item  the permutational invariant quantum solver (PIQS \cite{Shammah_2018}) in v4.3.1 (2018), 
    \item the introduction of the \code{QObjEvo} class in v4.4 (2019) and
      \item   the first major update to the circuit simulator QuTiP-QIP in v4.5 (2020).   
          
\end{itemize}

QuTiP v4.6 (2021) saw:
\begin{itemize}
  \item further improvements to QuTiP-QIP, 
  \item OpenQASM support, 
  \item and the release of binary wheels on pip, which made continued support for Windows and other platforms much easier. 
\end{itemize}

  Finally, QuTiP v4.7 (2022) brought:
  \begin{itemize}
  \item a major update to the HEOM solver and 
  \item the introduction of a Krylov subspace solver.  
  \end{itemize}
  In February 2023, the first alpha pre-release version of QuTiP v5 was published. 

As described earlier, version 5 is a substantial new release. It includes deep and far-reaching changes to many of the core components of QuTiP. To avoid bugs or errors plaguing users, or incompatibility with old code, the pre-alpha and alpha development stages stretched over all of 2023.  In March 2024, a large QuTiP developer's workshop was held in Franco Nori's group at RIKEN and, during this workshop, QuTiP v5 was finally fully released. It represents a substantial reinvention of what QuTiP can achieve and validates the success of the academic support of open-source science, of programs like JST Moonshot and GSoC, and of non-profit organizations such as NumFOCUS and the Unitary Foundation.






\section{Core features}

\subsection{Quantum objects and the data layer}

At its core, QuTiP is a library for manipulating arbitrary quantum objects and for solving the time evolution of both open and closed quantum systems.
Its aim is to remove burdens from the researcher and to give them interactive programmatic access to descriptive objects and solvers. 

During its history, it has generally aspired to versatility and ease of use rather than optimization or keeping up with state-of-the-art benchmarks.  However, as the community using quantum software has grown and demands for the support of high-performance computing platforms have risen, a fundamental change in QuTiP's concept of a quantum object was needed to remain competitive and useful.  Towards this goal, QuTiP v5's design enables the support of arbitrary data formats through a flexible and powerful data layer, which features dynamic conversion between data types. This means automatic conversion when objects of different type are combined together, and the option to choose the data type that is optimal for a particular task.

\begin{figure}[p!]
\centering
\includegraphics[width = 1.\columnwidth]{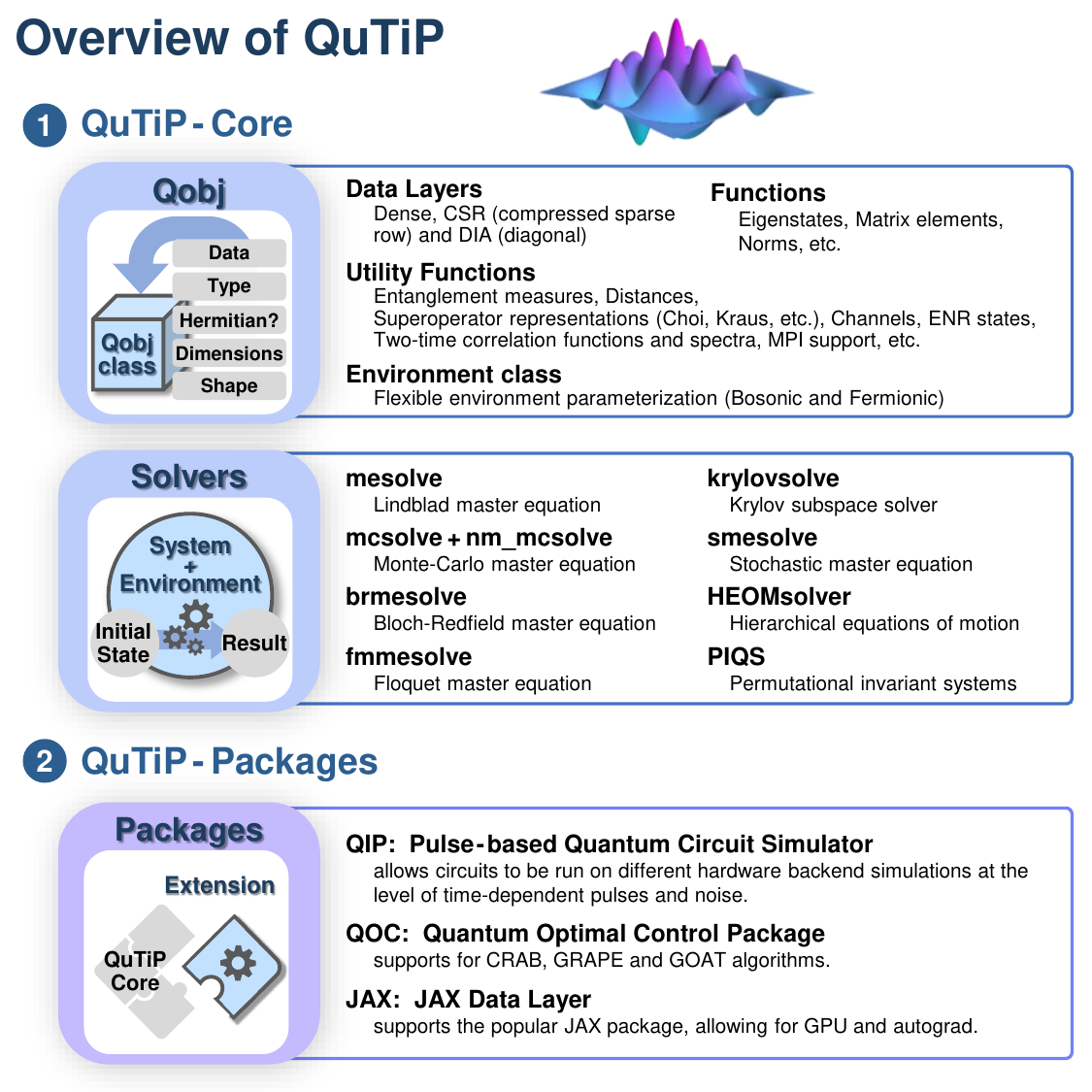}
\caption{A schematic overview of the QuTiP project, describing \code{Qobj} and its features/functions, solvers, and QuTiP sub-packages. For a complete list of \code{Qobj} methods and attributes see Table \ref{tab:qobj_methods}, for a list of libraries which QuTiP uses see Table \ref{tab:glossary_acronyms}, for a glossary of terms commonly used in QuTiP see Table \ref{tab:glossary_terms}, and for a list of state, operator, superoperator, entanglement measure and metric functions see Tables \ref{tab:qutip_states_functions}, \ref{tab:qutip_operators_functions}, \ref{tab:qutip_superoperators_functions}, \ref{table:entropy_functions} and \ref{table:metrics_functions}.}\label{schematic}
\end{figure}

\begin{table}[ht]
    \centering
         \renewcommand{\arraystretch}{1.3} 
    \begin{tabular}{@{}lp{9.8cm}@{}}\toprule
        \textbf{Method} & \textbf{Description} \\
        \midrule
        \code{qobj.to(x)} & Change the data layer to the format specified by the string \code{x}, which may be \code{"dense"}, \code{"csr"}, or \code{"dia"}.  If QuTiP-JAX is installed, \code{"jax"} and \code{"jaxdia"} are also available. \\

        \code{dtype=x}&  When creating quantum objects, most functions allow the user to specify the data-layer type \code{x} (allowed values described above). \\

         \code{qt.CoreOptions(default\_dtype=x)} & Setting this option will cause quantum objects created with most internal functions to use the same data-layer type \code{x} (allowed values described above) by default. \\

        \code{qobj.data\_as()} & Returns the raw data defining the quantum object in its current format.\\
\bottomrule
    \end{tabular}
    \caption{Summary of methods to change the data-layer in a Qobj}
    \label{tab:data_layer}
\end{table}

\begin{table}[ht]
    \centering
         \renewcommand{\arraystretch}{1.3} 
    \begin{tabular}{@{}lp{12cm}@{}}
        \toprule
        \textbf{Acronym} & \textbf{Description} \\
        \midrule
        NumPy & Numerical Python: A library for the Python programming language that provides support for large, multi-dimensional arrays and matrices, along with a collection of mathematical functions to operate on these arrays. \\
      
        SciPy & Scientific Python: A Python library used for scientific and technical computing, building on the capabilities of NumPy. It includes functions for optimization, integration, interpolation, eigenvalue problems, and other advanced mathematical operations. \\
 
        Matplotlib & A plotting library for the Python programming language, used for creating static, interactive, and animated visualizations. \\
   
        JAX & A library developed by Google for high-performance numerical computing that is particularly useful for machine learning and scientific computing, with automatic differentiation capabilities. \\
  
        Diffrax & Differential Equations in JAX: A library that provides numerical solvers for differential equations, compatible with JAX for high-performance computation. \\
   
        MKL & Math Kernel Library: A library developed by Intel that provides optimized mathematical routines. \\
  
        MPI & Message Passing Interface: A standardized and portable message-passing system used for parallel computing. \\
\bottomrule
    \end{tabular}
    \caption{Glossary of acronyms and terms for libraries used by QuTiP}
    \label{tab:glossary_acronyms}
\end{table}

\begin{table}[ht]
    \centering
         \renewcommand{\arraystretch}{1.3} 
    \begin{tabular}{@{}lp{12cm}@{}}
        \toprule
        \textbf{Term} & \textbf{Description} \\  
 \midrule
        CSR & Compressed Sparse Row: A SciPy data format for sparse matrices. \\
 
        ENR States & Excitation Number Restricted States: Quantum states where the total number of excitations across subsystems is restricted, reducing the Hilbert space dimension size drastically. \\
  
        Floquet & Floquet theory is a tool for the description of periodically driven systems. Some QuTiP solvers are based on Floquet theory. \\
 
        HEOM & Hierarchical Equations of Motion: A formalism used in quantum dynamics to describe non-Markovian open quantum systems. \\
      
        ODE & Ordinary Differential Equation: The QuTiP solvers allow users to choose from various numerical ODE integration methods. \\

        PIQs & Permutationally Invariant Quantum Solver \cite{Shammah_2018}: A solver in QuTiP for efficiently simulating systems that have permutational symmetry, reducing computational complexity. \\

        QuTiP-JAX & QuTiP-JAX: A module in QuTiP that leverages JAX, allowing the use of GPUs and automatic differentiation. \\

        QuTiP-QIP & QuTiP Quantum Information Processing: A module in QuTiP that provides tools for simulating quantum circuits, quantum algorithms, and other aspects of quantum information processing. \\

        QuTiP-QOC & QuTiP Quantum Optimal Control: A module in QuTiP for designing and optimizing quantum control pulses, allowing users to find control solutions for specific quantum dynamics tasks. \\   
 
        GRAPE & Gradient Ascent Pulse Engineering: A numerical algorithm used in quantum optimal control. It uses time discretization in order to identify the optimal control pulses for quantum systems. \\
  
        CRAB & Chopped Random Basis: A method for quantum optimal control that uses a randomly truncated basis to optimize control pulses. \\
   
        GOAT & Gradient Optimization of Analytic Controls: An optimal control method that uses analytical functions for the available controls. \\
        \bottomrule
    \end{tabular}
    \caption{Glossary of acronyms and terms for large subpackages and methods available in QuTiP}
    \label{tab:glossary_terms}
\end{table}

\subsubsection{The \textup{\code{Qobj}} class}
To understand this fundamental change, we have to introduce the cornerstone of QuTiP, the Python class \code{Qobj}, which provides an intuitive way to store and manipulate commonly used quantum objects.  In most instances, \code{Qobj}s are used to store quantum states (vectors), operators or super-operators, and manipulate them according to the rules of linear algebra. 

For example, a Pauli-Z matrix can be simply defined either using the built-in function \code{sigmaz()} or by constructing a \code{Qobj()} from a matrix: 

\inputpython{qobj.tex}{1}{7}
This \code{Qobj} is an operator, as indicated by its \code{type}, the \code{shape} of its matrix representation, and its dimensions (\code{dims}).  The dimensions allow us to keep track of subsystems and indirectly infer the type of the quantum object itself.  For example, in this case the object is an operator on a single sub-system, and can be thought of as a map which takes vectors to vectors.
\newpage 

Compare the output above to a tensor product of two Pauli operators:

\inputpython{qobj.tex}{9}{18}
Here, we used the \code{\&} operator introduced in version 5, which denotes the tensor product (instead, we could also have used the \code{tensor()} function like in earlier versions of QuTiP).  We see that the dimensions have the structure \code{[[2,2],[2,2]]}, which implies that the operator acts on vectors that live in the tensor product of two $2$-state subsystems.  In contrast, an operator acting on a single 4-state system would have the dimensions \code{[[4],[4]]}.  A more complex version of this sub-system labeling can be seen with superoperators, which map operators to operators, and will be explored later when the Lindblad master equation solver is introduced.

\begin{table}[ht]
     \renewcommand{\arraystretch}{1.25} 
    \centering
    \begin{tabular}{@{}lp{10cm}@{}}
        \toprule
        \textbf{Function} & \textbf{Description} \\
        \midrule
        \code{basis(N,n)} & Generates the vector representation of a Fock state, $\ket{n}$ in an $N$-dimensional Hilbert space. Also supports lists for construction of tensor spaces. \\
    
        \code{qutrit\_basis()} & Returns a list of basis states for a qutrit (three-level) system. \\
        
        \code{bra(x)} & Produces a bra state given a list or string of excitation numbers. For example, $x=\text{``01010''}$ produces the bra state $\bra{01010}$. \\
        
        \code{ket(x)} & Produces a ket state given a list or string, as with \code{bra}. \\
        
        \code{fock(N,n)} & Bosonic Fock (number) state. Same as \code{basis}. \\
        
        \code{coherent(N, $\alpha$)} & Generates a coherent state with eigenvalue $\alpha$ for a harmonic oscillator in an $N$-dimensional truncated Hilbert space. \\
        
        \code{spin\_state(j,m)} & Generates the spin state with the quantum numbers $j$ and $m$. \\
        
        \code{spin\_coherent(j, $\theta$, $\phi$)} & Generates the spin coherent state $\ket{\theta, \phi}$ for a spin-$j$ system. \\
        
        \code{projection(N,n,m)} & Generates the projection operator from state $m$ onto state $n$, i.e., $\ket{n}\bra{m}$, for a system with Hilbert space dimension $N$. \\
       
        \code{maximally\_mixed\_dm(N)} & Generates the maximally mixed density matrix for dimension $N$.\\
       
        \code{fock\_dm(N,n)} & Density matrix representation of the Fock state. \\
        
        \code{coherent\_dm(N, $\alpha$)} & Density matrix representation of the coherent state. \\
        
        \code{thermal\_dm(N,n)} & Density matrix for the thermal state of a harmonic oscillator in an $N$-dimensional truncated Hilbert space. \\
     
        \code{singlet\_state()} & Returns the singlet state $\ket{S}=\frac{1}{\sqrt{2}}\bigl( \ket{01}-\ket{10} \bigr)$. \\
       
        \code{triplet\_states()} & Returns a list of the triplet states $\ket{11}$, $\frac{1}{\sqrt{2}}\bigl( \ket{01}+\ket{10} \bigr)$ and $\ket{00}$. \\
      
        \makecell{\code{bell\_state(}\\ \code{    ['00', '01', '10', '11'])}} & Returns the selected Bell state, 
        $\ket{B_{00}} = \frac1{\sqrt2}\bigl( \ket{00} + \ket{11} \bigr)$, 
        $\ket{B_{01}} = \frac1{\sqrt2}\bigl( \ket{00} - \ket{11} \bigr)$,
        $\ket{B_{10}} = \frac1{\sqrt2} \bigl( \ket{01} + \ket{10} \bigr)$, 
        $\ket{B_{11}} = \frac1{\sqrt2} \bigl( \ket{01} - \ket{10} \bigr)$.\\
        
        \code{ghz\_state(N)} & Produces the $N$-qubit GHZ-state, e.g., $\frac1{\sqrt2}\bigl( \ket{0000} + \ket{1111} \bigr)$ for 4 qubits. \\
        
        \code{w\_state(N)} & Returns the W  state of $N$ qubits, $\frac1{\sqrt N}\bigl( \ket{100 \cdots 0} + \ket{010 \cdots 0} + \dots + \ket{00 \cdots 1} \bigr)$. \\
       \bottomrule
    \end{tabular}
    \caption{List of commonly used functions to create pre-defined states}
    \label{tab:qutip_states_functions}
\end{table}

\begin{table}[ht]
    \centering
         \renewcommand{\arraystretch}{1.2} 
    \begin{tabular}{@{}lp{9.6cm}@{}}
        \toprule
        \textbf{Function} & \textbf{Description} \\
        \midrule
        \code{commutator(A,B)} & Computes the commutator (or anti-commutator, given an additional optional argument) of two operators $A$ and $B$. \\
      
        \code{identity(dimensions)} & Returns the identity operator for a single or multipartite system described by the integer or list \code{dimensions}. \\
   
        \code{qeye(dimensions)} & Alias for \code{identity(dimensions)}. \\
  
        \code{qeye\_like(qobj)} & Generates the identity operator with the same dimensions and type as the reference quantum object \code{qobj}. \\
    
        \code{create(N)},  \code{destroy(N)} & Generates the creation (raising) or annihilation (lowering) operator for an $N$-dimensional Fock-space. \\
     
        \code{momentum(N)}, \code{position(N)} & The momentum or position operator of a single $N$-dimensional harmonic oscillator, i.e., $p=\frac{i}{\sqrt{2}} \bigl(a-a^{\dagger}\bigr)$ or $x=\frac{1}{\sqrt{2}}\bigl(a+a^{\dagger}\bigr)$.\\
   
        \code{num(N)} & Generates the number operator of a single $N$-dimensional harmonic oscillator, i.e., $n = a^{\dagger}a$. \\

        \code{displace(N,$\alpha$}) & Generates the displacement operator for a distance $\alpha$ in a single $N$-dimensional Fock space, i.e., $D(\alpha) = \exp\bigl( \alpha a^{\dagger} - \alpha^* a \bigr)$.\\
       
        \code{squeeze(N,z)} & Generates the single-mode squeezing operator in a single $N$-dimensional Fock space, i.e., $S(z) = \exp\bigl[ \bigl( z^* a^{2} - z a^{\dagger 2} \bigr) / 2 \bigr]$. \\
 
        \code{squeezing(a1,a2,z)} & Generates the two-mode squeezing operator for two harmonic oscillators predefined with annihilation operators \code{a1} and \code{a2}, i.e., $S_2(z) = \exp\bigl[ \bigl( z^*a_1a_2 - za_1^\dagger a_2^\dagger \bigr) / 2 \bigr]$. \\
     
        \code{fcreate(n,m)}, \code{fdestroy(n,m)} & Generates the fermionic creation or annihilation operator using the Jordan-Wigner transformation. Returns the $m$-th fermionic operator out of $n$ total fermions.\\
       
        \code{sigmax()}, \code{sigmay()}, \code{sigmaz()} & Generates the Pauli-x, y or z operator. \\

        \code{sigmam()}, \code{sigmap()} & Generates the lowering or raising operator for Pauli spins. \\
 
        \code{spin\_Jx(j)},... & Generates the spin-\code{j}, \code{x}, \code{y} or \code{z} operator. \\

        \code{spin\_Jm(j)}, \code{spin\_Jp(j)} & Generates the spin-$j$ lowering or raising operator. \\
 
        \code{jmat(j, which)} & Alias for \code{spin\_Jx(j)}, \code{spin\_Jy(j)}, \code{spin\_Jz(j)}, \code{spin\_Jm(j)} or \code{spin\_Jp(j)}, depending on \code{which} $\in$ \code{"x","y","z","-","+"}. \\
        \bottomrule
    \end{tabular}
    \caption{List of commonly used functions to create pre-defined operators}
    \label{tab:qutip_operators_functions}
\end{table}

In previous versions of QuTiP, the data actually defining the quantum object, a vector or matrix of complex numbers, was usually represented in terms of SciPy's implementation of the ``Compressed Sparse Row'' (CSR) matrix format for complex numbers.  This was a convenient format for many problems in open quantum systems, where density operators and superoperators are often naturally sparse, since the CSR format provides a fast matrix-vector product. However, for studying the dynamics of small quantum systems or performing computations on specialized hardware like GPUs, other types of data formats are preferred.

QuTiP v5 introduces a new data layer which allows for custom data formats.  The standard formats included and available in v5 are \code{CSR}, \code{Dense} and \code{Dia} (diagonal sparse).  For example, the operator we defined earlier was constructed from dense data provided by the user and is hence in the \code{Dense} format, as indicated by the \code{dtype} in the output above. It can be converted to another format with \code{.to()}:
\newpage 
\inputpython{qobj.tex}{20}{23}
Here, the parameter \code{fortran} refers to the layout of the dense data in the computer memory, and the parameter \code{nnz} to the number of non-zero entries in the sparse matrix.

When combining data formats, such as multiplying a sparse matrix with a dense vector, QuTiP performs conversions automatically. When new data formats are added, it is sufficient to define conversion methods between that new data format and only one existing format. QuTiP will then automatically use a graph of conversion methods to allow operations between that new format and any existing one. However, converting between data formats is generally not numerically optimal, and it is good practice to restrict oneself to the format that is most useful for the problem at hand.  Specific examples will be provided later. 

The true power of this approach lies in allowing new flexible data formats to be developed in the future. As an example, together with QuTiP v5 we are also releasing QuTiP-JAX \cite{qjax}, which takes advantage of the powerful JAX library for GPU-based performance enhancements and automatic differentiation~\cite{jax2018github}. Data formats for CuPy \cite{cupy}, TensorFlow \cite{tflow} and tensor networks \cite{tnetwork} have also been explored, and alpha versions of these data layer implementations are available. They will be expanded upon in future releases.  We refer to Table \ref{tab:data_layer} for a summary of methods to convert between data layer formats.

Note that the Array API consortium \cite{arrayconsort} aims to solve a similar problem (to standardize functionality available across Python libraries and frameworks), but our custom approach to this problem offers several benefits. Firstly, it enables incremental development of new data formats (if some solver or function does not support that new format, the dispatcher will convert to a format which is supported).  Secondly, it enables interaction between different data formats; for example, it supports sparse-dense matrix-vector multiplication, which can be an optimal choice for some solvers. Finally, both of these features are powered by a multiple-dispatch system which, to our knowledge, is a unique innovation. For example, it goes beyond Julia's multiple-dispatch system, which has ``promote'' rules that cannot handle the arbitrary heterogeneous data inputs of QuTiP's implementation.

In addition to storing the data describing a quantum object, the \code{Qobj} class has a large range of built-in utility functions to calculate common properties of quantum systems. For example, \code{eigenstates()} computes the eigenvalues and eigenstates of an operator, \code{expm()} takes the matrix exponential, \code{norm()} finds the norm of states and operators, and so on.  We provide a full list of these functions in Table \ref{tab:qobj_methods} and lists of commonly used functions to create states, operators and superoperators in Tables \ref{tab:qutip_states_functions}, \ref{tab:qutip_operators_functions} and \ref{tab:qutip_superoperators_functions}. Examples of commonly used utility functions to calculate entropies, entanglement measurements, and distances between states are given in Tables \ref{table:entropy_functions} and \ref{table:metrics_functions}.


\subsubsection{The \textup{\code{QobjEvo}} class}
The \code{QobjEvo} class provides a useful extension of the \code{Qobj} class to describe time-dependent quantum objects and to optimize their use by QuTiP's solvers. As we will demonstrate later, when calling a solver with a time-dependent Hamiltonian or time-dependent bath operators, we typically specify the time-dependent operator by a list of tuples. Each tuple defines an operator and its time-dependent prefactor in the form of a Python function, an array or a string. This list of tuples is then internally converted into the \code{QobjEvo} class, which involves optimization in preparation for their use by the solvers.  

However, \code{QobjEvo} objects can also be manually created and manipulated by the user, and they provide a flexible framework for dealing with multiple time-dependent systems that have their time-dependence specified in different ways. For example, they can be instantiated from continuously defined functions or from discrete time-dependent data. In the latter case, times in between data points, where the dependence is undefined, are filled in using a cubic spline interpolation.  Different \code{QobjEvo} objects can be added, multiplied, etc., with each other and with constant \code{Qobj} and scalar objects.  They also support many of the utility methods available to \code{Qobj}s like \code{dag()} (Hermitian conjugation) or \code{conj()} (complex conjugation). Further, they support tensor products with other objects and they can be converted into superoperators. We will demonstrate the utility of these features later.


\subsection{Solvers}

Armed with the \code{Qobj}  and \code{QobjEvo} classes, one can simulate a large range of open quantum system dynamics using the solvers provided by QuTiP.  These simulations are, for the most part, initial-value ordinary differential equation problems: given an initial state for a system, a Hamiltonian (possibly time-dependent) describing its energy and interactions, and an environment described for example in terms of rates or coupling strengths (depending on the solver), QuTiP uses numerical integration (either a custom method or one provided by SciPy) to find the evolution of the system as a function of time. 

These solvers are also employed as needed in other QuTiP packages like QuTiP-QIP or QuTiP-QOC. For general tasks, one should consider the effects of the environment on the system under study and choose the solver based on the level of approximation one wishes to take. This concept, core to QuTiP and to open quantum systems in general, will be developed step by step in the following sections. A brief guide to the solvers, their fields of application and their memory usage is provided in Table \ref{tab:qutip_solvers}.



\subsubsection{A new solver class}

A convenient change in QuTiP v5 is the introduction of a unified class interface to control how a user interfaces with the solvers.  Using this class interface is optional, but useful when reusing the same Hamiltonian data with different initial conditions, time steps or options.  This procedure usually provides only a minor numerical advantage, but the speed-up can be significant if the solver is reused many times.

When a solver is instantiated, one first supplies only the Hamiltonian and the operators defining the bath (e.g., collapse operators for a Lindblad master equation). Then, the initial condition and time steps are passed to the \code{Solver.run()} method, which performs the time evolution. Alternatively, one can also use the \code{Solver.start()} and \code{Solver.step()} methods in order to manually control the spacing of the time steps during the simulation.

We will first illustrate the usage of the solvers with an elementary example that builds upon the concepts introduced in the section on the \code{Qobj} class. We consider two interacting qubits, which are not interacting with an environment and thus are entirely defined by their Hamiltonian:
\begin{equation}
 H = \frac{\epsilon_1}{2} \sigma_z^{(1)}+ \frac{\epsilon_2}{2} \sigma_z^{(2)} +g \sigma_x^{(1)} \sigma_x^{(2)}  \, .
 \label{H0}
\end{equation}
\newpage
 
\inputpython{solver.tex}{4}{22}
The Hamiltonian is an operator with the same dimensions that we saw earlier, acting on the tensor product of two $2$-state Hilbert spaces. We can see that the sparse CSR data format is used now, since we used the built-in functions \code{sigmaz()} and \code{qeye()} to construct the operator.

The dynamics of a pure state of this system obeys the Schr\"odinger equation
\begin{equation}
i\hbar\frac{d}{dt} \ket\psi = H \ket\psi \, .
\label{schrodinger}
\end{equation}
Here, the Hamiltonian $H$ is the matrix in the code snippet above. The initial condition $\left|\psi(t=0)\right>$ and the final time $t$ must be provided to the the solver at run time. For example, with the traditional approach (that is, without using the class interface), we would use the Schr\"odinger equation solver \code{sesolve()} with:

\inputpython{solver.tex}{24}{26}
Note that we have used the function \code{basis(N,n)} to construct the initial condition. This function creates a vector in an $N$-state Hilbert space with a $1$ as the amplitude for the state $|n\rangle$ (note that for historical reasons, the operator \code{sigmaz()} is defined such that \code{basis(2,0)} is the excited state and \code{basis(2,1)} is the ground state, a convention which sometimes confuses users). Alongside the standard rules of linear algebra and the use of the \code{tensor} function (or the equivalent \code{\&} operator), this allows us to create any desired state.

With the new solver class interface, we can perform the same calculation with the following:

\inputpython{solver.tex}{28}{29}
The result objects returned from both methods are equivalent.
Since we did not specify any particular observables to evaluate, the solver returns the state vector of the system at all time points specified in the \code{tlist} ($100$ equally spaced steps between $t=0$ and $t=40$).

Finally, the aforementioned manual stepping interface would be used as follows:

\inputpython{solver.tex}{31}{37}
This interface is particularly useful if the Hamiltonian depends on external control parameters such as field strengths. Such parameters can be updated in each step using the optional parameter \code{args}.

\begin{table}[t!]
    \centering
     \renewcommand{\arraystretch}{1.7} 
    \begin{tabular}{@{}lp{10cm}c@{}}
        \toprule
        \textbf{Solver} & \textbf{When to use} & \textbf{Memory} \\
        \midrule
        \code{sesolve} & Closed system evolution following Schrödinger's equation. &  $d^2$ \\
      
        \code{mesolve} & Describes completely positive (CP) Lindblad evolution. Used for Markovian environments with a flat spectral density and typically at weak system-environment coupling.\par This solver can also be used to simulate arbitrary quantum master equations with superoperator generators. & $d^4$ \\
    
        \code{brmesolve} & Also used for Markovian environments. In contrast to \code{mesolve}, no secular approximation is made, leading to non-CP dynamics.\par This solver can automatically diagonalize time-dependent system Hamiltonians at each time step, as required by the Born-Markov approximation. & $d^4$ \\

        \code{mcsolve} & Monte Carlo Wavefunction method. Converges to \code{mesolve} for large number of trajectories $N_T$. Use to simulate statistical transport properties, or if \code{mesolve} consumes too much memory. & $N_T\, d^2$ \\

        \code{nm\_mcsolve} & Like \code{mcsolve}, but can be applied to non-CP master equations. & $N_T\, d^2$ \\

        \code{ssesolve} & To simulate systems subject to continuous (homodyne or heterodyne) measurement. Can include measurement feedback. & $N_T\, d^2$ \\

        \code{smesolve} & Like \code{ssesolve}, supports additional deterministic decay channels. & $N_T\, d^4$ \\

        \code{heomsolve} & Open quantum systems with Gaussian, bosonic or fermionic environments at arbitrary coupling strength and temperature.\par In the right column, $N_c$, $N_R$ and $N_I$ are the hierarchy cutoff and the number of real and imaginary exponents, see Sec.~\ref{subsec:heom}. & $d^4 \binom{N_c + N_R + N_I}{N_c}^2$ \\

        \code{FloquetBasis} & Periodically driven closed systems. One driving period is discretized into $N_t$ time steps. & $N_t d^2$ \\
\bottomrule
    \end{tabular}
    \caption{Overview of most solvers in QuTiP. The last column is the dimensionality of the dynamical generator, which (for dense generators) asymptotically equals the space complexity, that is, the memory usage of the simulation. Here, $d$ is the Hilbert space dimension.}
    \label{tab:qutip_solvers}
\end{table}
\subsubsection{Solver and integrator options}
The behavior and the output of the Schrödinger equation solver, and of all other solvers, can be finely controlled by making use of the optional \code{options} argument. In contrast to earlier versions, the options in QuTiP v5 are now specified by a Python dictionary instead of a custom class. This change increases the flexibility for future extensions and allows different solvers to provide different sets of options more easily.

There is a large range of options available; a full list is provided in the online documentation for each solver.
Frequently used options include \code{store\_states}, determining whether the system state at each time in the provided \code{tlist} should be included in the output, and \code{store\_final\_state}, determining whether the final state of the evolution should be included. These states are then included in addition to the computed values of any requested observables, see below. Other frequently used options are \code{method}, specifying the ODE integration algorithm, and specific options for that algorithm such as the desired numerical precision and the maximum steps used in the solver:
\newpage
\inputpython{solver.tex}{39}{54}

In addition to using the \code{store\_states} option, we have here shown how to increase the precision (absolute tolerance) \code{atol}, the maximum number \code{nsteps} of integration steps between two points in the \code{tlist}, and the maximum allowed integration step \code{max\_step} of the default Adams ODE integration method. The \code{max\_step} option is often important in time-dependent problems with periods of idling interspersed with short pulses; without setting a maximum time step for the solver to take, these short pulses might be ignored when the ODE solver takes too large time steps.

\subsubsection{\textup{\code{mesolve}} part 1: A master equation solver for Lindblad dynamics and beyond} \label{mesolvept1}

While the time-dependent Schrödinger equation in principle describes the dynamics of any quantum system, it is often impossible to solve in practice when the number of constituent systems and the dimensionality of the Hilbert space become too large (this issue is known as the exponential explosion problem of quantum mechanics).  This is problematic when we want to consider how a given finite quantum system is influenced by a large, perhaps even continuous, environment.  Decades of research have lead to tractable solutions to this problem in many parameter regimes, and in QuTiP we aim to make many of these solutions readily and easily available to all.

Of these solutions, master equations are by far the most common.  
The term master equation generally refers to a first-order linear differential equation for $\rho(t)$, the reduced density operator describing the state of the finite (open) quantum system. Such equations can be derived in numerous ways, for example from various approximation schemes to microscopic system-bath models, from Hamiltonian-level stochastic noise approaches, or from abstract mathematical considerations about the general properties of completely positive trace-preserving maps of quantum states.  In the latter context, the master equations take on a particularly nice form and are called Lindblad (or Gorini–Kossakowski–Sudarshan–Lindblad) master equations. It is this form which is the most commonly used, and by default implemented in QuTiP's \code{mesolve()} function. However, \code{mesolve()} can also be applied to master equations that do not have the Lindblad form, as we will show later.

The Lindblad master equation is an equation of motion for the density operator $\rho(t)$ of the open quantum system. It has the following form:
\begin{align} \label{lindblad}
\dot\rho(t) = -\frac{i}{\hbar}[H(t),\rho(t)] + \sum_n \frac{1}{2} \left[2 C_n \rho(t) C_n^\dagger - \rho(t) C_n^\dagger C_n - C_n^\dagger C_n \rho(t)\right] \, .
\end{align}
The use of a density operator instead of a state vector allows us to consider classical uncertainty in the description of the system. The density operator
\begin{equation}
\rho = \sum_k p_k \ketbra{\psi_k}{\psi_k} \, ,
\end{equation}
where $\sum_k p_k = 1$, describes a classical mixture of possible quantum states $\ket{\psi_k}$ occurring with probabilities $p_k$. For example, this kind of state describes a system in thermal equilibrium, where $\ket{\psi_k}$ are then the eigenstates of the system's Hamiltonian and $p_k=\exp[-\hbar \omega_k / (k_\mathrm{B} T)] / Z$. Here, $k_\mathrm{B} T$ is the thermal energy, $\hbar\omega_k$ the eigenenergy, and $Z$ the partition function.  This particular state often arises as the long-time steady-state of master equations obeying the detailed balance condition (see below). 

In addition to the system Hamiltonian, we see that the evolution \eqref{lindblad} depends on so-called collapse operators, or jump operators, $C_n=\sqrt{\gamma_n}A_n$.  Here, $\gamma_n$ can generally be understood as rates describing the frequency of transitions between the states connected by the operators $A_n$. A standard way of deriving the Lindblad equation \eqref{lindblad} from a microscopic description is using the Born-Markov and secular approximations; in this case, $A_n$ is related to the environment coupling operators of the microscopic description, and $\gamma_n$ is related to the square of the coupling strength. However, as mentioned before, one can also arrive at the Lindblad equation by considering the most general possible time evolution for $\rho$ that is completely positive and trace preserving. This requirement means roughly that $\rho$ will represent a physical state at all times, having eigenvalues that are positive and sum to one (as needed for a physical probability distribution). From the latter, mathematical point of view, $A_n$ could be arbitrary operators on the system Hilbert space and $\gamma_n$ arbitrary positive rates.


In QuTiP, the master equation solver \code{mesolve()} follows mostly the same syntax as the Schrödinger equation solver discussed earlier. The main difference is that collapse operators $C_n$, defining the dissipation due to contact with an environment, can be specified. Using the same Hamiltonian, time steps, and initial condition as before, we can add dissipation on both qubits by defining the following list of collapse operators:
$C_1 = \sqrt{\gamma} \sigma_-^{(1)}$ and $C_2 = \sqrt{\gamma} \sigma_-^{(2)}$, where $\sigma_-^{(i)}$ is an operator which takes the qubit $(i)$ from its excited state to its ground state.

\inputpython{mesolve.tex}{15}{22}
We have also provided a list of expectation operators in the argument \code{e\_ops}. This argument tells the solver that we want to evaluate and return the expectation values $\ex{E_n} = \tr[E_n \rho(t)]$ for each observable $E_n$ in the list \code{e\_ops} at every time step in \code{tlist}.  If the \code{e\_ops} argument is provided, the list of system states will not be stored unless asked for by providing the flag \code{store\_states} in \code{options}.

In this example, the collapse operators were introduced phenomenologically and act locally, or independently, on each qubit.  However, different choices of collapse operators are possible; one can derive a variety of master equations from microscopic models under different types of approximations. For example, assuming that the qubits are interacting more strongly with each other than with the bath, one arrives at a so-called global master equation under the standard Born-Markov-secular approximations. The global master equation involves collapse operators that act like annihilation and creation operators on the total, coupled, eigenstates of the interacting two-qubit system, 
\begin{equation}
A_{ij} = \ketbra{\psi_i}{\psi_j} \, ,
\end{equation}
and rates
\begin{equation}
\gamma_{ij} =\abs{ \bra{\psi_i} d \ket{\psi_j} }^2 S(\Delta_{i,j})\, .
\end{equation}
Here, the states $\ket{\psi_i}$ are the eigenstates of $H$ and $\Delta_{i,j} = E_j-E_i$ are differences of eigenenergies. Furthermore, $d$ is the coupling operator of the system to the environment.  The power spectrum 
\begin{equation} \label{eq:power_spectrum_definition}
S(\omega) = 2J(\omega) [ n_{th}(\omega)+1 ] \theta(\omega) + 2J(-\omega) [n_{th}(-\omega)]\theta(-\omega) \, , 
\end{equation}
depends on details of the environment like its spectral density $J(\omega)$ and its temperature through the Bose-Einstein distribution $n_{th}(\omega)$. Here, $\theta$ is the Heaviside function. These rates now obey the detailed balance condition
\begin{equation}
    \gamma_{ij} / \gamma_{ji} = \exp[ \Delta_{i,j} / (k_\mathrm{B} T)]\, .
\end{equation}

\begin{figure*}%
\includegraphics[width = 0.5\linewidth]{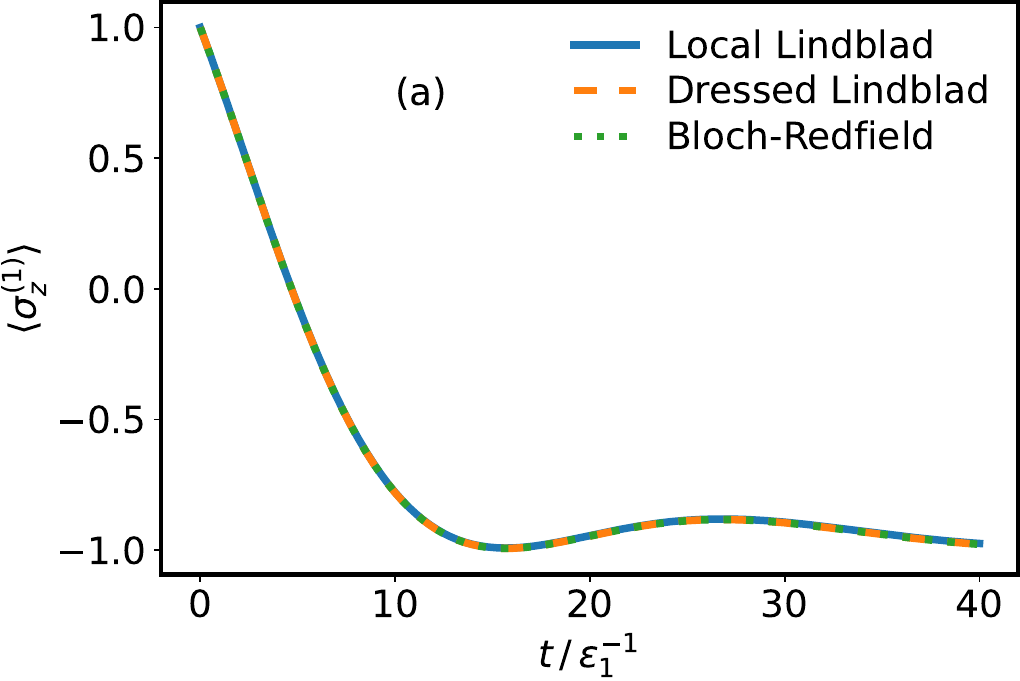}%
\includegraphics[width = 0.5\linewidth]{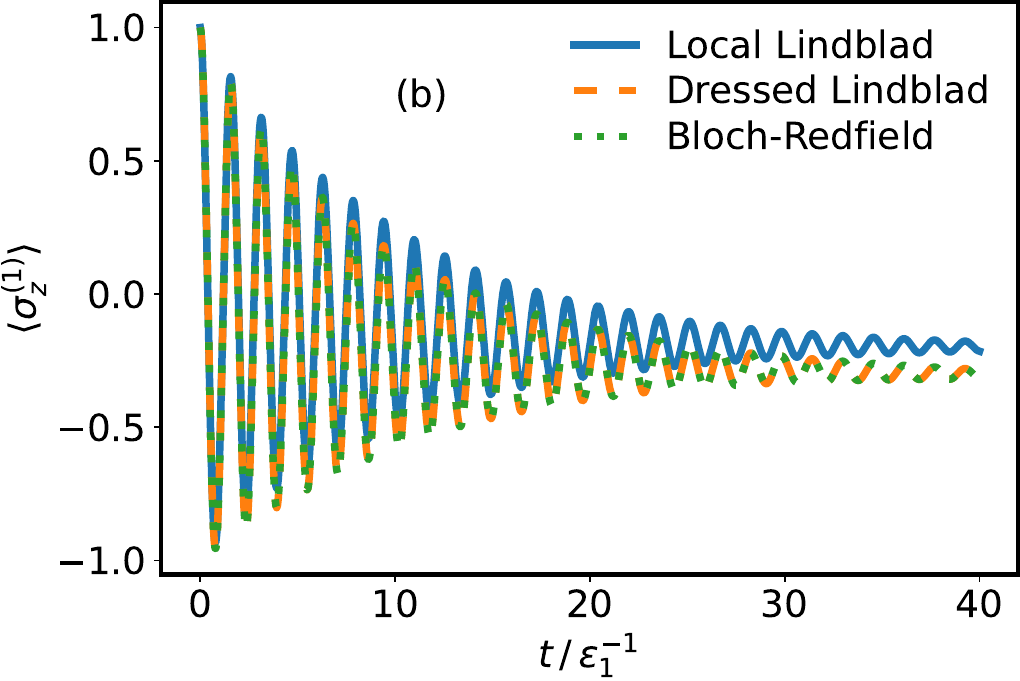}%
\caption{For the example problem of two interacting qubits, we compare the output of two \code{mesolve()} simulations. One uses local collapse operators acting on the bare states of each qubit (local Lindblad equation), and the other uses dressed collapse operators acting on the global eigenstates (global Lindblad equation). In addition, we include a simulation performed with the Bloch-Redfield solver (\code{brmesolve()}). Panel (a) shows weak coupling between the qubits ($g=0.1\epsilon_1$), where the local Lindblad description is sufficient from the perspective of detailed balance. Panel (b) shows the dynamics with strong coupling ($g=2\epsilon_1$), where local collapse operators predict a steady state that is not compatible with detailed balance.
In both cases, the qubits are resonant $\epsilon_1=\epsilon_2$.}\label{fig1}
\end{figure*}

Assuming just a flat spectral density $J(\omega) = \gamma/2 $ and zero temperature gives $S(\omega) = \gamma \theta(\omega)$, i.e., only spontaneous decay can occur, no stimulated decay or excitation can happen.
In the following example, we implement this zero-temperature environment manually for our two-qubit model using \code{mesolve()}. We see that the long-time evolution leads to a state that is close to the coupled ground state of the two-qubit Hamiltonian:

\inputpython{mesolve.tex}{47}{70}
In \fig{fig1}, we show the results of the local and dressed (global) Lindblad simulations and compare the results with the Bloch-Redfield solver.
The Bloch-Redfield solver, which will be explained later, is designed to automatically construct a weak-coupling master equation from a given bath power spectrum, allowing for time-dependent system Hamiltonians and arbitrary degrees of the secular approximation. If the secular approximation is fully implemented, using the Bloch-Redfield solver is equivalent to the global master equation that we constructed above manually.
We find that when the qubit-qubit coupling is small, the results from the local and global master equations both agree well with the Bloch-Redfield solver. For large qubit-qubit coupling,  the local master equation produces deviating results from the global/Bloch-Redfield approach.

For advanced applications of QuTiP, it is important to understand how a Lindblad equation is constructed internally.  In order to start a simulation, the user must first pass a Hamiltonian and collapse operators to QuTiP, which are operators, and an initial condition, which may be given as a vector (for a pure state) or an operator (for a mixed state).
However, internally, QuTiP uses the superoperator representation to solve the Lindblad equation \eqref{lindblad}.  In short, this means that the density matrix representing the system state is stacked, in column order, into a single large vector.  The right-hand-side of the Lindblad equation, called the Liouvillian or the Lindbladian, is a superoperator acting on this new larger vector space. It can thus be represented as a large matrix, which multiplies the new larger state vector only from the left. 

Formally, this representation takes advantage of the isomorphism between operators acting on a Hilbert space and vectors in an enlarged space consisting of two copies of the original Hilbert space, $\mathcal{L}(\mathcal{H}) \cong \mathcal{H} \otimes \mathcal{H}$.  One can visualize this as mapping the component of a density operator corresponding to the basis projector $\ket{\psi_i} \bra{\psi_j}$ to a component in a new vector on the double space corresponding to the basis vector $\ket{\psi_j}\otimes \ket{\psi_i}$.  Then, operators acting on the original density operator element from the left and right, $A \ket{\psi_i} \bra{\psi_j} B$, become operators acting just from the left as $(B^{\dagger} \otimes A) (\ket{\psi_j}\otimes \ket{\psi_i})$.  

This construction makes the construction of the actual matrix describing the Lindbladian formally very simple, and it is usually numerically more straightforward to solve an ODE in a matrix-vector format, as opposed to a matrix-matrix format.  However, it comes with the downside that the memory cost increases quadratically due to the doubling of the Hilbert space. When this cost becomes prohibitive, one can consider employing the Monte-Carlo solver described in the next section. In future QuTiP versions, we plan to implement more memory-efficient alternative ways of storing the Lindbladian by avoiding vectorization (as done in the dynamiqs package \cite{guilmin2024dynamiqs}).

As noted before, QuTiP's master equation solver is not restricted to just solving Lindblad equations like \eqref{lindblad}, but can also be applied to any other master equation.  The Liouvillian superoperators for other master equations must be constructed by hand, for example using the built-in functions \code{spre()}, \code{spost()} and \code{sprepost()}, which convert operators on the original Hilbert space to operators in the double space discussed earlier (see Table \ref{tab:qutip_superoperators_functions} for a list of more superoperator-related functions). These Liouvillians can be then passed directly to \code{mesolve()} in place of the system Hamiltonian.  For example, the Lindbladian corresponding to the master equation of the earlier example can be constructed manually via:

\inputpython{mesolve.tex}{38}{42}

In QuTiP v5, the master equation solver can be used with a new solver class interface, like we have seen earlier for the Schrödinger equation solver. It has also been augmented with new options for ODE integration methods: Verner's ``most efficient'' Runge-Kutta methods of order 7 and 9 are available as \code{vern7} and \code{vern9}, and an approach based on diagonalization for time-independent systems is available as \code{diag}.  Furthermore, the solver has been updated to take advantage of the new data layers. In section \ref{sec:me_jax_gpu}, we will demonstrate how to use the QuTiP-JAX data layer to take advantage of custom data layers to enhance \code{mesolve()} using GPUs.

\subsubsection{\textup{\code{mesolve}} part 2: Time-dependent systems}\label{section:medriven}

As mentioned in the earlier summary of the \code{QobjEvo} class, QuTiP and most of its solvers support time-dependent quantum objects. The user may specify the time dependence in a variety of ways: as a Python function, as a string (which will be compiled into machine code at the first usage), or as discrete time-dependent data which will be interpolated with cubic splines by QuTiP.

Usually, the user does not need to deal with \code{QobjEvo} objects themselves, and can just directly provide the time-dependence to the solver. Consider a standard example of a driven qubit with the time-dependent Hamiltonian
\begin{equation}
H = \frac{\Delta}{2} \sigma_z + \frac{A}{2} \sin (\omega_d t) \sigma_x \, ,
\end{equation}
which experiences dissipation through contact with a zero-temperature bath with the coupling rate $\gamma$. We model the influence of the bath by a Lindblad equation with the collapse operator $\sigma_-$ in the undriven basis. The assumption that the noise acts in the undriven basis is valid when the drive amplitude $A$ is much smaller than the natural system frequency $\Delta$.

If we assume in addition that the driving is close to resonant, $\omega_d \approx \Delta$, we can perform the rotating wave approximation to find, in a rotating frame,
\begin{equation} \label{eq:mesolve:H_RWA}
H_{\mathrm{RWA}} = \frac{\Delta-\omega_d}{2} \sigma_z + \frac{A}{4} \sigma_x \, ,
\end{equation}
which reduces the problem to an undriven one.  This is a useful consistency check on the driven results for this simple example.

Defining the driven setup in QuTiP is easy; we first enter the driving field as a Python function:

\inputpython{mesolvedriven.tex}{20}{21}
Then the undriven and driven parts of the Hamiltonian can be defined with:

\inputpython{mesolvedriven.tex}{25}{27}
This specification of the time-dependent Hamiltonian can be passed to \code{mesolve} as usual:

\inputpython{mesolvedriven.tex}{43}{45}
Note that, as mentioned, the collapse operator is constant in time, i.e., it acts on the undriven basis.
Alternatively, we could have entered the driving field as a string:

\inputpython{mesolvedriven.tex}{226}{226}
or as time-dependent data:

\inputpython{mesolvedriven.tex}{228}{228}
In the latter case, one has to take care that the spacing of the times in \code{tlist} is sufficiently smaller than the period length.

To check the validity of a given model, it is useful to compare to other solvers. In this case, we will compare to the Bloch-Redfield solver and to the solver for the hierarchical equations of motion (HEOM).  More details on these will be given later.

For the Bloch-Redfield solver, we need to specify the bath's power spectrum, which captures how it affects the system at different frequencies. As in the previous example, we will use a white-noise bath at zero temperature. For the reader's convenience, the following code snippet, which defines the power spectrum and invokes the Bloch-Redfield solver, demonstrates also how a finite-temperature environment could be handled:
\newpage
\inputpython{mesolvedriven.tex}{88}{104}
Here, \code{H} is the same time-dependent system Hamiltonian that was defined before.

\begin{figure*}%
\includegraphics[width = 0.5\linewidth]{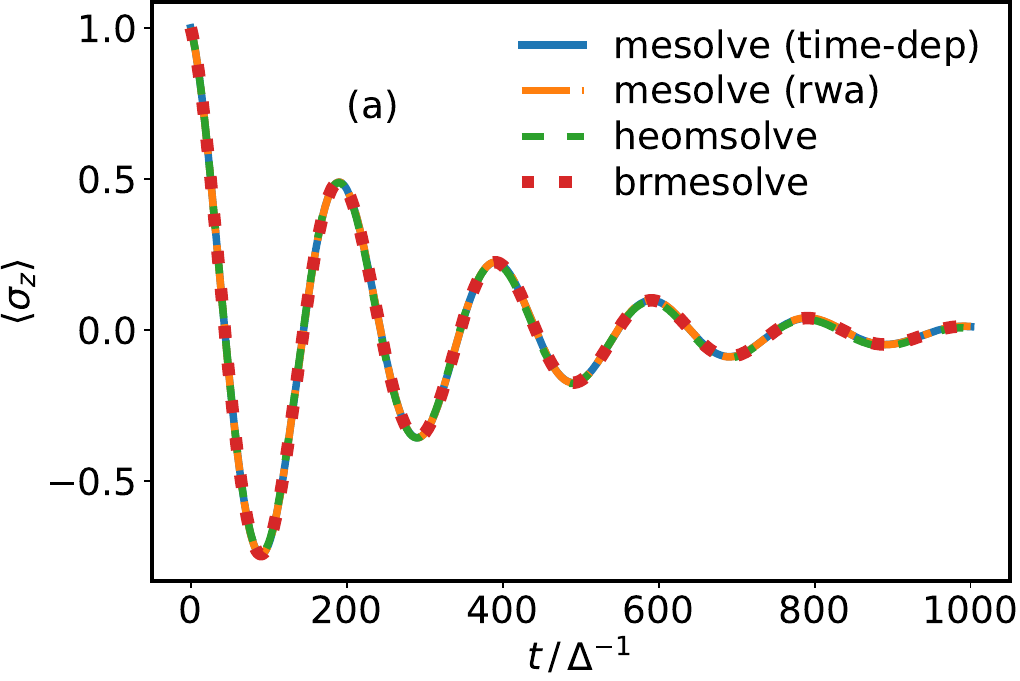}%
\includegraphics[width = 0.5\linewidth]{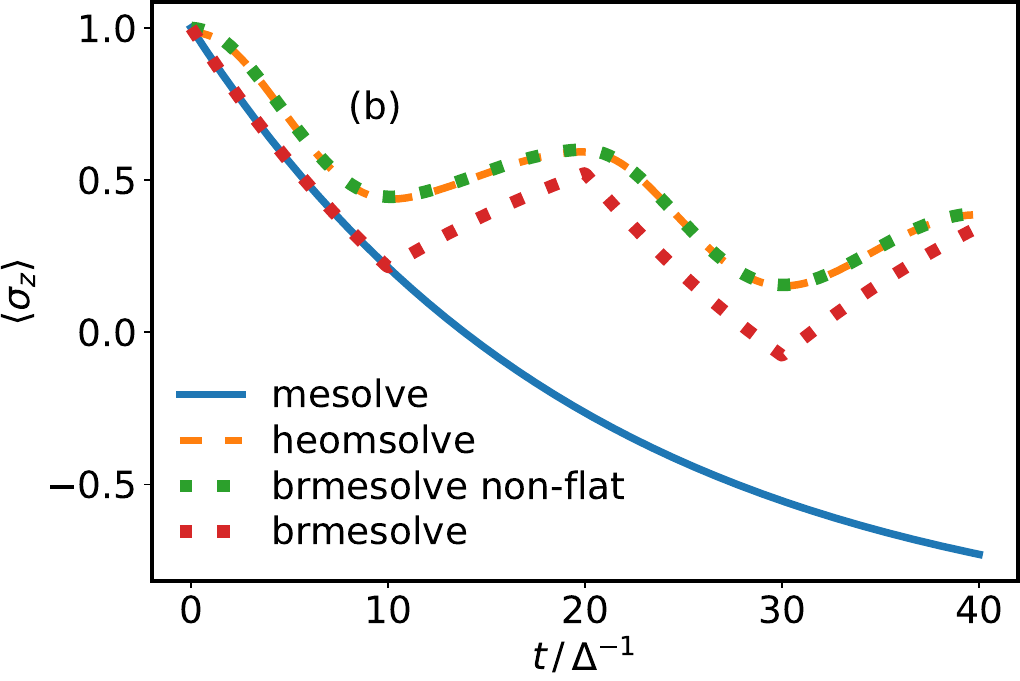}%
\caption{In panel (a), we show the dynamics of a resonantly driven qubit simulated with \code{mesolve}, \code{brmesolve} and the HEOM solver. We also compare the results to a simulation of the time-independent RWA Hamiltonian, Eq.~\eqref{eq:mesolve:H_RWA}.  The bath is chosen to have zero temperature and a flat spectral density $J(\omega) = \gamma$ with a rate $\gamma = 0.005 \Delta /(2\pi)$. The drive is assumed to be weak ($A=0.01 \Delta$) and on resonance ($\omega_{d} = \Delta$).  In panel (b), we show the second example, where the energy of a qubit is adiabatically, sinusoidally modulated between positive and negative values with frequency $\omega_d = 0.05 \Delta$ and amplitude $A = \Delta$. The qubit is in contact with a flat zero-temperature bath with rate $\gamma = 0.05 \Delta / (2\pi)$. The naive \code{mesolve} implementation with a single collapse operator that is not sensitive to the system's energy predicts just exponential decay. The Bloch-Redfield and HEOM solvers predict reversal of the decay when the energy changes sign, as a true zero-temperature bath can only remove energy from the system. The \code{heomsolve} result differs slightly from the \code{brmesolve} result since the spectral density used in \code{heomsolve} is not flat.  When the same spectral density used in the HEOM method is used explicitly in the Bloch-Redfield solver both approaches agree. This is shown with the green dotted line  which we called \code{brmesolve non-flat}, which uses the same spectral density as HEOM. 
}\label{medriven}
\end{figure*}

For the HEOM solver, the process is much more involved; we need to specify a multi-exponential decomposition of the correlation functions of the bath. However, the correlation functions are non-exponential at zero temperature, requiring us to apply a multi-exponential fit in order to use the HEOM approach. A full example of this process is given in the complete code example available on GitHub \cite{codeexamplesstatic}.  In \fig{medriven}a, we show a comparison of the resonantly driven qubit simulated with all three time-dependent solvers as well as the time-independent rotating wave approximation. As expected, all results agree with each other very well.

A second example, to illustrate where using naive local-basis collapse operators can fail, is that of a single qubit whose energies are adiabatically switched between positive and negative values,
\begin{equation}
H = \frac{\Delta}{2} \sin(\omega_d t) \sigma_z \, .
\end{equation}
 When the drive is slow, we expect the bath to be able to respond to this change. It should therefore always induce transitions from the higher-energy state to the lower-energy one, extracting energy from the system. As we see in \fig{medriven}b, a feature of the Bloch-Redfield solver and the HEOM method is that they can capture this switching effect automatically. The naive approach, however, using a single constant collapse operator $\sigma_-$ fails and is insensitive to the drive. This latter approach could be easily improved by defining more correct collapse operators, but we included the most simple choice of collapse operator here in order to demonstrate a potential pitfall of using phenomenological Lindblad equations, and to demonstrate the utility of QuTiP's more advanced solvers in validating or invalidating one's choice of approximate master equation.

\subsubsection{\textup{\code{mesolve}} part 3: JAX and GPU acceleration with Diffrax} \label{sec:me_jax_gpu}

The use of graphical processing units (GPUs) to accelerate the solution of numerical tasks has become ubiquitous in the last 10 years, with applications ranging from mining cryptocurrency to training large language models. Historically, using GPUs for science required custom algorithm implementations in Cuda or OpenCL and was labour intensive, limiting their use to only the most demanding of applications.  However, in recent years, the use of GPUs also for scientific tasks has greatly increased due to the availability of off-the-shelf packages like TensorFlow, CuPy, or JAX. With the flexible data layer in QuTiP, these packages can be slotted in and used with minimal overhead for the user. 

After some initial experimentation with CuPy \cite{cupy} and TensorFlow \cite{tflow}, development has focused on a QuTiP-JAX \cite{qjax} data layer because of its powerful auto-differentiation capabilities and its mass adoption by the broader machine learning community.  We will discuss the auto-differentiation aspects later, and in this section just provide an example of how QuTiP-JAX can be used for solving open quantum system dynamics on a GPU.  


Before continuing, it is important to consider in which circumstances GPUs would actually accelerate such a task. GPUs tend to shine when evaluating many small matrix-vector problems in parallel, so one expects it to help substantially when integrating the behavior of a small system for many different parameters, such as for qiskit-dynamics~\cite{2024qiskit}, or repeated multiplication of small matrices to the subspace of a large quantum state, as demonstrated in various quantum circuit simulators such as yao.jl~\cite{YaoFramework2019}. However, when solving a single ODE of a large system, whose integration involves repeated matrix-vector products with very large matrices, the potential advantage is less clear. The sequential nature of an ODE integrator makes it hard to parallelize. Although we expect that, for large matrices, it could be beneficial to parallelize individual matrix-vector products themselves at the level of column-row products, this optimization would require substantial modification of lower-level libraries. Here, we will show, with a practical example, that there is a cross-over in the system size where it becomes advantageous to use a GPU in QuTiP nevertheless.

In order to use QuTiP-JAX, one must install it as well as the JAX package in addition to the core QuTiP package. It can then be imported as follows:

\inputpython{JAX-mesolve.tex}{3}{5}
Using \code{import jax.numpy as jnp} is convenient as JAX mirrors all of the functionality of NumPy with equivalent functions that can be used with JAX (for the purposes of auto-differentiation, for example). The import of the \code{qutip\_jax} module enables, as a side effect, the data layer formats \code{'jax'} and \code{'jaxdia'}.  The former is a dense format, while the latter is a custom sparse diagonal format.  An experimental CSR format is available within JAX, but currently not yet supported by QuTiP.  

\inputpython{JAX-mesolve.tex}{7}{11}

To use the JAX data-layer within the master equation solver we need to use the \code{diffrax} ODE integrator, which we select by passing the option \code{"method":"diffrax"} to the solver.
A short-cut to this can be done with the new \code{qutip\_jax.set\_as\_default()} functionality, which enables us to switch all QuTiP objects' data types to JAX and change the solver method to \code{diffrax} in a single step:

\inputpython{JAX-mesolve.tex}{75}{76}
The following code can be used to revert this change:

\inputpython{JAX-mesolve.tex}{79}{79}

Apart from this, manipulating QuTiP-JAX objects and solving problems with the master equation solver proceeds as with any other data format.  To give a concrete example, and demonstrate an advantage of using GPUs, we consider a 1D Ising spin chain model:
\begin{equation}
    H = \sum_{i=1}^N g_0\, \sigma_z^{(n)} - \sum_{n=1}^{N-1} J_0\, \sigma_x^{(n)} \sigma_x^{(n+1)} \, .
\end{equation}
Here, $N$ is the number of spins, $g_0$ the level splitting and $J_0$ a coupling constant. The end of the spin chain is in contact with an environment, modeled as a Lindblad dissipator with the collapse operator $\sigma_x^{(N-1)}$ and the coupling rate $\gamma$. This model can be defined and solved in QuTiP as follows:

\inputpython{JAX-mesolve.tex}{13}{48}
Note that we used the \code{qt.CoreOptions(default\_dtype=data\_type)} context manager so that objects  created with most internal functions use the same data format (overriding their default behavior). Note however that this currently does not override the behavior of every function which can create a \code{Qobj} in QuTiP (e.g., \code{PIQs} and \code{HEOMSolver} do not support this option).

With this definition, we can now compare solving the same problem with the default (sparse) data format on the CPU and with JAX on the GPU. For example, to run the solver on the GPU one can do the following:

\inputpython{JAX-mesolve.tex}{51}{71}

\begin{figure*}
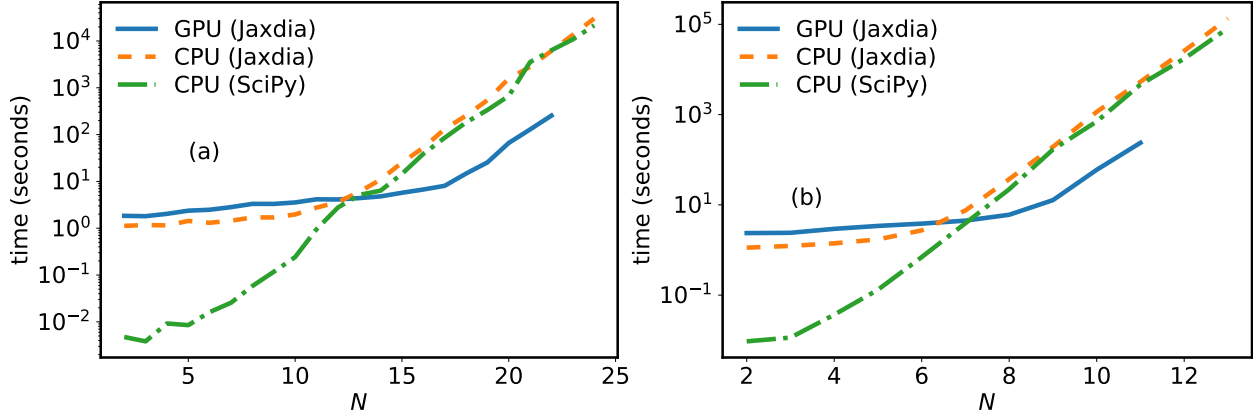
%
\includegraphics[width = 0.5\linewidth]{JAX-sesolvev3.pdf}%
\includegraphics[width = 0.5\linewidth]{JAX-mesolvev3.pdf}%
\caption{Benchmark of the time required to solve the dynamics of an Ising spin chain as a function of the number of spins $N$ using the JAX data layer. The simulation was performed on an NVIDIA A100 GPU with 80 GB of RAM. The CPU used for comparison was an AMD EPYC 7713 (64 cores).  Panel (a) shows a noiseless example for up to $22$ spins ($24$ on CPU) using \code{sesolve}. Panel (b) shows the same problem in the presence of noise for up to $11$ spins ($12$ on CPU) using \code{mesolve}. The Schrödinger equation solver is able to integrate more spins because of the memory cost of the superoperator constructed by \code{mesolve}.  In both cases, we see a crossover at a certain system size where the GPU solver becomes more performant, but the GPU memory limit is reached shortly after. 
 For the examples run on CPU we also differentiate between the default SciPy based solver and using JAX in CPU mode.  In all JAX examples we used the jaxdia data-layer. The dense Jax format runs out of memory much more quickly (not shown).}\label{fig2}
\end{figure*}

By comparing the performance of simulating the dynamics of the Ising spin chain with the standard QuTiP CPU-bound method (\code{adams}) and the equivalent \code{diffrax} method on a GPU, we observe a threshold in system size where the GPU outperforms the CPU calculation by up to two orders of magnitude.  Simulating systems of this size requires the use of the \code{jaxdia} format to avoid memory limitations of the dense \code{jax} format. However, even with the \code{jaxdia} format and using a state-of-the-art graphics card with 80 gigabytes of RAM, the memory limit is reached already at 11 spins. For the same model without dissipation, which can be solved with the Schrödinger equation solver, 22 spins are possible.  To go beyond these limits and harness large-scale high-performance computing requires us to distribute the ODE integration across multiple GPU nodes. This is a challenging task that we plan to explore in future evolutions of QuTiP-JAX.

\subsubsection{\textup{\code{steadystate}}: A steady-state solver for master equations}

When solving the dynamics of a closed Hamiltonian system with an initial condition that is not an eigenstate of the system, one will typically observe persistent oscillations that never decay.  However, when considering an open system described with a master equation $\dot \rho(t) = \mathcal L \rho(t)$, where $\mathcal L$ is the Liouvillian superoperator, the dissipation terms will gradually suppress coherences (in certain bases) and push the system in the long time limit into a steady state. The steady state satisfies the equation
\begin{equation}
\frac{d \rho(t\rightarrow \infty)}
{dt} = \mathcal{L}\rho(t\rightarrow \infty) = 0 \, , \label{ss}
\end{equation}
i.e., it does no longer change with time. The steady state is usually unique, but there can be multiple possible steady states under certain conditions (dark states which are not affected by the dissipation, or non-connected subspaces). We will here focus on the case where a unique steady state exists and mention some caveats about the degenerate case at the end of this section. One should also note that the discussion above applies only to time-independent systems. Systems with time-dependent driving typically have persistent oscillations around some fixed point in the long-time regime, which can be found either by explicitly solving the dynamics of the master equation or by using QuTiP's \code{steadystate\_floquet()} function. This function will not be discussed further here, and we refer to the online documentation for more information.

For time-independent problems described by Lindblad or other master equations, QuTiP provides the \code{steadystate()} function which supports various methods to solve the steady-state condition \eqref{ss}.  Like \code{mesolve()}, \code{steadystate()} takes as inputs either a Hamiltonian and collapse operators, or a Liouvillian superoperator constructed manually by the user.

The default solution method is called \code{direct} and directly solves \eqref{ss} as a linear equation, using normalization as an extra condition 
    to obtain a uniquely solvable set of linear equations.
Using the function's \code{solver} parameter, the user can select from a variety of linear equation solvers. For dense Liouvillians, there are the solvers \code{solve} and \code{lstsq} based on NumPy functions, and for sparse problems, many SciPy sparse linear equation solvers can be used: \code{spsolve}, \code{gmres}, \code{lgmres}, and \code{bicgstab}.  Some of these use iterative algorithms that become useful when the memory cost of finding the full exact solution is too high.  Finally, there is an \code{mkl\_spsolve} sparse solver which uses Intel's Math Kernel Library. This can offer substantial performance benefits, particularly on Intel CPUs, but requires that the necessary libraries are installed on the system.

The alternative method \code{power} also solves the linear equation \eqref{ss}, but it uses an additional iterative inverse power step \cite{nation2015steadystate} which starts by assuming $\rho^{(0)}_{ss} = \mathbb{1}$ and then solves $\mathcal L \rho^{(n)}_{ss} = \rho^{n-1}_{ss}$ until $\mathcal L\rho^{(n)}_{ss} < \varepsilon_{\mathrm{tol}}$ for some small tolerance $\varepsilon_{\mathrm{tol}}$ close to zero. With this method, the same solvers can be used as with the \code{direct} method discussed previously. Two further methods are available: \code{eigenvalue}, which finds the zero eigenvector of $\mathcal{L}$ iteratively, and \code{svd}, which use a dense singular-value decomposition of the Liouvillian.

Generally speaking, the direct method with either an exact linear equation solver or an iterative one (for large system sizes) are the most commonly employed methods. The optimal choice for a given situation tends to be problem specific. For more details, a comprehensive analysis and benchmark of some of these methods for common quantum problems is provided in \cite{nation2015steadystate}, alongside a deeper explanation of the origin of the different approaches.  Finally, in the situation where multiple steady states exist, the different solvers can often produce very different results; some may fail, others may produce linear combinations of possibilities, and so on. So far, QuTiP does not support an automated approach to dealing with this issue; therefore, the onus is on the user to understand the connectivity and properties of their model, or to check their results using long-time propagation of the dynamics with \code{mesolve()}.

\subsubsection{\textup{\code{mcsolve}}: A Monte Carlo solver for quantum trajectories}

The Lindblad master equation described in \eqref{lindblad} describes the ensemble-averaged dynamics of a quantum system in contact with an environment. In other words, it describes the expected results averaged over many repetitions of the same ``experiment''.  Interestingly, this equation can be unravelled in terms of trajectories of possible outcomes of single experiments. Each trajectory is assigned a classical probability, which is related to the likelihood of the environmental behavior that is necessary to produce that trajectory.

The trajectory unravelling thus provides a theoretical way of describing the fluctuation of results between multiple realizations of the experiment, and it sheds light on the physical interpretation of the Lindblad master equation. Further, it is also interesting from the point of view of numerical efficiency, because we can sometimes unravel the master equation in such a way that the density operator is written as an ensemble average of pure states which follow a non-linear stochastic Schr\"odinger equation \cite{mc1}. Therefore, it is not necessary to store the full super-operator form of the Lindbladian generator in memory, and one can work with regular states instead of density operators.  For large systems, this approach can save both computing time and memory, in particular if many trajectories can be simulated in parallel.

The Monte Carlo solver in QuTiP is based on the Monte Carlo wave function (MCWF) technique, which is an unravelling of the Lindblad equation in terms of pure states following quantum jump trajectories. These trajectories consist of periods of coherent, deterministic time evolution  described by the effective non-Hermitian Hamiltonian
\begin{equation} \label{heff}
H_{\mathrm{eff}} = H - \frac{i}{2} \sum_n C_n^{\dagger}C_n 
\end{equation}
interspersed with discrete quantum jumps. Here, $H$ and $C_n$ are the Hamiltonian and the collapse operators from the Lindblad equation \eqref{lindblad}.  The non-Hermitian evolution does not conserve the norm of the wave function, but causes it to continuously decrease towards zero.  The norm represents the survival probability, that is, the probability for the system to still evolve along that trajectory without having undergone a quantum jump (some new insights on this unraveling can be found in \cite{PhysRevResearch.4.023036, PhysRevA.104.062212}). The decrease of the norm of the state within a time step hence represents the probability of the system undergoing a quantum jump in that time step. 

At a certain time, randomly chosen by considering the survival probability, a quantum jump occurs and the system undergoes a transition described by one of the collapse operators $C_n$. The probability that the jump at time $t$ is described by the $n$-th collapse operator is proportional to $\bra{\psi(t)} C_n^\dagger C_n \ket{\psi(t)}$, where $\ket{\psi(t)}$ is the trajectory state before that jump.  This process can be interpreted as the environment ``measuring'' the system, such that the system is projected into the ``target state'' associated with the jump, and then re-normalized:
\begin{equation}
    \ket{\psi(t)} \quad\to\quad \bigl( \bra{\psi(t)} C_n^\dagger C_n \ket{\psi(t)} \bigr)^{-1/2}\, C_n \ket{\psi(t)}\, .
\end{equation}
The record of these jumps can in principle be obtained by an experimenter monitoring the environment and marking when they see the transition described by the collapse operator. For example, this transition might be the system emitting a photon into the environment, or charge moving between discrete states.


\paragraph{Basic example} --- The actual implementation of this method within QuTiP follows the algorithm outlined in \cite{mc1,mc2,mc3} and is described in the documentation. It is invoked with the function \code{mcsolve()}, which can be used exactly like \code{mesolve()} and returns an average over 500 trajectories by default. The number of trajectories can be adjusted using the parameter \code{ntraj}.

\inputpython{mcsolve.tex}{23}{24}

We used the option \code{"map":"parallel"} to automatically take advantage of multiple CPU cores to simulate different trajectories in parallel, and the option \code{"keep\_runs\_results":True}, which stores the results of the individual trajectories in addition to the ensemble average.  When the latter option is enabled, the \code{runs\_expect} property of the result object returns, for each expectation operator, a list of trajectories. Each of these trajectories is given as a time series for the expectation value of that operator, with the entries of the time series corresponding to the times in the \code{tlist} given to the solver. The average of the expectation values over all generated trajectories are returned in the \code{average\_expect} property, and their standard deviation is in \code{std\_expect}. These two properties are available also if the \code{keep\_runs\_results} option is not enabled.

Furthermore, if the option \code{"store\_states":True} is used, the \code{runs\_states} property returns the states for all trajectories (if requested), while \code{average\_states} contains their average. Given a sufficiently large number of trajectories, the states in the \code{average\_states} property of the Monte Carlo result will approximate the states returned by the regular master equation solver.

\paragraph{Changes in v5} --- Finally, in QuTiP v5, the result object gained a new \code{photocurrent} property, which saves the expectation value of the photocurrent.
This property replaces the \code{photocurrent\_sesolve()} stochastic methods from earlier QuTiP releases.

The finite sampling used by the Monte Carlo solver means that sometimes care must be taken to obtain a result close to the true ensemble average. The solver provides several methods to decide at which point to stop generating more trajectory samples. As mentioned previously, the option \code{ntraj} controls the maximum number of trajectories; calling \code{mcsolve} with \code{ntraj=1000} and no further options will thus use 1000 trajectories instead of the default 500. In addition, a maximum computation time can be specified; for example, using the option \code{timeout=60} will stop the solver after 60 seconds even if the requested number of trajectories has not been reached yet. 
Finally, the \code{target\_tol} option interrupts the computation when the statistical error of the results 
reaches the given target value. For more fine-tuning, target values can be given for both the absolute and the relative error, and it is also possible to use different target values for the different expectation operators.
How to specify these tolerances is explained in detail in the documentation.

In QuTiP v5, a class interface has been added to the Monte Carlo solver in analogy to the class interfaces for the other solvers discussed above.
Another addition in version 5 is an improved sampling algorithm, which can be enabled in the options with \code{"improved\_sampling":True}. This algorithm is particularly useful when the dissipation rates are very small. In this case, with the default algorithm, many of the simulated trajectories may end up containing no jumps and only deterministic evolution; all of these trajectories are thus identical. To avoid these redundant computations, the improved sampling algorithm only runs the no-jump trajectory once and includes it in the final statistics with an appropriate weighting.

Another improvement in the Monte Carlo solver in QuTiP v5 is the support for mixed initial states. Given a mixed initial state, the solver runs trajectories for each pure initial state in the mixture, and correctly weights the results in the final averaging. Users have the option to manually control the number of trajectories used for each initial state, or to select this number automatically based on the requested total number of trajectories. When combined with the improved sampling option, one no-jump trajectory will be generated for each initial state. Future improvements to the solver may incorporate recent advances in optimizing the Monte Carlo method based around the waiting-time distribution~\cite{mcnew}.

\begin{figure*}
{\centering \includegraphics[width = 0.7\columnwidth]{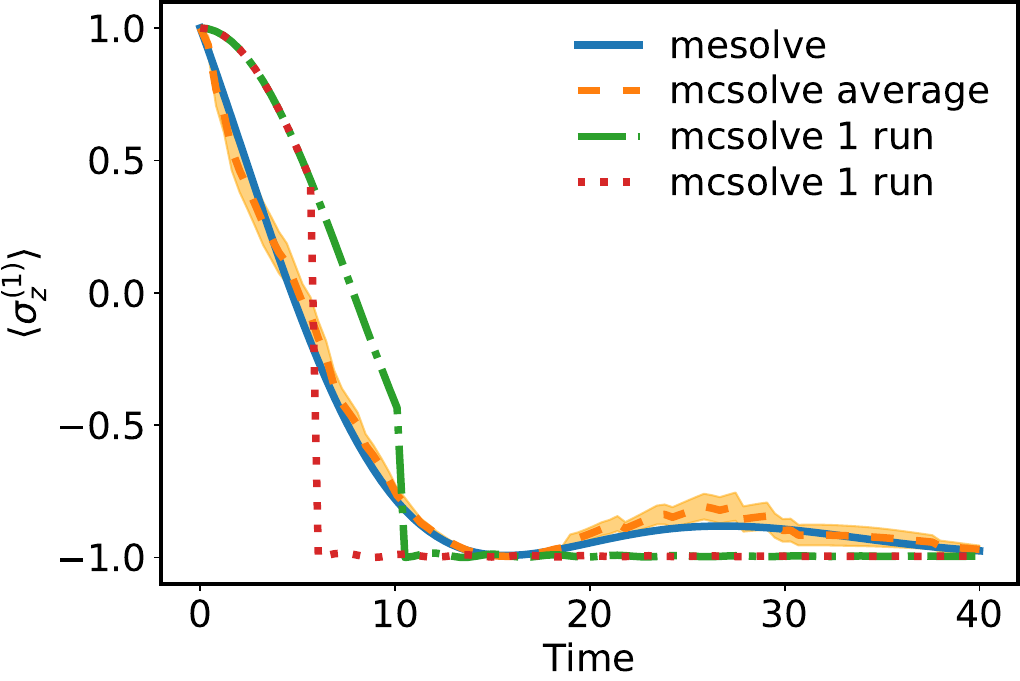}\par }
\caption{Here we show the same example as \fig{fig1} evaluated with the Monte-Carlo solver \code{mcsolve}.  The average behavior is compared to the result from \code{mesolve}, alongside two example trajectories showing discrete quantum jumps.  The yellow shaded region indicates the convergence error based on the standard deviation of the trajectories and the number of trajectories, $\sigma_{\mathrm{err}} = \sigma / \sqrt{N_{\mathrm{traj}}}$. Here, we use a smaller number of trajectories $N_{\mathrm{traj}}=100$ than default, to amplify the error and make it more visible in this plot. }\label{fig3mc}
\end{figure*}

\subsubsection{\textup{\code{nm\_mcsolve}}: A Monte Carlo solver for non-Markovian baths} \label{sec:nmmc}

If the time evolution of an open quantum system exhibits non-Markovian effects, the dynamics of its reduced density matrix cannot be described by a Lindblad master equation.
However, by applying the time-convolutionless (TCL) projection operator technique, it is still possible to write the dynamics in the time-local form \cite{breuer2002theory, BreuerPhysRevA2004, PiiloPhysRevLett2008, DonvilOpenSystInfDyn2023}
\begin{align}
    \dot \rho(t) &= -\frac{i}{\hbar} \comm{H(t)}{\rho(t)} + \sum_n \gamma_n(t)\, \mathcal D_n[\rho(t)]
        \qq{with} \nonumber\\
    \mathcal D_n[\rho(t)] &= A_n \rho(t) A_n^\dagger - \frac 1 2 \bigl[ A_n^\dagger A_n \rho(t) + \rho(t) A_n^\dagger A_n \bigr]\, . \label{eq:nmmc:pseudolindblad}
\end{align}
Here, $H(t)$ is a system Hamiltonian and $A_n$ are jump operators but, in contrast to a Lindblad equation, the coupling rates $\gamma_n(t)$ may be negative at some or all times.

In the Monte Carlo wave function (MCWF) method implemented in QuTiP's \code{mcsolve}, quantum jumps occur with probabilities proportional to the coupling rates.
In the present case, these probabilities would be negative; thus, the MCWF cannot be immediately applied to master equations like \eqref{eq:nmmc:pseudolindblad}.
This problem can be navigated by mapping the dynamics to an equivalent Lindblad master equation and applying the MCWF method to that Lindblad equation.
For example, it is known in general that time-local quantum master equations can always be mapped to Lindblad equations on the double Hilbert space \cite{BreuerPhysRevA1999, HushPhysRevA2015, MenczelArXiv240111830Quant-Ph2024}.

The non-Markovian Monte Carlo solver \code{nm\_mcsolve()}, which was added to QuTiP in version 5, follows a similar approach.
By introducing a trajectory weighting called the ``influence martingale'', a master equation of the type \eqref{eq:nmmc:pseudolindblad} can be mapped to a Lindblad equation on the same Hilbert space, as shown in Refs.~\cite{DonvilNatCommun2022, DonvilNewJPhys2023, DonvilOpenSystInfDyn2023}. 
The formalism requires that the jump operators satisfy a completeness relation of the form $\sum_n A_n^\dagger A_n = \alpha\, \mathbb 1$ for some scalar $\alpha > 0$.
The function \code{nm\_mcsolve()} automatically ensures that this relation is satisfied by adding, if necessary, an additional jump operator with zero coupling rate.
It then calculates the shift function
\begin{equation}
    s(t) = 2\, \abs*{ \min\bigl\{ 0, \gamma_1(t), \gamma_2(t), \dots \bigr\} }\, ,
\end{equation}
which ensures that the shifted rates $\Gamma_n(t) = \gamma_n(t) + s(t)$ are non-negative.
Finally, it uses the regular MCWF method to generate trajectories $\ket{\psi(t)}$ for the completely positive Lindblad equation
\begin{equation}
    \dot \rho'(t) = -\frac{i}{\hbar} \comm{H(t)}{\rho'(t)} + \sum_n \Gamma_n(t)\, \mathcal D_n[\rho'(t)] \, ,
\end{equation}
such that $\rho'(t) = \mathbb E\{ \ketbra{\psi(t)} \}$, where $\mathbb E$ denotes averaging over the trajectory ensemble.
The original state can then be reconstructed through the average $\rho(t) = \mathbb E\{ \mu(t) \ketbra{\psi(t)} \}$, where
\begin{equation} \label{eq:influence_martingale}
    \mu(t) = \exp\Bigl[ \alpha \int_0^t s(\tau)\, \dd\tau \Bigr]\, \prod_k \frac{\gamma_{n_k}(t_k)}{\Gamma_{n_k}(t_k)}
\end{equation}
is the influence martingale.
The product runs over all jumps on the trajectory with jump channels $n_k$ and jump times $t_k < t$.

We note that the technique described above, and the non-Markovian Monte Carlo solver, are not limited to completely positive dynamics and apply to any master equation of the form \eqref{eq:nmmc:pseudolindblad}.
Examples for non-positive dynamics described in this form can be found, for example, in the study of Redfield equations \cite{DavidovicQuantum2020} or of unitary evolution subject to classical noise \cite{GneitingPhysRevB2020, GroszkowskiQuantum2023}.

\begin{figure*}
    \includegraphics[width=\textwidth]{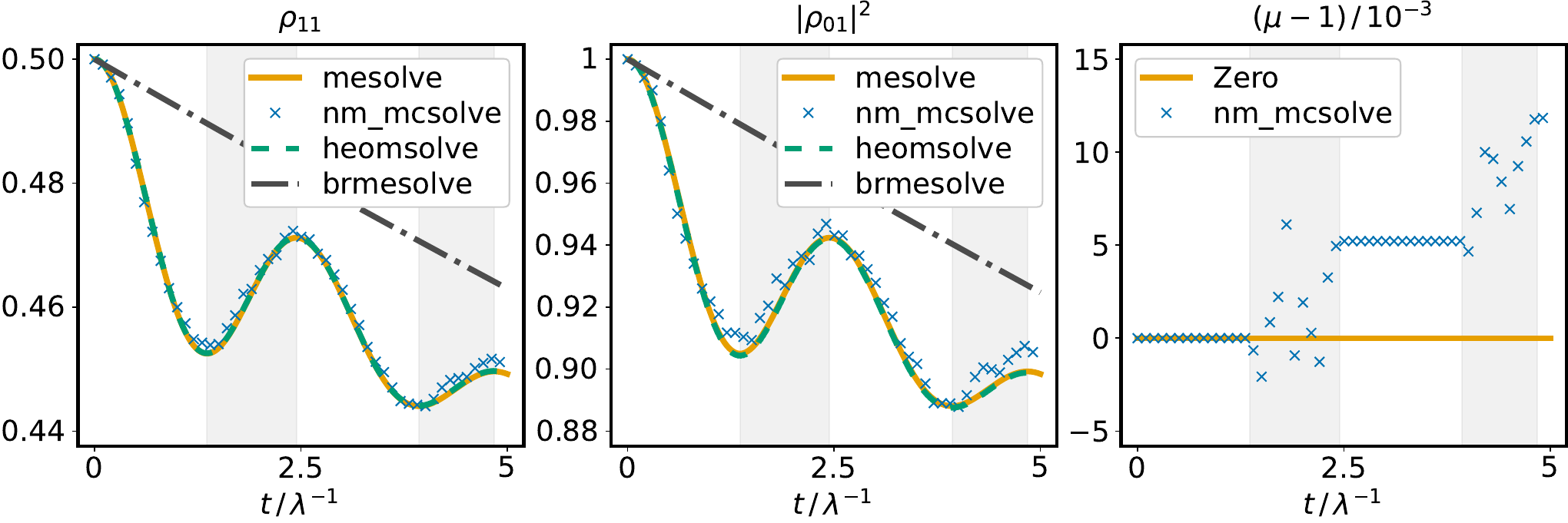}
    \caption{
        Damped Jaynes-Cummings Model.
        The figure shows results from the simulations described in Sec.~\ref{sec:nmmc}, obtained with four different solvers available in QuTiP, including the non-Markovian Monte Carlo solver.
        In all panels, the gray background indicates times where $\gamma(t)$ is negative.
        We used $\Gamma = 0.3\lambda$, $\Delta = 8\Gamma$, and $\omega_0 = 100\lambda + \Delta$.
    }
    \label{fig:nmmc}
\end{figure*}

For an illustrating example, we consider the damped Jaynes-Cummings model, which describes a two-level atom coupled to a damped cavity mode.
The cavity mode can be mathematically eliminated, leaving us with the two-level atom coupled to an effective environment with the power spectrum \cite{breuer2002theory}
\begin{equation} \label{eq:nmmc:power_spectrum}
    S(\omega) = \frac{\lambda \Gamma^2}{(\omega_0 - \Delta - \omega)^2 + \Gamma^2} \, .
\end{equation}
Here, $\lambda$ is the atom-cavity coupling strength, $\omega_0$ the atom transition frequency, $\Delta$ the cavity detuning and $\Gamma$ the spectral width.
Under the rotating wave approximation and at zero temperature, the dynamics of the two-level atom can be shown to follow the exact master equation \cite{breuer2002theory}
\begin{equation} \label{eq:nmmc:tcl}
    \dot \rho(t) = \frac{A(t)}{2i\hbar} \comm{\sigma_+ \sigma_-}{\rho(t)} + \gamma(t)\, \mathcal D_-[\rho(t)] \, ,
\end{equation}
where $\mathcal D_-$ is the dissipator for the Lindblad operator $\sigma_-$.
Further, $\sigma_\pm$ are the ladder operators for the atom, $\rho(t)$ is the atom state in the interaction picture, and $\gamma(t)$ and $A(t)$ are the real and imaginary parts of the expression
\begin{equation}
    \gamma(t) + i A(t) = \frac{2\lambda\Gamma \sinh(\delta t / 2)}{\delta \cosh(\delta t / 2) + (\Gamma - i\Delta) \sinh(\delta t / 2)} \, ,
\end{equation}
with 

\begin{equation}
    \delta = [(\Gamma - i\Delta)^2 - 2\lambda\Gamma]^{1/2}.
\end{equation}
Depending on the choice of system parameters, the coupling rate $\gamma(t)$ may become negative at some times.

We now apply a variety of QuTiP's solvers, including \code{nm\_mcsolve()}, to this example.
We first define the system Hamiltonian and initial state,

\inputpython{nm_mcsolve_example.tex}{224}{226}
and the functions $\gamma(t)$ and $A(t)$ as Python functions \code{gamma} and \code{A}.
Despite $\gamma(t)$ being negative at some times, the exact master equation \eqref{eq:nmmc:tcl} can be integrated using \code{mesolve()}.
However, the Liouvillian superoperator must be constructed by hand, since, by definition, dissipation terms added through the \code{c\_ops} parameters cannot have negative rates:

\inputpython{nm_mcsolve_example.tex}{230}{232}

We can now easily run the non-Markovian Monte Carlo simulation.
Instead of a list of collapse operators \code{c\_ops}, this solver takes a list of pairs of jump operators and corresponding rates, which makes negative rates possible.
Otherwise, it accepts mostly the same parameters as \code{mcsolve()}:

\inputpython{nm_mcsolve_example.tex}{236}{238}

The solvers used so far were based on the exact master equation \eqref{eq:nmmc:tcl}.
Additionally, we consider methods that will be discussed below, that apply directly to a spin-boson model with the given power spectrum \eqref{eq:nmmc:power_spectrum} and corresponding free reservoir auto-correlation function
\begin{equation} \label{eq:nmmc:cfct}
    C(t) = \frac{\lambda \Gamma}{2} \exp[-i (\omega_0 - \Delta) t - \lambda \abs{t}] \, .
\end{equation}
We define the system Hamiltonian and system coupling operator $Q = \sigma_+ + \sigma_-$ in the Schrödinger picture,

\inputpython{nm_mcsolve_example.tex}{242}{243}
where $\omega_0$ was chosen much larger than $\Delta$ to ensure validity of the rotating wave approximation.
The HEOM solver can be applied after decomposing the correlation function \eqref{eq:nmmc:cfct} into its real and imaginary parts:

\inputpython{nm_mcsolve_example.tex}{247}{253}
Note that the parameter \code{omega\_c} appearing here is $\omega_c = \omega_0 - \Delta$.

Finally, we compare with the Bloch-Redfield solver, which takes as its input the power spectrum \eqref{eq:nmmc:power_spectrum}:

\inputpython{nm_mcsolve_example.tex}{257}{260}
For comparison, the results obtained with the HEOM solver and the Bloch-Redfield solver must be transformed into the interaction picture:

\inputpython{nm_mcsolve_example.tex}{264}{266}

The results of these simulations are shown in Fig.~\ref{fig:nmmc}.
Using only 1000 trajectories, the MCWF simulation reproduces the exact solutions obtained with \code{mesolve()} and \code{heomsolve()} well.
The Bloch-Redfield equation produces a very different picture at the short time-scale considered here, showing that we are deep in the non-Markovian regime.
The small deviations between \code{mesolve()} and \code{heomsolve()} stem from the rotating wave approximation.

Figure~\ref{fig:nmmc} shows that whenever $\gamma(t)$ is negative, coherence is restored in the atom state.
It also shows the average value of the influence martingale \eqref{eq:influence_martingale}, which is stored in the \code{mc\_solution.trace} field.
The average influence martingale is an estimator for $\tr \rho(t) = \mathbb E\{ \mu(t) \}$.
We see that it is constant when $\gamma(t)$ is positive but fluctuates otherwise.
Its deviation from the exact value $\tr \rho(t) = 1$ can be used as an indicator for how well the Monte Carlo simulation has converged.

\subsubsection{\textup{\code{brmesolve}}: Bloch--Redfield master equation solver}

Earlier, in Section \ref{mesolvept1}, we discussed a master equation in Lindblad form, which described transitions between system eigenstates with rates proportional to the power spectrum of an environment. Such master equations can be derived microscopically from a system-bath model, where the power spectrum encodes the frequency-dependent coupling strength and the temperature of the environment.  The approximations used to derive the master equation in this form are called the Born-Markov-Secular approximations.  The Born and Markov approximations capture the weak-coupling and memoryless nature of the environment, and the secular approximation assumes that certain high frequency terms can be discarded.  

A master equation derived through these approximations is called a Bloch-Redfield master equation, and it can be conveniently constructed in QuTiP using the \code{brmesolve()} solver.  A useful feature of this solver is that the secular approximation can be relaxed to a specified degree, leading to a so-called non-secular and non-Lindblad equation of motion. Despite not having strict Lindblad form, this generalized master equation is still capable of describing the reduced state of the system in certain parameter regimes.  Being able to soften the secular approximation can be important if the bath has a certain structure, or if the dissipation rates or the temperature of the bath are large.

A full derivation can be found in the literature \cite{breuer2002theory}, and a short discussion of its most important steps is included in the QuTiP documentation. The final equation implemented in QuTiP can be written as

\begin{equation}
  \frac{d}{dt}\rho_{ab}(t)
    =
    -i\omega_{ab}\;\rho_{ab}(t)
    +
    \sum_{c,d}^{\mathrm{sec}}\frac{1}{2}R_{abcd}\;\rho_{cd}(t) \, .\label{br}
\end{equation}
The indices $a,b$ refer to matrix-elements of operators in the eigenbasis of the system Hamiltonian $H_{sys}$ with eigenenergies $\omega_m$ and $\omega_{ab}=\omega_a-\omega_b$. The system is coupled to one or more baths labelled by the index $\alpha$ through operators $A^{\alpha}$, with matrix elements $A^{\alpha}_{ab}$ in the Hamiltonian eigenbasis.  The bath is fully described by its power spectrum $S_{\alpha}(\omega)$, which we defined in \eqref{eq:power_spectrum_definition}.

Given these definitions, the Bloch-Redfield tensor can be written as

\begin{align} R_{abcd} =  -\sum_{\alpha,\beta}
 \left\{
 \delta_{bd}\sum_nA^\alpha_{an}A^\beta_{nc}S_{\alpha\beta}(\omega_{cn})
 -
 A^\beta_{ac} A^\alpha_{db} S_{\alpha\beta}(\omega_{ca})
 \right. \nonumber\\
+
 \left.
 \delta_{ac}\sum_n A^\alpha_{dn}A^\beta_{nb} S_{\alpha\beta}(\omega_{dn})
 -
 A^\beta_{ac}A^\alpha_{db} S_{\alpha\beta}(\omega_{db})
 \right\},\label{Rt} \end{align}

The sum in \eqref{br} refers to the degree to which the secular approximation is applied in the derivation of \eqref{Rt}.  Making the secular approximation corresponds to removing certain terms in the sums over $c$ and $d$ (which arise from fast oscillating terms in the interaction picture). 

In QuTiP, the Bloch-Redfield tensor can be calculated with the function \code{bloch\_redfield\_tensor()} and used directly in \code{mesolve()}, akin to the manual Lindblad construction we showed earlier. These two steps can be combined by using the \code{brmesolve()} solver interface directly.  The secular approximation can be controlled using the \code{sec\_cutoff} parameter, with \code{sec\_cutoff=-1} implementing the complete non-secular tensor, and any positive float giving a partial secular approximation (neglecting terms oscillating with frequencies $|\omega_{ab} - \omega_{cd}|$ larger than this value).  The default value of $0.1$ will give rise to a secular equation of motion in most cases.

The other powerful and important feature of this solver is its support for time-dependent Hamiltonians, which we demonstrated earlier in Sec.~\ref{section:medriven}. The assumption used here is that the environment always sees the system in its `instantaneous eigenbasis' at any given time, so the above equation of motion applies, but with the eigenbasis used to construct the Redfield tensor constantly changing. This feature of the time-dependent Bloch-Redfield equation is very challenging numerically because the Hamiltonian must be diagonalized at every instance in time.  The solver \code{brmesolve()} is optimized for this diagonalization, but it remains a challenging numerical problem.

An example of using \code{brmesolve()} is shown in \fig{fig1} (see also \fig{medriven} and \fig{fig:nmmc}), alongside the results from local and global Lindblad equations solved with \code{mesolve()}. To define a problem for the \code{brmesolve()} solver, we must provide a list of system operators that couple to an environment, and the power spectrum of that environment. The example in \fig{fig1} is constructed with:
\newpage
\inputpython{mesolve.tex}{75}{82}

The example for a driven problem, in Sec.~\ref{section:medriven}, was defined there already, and demonstrated in \fig{medriven}.

Finally, we note that the definition in Eq.~\eqref{Rt} allows for non-Hermitian coupling operators, as required by interacting Fermionic systems which might obey an interaction Hamiltonian in the form $H_I = \sum_\alpha A^\dagger_\alpha \otimes B_\alpha + A_\alpha \otimes B^\dagger_\alpha$. In QuTiP 5 this is now supported through the use of a \code{FermionicEnvironment} to describe the bath, or \code{brcrossterm()} to manually build custom interactions. An explicit example of how to use the Bloch-Redfield solver with the environment class is described in \ref{subsec:heom}.

\subsubsection{Floquet methods: the Floquet basis}

In previous sections we have seen different methods to solve time-dependent Hamiltonians, each of them suitable for different scenarios.
Here, we discuss Hamiltonians with a periodic time-dependence. In such cases, a natural approach is using the Floquet theorem to tackle the problem, similar to how using the Bloch theorem (a particular application of the Floquet theorem) greatly simplifies problems with spatial periodicity.

Literature on Floquet theory~\cite{Floquet1883} and its application to quantum systems is extensive~\cite{Shirley1965,Grifoni1998, Creffield2003}. Here, for completeness, we will only introduce the main results when applied to time-periodic Hamiltonians. 
Let $H$ be a Hamiltonian periodic in time with period $T$, such that
\begin{equation}
H(t) = H(t+ T) \, .
\end{equation}
The system state follows the time-dependent Schr\"odinger equation $i\hbar\frac{\partial}{\partial t} \ket{\psi(t)} = H(t) \ket{\psi(t)}$.
The Floquet theorem states that there exist state solutions defined as
\begin{align}\label{eq:FloquetStates}
\ket{\psi_{\alpha}(t)} = \exp (-i\epsilon_{\alpha} t / \hbar) \ket{\Phi_{\alpha} ( t)} \, ,
\end{align}
where $\epsilon_{\alpha}$ are the quasi-energies and $\ket{\Phi_{\alpha} (t)} = \ket{\Phi_{\alpha} (t +T)}$ the Floquet modes. Then, any solution $\ket{\psi(t)}$ of the time-dependent Schrödinger equation can be written as a linear combination of the Floquet states such that
\begin{equation}
\ket{\psi( t)} = \sum_{\alpha} c_{\alpha} \ket{\psi_{\alpha}(t)} \, ,
\end{equation}
where the constants $c_{\alpha}$ are determined by the initial conditions.

By inserting \eqref{eq:FloquetStates} into the Schrödinger equation, one can define the Floquet Hamiltonian,
\begin{equation}
H_F (t) \equiv H(t) - i\hbar \frac{\partial}{\partial t}\, ,
\end{equation}
which converts the time-dependent Schr\"odinger equation into a time-independent problem,
\begin{equation}\label{eq:H_F}
H_F (t) \ket{ \Phi_{\alpha}(t) } = \epsilon_{\alpha} \ket{ \Phi_{\alpha}(t) } \, .
\end{equation}
Once we obtain the Floquet modes at $t \in [0, T]$, using Eq.~\eqref{eq:FloquetStates} we immediately know $\ket{\psi(t)}$ at any large $t$. 



We now turn our attention to the application of Floquet theory in QuTiP. 
We first analyse the computational performance of using the Floquet basis compared to the Schrödinger equation solver \code{sesolve()}.
For this example, we use the Hamiltonian of a two-level system driven by a periodic function,
\begin{equation}\label{eq:H_TLS_d}
H= -\frac{\epsilon}{2} \sigma_z -\frac{ \Delta}{2} \sigma_x +\frac{A}{2} \sin (\omega_d t) \sigma_x \, ,
\end{equation}
where $\epsilon$ is the energy-splitting, $\Delta$ the coupling strength and $A$ the drive amplitude.
As discussed earlier, this Hamiltonian can be defined in QuTiP as follows.

\inputpython{p_Floquet_speed_test.tex}{28}{31}

\begin{figure}
\centering
\includegraphics[width=\textwidth]{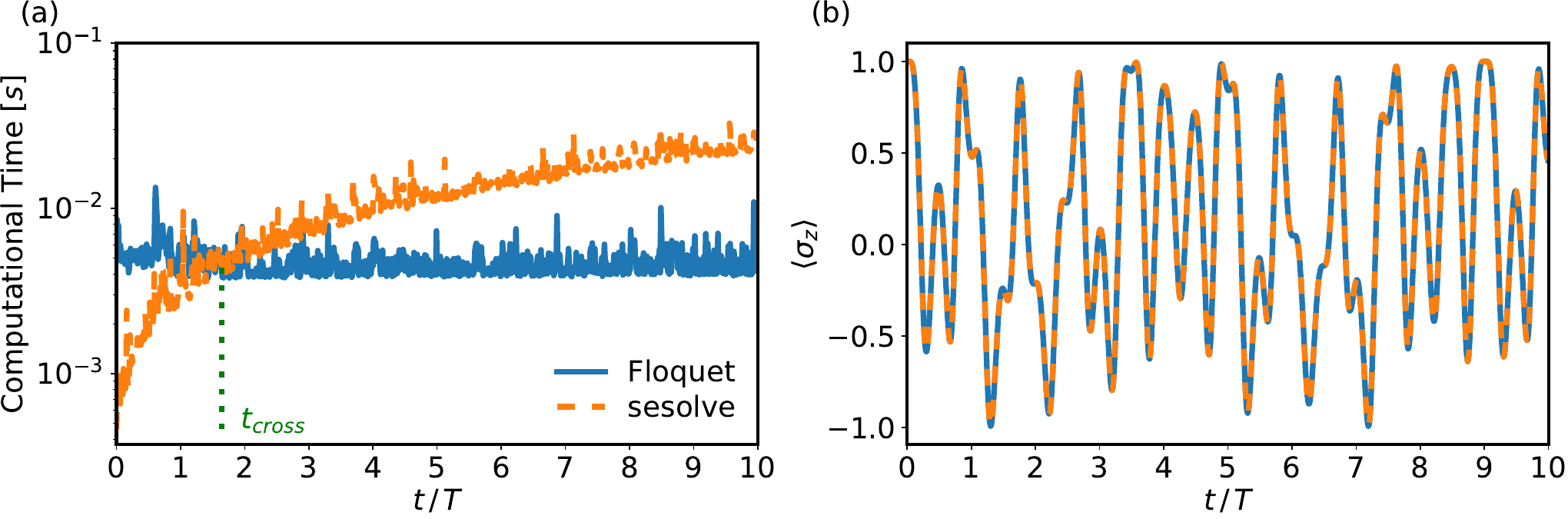}
\caption{Two-level system driven periodically (Eq.~\eqref{eq:H_TLS_d}) with energy splitting $\epsilon=2\pi$, coupling strength $\Delta=0.2 \times 2\pi$, drive amplitude $A=2.5 \times 2\pi$ and drive frequency $\omega_d = 2\pi $. (a) Time to evolve the state from the initial state until the given time, using the Floquet method (solid blue line) and \code{sesolve()} (orange dashed line). The dotted green line indicates the normalised time, $t_{\mathrm{cross}}$, at which Floquet basis and \code{sesolve()} take the same computational time to evolve the state. For times beyond this, direct integration with \code{sesolve()} is less efficient than using the Floquet basis. Note that intermediate results such as the computed Floquet basis or the evolved state are not reused between data points. (b) Expected value of $\sigma_z$ as a function of time, computed using the Floquet method and \code{sesolve()}.}
\label{fig:Floquet_comp_time}
\end{figure}

Figure~\ref{fig:Floquet_comp_time}(a) shows the speed comparison between the Floquet approach and direct integration of the Schrödinger equation. One can see that the computational time needed using the Floquet method is, on average, independent of the time until which we want to evolve our state. Because we always need to compute at least one full period to use the Floquet basis, the overhead of the Floquet method is disadvantageous compared to \code{sesolve()} at short times. However, at large stroboscopic times it becomes more efficient to use the Floquet basis, since we take advantage of the system periodicity in that way.
In \code{QuTiP} we could implement this using the \code{fsesolve()} solver, or manually as follows:

\inputpython{p_Floquet_speed_test.tex}{115}{119}
Figure~\ref{fig:Floquet_comp_time}(b) shows the agreement between the two methods at arbitrary stroboscopic times.


A good example to analyze the dimensional scaling of the Floquet implementation is to study a one-dimensional Ising spin chain under a periodic drive. The Hamiltonian of this system can be expressed as
\begin{equation}\label{eq:H_Ising_d}
H = g_0 \sum_{n=1}^{N} \sigma_z^{(n)} - J_0 \sum_{n=1}^{N-1} \sigma_x^{(n)} \sigma_x^{(n+1)} + A \sin(\omega_d t) \sum_{n=1}^{N} \sigma_x^{(n)}  \, ,
\end{equation}
where $g_0$ is the level splitting, $J_0$ is the coupling constant between nearest-neighbour spins and $A$ is the drive strength.
In QuTiP, this Hamiltonian can be implemented similarly to the Ising Hamiltonian without drive used in Sec.~\ref{sec:me_jax_gpu}, except that here we additionally include the driving term.
%
%
%

\begin{figure}
\centering
\includegraphics[scale=0.4]{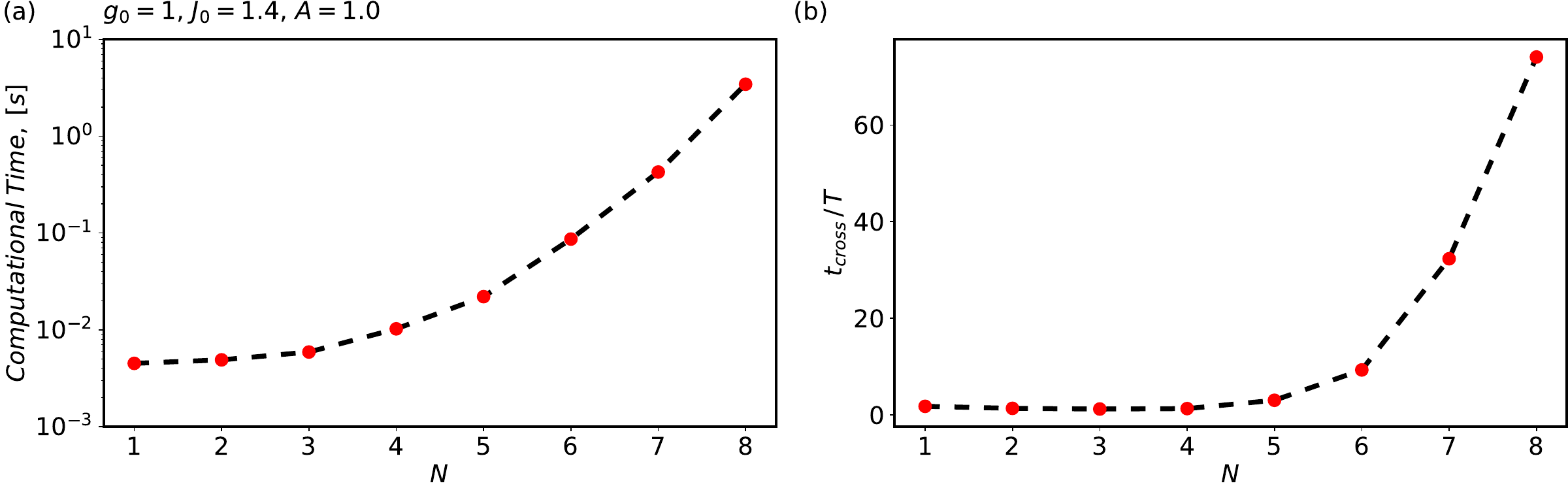}
\caption{Numerical study of the performance of Floquet basis and \code{sesolve} depending on the dimension of the system $N$. (a) Computational time needed to reach the $t_{\mathrm{cross}}$ depending on $N$. (b) Normalised time $t_{\mathrm{cross}}/T$ at which Floquet basis and \code{sesolve} require similar computational time versus $N$. Discrete data points are shown as red dots, while dashed lines are included to facilitate visualization of the overall trend.}
\label{fig:crosst_vs_N}
\end{figure}

In Fig.~\ref{fig:crosst_vs_N}, we compare the performance of \code{FloquetBasis()} and \code{sesolve()} when studying the Ising spin chain under a periodic drive, Eq.~\eqref{eq:H_Ising_d}. Particularly, we are interested in comparing these two methods depending on the number of particles in the system $N$.
Figure~\ref{fig:crosst_vs_N}(b) shows, depending on the number of particles $N$, the crossover time $t_{cross}$ at which the Floquet method becomes more performant than the direct \code{sesolve()} method.
Figure~\ref{fig:crosst_vs_N}(a) shows the computational time required to evolve the system until that crossover time.

In addition to the Floquet basis transformation and the \code{fsesolve()} solver mentioned before, QuTiP includes the solver \code{fmmesolve()} which is capable of analyzing time-periodic Hamiltonians which are affected by a dissipative bath, i.e., quasi-time-periodic Hamiltonians. We refer the reader to the QuTiP documentation for examples using this solver. When applying it, the user must proceed with caution since a generalized Floquet method can give unphysical results if the dissipation rate is too large compared to the pure eigenenergies of the periodic Hamiltonian. In a future work, we plan to discuss the intricacies of generalized Floquet methods and this solver in particular. Moreover, another solver \code{flimesolve()}, which also uses Floquet theory to approach open quantum systems, is currently being developed \cite{fenton}.

\subsubsection{\textup{\code{smesolve}}: Stochastic master equation solver}

When modelling an open quantum system, classical stochastic noise can be used to simulate a large range of phenomena. For example, classical noise can be used as a random term in the Hamiltonian as a means to simulate a classical environment randomly changing some system property in each run of an experiment. Another example is the Monte Carlo solver discussed earlier, where classical randomness is used to simulate the random chance of a quantum jump occurring in a dissipative Lindblad process. 

In the \code{smesolve()} solver that we will discuss in this section, noise appears because of a continuous measurement. The solver allows us to generate the trajectory evolution of a quantum system conditioned on a noisy measurement record.  Historically, this type of solver was used by the quantum optics community to model homodyne (single quadrature) and heterodyne (two-quadrature) detection of light emitted from a cavity. However, the solver is quite general, and can in principle be also applied to other types of problems.

To demonstrate its use we will focus on the standard example of a stochastic master equation describing an optical cavity whose output is subject to homodyne detection. 
The cavity obeys the general stochastic master equation,
\begin{equation}
d \rho(t) = -i [H,\rho(t)]\, dt + \mathcal D[a]\rho(t)\, dt + \mathcal{H}[a]\rho\, dW(t)
\end{equation}
with $\mathcal D[a] \rho = a\rho a^{\dagger} - \frac{1}{2}a^{\dagger}a \rho - \frac{1}{2}\rho a^{\dagger}a$ being the Lindblad dissipator and $H=\Delta a^{\dagger}a$ the Hamiltonian, which together capture the deterministic part of the system's evolution. The term $\mathcal{H}[a] \rho = a\rho + \rho a^{\dagger} - \mathrm{tr}\left[a\rho + \rho a^{\dagger}\right]$ represents the stochastic part which captures the conditioning of a trajectory through continuous monitoring of the operator $a$.
Here, $dW(t)$ is the increment of a Wiener process obeying $\mathbb E[dW]=0$ and $\mathbb E[dW^2]=dt$.

\begin{figure}
    {\centering \includegraphics[width=0.5\columnwidth]{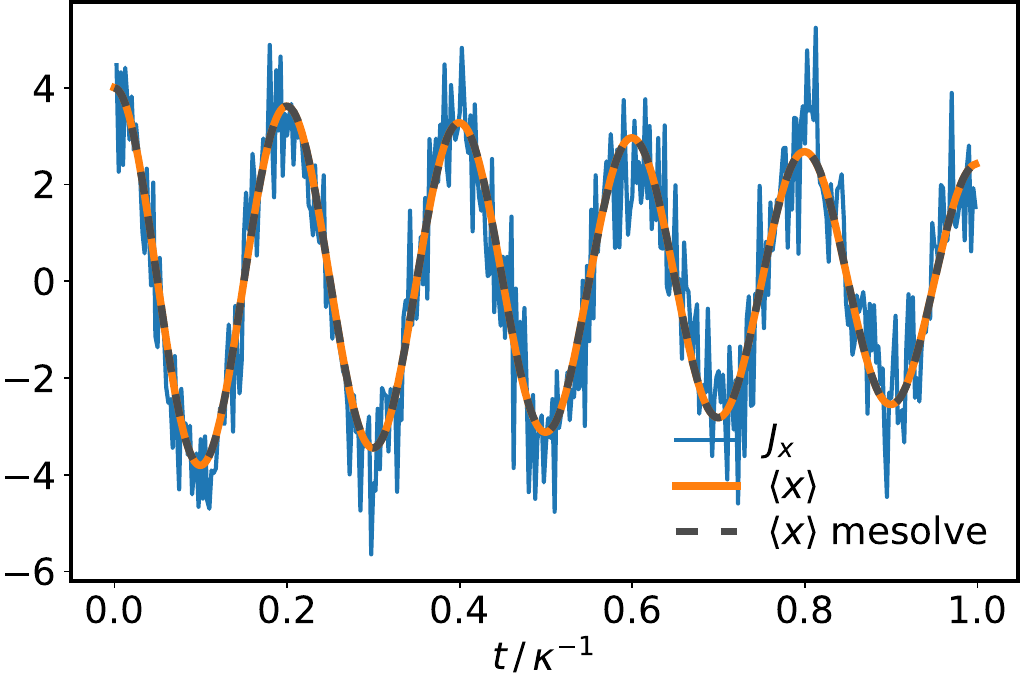}\par}
    \caption{An example of solving the stochastic master equation for a dissipative cavity, with decay rate $\kappa$ and Hamiltonian  $H=\Delta a^{\dagger}a$,  undergoing homodyne monitoring of the output field. Here we plot the averaged homodyne current $J_x=\ex{x} + dW/dt$, the average system behavior $\ex{x}$ for $50$ trajectories, and the average result for $\ex{x}$ predicted  by \code{mesolve} for the same model (without resolving conditioned trajectories). In this example the conditioned trajectories are only weakly affected by the noise, while the homodyne current remains noisy. Parameters used were $\Delta = 10\pi \kappa$, and initial condition of a coherent state with displacement $\alpha=2$.
    }
    \label{fig:smme}
\end{figure}

This equation can be easily implemented in \code{smesolve()} with
\newpage
\inputpython{smesolve.tex}{23}{26}
We have chosen here an initial coherent state, and collect expectation values of the operator $x=a+a^{\dagger}$.  The parameter \code{sc\_ops} indicates which operators are being monitored, and \code{stoc\_solution.expect} returns a list of the averaged results, while \code{stoc\_solution.measurement} returns a list of lists of individual measurement results of the operator $J_x(t) = \ex{x} + dW/dt$. The optional \code{c\_ops} can be a standard list of collapse operators describing additional, unmonitored, baths. The results of this simulation are shown in \fig{fig:smme}. Interestingly, in this example the conditioned system trajectories are only weakly affected by the noise, making it a 
  useful example for tests of this solver.

\subsubsection{\textup{\code{HEOMSolver}}: Hierarchical Equations of Motion} \label{subsec:heom}

While the other solvers in QuTiP largely rely on either perturbative approximations and/or assumptions about the Markovianity of the environment a system is coupled to, the hierarchical equations of motion (HEOM) are a numerically exact method \cite{tanimura_numerically_2020,Tanimura1989,lambert2020bofinheom} to solve the dynamics of an open quantum system, under a minimal set of assumptions. It originated in the field of physical chemistry, where it was used to solve problems related to electronic energy transport in photosynthetic light-harvesting. It has recently found utility in a broad range of other fields, from quantum information to quantum electronics, and it is now used as a benchmark for developing other methods \cite{10.1063/5.0221182}. In QuTiP we provide a solver for both bosonic and fermionic environments, which in an upcoming release will support arbitrary spectral densities and correlation functions via a built-in fitting procedure. 

The minimal assumptions that the HEOM method relies on are that the bath is Gaussian, initially in an equilibrium thermal state, and that the bath operator which couples to the system is linear. Using the Feynman-Vernon influence functional, one can show that, under these assumptions, the influence of the environment is fully characterized by its second order correlation function. For a bosonic environment, this correlation function can be expressed as (see \cite{PhysRevResearch.5.033011} for an explanation of the fermionic bath case, and \cite{lambert2020bofinheom,PhysRevResearch.5.043177} for applications)

\begin{equation}
    C(t)= \int_{0}^{\infty} d\omega\, \frac{J(\omega)}{\pi} \left( \coth\left({\frac{\beta \omega}{2}}\right) \cos\left({\omega t}\right) - i \sin\left({\omega t}\right)\right)\, .
\end{equation}






The derivation of the HEOM is also based on the Feynman-Vernon influence functional. The HEOM relies on the assumption that the correlation function can be written as a sum of decaying exponentials like


\begin{subequations}
\begin{align}
      C(t)&= C_{R}(t)+ C_{I}(t) \qquad \qquad\text{with} \\ C_{R}(t) &=  \sum_{k=1}^{N_{R}} c_{k}^{R} \exp(-\gamma_{k}^{R} t) \qquad \; \;\text{and}  \\
      C_{I}(t)&= \sum_{k=1}^{N_{I}} c_{k}^{I} \exp(-\gamma_{k}^{I} t)
\end{align}\label{eq:exponentially_decayin}
\end{subequations}
(though some variants of the technique have generalized this assumption \cite{Rahman_Kleinekathöfer_2019}).
By taking repeated derivatives of the influence functional, in conjunction with the assumption \eqref{eq:exponentially_decayin}, one arrives at the  coupled HEOM differential equations
\begin{align}
    \dot{\rho}^{n}(t) &=  -i [H_{S},\rho^{n}(t) ]  - \sum_{j=R,I}\sum_{k=1}^{N_{j}} n_{jk} \gamma_{k}^{j} \rho^{n}(t) - i \sum_{k=1}^{N_{R}} c_{k}^{R} n_{Rk} \{Q, \rho^{n_{Rk}^-}(t)\} \nonumber \\  &+ \sum_{k=1}^{N_{I}} c_{k}^{I} n_{Ik}\{Q, \rho^{n_{Ik}^-}(t)\}  - i  \sum_{j=R,I}\sum_{k=1}^{N_{j}} [Q,\rho^{n_{jk}^+}(t)]\, ,
\end{align}
where $n=(n_{R1},n_{R2},\hdots ,n_{R N_{R}},n_{I1},n_{I2},\hdots,n_{I N_{I}})$  is a multi-index label of non-negative integers $n_{jk}$, and $n_{jk}^-$ ($n_{jk}^+$) denotes the multi-index with the selected entry reduced (increased) by one. Further, $Q$ is the generic system operator which couples to the bath. In practice, the \emph{a priori} infinite hierarchy is truncated to $n_{jk} \leq N_{c}$, for a suitably chosen cutoff $N_{c}$. While the label $n=(0,0,\hdots,0)$ corresponds to the system density matrix, operators $\rho^n(t)$ with $n \neq (0,0,\hdots,0)$  are referred to as auxiliary density operators (ADOs), and encode correlations between system and bath.

To choose the value of the cutoff $N_c$, one typically starts with a small value which is then increased step by step until convergence is found. Heuristic arguments indicate that a lower bound is given by \cite{IshizakiProcNatlAcadSci2009}
\begin{equation}
    N_c \gtrsim \frac{\omega_S}{\min_{k,j} \Re[\gamma_k^j]} ,
\end{equation}
where $\omega_S$ is the largest system frequency and $\Re$ denotes the real part.
In future versions of QuTiP, we are planning to implement more efficient cutoff mechanisms, where ADOs are kept or discarded according to an importance criterion \cite{PhysRevB.88.235426, PhysRevB.94.121303, Huang2023}.

The implementation of the HEOM in QuTiP is explained in greater detail in \cite{lambert2020bofinheom}. Following the release of QuTiP v5, it is currently being enhanced to be more compatible with the other solvers (\code{mesolve()} and \code{brmesolve()}) with a generic environment class that allows us to quickly compute the power spectrum. The logic of this new environment class, and its functionality, are described in Fig.~(\ref{fig:environment_class}) and Tables (\ref{tab:predefined}), (\ref{tab:custom}) and (\ref{tab:comparison}).

\paragraph{Basic example} --- To demonstrate how to use the HEOM solver in practice, let us consider the evolution of a qubit in a thermal bosonic environment with the Hamiltonian
\begin{align}
     H = \frac{\omega_{0}}{2} \sigma_{z} + \frac{\Delta}{2} \sigma_{x} + \sum_{k} w_{k} a_{k}^{\dagger} a_{k} + \sum_{k} g_{k} \sigma_{z} (a_{k}+a_{k}^{\dagger}) \, . \label{Hheom}
\end{align}
In the continuum limit, one can describe the couplings through the spectral density 
\begin{align}
    J(\omega) = \pi \sum_{k} |g_{k}|^{2} \delta(\omega-\omega_{k}) \,.
\end{align}
Typically, bosonic HEOM solvers use either the overdamped Drude-Lorentz spectral density or the underdamped Brownian motion spectral density. Let us for example consider the underdamped spectral density

\begin{align}
 J(\omega) = \frac{\lambda^{2} \Gamma \omega}{(\omega_{c}^{2}-
\omega^{2})^{2}+ \Gamma^{2} \omega^{2}} \, .
\end{align}

We begin by initializing a bath, based upon the parameters in the spectral density:

\inputpython{heom_example.tex}{27}{28}
%
The \code{env} object contains the exact information about the bosonic bath, while the \code{bath} object is an approximated version of \code{env} that is specifically designed for the HEOMSolver, which requires as input an exponential decomposition of the bath correlation functions as described above. For this particular spectral density, Eq.\ \ref{eq:exponentially_decayin} can be realized via the Matsubara or Pad\'{e} decompositions. These decompositions rely on a truncation parameter $N_k$ that introduces an approximation to the true bath dynamics by truncating the Matsubara or Pad\'{e} decompositions into a finite number of exponentials. The QuTiP implementation makes it easy to see the impact of the approximation, as it is straightforward to compute the exact and approximated quantities that describe the bath (correlation function, spectral density and power spectrum)

\inputpython{heom_example.tex}{178}{181}



\begin{figure*}[h]
\includegraphics[width= \columnwidth]{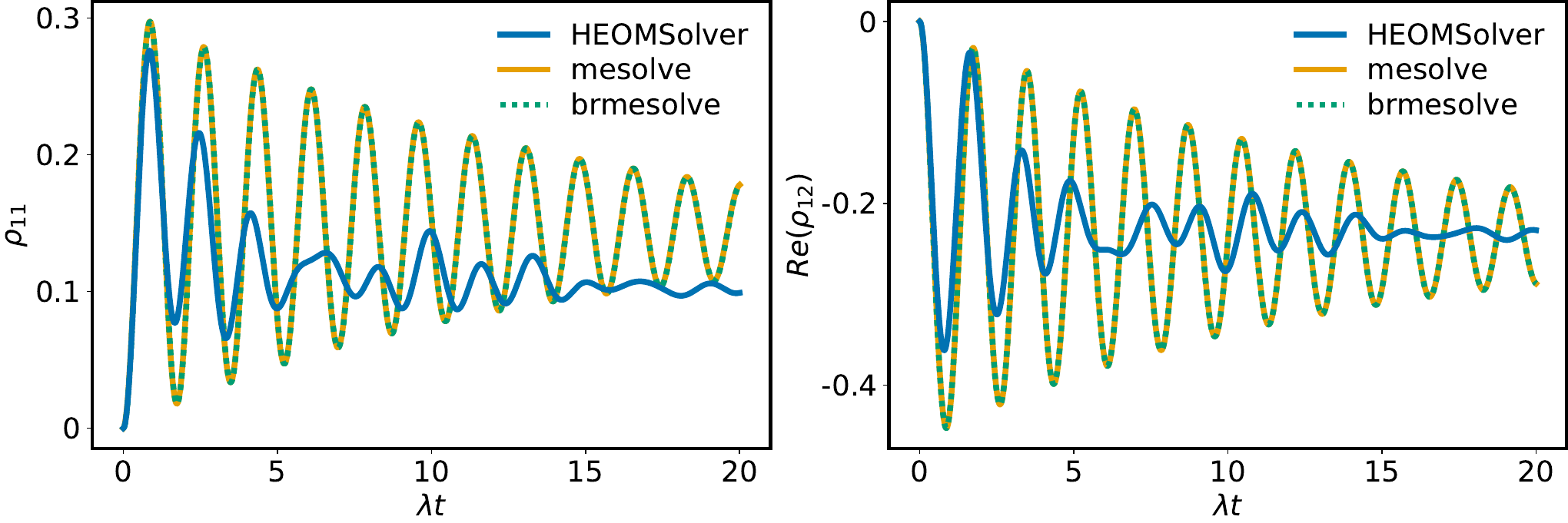}
    \caption{
       For the example of a standard spin-boson problem, we compare the output of the different master equation solvers available in QuTiP, namely, the Lindblad master equation solver (\code{mesolve()}), the Bloch-Redfield equation solver (\code{brmesolve()}) and the hierarchical equations of motion solver (\code{HEOMSolver}) for a spin coupled to a bosonic bath described by an underdamped Brownian motion spectral density. Left, (a), shows the evolution of the qubit population for the different approaches. Notice that both \code{brmesolve} and \code{mesolve} coincide, but they differ substantially from the HEOM result in this deeply non-Markovian regime. The right figure, (b), shows the dynamics of the coherence values. The parameters used to generate these figures are: $\lambda=0.5\, \Delta$, $\Gamma=0.1\, \Delta$, $T=0.5\, \Delta$, $N_{k}=5$, $N_{c}=6$ and $\omega_{0}=\frac{ 3 \Delta}{2}$.
    }\label{fig:correlation_heom1}    
\end{figure*}
\begin{figure*}[p!]\centering
\includegraphics[width = 0.8\columnwidth]{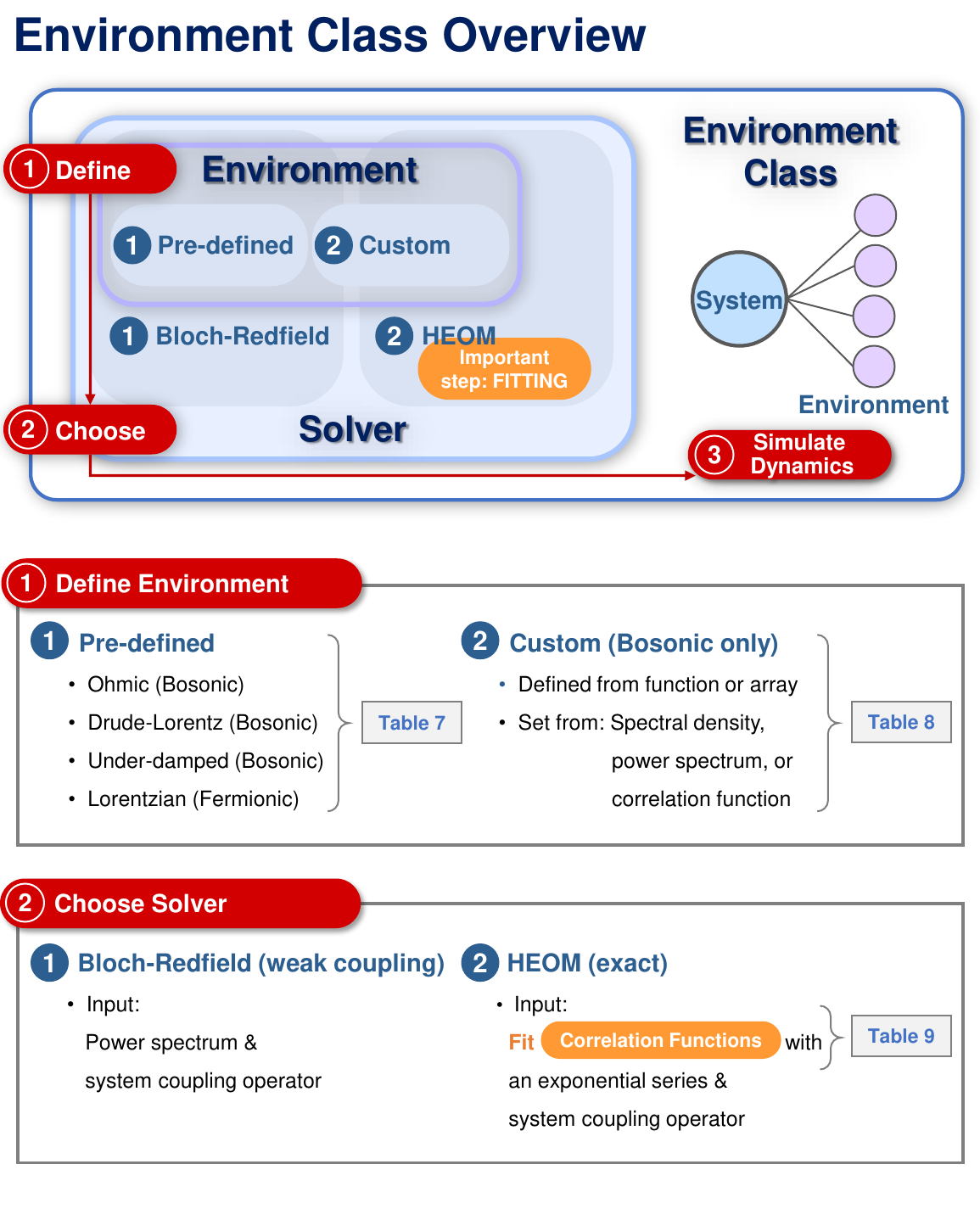}
\caption{An overview of the environment class, which is now supported by the HEOM and Bloch-Redfield solvers. Details on how predefined and custom environments can be made is provided in tables (\ref{tab:predefined}) and (\ref{tab:custom}), and the fitting methods that can be used to approximate the bath correlation function for use in the HEOM solver are described in table (\ref{tab:comparison})}\label{fig:environment_class}
\end{figure*}

Similar notation is used for both the power spectrum and spectral density. The power spectrum can be used as quick gateway to compare with the other available solvers. For example, we can solve the HEOM equations by using the bath object (in a tuple, alongside which system operator \code{Q} it couples to), the system Hamiltonian and the \code{max\_depth} parameter that specifies the cutoff $N_c$ of the hierarchy equations, with the \code{HEOMSolver} as follows:

\inputpython{heom_example.tex}{32}{35}
Simultaneously, we can compare this to a Bloch-Redfield solution also using the bath object properties,

\inputpython{heom_example.tex}{63}{65}
%
See \fig{fig:correlation_heom1} for a comparison of the results.

\paragraph{Ohmic bath with exponential cutoff} --- The QuTiP implementation of the HEOM allows for the simulation of more general spectral densities: the user can create an arbitrary \code{BosonicEnvironment} from the spectral density, correlation function or power spectrum. As the HEOM requires a decaying exponential representation of the correlation functions, we can either fit the spectral density with one with a known analytical decomposition of its correlation function,  or the correlation functions themselves, as explained in \cite{lambert2019modelling, lambert2020bofinheom}. Let us consider the simulation of a Ohmic bath, whose spectral density is given by

\begin{equation}
    J(\omega) = \alpha \omega \exp\left(-|\omega|/\omega_{c}\right) \, .
\end{equation}

\begin{table}[h]
\centering
     \renewcommand{\arraystretch}{2.5} 
\resizebox{\textwidth}{!}{
\begin{tabular}{@{}lcccc@{}}
\toprule
\textbf{\shortstack{\\Predefined \\ environment \\type}} &
\textbf{\shortstack{Spectral density \\ Function}} &
\textbf{\shortstack{Python \\ Function}} &
\textbf{\shortstack{\\Pad\'{e} \\or Matsubara\\ Available}} &
\textbf{\shortstack{Bosonic or \\ Fermionic}} \\ \midrule 
\textbf{Ohmic} & $J(\omega) = \alpha (\omega^s/\omega_c^{s-1}) \exp[-\omega / \omega_c]$  & \code{OhmicEnvironment(T, alpha, wc, s)} & \textcolor{red}{\faTimes}  & Bosonic  \\ 

\textbf{Drude-Lorentz} & $J(\omega) = 2 \lambda \gamma \omega/(\gamma^{2}+\omega^{2})$ & \code{DrudeLorentzEnvironment(T, lam, gamma)} & \textcolor{green}{\faCheck} & Bosonic  \\ 

\textbf{Under-Damped} & $J(\omega) = \lambda^{2} \Gamma \omega/[(\omega_0^{2}-
\omega^{2})^{2}+ \Gamma^{2} \omega^{2}]$ & \code{UnderDampedEnvironment(T, lam, Gamma, w0)} & \textcolor{green}{\faCheck} & Bosonic  \\ 

\textbf{Lorentzian} & $J(\omega) =\gamma W^2/[(\omega - \omega_0)^2 + W^2]$ & \code{LorentzianEnvironment(T, mu, gamma, W, w0)} & \textcolor{green}{\faCheck} & Fermionic  \\ \bottomrule
\end{tabular}}
\caption{Predefined environments: Table of predefined environments as given by their spectral density, and the internal QuTiP function to create them.  The fourth column indicates whether or not a Pad\'{e} or Matsubara series expansion of the bath correlation functions is available. All baths require an initial temperature (which can be zero, though this invalidates the series expansion for the correlation functions, and fitting should be used instead). For the Fermionic Lorentzian bath we also need to provide a chemical potential (defined via the \code{mu} parameter). }\label{tab:predefined}
\end{table}

\begin{table}[h]
\centering
     \renewcommand{\arraystretch}{1} 
\begin{tabular}{@{}p{4cm}p{6cm}p{5.5cm}@{}}
\toprule
\textbf{Defined from:} &
\textbf{Python Function} &
\textbf{Important optional arguments}\\ \midrule

\textbf{Spectral density} & \code{BosonicEnvironment} \code{.from\_spectral\_density(func)}  & 
\code{T}: temperature of the bath. \code{wlist}: array of frequencies (if array-based). \code{wMax}: maximum frequency used in Fourier-transform conversions.
\\
\midrule
\textbf{Power Spectrum} & \code{BosonicEnvironment} \code{.from\_power\_spectrum(func)} & As above\\ 
\midrule
\textbf{Correlation function} & \code{BosonicEnvironment} \code{.from\_correlation\_function(func)}&
\code{T}: temperature of the bath.  \code{tlist}: array of times (if array-based). \code{tMax}: maximum time used in Fourier-transform conversions. \\ 
\midrule
\textbf{Correlation function from exponential series (bosonic or fermionic)} 
& 
\code{ExponentialBosonicEnvironment} \code{(ck\_real, ck\_imag, vk\_real,vk\_imag)}
\code{ExponentialFermionicEnvironment} \code{(ck\_plus, vk\_plus, ck\_minus, vk\_minus)} &

\code{T}: temperature of the bath. \code{combine}: Boolean,  indicates whether common-frequency compression is used (bosonic case only). \code{mu}: Chemical potential (fermionic case only).\\ 

\bottomrule
\end{tabular}
\caption{Custom environments: Table of functions for defining custom environments from either their spectral density, power spectrum or correlation functions (only custom bosonic environments are currently supported in this way, though fermionic ones can be done manually as shown in the final row in the table). The parameter \code{func} can be a Python function or a list/array of points (requires a corresponding \code{tlist} or \code{wlist} optional parameter to be provided). Both \code{wMax} and \code{tMax} should be chosen such that the function or array used is negligible for higher values of $t$ or $\omega$.}\label{tab:custom}
\end{table}

\begin{table}[h]
\centering
     \renewcommand{\arraystretch}{1.2} 
\begin{tabular}{@{}lcccccccp{6cm}@{}}
\toprule
\textbf{Method} &
\textbf{\rotatebox{90}{\shortstack{Arbitrary  \\ Functions}}} &
\textbf{\rotatebox{90}{\shortstack{Allows \\Constraints}}} &
\textbf{\rotatebox{90}{\shortstack{No need of \\ Extra Input}}} &
\textbf{\rotatebox{90}{\shortstack{Optimization \\ Convergence}}} &
\textbf{\rotatebox{90}{\shortstack{Sampling \\ Insensitive}}} &
\textbf{\rotatebox{90}{\shortstack{Arbitrary \\ Temperatures}}} &
\textbf{\rotatebox{90}{\shortstack{Works on \\ Noisy data}}} &
\textbf{Recommended when...} \\ 
\midrule
\makecell[cl]{\textbf{NLSQ} \\ (ps,sd,cf)} & \textcolor{green}{\faCheck} & \textcolor{green}{\faCheck} & \textcolor{red}{\faTimes} & \textcolor{red}{\faTimes} & \textcolor{red}{\faTimes} & \textcolor{green}{\faCheck} & \textcolor{green}{\faCheck} & You have an idea about which  exponents  should be included \cite{lambert2019modelling}. \\ 

\textbf{AAA} & \textcolor{green}{\faCheck} & \textcolor{red}{\faTimes} & \textcolor{green}{\faCheck} & \textcolor{green}{\faCheck} & \textcolor{red}{\faTimes} & \textcolor{green}{\faCheck} & \textcolor{green}{\faCheck} & You need high-accuracy in the steady-state and the spectral density  is not too structured. \\ 

\textbf{Prony} & \textcolor{green}{\faCheck} & \textcolor{red}{\faTimes} & \textcolor{green}{\faCheck} & \textcolor{green}{\faCheck} & \textcolor{red}{\faTimes} & \textcolor{green}{\faCheck} & \textcolor{red}{\faTimes} & The correlation function is noiseless and long lived. \\ 

\textbf{Matsubara} & \textcolor{red}{\faTimes} & \textcolor{red}{\faTimes} & \textcolor{green}{\faCheck} & \textcolor{green}{\faCheck} & \textcolor{green}{\faCheck} & \textcolor{red}{\faTimes} & \textcolor{red}{\faTimes} & Doing high temperature simulations using the specific spectral densities it is available for.  \\ 

\textbf{Pade} & \textcolor{red}{\faTimes} & \textcolor{red}{\faTimes} & \textcolor{green}{\faCheck} & \textcolor{green}{\faCheck} & \textcolor{green}{\faCheck} & \textcolor{red}{\faTimes} & \textcolor{red}{\faTimes} & The same cases as the Matsubara method. This is  recommended over Matsubara whenever available.\\ 

\textbf{ESPIRA} & \textcolor{green}{\faCheck} & \textcolor{red}{\faTimes} & \textcolor{green}{\faCheck} & \textcolor{red}{\faTimes} & \textcolor{red}{\faTimes} & \textcolor{green}{\faCheck} & \textcolor{green}{\faCheck} & Looking for a general-purpose method.  \\ 

\textbf{ESPRIT} & \textcolor{green}{\faCheck} & \textcolor{red}{\faTimes} & \textcolor{green}{\faCheck} & \textcolor{green}{\faCheck} & \textcolor{red}{\faTimes} & \textcolor{green}{\faCheck} & \textcolor{green}{\faCheck} &The correlation function is  long lived. 
\\ 
\bottomrule
\end{tabular}
\caption{Once an environment is created (see Tables (\ref{tab:predefined}) and (\ref{tab:custom})), its correlations can then be approximated as an exponential series in a variety of ways. This is done by the class method \code{.approximate("type")} where \code{"type"} is a string defining the method desired, as listed in the first column of this table.  They take a variety of optional arguments to define fitting range or bounds, depending on the method used.
 Non-linear least squares fitting (NLSQ), the method used in this section, might be performed on the spectral density, the power spectrum, or the correlation function, as specified by the type strings "ps", "sd",  and "cf". ``Arbitrary Functions'' stands for the ability of approximating an arbitrary user-defined environment. ``Allows Constraints'' stands for the ability to limit the range of values the parameters in Eq. \ref{eq:exponentially_decayin}, which is often useful when working with HEOM as it may improve convergence. ``No need of Extra Input'' refers to the input that the algorithm needs to work properly: we consider a set of sampling points as basic input, and other pieces of information required as extra input. ``Optimization Convergence'' refers to the fact that the approximation does not get stuck in local minima preventing convergence, in some cases failing to provide any approximation. ``Sampling Insensitive'' refers to the fact that changing the sampling points does not change the approximation. ``Arbitrary Temperatures'' refers to the method converging at arbitrarily low temperatures. ``Works on Noisy data'' applies when the user defined spectral density comes from data that is noisy.}\label{tab:comparison}
\end{table}
Because this spectral density is frequently used in the literature, a special class, \code{OhmicEnvironment}, has been added for convenience. As mentioned, in order to use the HEOM solver, we must first choose between fitting either the correlation function or the spectral density. The code snippet below shows how easy it is to set up simulations using these different approaches:

\inputpython{heom_example.tex}{98}{105}

The first output of the approximation by fitting is a \code{BosonicEnvironment} object, while the second output provides the fit information. This is particularly useful when dealing with non-standard baths, where one often needs more exponents to accurately describe the bath even when dealing with high temperatures, as the fit information helps one decide how many exponents to take into consideration.

\begin{figure*}[h]
\includegraphics[width=\columnwidth]{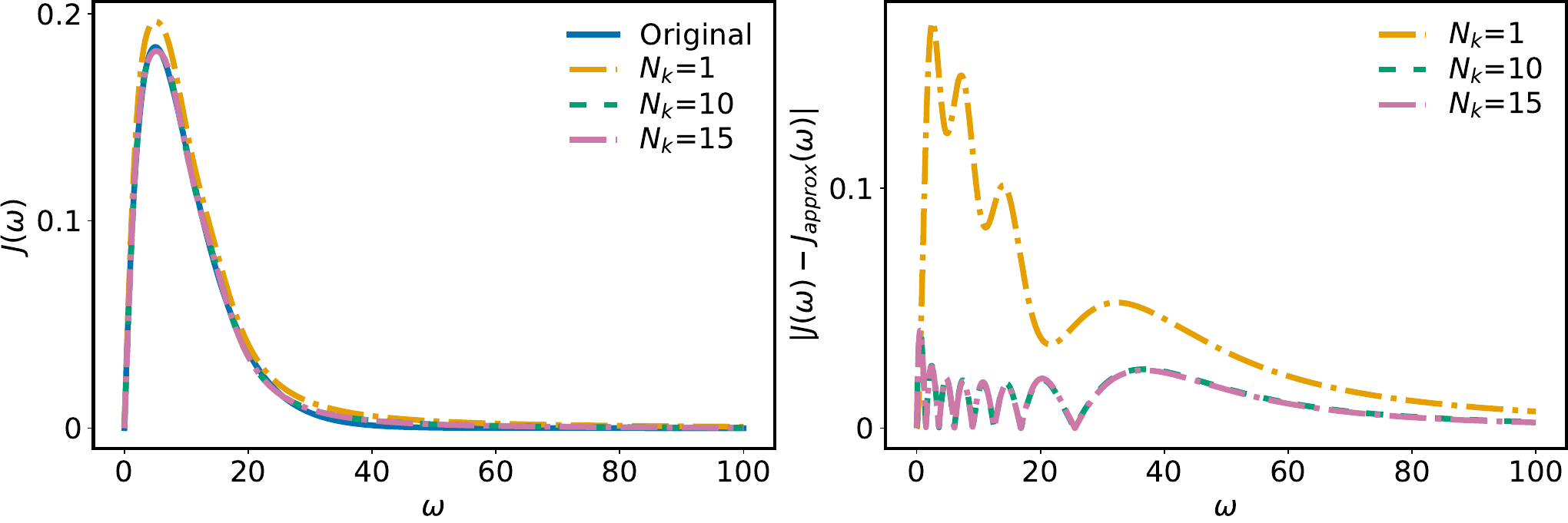}
\caption{
Approximation of the spectral density of an Ohmic Bath via fitting with three underdamped Brownian motion spectral density functions. Importantly, when fitting spectral densities for the HEOM method, we must still expand each spectral density's contribution to the total correlation functions in terms of exponentials. To gain insight of the effectiveness of both the fit and the correlation function expansion, we can then use that expansion, along with the temperature information of the bath, to reconstruct a total effective spectral density, which is what we show in this plot. In the figure, we can see that the number $N_k$ of exponents kept per underdamped mode has an effect on the approximation of the resulting effective bath spectral density. Ultimately one should aim to find the minimum number of exponents that make up a good approximation of the spectral density. The parameters of the original Ohmic bath are $\lambda = 0.1$, $\omega_c = 5 $, and $T = 1$.
}\label{fig:correlation_heom}    
\end{figure*}


The bath obtained from the fitting can be passed to the solver to quickly obtain its dynamics

\inputpython{heom_example.tex}{126}{128}

Figure \ref{fig:correlation_heom} shows an example of fitting the Ohmic spectral density with exponential cut-off with a set of underdamped Brownian motion spectral densities (sometimes called the Meier-Tannor fitting approach).

\paragraph{Zero temperature and the localization-delocalization phase transition} --- One of the benefits of including these fitting routines is to be able to simulate situations where the structure of the spectral density has non-trivial effects. An example of such a situation is the localization-delocalization transition in the spin-boson model \cite{xu_taming_2022,Wenderoth_Breuer_Thoss_2021,Zhou_Shao_2008,Wang_Thoss_2008}. When the temperature of the bath is $T=0$, one expects the steady state of the system to be 
 delocalized $\langle \sigma_z(t\rightarrow \infty) \rangle = 0$. However, when the coupling to the bath is increased the steady state goes from a completely delocalized state to what appears to be a localized state for long-times $\langle \sigma_z(t\rightarrow \infty) \rangle \neq 0$. To demonstrate this, we follow \cite{Zhou_Shao_2008} and choose an Ohmic spectral density with a polynomial cutoff of the form 
\begin{align}
    J(\omega)= \frac{\pi \alpha \omega}{2(1+(\frac{w}{wc})^{2})} \, .
\end{align}
Due to this effect being present at $T=0$, we fit the correlation function rather than the spectral density. Obtaining the correlation function is straightforward once we create a \code{BosonicEnvironment}.

\inputpython{heom_transition.tex}{100}{105}

\begin{figure*}[t]
\centering
\includegraphics[width=\columnwidth]{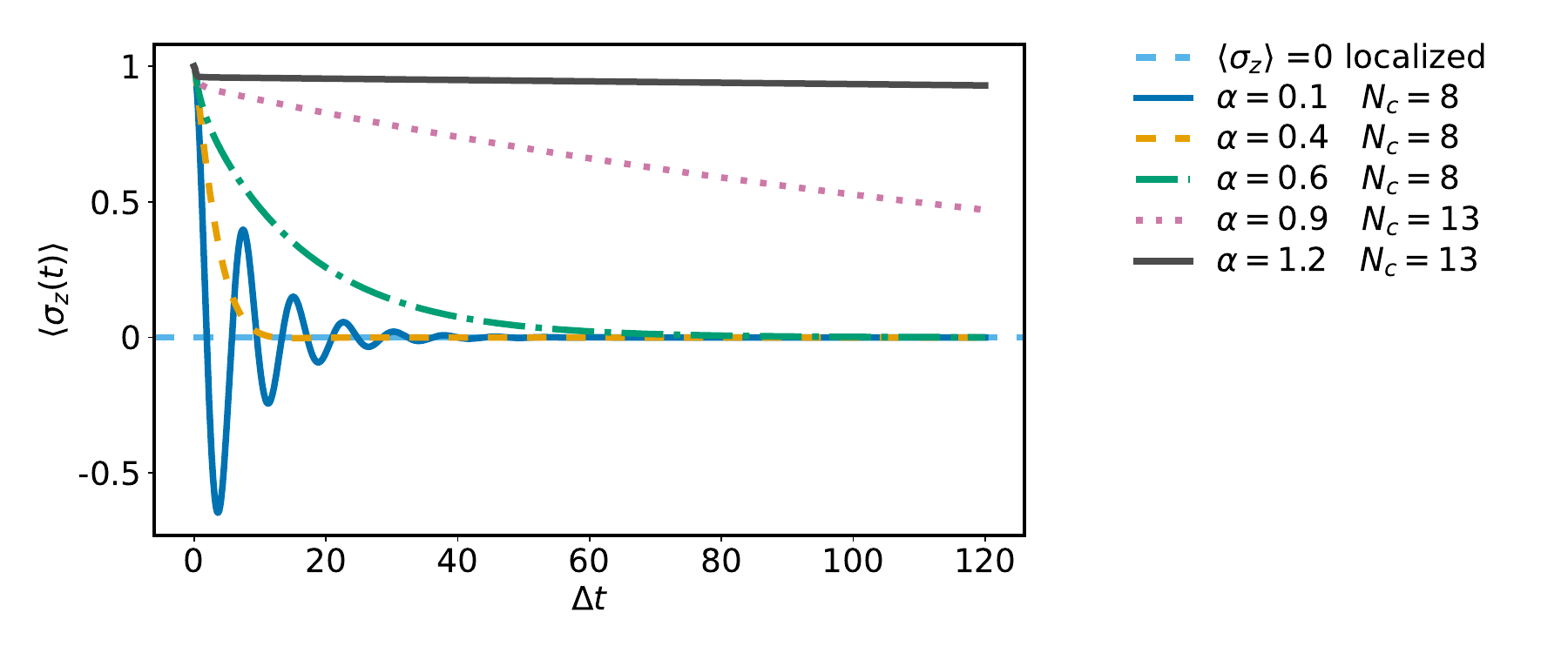}
    \caption{This figure shows the spin-boson localization-delocalization phase transition. For the simulation, we used $\omega_{0}=0$, $\omega_{c}=10\, \Delta$ and $T=0$. The correlation function was fit with two exponents for the real part and one for the imaginary part. 
    }\label{fig:transition_heom}    
\end{figure*}

For this example, let us consider the same system Hamiltonian as before, with $\omega_{0}=0$. Figure \ref{fig:transition_heom} shows how by increasing the coupling to the bath, the system goes from a localized state to a delocalized state. This is the so called spin-boson localization-delocalization phase transition \cite{xu_taming_2022}. 

Apart from the method used above, the QuTiP environment class provides several different ways to decompose arbitrary environments into a decaying exponential form as in Eq.~\eqref{eq:exponentially_decayin}, as needed for simulation with the HEOMSolver. Table~\ref{tab:comparison} provides a summary of the methods available. For details about the methods, we refer to the documentation or Refs.~\cite{Takahashi_Rudge_Kaspar_Thoss_Borrelli_2024,suárez2025makingnonmarkovianmasterequations}. At present, most of these methods only work for bosonic environments; however, fermionic environments will be supported in the near future.




\subsubsection{Visualization of solver results}

QuTiP comes with a range of functions to visualize the results returned by its solvers. In addition to automated functions for plotting expectation values with Matplotlib, QuTiP provides utility functions to calculate important and commonly used representations of quantum states, including the Bloch-sphere for two-level systems and pseudo-probability functions, such as the Wigner and Husimi functions, for harmonic systems like cavities.

In this section, we present a few examples to illustrate the utility and features of these plotting functions. Many more examples can be found in the tutorial notebooks described in \ref{app:notebooks}.

For both learners and researchers, using the Bloch sphere to visualize the overlap of a qubit state with the Pauli matrices in three-dimensional space can be useful to help understanding how the state evolves, particularly when combined with QuTiP's animation features.  In QuTiP, a Bloch sphere is instantiated with \code{b = Bloch()}.  Points can be added to it using coordinates, as in \code{b.add\_points([1/np.sqrt(3), 1/np.sqrt(3), 1/np.sqrt(3)])}, and the sphere plotted with \code{b.render()}.  Vectors can be added in a similar way, using  \code{b.add\_vectors([0, 1, 0])}.  

\begin{figure}[t]
\includegraphics[width = 0.33\columnwidth]{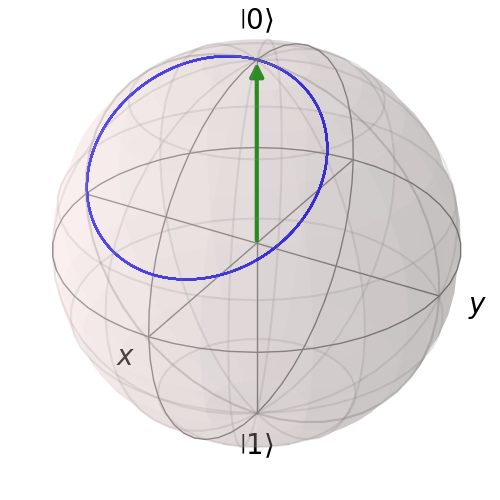}
\includegraphics[width = 0.33\columnwidth]{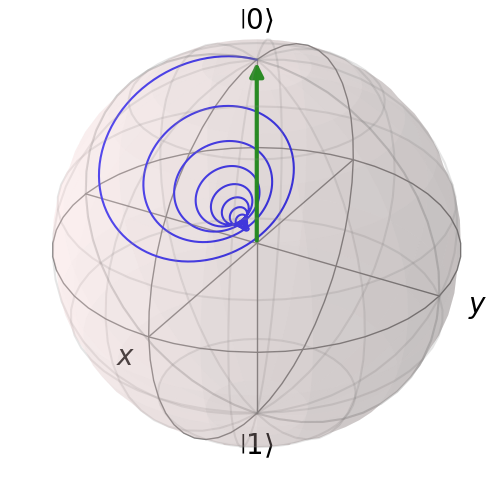}
\includegraphics[width = 0.33\columnwidth]{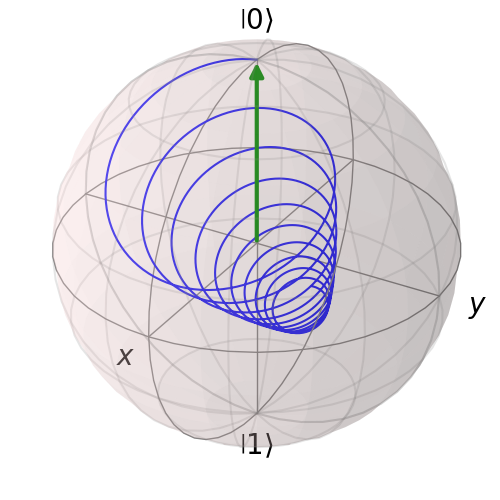}
\caption{The left figure shows the Bloch sphere representation of the dynamics of a qubit undergoing unitary evolution under the Hamiltonian 
$H = \Delta [\cos(\theta)  \sigma_z + \sin(\theta)  \sigma_x  ]$ with $\theta = 0.1\, \pi$ for an initially excited qubit. The middle figure shows the dynamics of the same system with added dephasing, with a rate $\gamma_p = 0.5\, \Delta / (2\pi)$. The right figure shows the same model but with added relaxation at the rate $\gamma_r = 0.5\, \Delta/(2\pi)$.}\label{bloch2}
\end{figure}

More commonly, one wants to visualize the output of solvers, which is provided in the form of state vectors or density operators. These objects can be added to a sphere using the method \code{b.add\_states(state)}.  The tutorial notebook \code{0004\_qubit-dynamics}, briefly summarized in \ref{app:notebooks}, demonstrates this feature using the example of a driven qubit, both with and without noise, and the result is shown in \fig{bloch2}.

Pseudo-probability functions are also commonly used to visualize data arising from continous variable systems. In QuTiP, such systems are truncated on finite dimensional Fock spaces, but these pseudo-probability functions can still be used, given sufficient truncation. The Wigner function, for example, helps visualize the probability of the position and momentum quadratures of cavities, and famously contains negative probabilities for non-classical states.

\begin{figure}[t]%
\includegraphics[width = 0.5\linewidth]{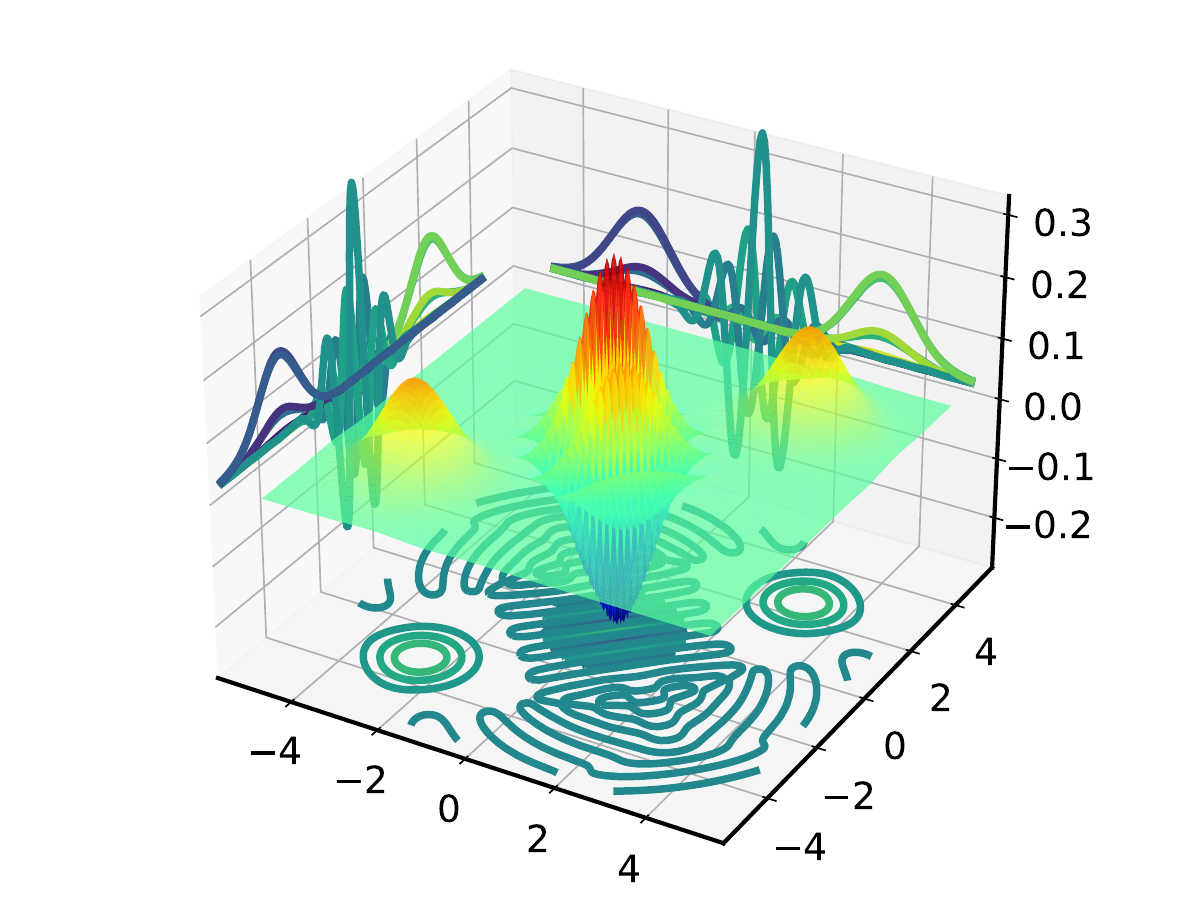}%
\includegraphics[width = 0.5\linewidth]{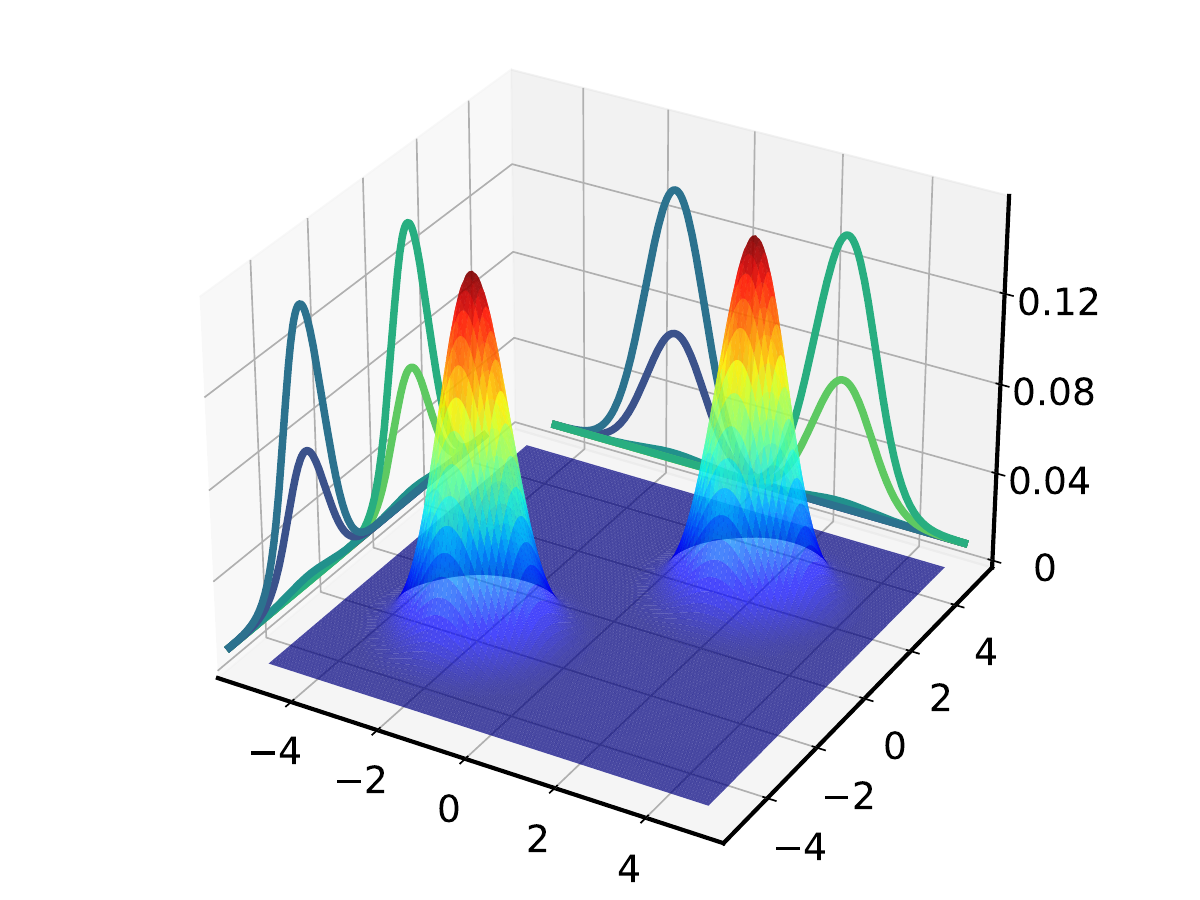}%
\caption{Left figure shows the Wigner function, a pseudo-probability distribution, of a cavity prepared in a Schr\"odinger cat state $\psi=\frac{1}{\sqrt 2}\left(\ket{\alpha_1} + \ket{\alpha_2}\right)$, with $\alpha_1 = -2.0 - 2j$ and $\alpha2 = 2.0 + 2j$. The right figure shows instead the Husimi-$Q$ function for the same state.
}\label{wigner}
\end{figure}

We can demonstrate the visualization of Wigner functions straightforwardly with another common example; a cavity prepared in a ``Schrödinger cat'' state, i.e., a superposition $\psi=\frac{1}{N} \left(\ket{\alpha_1} + \ket{\alpha_2}\right)$ of two coherent states (where $N$ is a normalization factor). The Wigner function, shown in Fig.~\ref{wigner}, clearly shows the expected negative values for such a highly non-classical state. The non-classicality is less apparent in the Husimi-Q function, also shown in Fig.~\ref{wigner}, which is non-negative by definition.

New in QuTiP v5 is a suite of tools to automate the animation of many of these commonly used functions.  These can be explored in the tutorial \cite{animation_tutorial}, and they include customized methods for the:
\begin{itemize}
\item Wigner function (\code{anim\_wigner} and \code{anim\_wigner\_sphere}), 
\item Hinton plots (\code{anim\_hinton}), 
\item sphere plots (\code{anim\_sphereplot}), 
\item histograms (\code{anim\_matrix\_histogram}), 
\item Fock state distributions (\code{anim\_fock\_distribution}), 
\item spin distributions (\code{anim\_spin\_distribution}), 
\item Qubism plots (for ploting the states of many qudits, \code{anim\_qubism}),  and 
\item Schmidt plots (for plotting matrix elements of a quantum state, \code{anim\_schmidt}). 
\end{itemize}
In addition, there is a new option (\code{qutip.settings.colorblind\_safe}) to choose plotting colors from a pallette of colorblind safe colors.

\subsection{Additional features in QuTiP v5}

In addition to the new data layer and solver features described earlier, there are other new features in QuTiP v5, and some of the previously existing features have received important updates. Below, we describe in detail three such new or updated features: excitation number restricted states, which are crucial for the simulation of large composite systems, and now have improved back-end support through a new dimensions class, the option of using the JAX auto-differentiation functionality with the new JAX data layer, and support for the Message Passing Interface (MPI) in the parallelization of various solvers, which enables the easy use of super-computing resources.

\subsubsection{Excitation number restricted states}\label{enr}

When modeling many interacting quantum systems, QuTiP does not by default apply any approximation
apart from truncating the individual systems' Hilbert spaces. For example, when modelling
the discrete Fock-space representation of a quantum harmonic oscillator, we normally truncate the states at some finite number of excitations. This is clearly demonstrated when defining the commonly used Jaynes-Cummings model from quantum optics, which describes a single two-level atom interacting with a single-mode cavity under the rotating wave approximation:

\inputpython{enrstates.tex}{11}{21}

In this case the Hilbert space of the cavity is truncated at two Fock states, so the total number of states is four. However, the Hamiltonian conserves the total number of excitations in the coupled system. With an initial condition containing at most a single excitation, the double-excitation state, where the atom is excited and a photon is in the cavity, is therefore decoupled from the dynamics.  Using excitation number restricted (ENR) states, we can truncate the total Hilbert space to exclude the double-excitation state:

\inputpython{enrstates.tex}{28}{38}

Here, \code{N\_exc=1} is the maximum number of excitations we wish to consider across the whole Hilbert space. The function \code{enr\_destroy(dims, N\_exc)} returns a list of annihilation operators for each subsystem in \code{dims} which only act on a reduced space including states with up to that total excitation number.  In this example, the restricted Hilbert space is spanned by $\ket{0,0}$, $\ket{0,1}$ and $\ket{1,0}$.  The first annihilation operator can be thought of as the operator $\ket{0,0}\bra{1,0}$, and the second one as $\ket{0,0}\bra{0,1}$.  With these constructions, we can then recreate the full Hamiltonian and dynamics of the Jaynes-Cummings model, omitting the unimportant double occupation state.

The power of this approach lies in situations with many subsystems, where one only needs to consider a limited number of excitations. One of the core QuTiP notebooks demonstrates this well, with a large chain of coupled Jaynes-Cummings models \cite{enr_chain}.  But it can also be used as a powerful truncation tool in situations where the Hamiltonian is not necessarily excitation-number conserving (see \cite{feist}).

It is important to note that since ENR states essentially compress the normal tensor structure of states and operators onto one single reduced Hilbert space, annihilation and creation operators of different subsystems no longer commute. Hence, care must be taken when representing operators on the ENR state space.  Typically, when constructing Hamiltonians, one should order annihilation operators to the right and creation operators to the left. In addition, ENR states require the use of a range of custom functions, like \code{enr\_fock()}, and many standard utility functions in QuTiP will fail when used with them.  In v5 the addition of unique dimension objects for ENR states potentially allows these issues to be resolved, but general compatibility is still ongoing work.

To demonstrate the utility of ENR states in a complex problem, we now consider an important example from the literature; that of a qubit interacting with a one-dimensional waveguide \cite{zueco} truncated by a mirror.  Generally, this type of problem involving time-delayed feedback is difficult to model numerically \cite{Zhang2017}.  One common method to capture the finite time delay of photons reaching the mirror and returning to the system is discretization of modes in the waveguide.  This discretization can be performed in multiple ways (see \cite{zueco} and \cite{arne}), but the approach taken in \cite{stephenhughes1,PhysRevA.106.013714} is  particularly amenable to using ENR states. In this approach, spatial discretization and temporal discretization of the waveguide are done hand-in-hand, and open ends of the wave guide are truncated using a Monte Carlo-like measurement step. This procedure still requires modelling a large number of waveguide modes, or ``boxes'', which the authors of \cite{stephenhughes1} were able to achieve using an ENR-like truncation of their basis states (albeit done manually, not using QuTiP).

It is relatively straightforward to implement this procedure in QuTiP. However, following \cite{stephenhughes1}, we implement the time evolution manually using a product of propagators for small discrete time steps rather than using one of the standard QuTiP solvers directly. We refer readers to \cite{stephenhughes1} for a complete description, but succinctly, one starts with the Hamiltonian for the system coupled to a discrete-frequency waveguide:
\begin{align}
H = \frac{\epsilon}{2} \sigma_z + \sum_{\alpha \in \{L, R\}} \sum_{k={0}}^{N-1} \omega_k b_{k,\alpha}^{\dagger}b_{k,\alpha} 
+  \sqrt{\frac{2\pi}{L_0}} \sum_{\alpha \in \{L, R\}} \sum_{k={0}}^{N-1} \kappa_{\alpha}(\omega_k)\left[ \sigma_+ b_{k,\alpha} + \text{H.c.} \right] \, .
\end{align}
Here, it was assumed that a mirror truncates the left side of the waveguide at some finite distance $L_0/2$, and the sums run over left- and right-moving modes and over their discretized frequencies $\omega_k$. The coupling terms are $\kappa_L(\omega) = \sqrt{\gamma_L/(2\pi)}$ and $\kappa_R(\omega)=\sqrt{\gamma_R/(2\pi)} e^{i\phi} e^{i\omega\tau}$, where $\gamma_{\alpha}$ are coupling constants, $\tau = L_0 / c$ is the total travel time of photons to the mirror and back (with $c$ being the speed of light in the waveguide), and $\phi$ is an additional phase change incurred from the reflection at the mirror. In other words, a photon can propagate to the left, hit the mirror, and then return as a right propagating photon that then interacts with the system with an accumulated phase $(\omega \tau + \phi)$.  

The key step is to transform the discrete frequencies into spatially discretized modes with the discrete Fourier transform
\begin{equation}
B_{n,\alpha} =(1/\sqrt{N})\sum_{k=0}^{N-1} b_{k,\alpha}\exp[(\pm)_\alpha i\omega_k n \Delta t]\, ,
\end{equation} 
where $\Delta t=L_0/N$ is the time-domain sampling corresponding to the spatial discretization of the total travel length $L_0$. The number of modes is $N$ and, assuming linear dispersion, their frequencies are $\omega_k = 2\pi k/L_0$ (setting now $c=1$). Finally, the sign in the exponent is ``$+$'' for right-moving and ``$-$'' for left-moving modes.

Under this transformation the model becomes one where photons emitted in a time interval $\Delta t$ are then moved, conveyor-belt-like, through these discrete modes (``boxes'') until they hit the mirror at time $\tau/2 \equiv N \Delta t/2$, and then return in right-moving boxes until they again interact with the system at time $\tau$.  Photons emitted into the right side of the waveguide never return, and can be accommodated by projecting, at each time step, the system onto the appropriate state depending on whether a photon is observed in the right-most-box or not. As previously mentioned, this projection technique has formal similarity with the Monte-Carlo wavefunction method.

\begin{figure}[t]
\centering \includegraphics[width = 0.6\columnwidth]{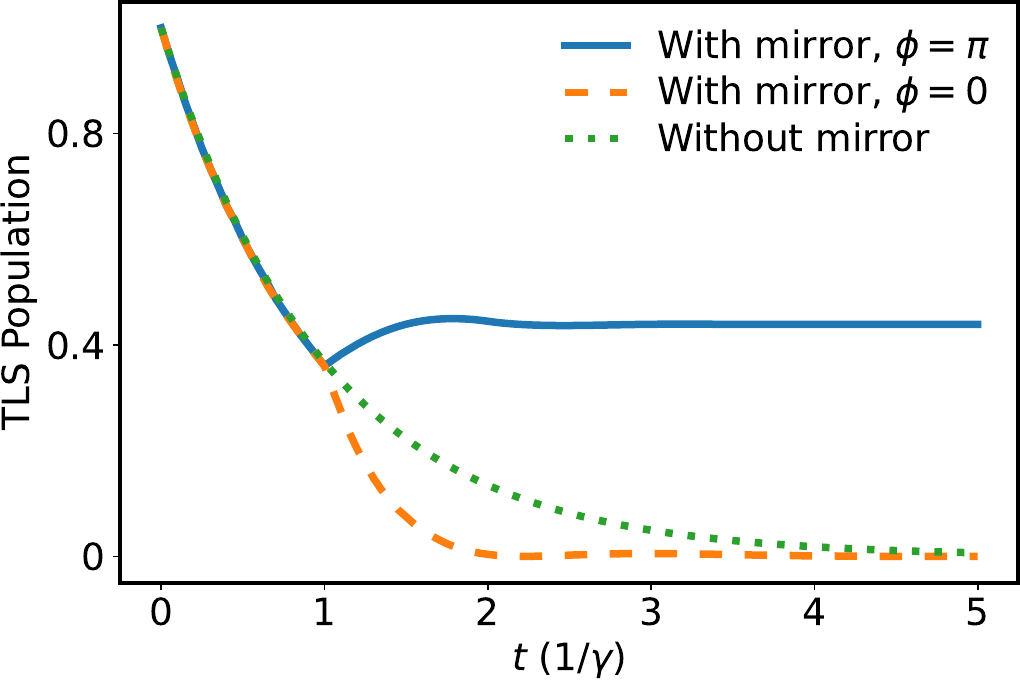} 
\caption{The dynamics of a two-level system, or qubit, interacting with a waveguide truncated at one end by a mirror. Here, we chose $\epsilon = 0$ and used $N=21$ discrete waveguide modes, $\Delta t = L_0/N$, with $L_0 = \gamma c = 1$, with $\gamma=\gamma_L+\gamma_R$ so that the roundtrip time $\tau = 1/\gamma$. The orange curve shows the case with a mirror reflection with phase $\phi=\pi$, and the green dashed curve shows mirror reflection with phase $\phi=0$, both averaged over $4000$ trajectories. The red dotted curve shows the exponential decay expected from an open waveguide without mirror, and the blue dashed line shows the expected population at the time the first emitted photon returns and interacts with the qubit again. }\label{waveguide-enr}
\end{figure}

\begin{figure}[t]
\centering \includegraphics[width =\columnwidth]{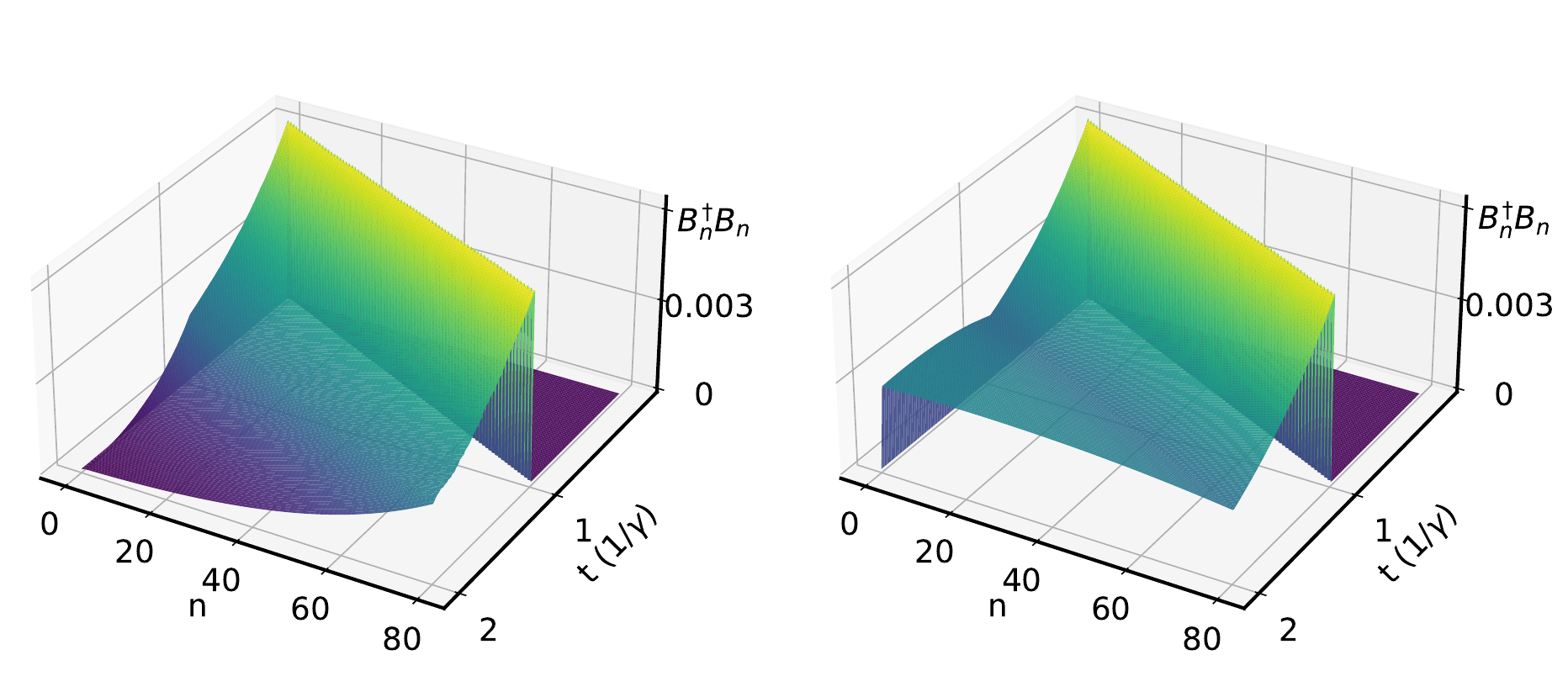} 
\caption{Occupation of the waveguide modes  $B_n^{\dagger}B_n$, where we we have increased the number of modes to $80$.  The time steps only extend to $\gamma t=2$, so that in the left figure, for $\phi=0$, we can see the overall loss of population in the waveguide modes after around $\gamma t =1$, the round-trip time, while in the right figure, for $\phi=\pi$, we see the saturation of the occupation.}\label{waveguide-enr2}
\end{figure}

Because the interaction between the qubit and the waveguide is assumed to be weak, and in a rotating wave approximation, we can model the whole setup including the qubit and $N$ modes efficiently using ENR states.  In \fig{waveguide-enr} we show, reproducing \cite{stephenhughes1}, the dynamics of the excited state population of the qubit with two different choices of phase, $\phi=0$ and $\phi=\pi$, demonstrating the effect of interference with the returning photons.  Here we used $N=21$, and only a single excitation, which with ENR states can be presented with a very small state space of just $23$ states (to be compared to $2^{22}$ states needed for a brute force calculation with a Fock space truncation of each mode of just $2$, without ENR states). Figure \ref{waveguide-enr2} shows the occupation of the waveguide modes as a function of time, illustrating the linear transportation of the excitation through the waveguide and the effect of the reflection phase $\phi$ on the waveguide populations.





\subsubsection{Automatic Differentiation: JAX}

In an earlier section, we showed how QuTiP's new JAX data layer can be used to run calculations on the GPU.
An additional feature of using the JAX data layer is access to automatic differentiation, or auto-differentiation.  In problems where derivatives are important, like optimization, we must often resort to numerical approximations, e.g., finite difference, to evaluate them. When higher order derivatives are required, such approximations can be numerically costly and inaccurate.  Auto-differentiation relies on the concept that any numerical function is, at a low-level, expressible in elementary analytical functions and operations. This can be exploited, via the chain rule, to give access to the derivative of almost any higher-level function.  

The JAX library makes it possible to conveniently and easily use auto-differentiation for a variety of applications. For example, in the \texttt{QuTiP-QOC} library we take advantage of this feature to find derivatives of a control objective with respect to the control parameters in order to find the optimal pulse shape implementing a complex operation.

A full explanation and set of examples of automatic differentiation is beyond the scope of this work, but we will showcase two basic examples here: one arising from the field of counting statistics, and the other relevant to Hamiltonian control.  In addition, the auto-differentiation capabilities of QuTiP-JAX will also be used in the section on the optimal control package QuTiP-QOC.

\paragraph{Counting statistics of an open quantum system.} -- In the first example, we have an open quantum system in contact with an environment, and there is a measurement device which keeps track of excitations flowing between system and environment.  

The measurement results of this process can be expressed in a variety of ways, but considerable insight can be gained by thinking about the statistics of such events; the mean, variance, skewness, and so on, of the probability distribution describing the number of excitations $n$ that have been exchanged by a certain time $t$. This distribution $P_n(t)$ is called the full counting statistics, and many common experimental observables such as current or shot noise can be extracted from its properties.

When the interaction between a system and its environment is described by a Lindblad master equation, we can obtain this distribution from a slightly modified definition of the density operator and said master equation.  Succinctly, one introduces the ``tilted'' density operator $G(z,t) = \sum_n e^{z n} \rho^{n}(t)$, where $ \rho^{n}(t)$ is the density operator of the system conditioned on $n$ jumps, or exchanges, occurring by time $t$ and $\Tr[ \rho^{n}(t)] = P_n(t)$. It obeys the tilted equation of motion, which reads (in the case where the contact with the environment is just through a single jump operator $C$)
\begin{align}\label{Glindblad}
\dot G(z,t) = -\frac{i}{\hbar}[H(t),G(z,t)] + \frac{1}{2} \left[2 e^{z} C \rho(t) C^\dagger - \rho(t) C^\dagger C - C^\dagger C \rho(t)\right]\, .
\end{align}
The dummy variable $z$ introduced here is conventionally called a counting field.
For $z=0$, we obtain $G(0,t)=\rho(t)$, and Eq.~\eqref{Glindblad} becomes the Lindblad equation \eqref{lindblad}. However, having access to $G(z,t)$ gives us access to the moments of $P_n$ through derivatives
\begin{equation}
\langle n^{m}\rangle(t) = \sum_n n^{m} \Tr[\rho^{n}(t)]=\frac{d^m}{dz^m}\Tr[G(z,t)]|_{z=0} \, .
\end{equation}

Normally, obtaining these derivatives would involve taking finite differences of Eq.~\eqref{Glindblad},  or writing explicit dynamic equations of motion for the moments themselves and solving them simultaneously.  With JAX and auto-differentiation we can obtain them explicitly, as shown in the following example.
We model a system with two levels, described by the annihilation operator $d$.  This system is coupled to two reservoirs via the rates $\Gamma_L$ and $\Gamma_R$. One of the couplings (governed by the rate $\Gamma_R$) is modified with a counting field as shown in Eq.~\eqref{Glindblad}; this is the channel whose counting statistics we monitor.  This simple model is often used to study the basic dynamics of charge being transported through a single quantum dot.
\newpage
\inputpython{JAX-countstats.tex}{15}{46}

The code above manually constructs the Lindbladian, with the counting field on the appropriate jump term, and then solves the master equation using \code{mesolve()}. The solution is a function of two parameters, time and the counting field.  We can then find the first and second counting-field derivatives of the state at any time.  In the long-time limit, analytical expressions for the current $I= \ex{n(t)}/t $ and the shot noise (variance) $S = \left(\ex{n^2(t)} - \ex{n(t)}^2\right)/t$ are well known; we compare them to the output of the our numerical functions below. Note that JAX's auto-differentiation capabilities function both on CPU and GPU; we used the command \code{jax.default\_device} to run our calculations on a GPU with no further modifications to the code required.

\inputpython{JAX-countstats.tex}{49}{61}

\paragraph{Derivatives of a driven Rabi model.} --- To further demonstrate the utility of automatic differentiation, we show in the next example how to take derivatives with respect to Hamiltonian parameters. We consider the driven Rabi model, which describes the interaction of a two-level quantum system (qubit) with an external driving field. The system is described by the time-dependent Hamiltonian

\begin{equation}
H(t) = \frac{\hbar \omega_0}{2} \sigma_z + \frac{\hbar \Omega}{2} \cos(\omega t)\sigma_x \, ,
\end{equation}
where $\omega_0$ is the qubit energy splitting, $\Omega$ is the Rabi frequency, $\omega$ is the driving frequency, and $\sigma_{x/z}$ are Pauli matrices. In the presence of dissipation, the system's evolution is governed by the Lindblad master equation, which includes energy relaxation via collapse operators. We include the dissipation rate $\gamma$ with the collapse operator $C = \sqrt{\gamma} \sigma_-$.

As we will discuss in the section on QuTiP-QOC, in quantum control and optimization tasks it is crucial to understand how physical observables depend on system parameters. Automatic differentiation provides an efficient and accurate way to compute gradients with respect to parameters such as the driving frequency $\omega$. This allows for gradient-based optimization techniques to be applied, bypassing the inefficiencies and inaccuracies of finite differences.

In this example, we compute the gradient of the population of the excited state $P_e(t) = \langle e | \rho(t) | e \rangle$ with respect to $\omega$. By using JAX's auto-differentiation tools in combination with QuTiP's Monte Carlo solver \code{mcsolve()}, we can compute the exact gradient of the final excited-state population with respect to the driving frequency. The example only showcases the ability to compute the derivative; this ability could then be employed to aid with optimization tasks, such as achieving maximum excitation of the qubit.

Below is the implementation in Python using QuTiP and JAX:
\newpage 
\inputpython{JAX-countstats.tex}{64}{99}

\subsubsection{MPI support for high performance computing}

It is in the nature of the trajectory solvers \code{mcsolve()}, \code{nm\_mcsolve()} and \code{smesolve()} that their simulations can be easily parallelized.
Previously, these functions accepted a \code{map\_func} argument which could be set to either \code{serial\_map} or \code{parallel\_map} to either simulate only one trajectory at a time, or multiple trajectories in parallel.
The function \code{parallel\_map} utilizes Python's multiprocessing module, which runs multiple processes on the same computer.

In QuTiP version 5, the \code{map\_func} argument has given way to the \code{"map"} option, which may be set either to \code{"serial"} or \code{"parallel"} in order to invoke \code{serial\_map()} or \code{parallel\_map()}, or to one of the new options \code{"loky"} or \code{"mpi"}.
The option \code{"loky"} is mostly equivalent to \code{"parallel"} (but may be more performant in some situations).
We will in the following discuss the remaining option, \code{"mpi"}.

The Message Passing Interface (MPI) is a standardized API facilitating parallel computations on multiple nodes of parallel computing architectures such as high performance computing clusters.
In Python, this API can be conveniently accessed through the \emph{MPI for Python} package \cite{DalcinJParallelDistribComput2005, DalcinJParallelDistribComput2008, DalcinAdvWaterResour2011, DalcinComputSciEng2021}.
When the option \code{"mpi"} is passed to one of QuTiP's trajectory solvers, an instance of this package's \code{MPIPoolExecutor} class is created that QuTiP will rely on for the parallelization.
It is strongly recommended to also pass the option \code{"num\_cpus"} which determines the number of worker processes to use.
The environment must therefore be configured to allow the application to use at least this number of processes, plus one (for the parent process).
The following code snippet demonstrates the use of this option with the Monte Carlo solver:
\newpage
\begin{python}
result = qt.mcsolve(
    H, initial_state, tlist, collapse, ntraj=NUM_TRAJECTORIES,
    options={"progress_bar": False, "map": "mpi", "num_cpus": NUM_WORKER_PROCESSES}
)
\end{python}

The exact procedure how to set up an environment in which \emph{MPI for Python} can successfully interact with an MPI implementation depends strongly on which versions of what MPI implementations and of what job schedulers are available on the cluster.
Providing a guide for this task goes beyond the scope of this text, and of the QuTiP project.
QuTiP simply assumes that the environment is set up correctly, and provides the additional option \code{"mpi\_options"} through which users may provide a dictionary with configuration options that will be passed directly to the constructor of the \code{MPIPoolExecutor}.

We note that the \code{MPIPoolExecutor} class is based on the \code{MPI\_Comm\_spawn} routine, which may not be available in all environments.
For environments where it is not, \emph{MPI for Python} provides a workaround where one replaces the normal script invocation ``\verb+python <FILE>+'' with
\begin{verbatim}
mpiexec -n <N> python -m mpi4py.futures <FILE>
\end{verbatim}
The \code{mpiexec} command will then immediately spawn the requested number $N$ of processes.
\emph{MPI for Python} handles the management of these processes and behaves as if the \code{MPIPoolExecutor} had spawned the worker processes.
The use of this workaround should not necessarily harm the performance, but it means that the full number of worker processes will live during the entire runtime of the application.


\section{QuTiP's other main packages}

In QuTiP v5, the structure of the QuTiP library has changed, as we have chosen to move large, independent, features into distinct sub-packages. Primarily, this choice was made to minimize the number of external dependencies of the core of QuTiP, increasing its maintainability, but it also enables the development of data layer ``plug-ins'' like QuTiP-JAX and other experimental ones.  In this section, we will review the main sub-packages, QuTiP-QOC and QuTiP-QIP.

\subsection{Optimal control: \textup{QuTiP-QOC}}
Quantum systems are generally sensitive to external perturbations. On the one hand, this sensitivity can be used to perform precise measurements or operations, but on the other hand, it makes the implementation of quantum devices difficult by introducing noise and errors.
Therefore, finding the optimal control fields that achieve a desired quantum operation under various objectives (e.g., minimum energy consumption or maximum robustness to noise) is a challenging and important problem. 
In practice, there are often constraints on the control fields, such as bounds on their bandwidth, amplitude or duration. 
These factors make quantum optimal control a complex and rich field of research, with diverse methods and applications.

Among the most frequently used techniques for finding optimal control functions are the GRadient Ascent Pulse Engineering (GRAPE) and Chopped RAndom Basis (CRAB) methods, which are both supported by the quantum optimal control package of QuTiP v4, QuTiP-QTRL \cite{crab, grape, qqtrl}. 
Along with the QuTiP v5 release comes the new family package QuTiP-QOC \cite{qqoc}, which includes both of these methods as well as two new ones.
Further, it introduces a general control framework to address pulse optimization in a customizable manner by providing keyword argument access to the underlying QuTiP and SciPy functions.
The new optimization techniques are the Gradient Optimization of Analytic conTrols (GOAT) algorithm \cite{MachnesGOAT} and the JAX OPTimization (JOPT), which is based on the auto-differentiation \cite{Goerz2022} capabilities of JAX and seamlessly integrates with QuTiP-JAX.

Both new techniques, GOAT and JOPT, work with analytic control functions and offer the possibility to treat the system evolution time as a variable optimization parameter.
Additionally, all pulse optimization routines now come with a global and local parameter search, making it easier to escape local minima. 
See Table~\ref{tab:qoc} for an overview of all supported features.

\begin{table}
    \centering
    \begin{tabular}{c|c|c|c|c|c}
        & analytic & local & global & time & multi-objective  \\ \hline
        GRAPE   & no & v4+ & v5 & no  & v5 \\
        CRAB    & v4+ & v4+ & v5 & no  & v5 \\
        GOAT    & v5 & v5 & v5 & v5 & v5 \\
        JOPT    & v5 & v5 & v5 & v5 & v5 \\
    \end{tabular}
    \caption{Support for parameterized analytic control functions, local and global parameter search, variable evolution time and multi-objective optimization in the QuTiP v4 (QuTiP-QTRL) and v5 (QuTiP-QOC) packages.}
    \label{tab:qoc}
\end{table}

\paragraph{Basic example of optimal control: optimizing the Hadamard gate} --- The following example shows how to use all of the above-mentioned methods to find optimal control parameters to implement a Hadamard gate on a single qubit. 
In general, the qubit might be subject to dissipation, captured in the Lindbladian formulation with the jump operator $\sigma_-$.
For simplicity we assume parameterized $\sigma_x, \sigma_y$ and $\sigma_z$ rotations for the control Hamiltonian
\begin{equation} \label{eq:qoc_hamil}
    H_\mathrm{c}(t) = c_x(t) \sigma_x + c_y(t) \sigma_y + c_z(t) \sigma_z \, ,
\end{equation}
where $c_x(t)$, $c_y(t)$ and $c_z(t)$ are independent control functions.
Furthermore, we model a constant drift Hamiltonian
\begin{equation}
    H_\mathrm{d} = \frac{1}{2} (\omega \sigma_z + \delta \sigma_x)
\end{equation}
with associated energy splitting $\omega$ and tunneling rate $\delta$. The amplitude damping rate for the collapse operator $C=\sqrt{\gamma}\sigma_-$ is denoted $\gamma$.
The total time evolution of the qubit is therefore assumed to have the form
\begin{equation}
    \dot \rho(t) = -\frac{i}{\hbar} [H_\mathrm{d} + H_\mathrm{c}(t), \rho(t)] + \frac{1}{2} \left[ 2 C \rho(t) C^\dagger - \rho(t) C^\dagger C - C^\dagger C \rho(t)\right] \, .
\end{equation}

By default, in the case of open system state transfer or map synthesis (as in this example), the optimization will minimize the trace distance to the target state or channel. In the case of closed system state transfer or gate synthesis objectives, it will minimize the overlap with the target vector or gate.
Even though not explicitly shown in the following example, all algorithms can be run with multiple objectives (list of \code{qoc.Objective} instances), where each objective can be supplied with an additional weight parameter.

\inputpython{qoc.tex}{4}{18}

\subsubsection{The GRAPE algorithm} 
The GRAPE algorithm, initially designed for nuclear magnetic resonance (NMR) pulse sequences, has applications in various physical systems, including superconducting qubits, and can be used to optimize noisy quantum devices in QuTiP \cite{grape, grape_sc, li2021pulselevel}. 
By minimizing an infidelity loss function that measures how close the final state or unitary transformation is to the desired target, the algorithm optimizes evenly spaced piecewise constant pulse amplitudes. 
The random or educated initial \code{guess} control pulse is updated iteratively according to the derivative of the loss function.

\inputpython{qoc.tex}{20}{48}

GRAPE concurrently updates the pulse amplitudes by calculating the derivatives using the discretized unitary forward and backward evolution operators. 
This approach bears similarities with the Krotov method that is also available as a QuTiP affiliated package, but instead updates the control intervals sequentially \cite{Goerz_2019_SciPost}. 
The QuTiP native Hamiltonian formulation makes it easy to define even multiple objectives in a common fashion for low effort comparisons with the Krotov library \cite{Goerz_2019_SciPost,PhysRevA.107.032609}.
The GRAPE algorithm is capable of accommodating various constraints on control fields, such as amplitude limits and offsets, for a range of default initial pulses (e.g., Gaussian, square, etc.), while striving to achieve the desired target infidelity, specified by the \code{fid\_err\_targ} keyword.

\subsubsection{The CRAB algorithm}
The CRAB algorithm has been applied to a range of challenging problems, like phase transitions in many-body systems, implementations of gates on transmon qubits, and entanglement generation for communication \cite{M_ller_2022, corti_robustness_2022}.
It is based on the idea of expanding the control fields in a random basis and optimizing the expansion coefficients $\vec{\alpha}$.
This has the advantage of using analytical control functions $c(\vec{\alpha}, t)$ on a continuous time interval, and is by default a Fourier expansion.
Instead of calculating the gradient with respect to individual time slots, the search space is reduced to the function parameters.
Typically, these parameters have one order of magnitude fewer dimensions and can efficiently be calculated through direct search algorithms (like Nelder-Mead).
The basis function is only expanded for some finite number of summands and the initial basis coefficients are usually picked at random.

The implementation in QuTiP-QOC also provides the possibility to balance multiple objectives in order to account for, e.g., variations in the control fields. Even though the optimization is performed over the function basis parameters, it is possible to initialize the overall pulse shape in the same way as with the GRAPE algorithm. 
On top of the original QuTiP-QTRL implementation, the method can now also exploit domain knowledge by providing an initial frequency (enabled through the \code{fix\_frequency} keyword) and amplitudes for the Fourier expansion.
\newpage
\inputpython{qoc.tex}{52}{78}

\subsubsection{The GOAT algorithm}
Similar to CRAB, this method also works with analytical control functions. By constructing a coupled system of equations of motion, the derivative of the (time ordered) evolution operator with respect to the control parameters can be calculated after numerical forward integration. In unconstrained settings, GOAT was found to outperform the previously described methods in terms of convergence and fidelity achievement \cite{riaz_optimal_2019}.
Our QuTiP implementation allows for arbitrary control functions provided together with their respective derivatives in a common Python manner.

\inputpython{qoc.tex}{82}{91}

For even faster convergence, it extends the original algorithm with the option to optimize controls with respect to the overall time evolution, which can be enabled by specifying the additional time keyword argument.
\newpage
\inputpython{qoc.tex}{93}{118}

\subsubsection{Integration with QuTiP-JAX}
As discussed earlier, QuTiP's new JAX backend provides state-of-the-art automatic differentiation capabilities.  Using the chain rule of differentiation for elementary computer operations, automatic differentiation is almost as exact as symbolic differentiation and is most commonly used in machine learning \cite{BaydinPR15}.
Recently it has also found its way into quantum optimal control \cite{Song22}.
Through the JAX backend, these capabilities can be used with the new control framework. 
As with QuTiP's GOAT implementation, any analytically defined control function can be handed to the algorithm. 
However, in this method, JAX automatic differentiation abilities take care of calculating the derivative throughout the whole system evolution. 
Therefore the user does not have to provide any derivatives manually. 
In the previous example, this simply means to swap the control functions with their just-in-time compiled version:

\inputpython{qoc.tex}{132}{150}

\begin{figure}[t]
\centering \includegraphics[width = \textwidth]{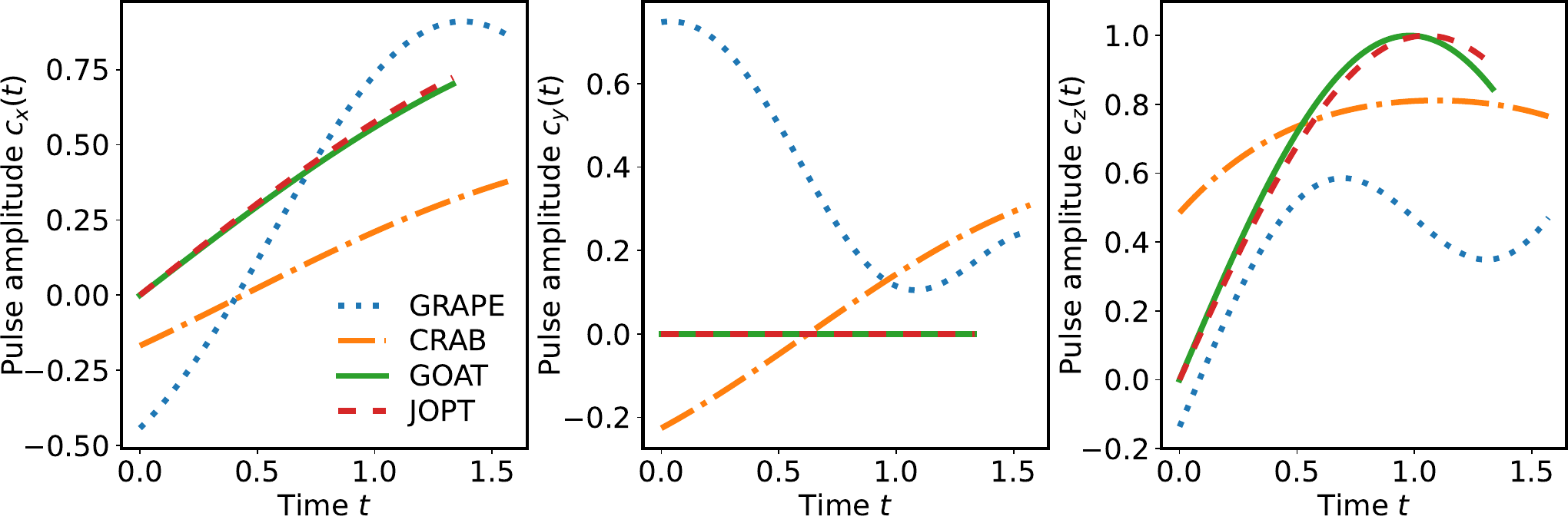}
\caption{Optimized pulse amplitudes implementing the Hadamard operator for a single qubit system with the control Hamiltonian \eqref{eq:qoc_hamil}.
The amplitudes were obtained through the various algorithms provided by the \texttt{QuTiP-QOC} package.
For this simple example, all algorithms quickly find pulse shapes with the requested fidelity (greater than $0.99$).
Note that the GOAT and JOPT curves end early because optimization was also performed over the final time.
}\label{qoc_pulse_plot}
\textbf{}\end{figure}

After running the global and local optimization, one can compare the results obtained by the various algorithms through a \code{qoc.Result} object, which provides common optimization metrics along with the \code{optimized\_controls} (see Fig.~\ref{qoc_pulse_plot}).

\subsection{Quantum circuits: \textup{QuTiP-QIP}}

Quantum circuits are a common conceptual and visual tool to represent and manipulate algorithms running on quantum computers. QuTiP supports this kind of language with the QuTiP-QIP \cite{qqip} family package, which offers several unique features. First, like the rest of QuTiP, it is one of the most popular fully independent and academically supported packages for its task. Second, it allows seamless integration of the unitaries representing a given circuit with the rest of QuTiP via the \code{Qobj} class. Finally, it integrates with both QuTiP-QTRL (soon QuTiP-QOC) and the various QuTiP open-system solvers to allow for pulse-level simulation of quantum circuits, including realistic noise.


The full scope of this package was recently demonstrated in \cite{li2021pulselevel}.  In this section, we first show the newly added circuit visualization feature, then walk through a new example to demonstrate some of the unique features of QuTiP-QIP.  To tie into the earlier examples in this paper, and the general theme of QuTiP of modeling open quantum system dynamics, our aim is to show: (i) How to construct a simple quantum circuit which simulates the dynamics of a given Hamiltonian (in this case, Eq.~\eqref{H0}). (ii) How to simulate the dynamics of an open system, i.e., a Lindblad equation, using ancillas to induce the correct noisy dynamics, Eq.~\eqref{lindblad}. (iii) How to then run both simulations on a hardware backend, termed a processor, that simulates itself the intrinsic noisy dynamics of a given quantum hardware implementation.
 
\subsubsection{Circuit Visualization}

In parallel to the QuTiP v5 release, a recent release of QuTiP-QIP introduced enhanced circuit visualization capabilities by extending the list of available renderers to include also Matplotlib-based and text-based renderers in addition to the previously available LaTeX renderer. This update makes circuit visualization more accessible and reduces (heavy and complex) reliance on external LaTeX dependencies. The Matplotlib renderer is particularly useful for generating visually appealing and highly customizable quantum circuits, while the text-based renderer is designed for quick, lightweight checks, ideal for development in command line interfaces or in environments with limited dependencies. Support for the LaTeX renderer remains available.

Additionally, the visualization of quantum circuits has been significantly streamlined through the use of the \texttt{draw} API. For instance, a quantum circuit can be defined as demonstrated below.
\newpage 
\inputpython{qip_renderer.tex}{3}{7}

\begin{figure}[ht]
\includegraphics[width = \columnwidth]{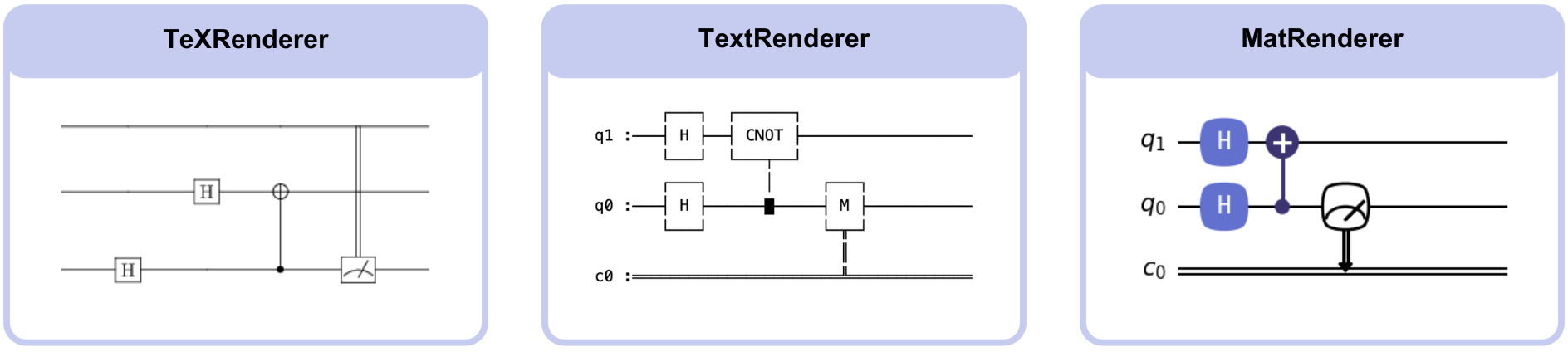}
\caption{The figure illustrates the visual output of three different quantum circuit renderers available in QuTiP-QIP: TeXRenderer (utilizes LaTeX for rendering), TextRenderer (utilizes ASCII characters for lightweight visualization), and MatRenderer (leverages Matplotlib for customizable graphics).}
\label{qip_renderer}
\end{figure}

Once defined, the circuit can be rendered using one of the following commands, depending on the desired output format as illustrated in \fig{qip_renderer}.
\inputpython{qip_renderer.tex}{10}{12}

Both new renderers utilize a layer-based system for gate placement, optimizing circuit compactness and enhancing the clarity of parallel computation steps during simulation. Customization options range from basic changes such as renaming gates or adjusting colors and titles to advanced features like modifying gate shapes, alignment, and spacing. The MatRenderer provides customization flexibility at both the gate and circuit levels, enabling users to tailor circuit representations to meet specific requirements.

\subsubsection{Simulating Hamiltonian dynamics}

In quantum simulation, one standard approach to simulating the dynamics of a quantum system is reducing the propagation of the Schrödinger equation into a discrete set of short time steps \cite{Buluta2009, Georgescu2014} (though analog simulations are also important in certain situations \cite{RevModPhys.84.1}). The propagator in one time step is approximated by ``Trotterization'',
\begin{equation} \label{trotter}
\psi(t_f) = e^{-i(H_A+H_B)t_f}\psi(0) \approx \left[e^{-iH_A dt }e^{-iH_B dt }\right]^d \psi(0) \, ,
\end{equation}
for sufficiently small time steps $dt = t_f/d$, which are repeated $d$ times (higher-order approximations can also be performed, which allows for larger time steps).

The Hamiltonians $H_A$ and $H_B$ here should be chosen such that the individual unitaries $e^{-iH_{A,B} dt}$ can be easily mapped to basic quantum gates in the circuit model.  When the simulated system can be represented as spin operators with some limited range of interactions, this mapping to qubits is straightforward \cite{nielsen2002quantum, lloyd96}, as with our simple example below. Alternatively, in a more general context, if the problem is represented as an abstract set of $N$ states, and $N\times N$ operators, various mappings are possible \cite{10.1145/780542.780546,10.5555/2481569.2481570,7354428, Low_2019}. A common approach is to map the $N$ states to the states described by  $n= \mathrm{log}_2(N)$ qubits. The $N\times N$ operators can then be mapped to strings of Pauli operators, for which efficient gate-mappings are known \cite{nielsen2002quantum}.  However, the number of strings required will be exponential in $n$ for dense problems, so some degree of sparsity or structure, and efficient access to matrix elements, is needed.

The example of two interacting qubits defined in Eq.~\eqref{H0} can be expressed in the above formula, Eq.~\eqref{trotter}, with
\begin{align}
 H_A &= \frac{\epsilon_1}{2} \sigma_z^{(1)}+ \frac{\epsilon_2}{2} \sigma_z^{(2)} \quad\text{and} \label{HA} \\
 H_B &= g \sigma_x^{(1)} \sigma_x^{(2)} \, . \label{HB}
\end{align}
In this case there is no difficulty with mapping the Hilbert space of the problem (two-level systems) being simulated to the circuit model (qubits). In more complex cases, various mappings exist for e.g.\ fermions, bosons or large spins, but they are not yet natively supported in QuTiP-QIP.

To implement \eqref{trotter}, we simply need to construct a circuit consisting of two qubits and a set of gates $A_1 = \operatorname{exp}[-i \epsilon_1 \sigma_z^{(1)} dt / 2]$, $A_2 = \operatorname{exp}[-i \epsilon_2\sigma_z^{(2)} dt / 2]$ and $B = \operatorname{exp}[-i g \sigma_x^{(1)} \sigma_x^{(2)} dt ]$ acting repeatedly ($d$ times) on a given initial condition. The gates one can use to represent these operations depend on the operations available for the underlying hardware. We use the predefined gates \code{RZ} defining a rotation around the $Z$ axis, and a combination of Hadamard gate and ZX rotation \code{RZX} gates to implement the \code{XX} interaction unitary, as these gates are native operations on the superconducting qubits processor \code{SCQubits}.

To define this circuit and its constituent gates we use the \code{QubitCircuit} class:

\inputpython{QuTiPqip2.tex}{43}{43}
Here we initialize a circuit with just two qubits. We can also choose to have classical bits as well, which can be used to store measurements and implement feedback. These are not needed here, however. We now add the gates which implement the Trotterization of \eqref{HA} for a small time step $dt = t_f/d$

\inputpython{QuTiPqip2.tex}{45}{46}
These \code{RZ} gates rotate both qubits around the $Z$ axis by an angle given by \code{arg\_value}.  The interaction between the two qubits, \eqref{HB}, can then be implemented with the gates:

\inputpython{QuTiPqip2.tex}{48}{50}

A full list of available gates can be found in the official documentation for QuTiP-QIP. In addition, custom gates can be easily implemented. One important point however is that, currently, not all gates can be decomposed into a set of native gates for a given hardware processor, as we will show below. Work is ongoing to have a gate decomposition graph for all predefined gates. In addition, a feature we will not demonstrate here is the simulation of measurements. Instead, we will just take the final output of the circuit and manipulate it as a standard quantum object.  
 
\begin{figure}[t]
{\centering \includegraphics[width = 0.6\columnwidth]{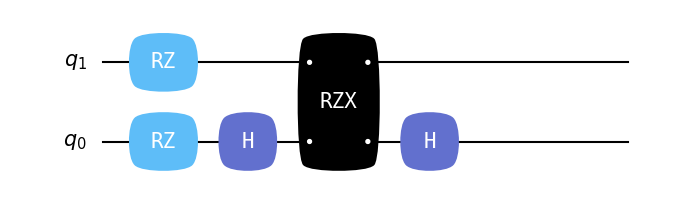} \par}
\caption{Circuit model of a single Trotterization step of \eqref{HA} and \eqref{HB}.}\label{qipcircuit}
\end{figure} 

We can directly visualize the circuit we just constructed, see \fig{qipcircuit}.  To construct a larger circuit to simulate more time steps we can keep appending the same circuit we defined above, $d$ times, until we simulate the full integration time $t_f$.  Alternatively, we can repeatedly run the above circuit, saving the output and using it as the input for the same circuit in the next iteration. For our purposes, the latter option is easier, as it allows us to save the state at all time steps:

\inputpython{qutipqip2.tex}{54}{60}

\begin{figure}[t]
{\centering \includegraphics[width = 0.6\columnwidth]{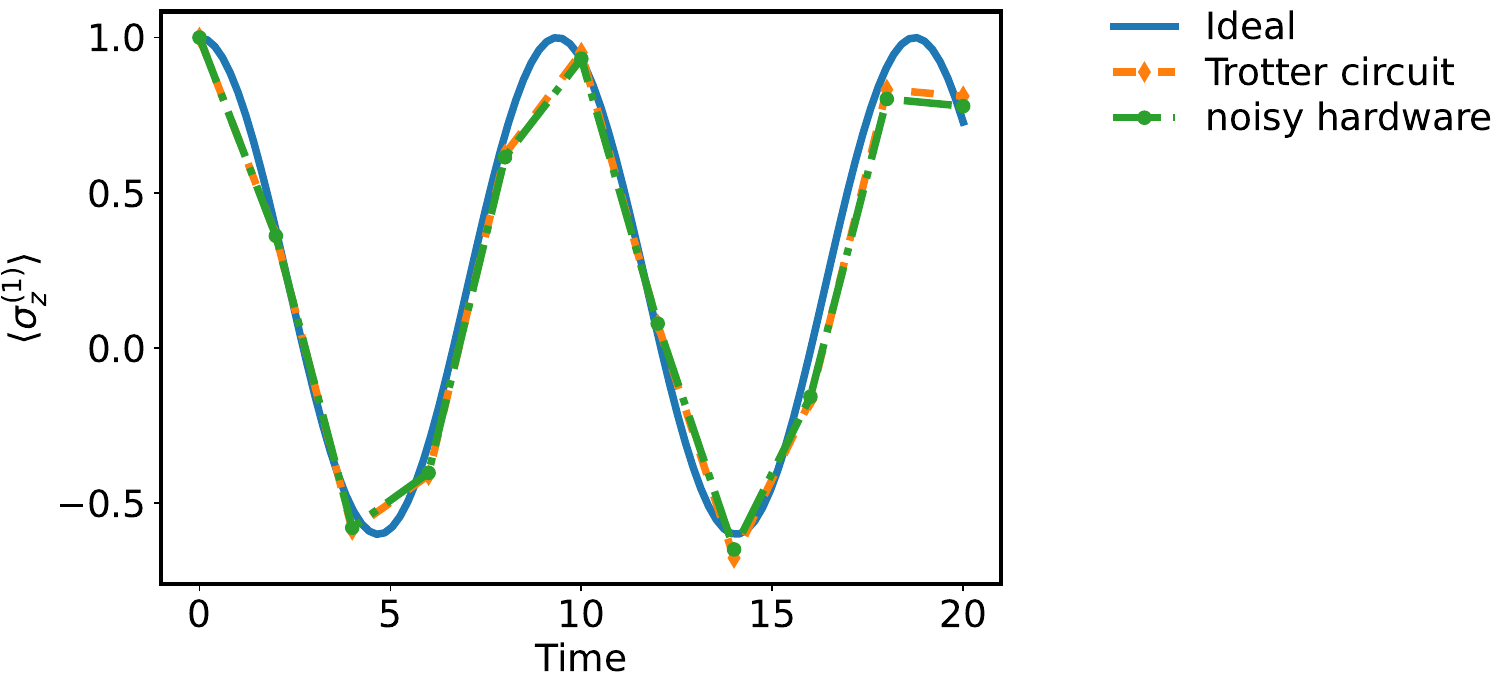} \par}
\caption{Results for the simulation of the closed system dynamics of the system described with the Hamiltonian \eqref{H0} for $g=0.3\, \epsilon_2$, $\epsilon_1=0.7\, \epsilon_2$. The circuit result, in blue, is the result of repeating the circuit described in \fig{qipcircuit} $d=t_f/dt$ times, with $t_f=20\, \epsilon_2^{-1}$, $dt=2\, \epsilon_2^{-1}$ and $d=10$.  In the processor simulation using a superconducting qubit backend, we set $T_1=T_2=2\times 10^5\, \epsilon^{-1}_2\, $ to see the influence of finite hardware noise on the simulation while still obtaining a reasonable approximation to the full simulation.}\label{qip1}
\end{figure}

The result of repeating this procedure $d$ times (equivalent to the \code{numsteps} parameter in the code example) is shown in \fig{qip1}. We compare the result to the exact integration solution and to the result from using a processor backend, which includes noise and pulse-level simulation.  Note that we chose the two qubits to be off-resonant to demonstrate the effect of finite Trotterization error (which would be negligible on resonance). We can thus see the trade-off between the Trotterization error and hardware errors: longer time steps in the simulation increase the Trotterization error but reduce the number of gates and, hence, the influence of hardware errors.

\begin{figure}[t]
{\centering \includegraphics[width = 0.7\columnwidth]{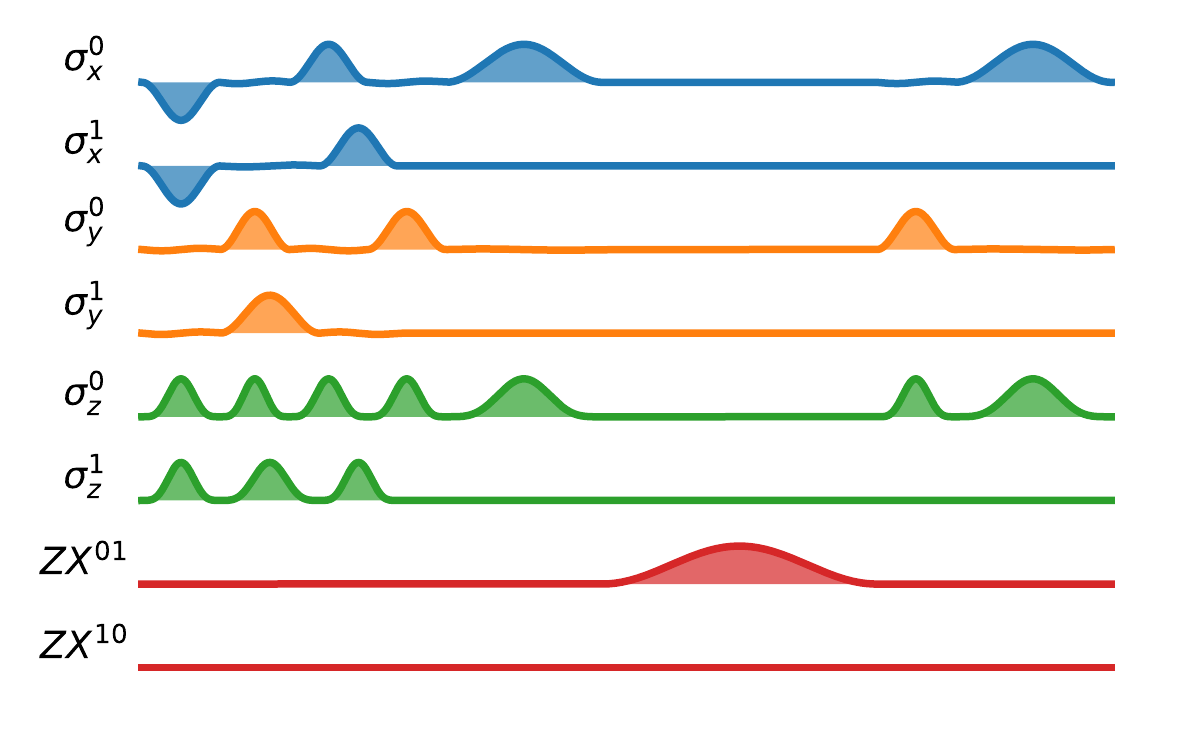} \par }
\caption{Pulse-level decomposition of the circuit in \fig{qipcircuit} as used by the superconducting qubits processor.}\label{qipgates}
\end{figure} 
To run a given quantum circuit on a processor, we simply initialize the desired processor and then load the circuit into it. For example, here we use the superconducting circuit processor:

\inputpython{QuTiPqip2.tex}{64}{67}

Similar to the direct simulation of the circuit, we define an initial condition and then run the processor with \code{processor.run\_state(init\_state)}. The output is now a \code{Result} object from the QuTiP solver being used. In this case, the solver is \code{mesolve()} as we specified finite $T_1$ and $T_2$ times when initiating the processor. The processor itself is defined internally by a Hamiltonian, available control operations and pulse shapes, $T_1$ and $T_2$ times and so on. It assumes a set of default parameters, but these can be overwritten upon initialization (as with $T_1$ and $T_2$ above).  We can see the pulse shapes used in the solver by calling \code{processor.plot\_pulses()}. An example of the compiled control pulse of the circuit in \fig{qipcircuit} is given in \fig{qipgates}.  A discussion of the full potential of processors, and how to implement custom ones, can be found in \cite{li2021pulselevel} and the package documentation.

\subsubsection{Simulating master equation dynamics}

We now demonstrate a more complex example where we can take advantage of QuTiP-QIP's easy integration with standard \code{Qobj} objects.  This example shows that, interestingly,  noise-free unitary quantum circuits can be used to simulate open noisy quantum systems. This type of simulation has a range of interesting applications, like dissipative state engineering \cite{dse1,dse2,dse3,dse4,dse5}, dissipative error correction \cite{PhysRevLett.131.050601}, and modelling many-body dissipative systems.

To realize the Lindblad master equation described in \eqref{lindblad}, we implement a recent proposal \cite{lindbladsim1, lindbladsim2}, which employs a single ancilla, and measurements/resets, for each Lindblad collapse operator. At $t=0$, we prepare the dilated state 
\begin{equation}
\ket{\psi_D(t=0)}=\ket{\psi(t=0)} \otimes \ket{0}^{\otimes K}    
\end{equation}
 for the $K$ total ancillas. At every time step $dt$, the system interacts with the ancillas for a time $\sqrt{dt}$, after which the ancillas are reset again to their ground-state.  The unitary operation part of the interaction between system $(i)$ and its associated ancilla for collapse operator $(k)$ is given by
\begin{equation}
U(\sqrt{dt}) = \exp[ -i\sqrt{\gamma_k}(\sigma_-^{(i)} \sigma_+^{(k)}+\sigma_+^{(i)} \sigma_-^{(k)})\sqrt{dt} ] \, .\label{ancilla}
\end{equation}

The unitary $U(\sqrt{dt})$ has a rather complex decomposition in terms of native gates for the backends used here, so instead we impose another Trotter approximation and write
\begin{align}
U(\sqrt{dt}) &= \exp[ -\frac{i}{2} (\sigma_x^{(i)} \sigma_x^{(k)}+\sigma_y^{(i)} \sigma_y^{(k)})\sqrt{\gamma_k dt} ] \nonumber\\
&\approx \exp[ -\frac{i}{2} \sigma_x^{(i)} \sigma_x^{(k)} \sqrt{\gamma_k dt}] \, \exp[ -\frac{i}{2} \sigma_y^{(i)} \sigma_y^{(k)}\sqrt{\gamma_k dt} ] ,
\end{align}
and then we decompose
\begin{equation}
\exp[ -\frac{i}{2} \sigma_x^{(i)} \sigma_x^{(k)} \sqrt{\gamma_k  dt} ] = H^{(i)}\, U_{\mathrm{RZX}}(\sqrt{\gamma_k dt})\, H^{(i)}
\end{equation}
as before, and 
\begin{align}
\exp[ -\frac{i}{2} \sigma_y^{(i)} \sigma_y^{(k)}\sqrt{\gamma_k dt} ] = U_{\mathrm{RX}}^{(i)}\left(\frac{\pi}{2}\right)^{(k)}\, U_{\mathrm{RZ}}^{(k)}\left(-\frac{\pi}{2}\right)\, 
U_{\mathrm{RZX}}(\sqrt{\gamma_k dt})\, U_{\mathrm{RX}}\left(-\frac{\pi}{2}\right)^{(i)}\, U_{\mathrm{RZ}}^{(k)}\left(\frac{\pi}{2}\right) \, .
\end{align}

\begin{figure}[t]
\includegraphics[width = \columnwidth]{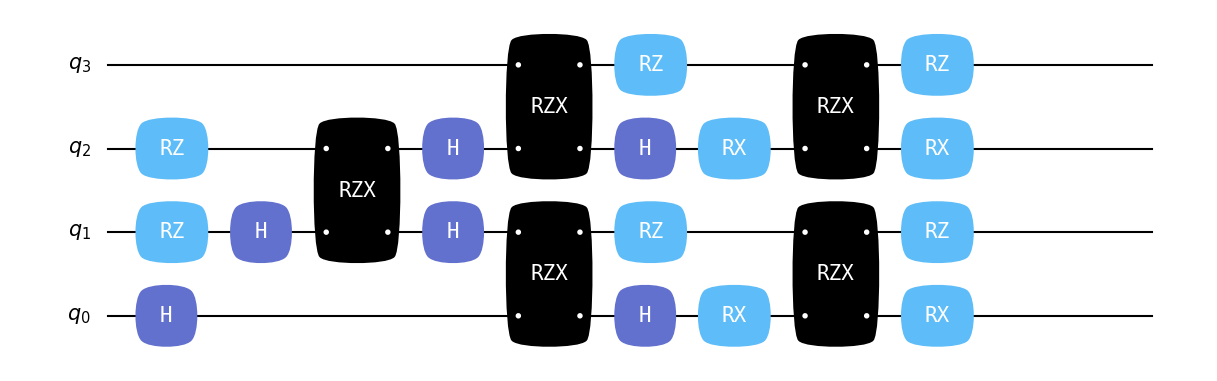}
\caption{Circuit model of a single Trotterization step of \eqref{HA} and \eqref{HB}, followed by the interaction with two ancillas implementing \eqref{ancilla}.}\label{qipcircuit2}
\end{figure}

\begin{figure}[t]
{\centering \includegraphics[width = 0.7\columnwidth]{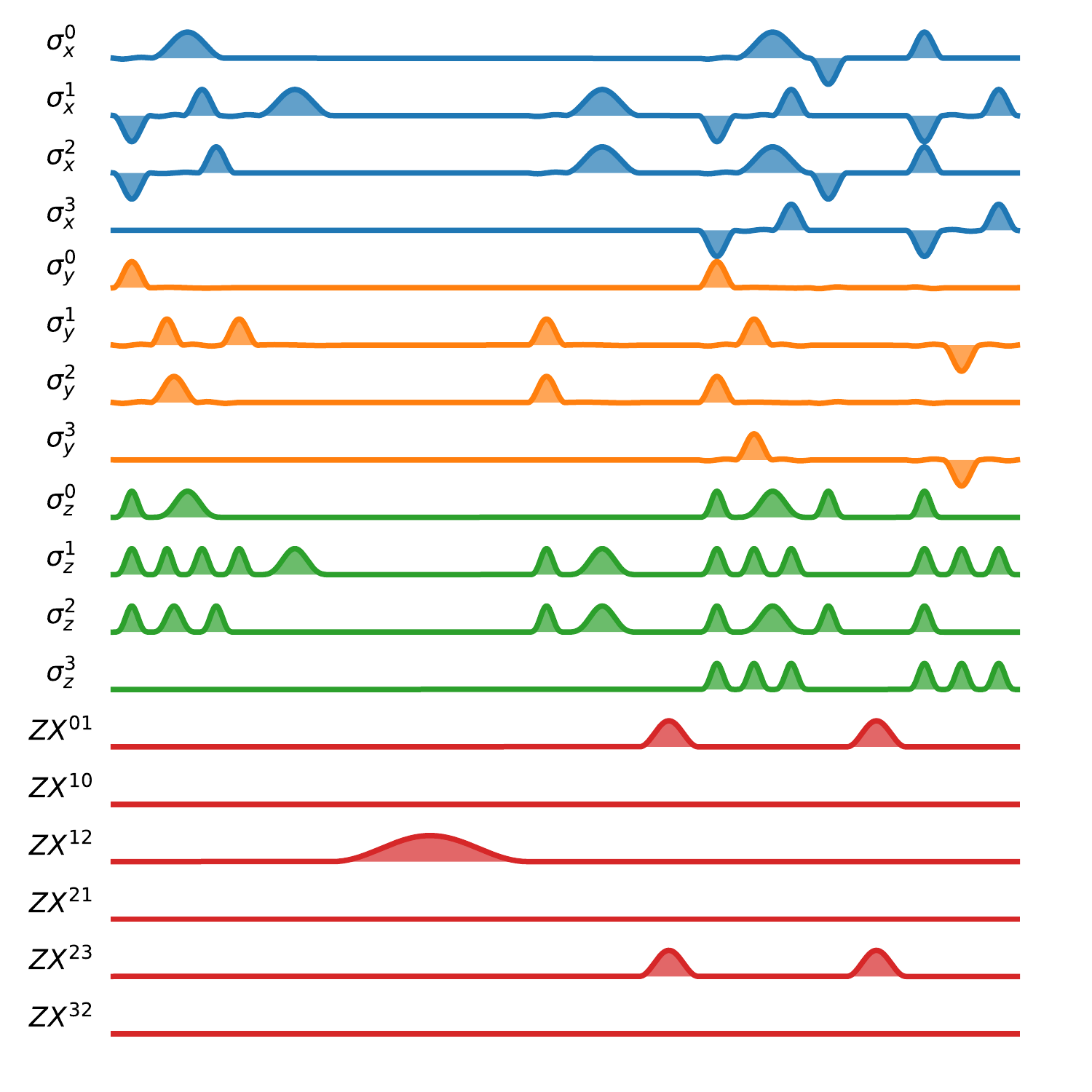} \par}
\caption{Pulse-level decomposition of the circuit in \fig{qipcircuit2} as used by the spin-chain processor.}\label{qipgates2}
\end{figure}

\begin{figure}[t]
{\centering \includegraphics[width = 0.6 \columnwidth]{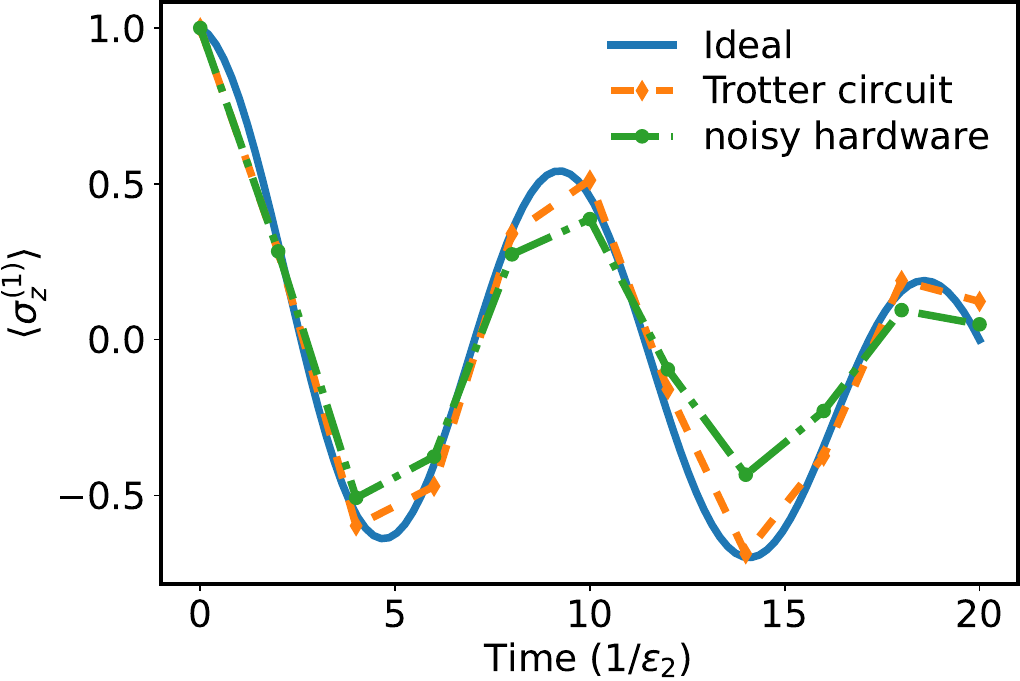} \par}
\caption{Results for the quantum simulation of open-system dynamics defined by the Hamiltonian \eqref{H0}.
The Hamiltonian is the same as used in \fig{qip1}. The Lindblad noise, acting on both qubits with rate $\gamma=0.03\, \epsilon_2$, can be simulated by an additional interaction with two ancillas. The magnitude of this interaction corresponds to simulating a time step $\sqrt{dt}$ for every full ``coherent'' time step $dt$ (see \cite{lindbladsim1, lindbladsim2}), then tracing out the ancillas and resetting them to their ground states. The Trotter circuit (orange) reproduces quite well the results of \code{mesolve} (blue). The green curve, showing the circuit run on a superconducting qubits processor, exhibits significantly more error than in \fig{qip1} because of the much longer circuit depth.}\label{qip2}
\end{figure}

After these operations have been applied, we trace out the ancillas, project them to their ground states, and use the new state
\begin{equation}
\ket{\psi(dt)} = \Tr_{\otimes K}\bigl[ \ket{\psi_D(dt)}\bra{\psi_D(dt)} \bigr] \otimes \ket{0}\bra{0}^{\otimes K}
\end{equation}
for the next time step. We show the circuit which implements a single time step in \fig{qipcircuit2}.
Note that to avoid unnecessary two-qubit operations between non-neighbouring states, we use the middle two qubits as the system qubits and the first and last one as ancillas.
The compiled pulse is shown in \fig{qipgates2}, which is much longer than the first one.
Therefore, the evolution will suffer more from the decoherence of the physical qubits.
In \fig{qip2}, we show the results of a standard \code{mesolve()} simulation, the Trotterized circuit simulation with ancillas, and the circuit simulation running on the same spin-chain backend as before.

\section{QuTiP's community}

QuTiP is an open source package that operates like an open-door community, with an active mailing list giving feedback on common problems and applications.  Community contributions to the QuTiP package are welcome, and only limited by the free time of the largely voluntary development and administration team.

\subsection{The admin team and governance}

The administration team, which reviews code, makes contributions, and decides on current and future development plans, is made up of members from across the globe. Regular volunteers or students who show a perspicacity are nominated by current administrators to join the team, which gives them authority to join monthly administration meetings and approve pull requests.  Other administration members are sometimes technical staff, directly hired by one of the academic institutions supporting QuTiP. Particular note should be given to support from JST Moonshot program, which supported several staff members in RIKEN, and Institute Quantique, which supported Eric Giguère in Sherbrooke.

The admin team relies on a governance body with five board members whose role is to resolve any serious conflicts and to give objective feedback on any large-scale decisions about QuTiP's future. Finally, we note that two members of the current administration team are students who graduated from Google Summer of Code's projects and decided to continue contributing to QuTiP voluntarily.  Such students have been a vital part of the development of QuTiP v5, keeping QuTiP alive and flourishing.

\subsection{RIKEN}

As described in the introduction, QuTiP was originally developed in the group of Franco Nori in RIKEN.  While the development team has since become global, RIKEN still serves as a focus for continued development by research scientists, technical staff, and intern students. Several recent important contributions of intern students include expansion of the HEOM method, the new QuTiP-QOC library, a general environment class, and an experimental tensor network data layer.  In addition, members of the RIKEN team serve as mentors on GSOC projects and assist in the development and management of new features across QuTiP.  

\subsection{NumFOCUS and Google Summer of Code} \label{subsec:gsoc}

QuTiP is affiliated to two non-profit organizations that provide support for development in various ways: NumFOCUS and the Unitary Foundation (the latter will be discussed in a later section).  NumFOCUS has a long history of supporting open source projects, and primarily acts as an umbrella organization which enables QuTiP to apply to Google Summer of Code as a mentoring organization.  It also supports development with periodic mini-grants. Their support for Google Summer of Code has been very useful, and enabled the QuTiP developers to mentor 2-3 students every year for the last few years. 

\subsection{The Unitary Foundation}

QuTiP is also affiliated with the Unitary Foundation, which focuses on the support and development of open source software for quantum computing applications. The Unitary Foundation has supported QuTiP development through micro-grants and Hackathon bounties, both of which have helped nascent volunteers to contribute to QuTiP and substantially raised the visibility of QuTiP in this community.  In addition, they operate a large Discord server through which people can interface with the QuTiP admin team.


\subsection{Packages that use QuTiP}
One of the most successful ingredients of QuTiP has been the flexibility and intuitive features of the \code{qutip.Qobj} class. Based on its use for the representation of quantum operators and quantum states, several other packages have formed a constellation of libraries.  To date, 733 GitHub repositories and 83 released packages depend on QuTiP.  A small selection of popular examples are:
\begin{itemize}
\item Krotov~\cite{Goerz_2019_SciPost}: As mentioned in the optimal control section, this QuTiP-affiliated library provides access to the Krotov method for optimizing control of quantum systems. 
\item SCqubits~\cite{Groszkowski2021}: A powerful and popular library for computing the energy spectra of common superconducting qubit designs, that interfaces with QuTiP to perform time-evolution simulations.
\item Sqcircuit~\cite{Rajabzadeh2023}: An alternative to SCqubits that also models superconducting circuits, but has the added feature of using an efficient choice of basis to perform diagonalization of complex circuits.
\item Dynamiqs~\cite{guilmin2024dynamiqs}: An open systems solver deeply integrated with JAX.  While lacking some of QuTiP's features, it has a robust and focused integration with the JAX library, and is compatible with QuTiP \code{Qobj}.
\item Qibo~\cite{qibo}, a hardware focused middleware API that supports a QuTiP backend.
\end{itemize}

Several commercially developed packages use QuTiP as well, including:

\begin{itemize}
    \item  NVIDIA CUDA-Q~\cite{cudaq}, which, for example, uses it for Bloch sphere visualization.
\item Qiskit~\cite{2024qiskit}, which also uses QuTiP's Bloch-sphere visualization and its parallel map function.
\item Pasqal's Pulser package~\cite{pulser} which supports a QuTiP backend.
\end{itemize}



\section{Future development}
Historically, QuTiP has been mostly used as an academic tool for research and education. It allows quick implementation and experimentation with theoretical models of closed and open quantum systems, and has made research both faster and more reproducible for thousands of scientists around the world. With QuTiP v5, the focus has been on maintaining that utility while also allowing for future expansion in terms of state-of-the-art performance and use of high-performance distributed computing.  Future development plans will continue in this regard.

In addition, recent developments in the field of quantum simulation software suggest another goal for QuTiP's future direction.
QuTiP, and in particular QuTiP-QIP, remain one of the few academically independent platforms for simulating quantum circuits and quantum computers in general.  Other powerful platforms with comparable feature sets have arisen from industrial efforts. Nevertheless, there is still an important role for QuTiP with its promise to be continually maintained and developed, and fully open-source.

In this regard, there is a strong incentive to improve QuTiP's support both for simulating quantum computing systems and for supporting cloud computing services as backends. In the future, QuTiP could serve as a system-agnostic and independent platform for development, with which one could access and use many hardware platforms (both industrial and academic). Initial efforts to this end came from backend support added by IonQ developers for their hardware, and we wish to continue in this direction by supporting as many hardware platforms as possible. On the question of academic independence, we note that some members of the administration team, present and alumni, have moved into industrial positions with companies like IBM, Zurich Instruments, and others. However, a robust board of mentors is in place to prevent bias.

In addition to these two goals, state-of-the-art performance and cloud-hardware support, we must also consider issues of maintainability. In that aspect, our strategy of QuTiP-family subpackages is working well, and we will continue with it in the future. However, there is an inevitability that some features that are not part of QuTiP-core may become abandoned or not well maintained, simply due to administrators or developers moving onto new tasks and fluctuations in academic funding support.  Finding solutions to this ongoing problem, common to many open-source software packages, is one of the important future goals of the QuTiP administration team.

In addition to these broad strategic goals, in our recent workshop the following features were discussed for future releases:

\subsection{QuTiP's role in the quantum computing revolution}
\paragraph{Hamiltonian library} --- A library of common models (Hamiltonians, master equations, and so on) for common physical systems people encounter would be useful both for academics and educators. At first, we plan to include them within the QuTiP core as Python functions and then consider later how they might be exported in a common format for use in other libraries.

\paragraph{Simulating QPUs} --- There are broad plans to extend the list of ``processors'', or quantum processing units, available to QuTiP-QIP to include a larger range of physical systems \cite{buluta2011natural,Gu2017,Kockum2019,Cheng2023}.  This extension would tie in with the Hamiltonian library specified above, and the longer-term goal to support physical hardware QPUs through cloud API services.

\paragraph{QIP Tutorials} --- In addition to predefined processors or QPUs, we plan to build a set of tutorials for QuTiP-QIP, showing explicitly how to create models of different types of quantum processors. 

\subsection{QuTiP's role in fundamental scientific research}
\paragraph{More and faster solvers} --- We want to always support the state of the art in open quantum systems, identifying and incorporating methods which provide utility for a broad range of problems, and overall continually work to make all solvers faster and more efficient.  New solvers currently being developed or considered are: methods to iteratively construct equations of motions of operators  (sometimes called a ``cumulant expansion''), and a non-Markovian master equation expansion based on time-convolutionless truncation of bath properties~\cite{GroszkowskiQuantum2023} (confusingly, sometimes also called a ``cumulant expansion'').  Additionally, currently under active development is a new Floquet master equation solver which features flexible levels of truncation of Floquet frequencies, a Dyson-expansion based solver for efficient simulation of quickly driven systems  \cite{dysolve}, and new Krylov-space solvers.

\paragraph{GPU support, high-performance computing, and qutip-cuquantum} --- Multiprocessing in QuTiP currently takes three forms:  
\begin{itemize} 
\item Support for GPU through custom data layers like JAX and the in-development cuquantum data layer (see below), 
\item parallel use of multiple cores on a single CPU or node, instantiated either by explicitly using functions like \code{parallel\_map()} or implicitly in natural multi-task problems like the stochastic and Monte Carlo solvers, and 
\item the use of multiple cores by underlying libraries and methods such as Intel-MKL, which can help speed up ODE solving and finding steady states. 
\end{itemize}
Historically, we also supported OpenMP, but this support was removed with version 5 due to lack of use and heavy maintenance.  Recently, support for MPI was added, extending point (ii) above to multiple nodes. 

A very recent project in development (in collaboration with NVIDIA) is support for multiple GPUs with the \code{qutip-cuquantum} data layer \cite{qutip-cuquantum}, which takes advantage of the powerful cuquantum library from NVIDIA themselves \cite{cuquantum}. By incorporating a symbolic representation of the problem being modeled, dynamics simulations using both \code{mesolve} and \code{sesolve} can be distributed across multiple GPUs, allowing for both speed-ups and larger scale simulations over that available via single-core CPU and single-GPU simulations. Expanding support for such symbolic representations across more parts of QuTiP is a planned core feature of our next release, QuTiP v6.

\paragraph{Tensor-network data layer} --- Approximate or exact truncation methods (e.g., matrix product states (MPS), matrix product operators (MPOs), and tensor networks (TN)) that take advantage of the
algebraic structure of quantum operators for specific problems (like one-dimensional spin chains) have proven very successfully in more efficiently analysing and simulating certain many-body problems (both closed and open) numerically.  Motivated by this progress in the community, early in the development of version 5, we began exploring a tensor-network data layer, which exists currently in a very early alpha form.  However, QuTiP has historically focused on providing solvers for arbitrary systems and its low-level data
layer operates on two dimensional matrices. The data layer is therefore not the correct level for describing these algebraic structures. Work is underway to extend QuTiP with a symbolic operator
description (akin to the early version being used in \code{qutip-cuquantum}), which will allow these richer structures to be captured
and for solvers to use this to efficiently compute solutions for
operators which have additional structure. Importantly, it will also enable easier integration
with existing packages for these methods, as well as developing our own implementations. We expect this to be a
useful extension not only for TN, MPS and MPOs but also for
structuring problems for GPUs and HPCs where matching the problem
description to the hardware provides significant performance benefits.

\paragraph{QuTiP beyond Python} --- QuTiP is, and always will be, a Python orientated package. However, the popularity of QuTiP means that it has led to the creation of QuTiP-like packages in other languages. One of these, written in Julia, QuantumToolbox.jl, was recently incorporated into the QuTiP organization (and includes a recent Julia version of the HEOM solver \cite{Huang2023}).  Being developed by Alberto Mercurio, Yi-Te Huang, and others, it aims to offer syntax compatibility with QuTiP, and support the powerful distributed computing capabilities Julia is offering. In addition, an older and alternative Julia package inspired by QuTiP is QuantumOptics.jl \cite{kramer2018quantumoptics}, which complements the efforts being made in QuantumToolbox.jl.

\section{Conclusion}

QuTiP remains to be one of the most popular academically independent and fully open-source toolkits for simulating open quantum systems.  Its wide adoption by industry, education, and research has also helped support its continued development, by attracting young and enthusiastic contributors.  The release of v5 of QuTiP, particularly the substantial changes to the data layer, enables QuTiP to remain relevant as the open-source quantum software community continues to grow.  The examples we provided in this work, demonstrating unique applications of both old and new solvers and features of QuTiP, also serve an important purpose in exemplifying unique and new ways QuTiP can be used by the community.

\section{Acknowledgements} 
Over the past five years, the development of QuTiP has greatly benefited from the generous support of the Japan Science and Technology Agency (JST) Moonshot R\&D Grant Number JPMJMS2061, and we thank them for their kind support.
We wish to thank a large range of past contributors to QuTiP, particularly Arne Grismo, Cassandra Granade, Michael H. Goerz, Anubhav Vardhan, Saumya Biswas, Sidhant Saraogi, Asad Raza, Felipe Bivort Haiek, Purva Thakre, Christian Staufenbiel, Xavier Spronken, Shreyas Pradhan, Trent Fridey, Yuji Tamakoshi and Alessia Parato. We also thank Mana Lambert for designing and contributing \fig{schematic}, Gavin Crowder for feedback on the the waveguide example in section (\ref{enr}), the NVIDIA cuquantum team for support in the development of the qutip-cuquantum data layer, and Maximilian Meyer-M\"olleringhof for assistance in adapting the code examples into tutorial notebooks.  F.N. is also supported in part by: the Japan Science and Technology Agency (JST) [via the CREST Quantum Frontiers program Grant No. JPMJCR24I2, the Quantum Leap Flagship Program (Q-LEAP)], and the Office of Naval Research (ONR) Global (via Grant No. N62909-23-1-2074).  N.~L.~is supported by the RIKEN Incentive Research Program and by MEXT KAKENHI Grant Numbers JP24H00816, JP24H00820. P.~M.\ performed this work as an International Research Fellow of the Japan Society for the Promotion of Science (JSPS). M.G.~is supported by the New Energy and Industrial Technology Development Organization (NEDO), project code JPNP16007. E.G. is supported by the Ministère de l’Économie, de l’Innovation et de l’Énergie du Québec. We also acknowledge the Information Systems
Division, RIKEN, for the use of their facilities.

\appendix
\newpage
\section{Tables}
Here we provide tables summarizing various useful methods, functions, and concepts in QuTiP.

\begin{table}[!hp]
    \centering
    \renewcommand{\arraystretch}{1.1}
    \begin{tabular}{@{}lp{10cm}@{}}
        \toprule
        \textbf{Method / Attribute} & \textbf{Description} \\
        \midrule
        \code{copy()} & Create a copy of the Qobj. \\
       
        \code{conj()} & Complex conjugate. \\
       
        \code{contract()} & Contract subspaces of the tensor structure that are 1D. \\
  
        \code{cosm()} & Matrix cosine of the Qobj. \\
   
        \code{dag()} & Adjoint (Hermitian conjugate) of the Qobj. \\
  
        \code{data\_as(format, copy)} & Retrieve the data in the desired format. \\
 
        \code{diag()} & Diagonal elements of the Qobj. \\

        \code{dnorm()} & Diamond norm of the Qobj. \\
    
        \code{dual\_chan()} & Obtain the dual channel of the Qobj. \\
   
        \code{eigenenergies()} & Eigenvalues of the Qobj. \\
 
        \code{eigenstates()} & Eigenvalues and eigenstates of the Qobj. \\

        \code{groundstate()} & Ground state eigenvalue and eigenvector. \\

        \code{expm()} & Matrix exponential of the Qobj. \\

        \code{full()} & Dense array representation of the Qobj data. \\

        \code{inv()} & Matrix inverse of the Qobj. \\

        \code{logm()} & Matrix logarithm of the Qobj. \\
    
        \code{matrix\_element(bra, ket)} & Matrix element between the specified bra and ket vectors. \\

        \code{norm()} & Norm of the Qobj. \\
     
        \code{overlap(other)} & Overlap between two Qobjs. \\

        \code{permute(order)} & Reorder the tensor structure of a composite Qobj. \\

        \code{proj()} & Projector for a ket or bra vector. \\
  
        \code{ptrace(sel)} & Partial trace over the specified subsystems. \\
 
        \code{purity()} & Purity of the Qobj. \\

        \code{sinm()} & Matrix sine of the Qobj. \\
   
        \code{sqrtm()} & Matrix square root of the Qobj. \\

        \code{tidyup(atol)} & Remove small elements (with tolerance atol) from the Qobj. \\

        \code{tr()} & Trace of the Qobj. \\
 
        \code{trans()} & Transpose of the Qobj. \\
   
        \code{transform(input)} & Perform basis transformation defined by the input matrix. \\

        \code{trunc\_neg()} & Remove negative eigenvalues. \\

        \code{unit()} & Normalizes the Qobj. \\

        \code{data} & The QuTiP-internal data-layer object storing the data. \\

        \code{dtype} & The data-layer type used for storing the data. \\

        \code{dims}, \code{shape}, \code{type} & Basic information about the Qobj, explained in the main text. \\

        \code{isherm}, \code{isunitary} & Indicates if the Qobj is a Hermitian / unitary operator. \\
   
        \code{isket}, \code{isbra}, \code{isoper}, & Indicates if the Qobj has the respective \code{type}. \\
        \code{issuper}, \code{isoperket}, \code{isoperbra} & \\
    
        \code{iscp}, \code{ishp}, \code{istp}, \code{iscptp} & Indicates if the Qobj is a map that is completely positive (CP) / hermiticity-preserving (HP) / trace-preserving(TP) / CP and TP. \\
        \bottomrule
    \end{tabular}
    \caption{Methods of the \code{Qobj} Class in QuTiP. We have omitted optional parameters for some of the methods; they can be found in the full online reference together with more detailed explanations.}
    \label{tab:qobj_methods}
\end{table}

\newpage

\begin{table}[!hp]
    \centering
     \renewcommand{\arraystretch}{1.3}
    \begin{tabular}{@{}lp{10cm}@{}}
        \toprule
        \textbf{Function} & \textbf{Description} \\
        \midrule
        \code{lindblad\_dissipator(a,b)} & Generates the Lindblad dissipator for the given Lindblad operators, $\mathcal{D}[a,b]\rho = a \rho b^\dagger -
\frac{1}{2}a^\dagger b\rho - \frac{1}{2}\rho a^\dagger b$. \\
  
        \code{liouvillian(H, c\_ops)} & Generates the Liouvillian superoperator for a given Hamiltonian and list of collapse operators. \\

        \code{spre(A)} & Generates the superoperator corresponding to left multiplication by the operator $A$. \\
  
        \code{spost(A)} & Generates the superoperator corresponding to right multiplication by the operator $A$.\\
     
        \code{sprepost(A,B)} & Generates the superoperator for left and right multiplication by operators $A$ and $B$ respectively. \\

        \code{operator\_to\_vector(op)} & Vectorizes an operator. \\

        \code{vector\_to\_operator(vec)} & Reshapes a vectorized operator back to its original matrix form. \\
\bottomrule
    \end{tabular}
    \caption{Table of commonly used functions for creating and manipulating superoperators}
    \label{tab:qutip_superoperators_functions}
\end{table}

\newpage

\begin{table}[!hp]
\centering
 \renewcommand{\arraystretch}{1.3}
\begin{tabular}{@{}lp{11cm}@{}}
\toprule
\textbf{Function} & \textbf{Description} \\ 
\bottomrule

\code{entropy\_vn(rho, base)} & Computes the von Neumann entropy of a density matrix, defined as
\[
S(\rho) = -\text{Tr}(\rho \log_b \rho) ,
\]
where $b$ is the logarithmic base (default is $e$). \\

\code{entropy\_linear(rho)} & Computes the linear entropy of a density matrix, given by
\[
S_L(\rho) = 1 - \Tr(\rho^2) .
\]
\\ 

\makecell{\code{entropy\_mutual(} \\ \code{~~~~rho, selA, selB} \\ \code{)}} & Calculates the quantum mutual information for a bipartite density matrix $\rho$, given by
\[
I(A:B) = S(\rho_A) + S(\rho_B) - S(\rho_{AB}) ,
\]
where $\rho_A$ and $\rho_B$ are the reduced density matrices. The parameters \code{selA} and \code{selB} specify the subspaces $A$ and $B$. \\ 

\makecell{\code{entropy\_conditional(} \\ \code{~~~~rho, selB} \\ \code{)}} & Computes the conditional entropy for a bipartite density matrix:
\[
S(A|B) = S(\rho_{AB}) - S(\rho_B) ,
\]
where $\rho_{AB}$ is the bipartite density matrix, and $\rho_B$ is the reduced density matrix of $B$. The parameter \code{selB} specifies the subspace $B$.\\ 

\makecell{\code{entropy\_relative(} \\ \code{~~~~rho, sigma, base} \\ \code{)}} & Calculates the relative entropy between two density matrices $\rho$ and $\sigma$, defined as
\[
S(\rho || \sigma) = \text{Tr}(\rho \log_b \rho) - \text{Tr}(\rho \log_b \sigma) ,
\]
where $b$ is the logarithmic base (default is $e$).
This quantifies the "distance" between the quantum states $\rho$ and $\sigma$. \\

\code{concurrence(rho)} & Calculates the concurrence of a two-qubit density matrix $\rho$, an entanglement measure:
\[
C(\rho) = \max\left(0, \sqrt{\lambda_1} - \sqrt{\lambda_2} - \sqrt{\lambda_3} - \sqrt{\lambda_4}\right) ,
\]
where $\lambda_i$ are the eigenvalues of $R = \rho (\sigma_y \otimes \sigma_y) \rho^* (\sigma_y \otimes \sigma_y)$. \\ 

\code{negativity(rho, subsys)} & Computes the negativity of a bipartite quantum state $\rho$, defined as
\[
\mathcal{N}(\rho) = \frac{\|\rho^{T_B}\|_1 - 1}{2} ,
\]
where $\rho^{T_B}$ is the partial transpose and $\|\cdot\|_1$ the trace norm.
The parameter \code{subsys} specifies for which subsystem to compute the negativity. \\ 
\bottomrule
\end{tabular}
\caption{Summary of entropy and entanglement measure functions in QuTiP}
\label{table:entropy_functions}
\end{table}

\newpage

\begin{table}[!hp]
\centering
 \renewcommand{\arraystretch}{1.3}
\begin{tabular}{@{}lp{11.3cm}@{}}
\toprule
\textbf{Function} & \textbf{Description} \\ 
\midrule

\code{fidelity(rho, sigma)} & Computes the fidelity between two quantum states $\rho$ and $\sigma$, defined as
\[
F(\rho, \sigma) =  \text{Tr} \left[ \sqrt{ \sqrt{\rho} \sigma \sqrt{\rho} } \, \right] .
\]
Fidelity measures how close two quantum states are to each other. \\ 

\code{tracedist(rho, sigma)} & Computes the trace distance between two density matrices $\rho$ and $\sigma$:
\[
D_{\text{tr}}(\rho, \sigma) = \frac{1}{2} \| \rho - \sigma \|_1 ,
\]
where $\| \cdot \|_1$ denotes the trace norm (sum of singular values). \\ 

\code{hilbert\_dist(A, B)} & Computes the Hilbert-Schmidt distance between two operators $A$ and $B$:
\[
D_{\text{HS}}(A, B) = \sqrt{ \text{Tr} \left[ (A - B)^\dagger (A - B) \right] } .
\]
This measures the ``distance'' between operators in Hilbert space. \\ 

\code{bures\_dist(rho, sigma)} & Calculates the Bures distance between two quantum states $\rho$ and $\sigma$:
\[
D_{\text{Bures}}(\rho, \sigma) = \sqrt{2 \left(1 - {F(\rho, \sigma)} \right)} ,
\]
where $F(\rho, \sigma)$ is the fidelity between $\rho$ and $\sigma$. \\ 

\code{bures\_angle(rho, sigma)} & Computes the Bures angle between two quantum states $\rho$ and $\sigma$:
\[
\theta_{\text{Bures}}(\rho, \sigma) = \arccos \left({F(\rho, \sigma)} \right) .
\]
This provides an angular measure of the distance between states. \\ 

\makecell{\code{average\_gate\_fidelity(} \\ \code{~~~~E, F)}} & Computes the average gate fidelity between two quantum channels $\mathcal{E}$ and $\mathcal{F}$, using the definition from \cite{PhysRevA.71.062310}. \\ 

\code{process\_fidelity(E, F)} & Calculates the process fidelity between two quantum channels $\mathcal{E}$ and $\mathcal{F}$, via their Choi matrices $\chi_{\mathcal{E}}$ and $\chi_{\mathcal{F}}$. Since QuTiP v5, this function uses the definition from \cite{PhysRevA.71.062310}. \\

\code{dnorm(E, F)} & Computes the diamond norm distance between two channels $\mathcal{E}$ and $\mathcal{F}$:
\[
\| \mathcal{E} - \mathcal{F} \|_{\diamond} = \sup_{\rho} \| (\mathcal{E} \otimes \mathbb{I})(\rho) - (\mathcal{F} \otimes \mathbb{I})(\rho) \|_1 ,
\]
where the supremum is over all density matrices \( \rho \) on an extended Hilbert space, and \( \mathbb{I} \) is the identity map. \\ 
\bottomrule

\end{tabular}
\caption{Summary of commonly used metric functions in QuTiP}
\label{table:metrics_functions}
\end{table}

\newpage

\section{Summary of tutorials and example notebooks} \label{app:notebooks}

QuTiP has historically provided a large array of tutorials and example notebooks \cite{tutorials,qnotebooks}.  With v5, initiated through the GSoC project of Christian Staufenbiel, these notebooks are going through an overhaul, to unify and modernize them, make them more compact with a switch to markdown, and include rudimentary tests of their functionality (which also acts as an additional check of QuTiP itself).

Here we provide a detailed summary of some of the updated notebooks for v5, including an abridged version of their content which is useful and complementary to the main text. 

\subsection{Time-evolution tutorials}

This set of tutorials can be seen as the core demonstration of QuTiP's functionality and features \cite{tutorials}.
\tocless\subsubsection{0001\_qobjevo}

In this tutorial we demonstrate the flexibility and power of the \code{Qobjevo} class for representing time-dependent objects.  This has largely been covered earlier in this work, but it is interesting to show some examples from that notebook.  To recap, when solvers in QuTiP are provided with a time dependent operator, typically in terms of a tuple containing the quantum object and a representation of the time-dependence as a function, string, or array, internally this is converted into \code{Qobjevo}, which compiles and optimizes the representation.

These objects can also be created and compiled manually with lists of objects and time-dependent coefficients
\code{[A0, [A1, f1], [A2, f2], ...]} where $A_k$ are quantum objects and $f_k$ either functions, strings, or arrays. Another flexible option is to use the \code{coefficient} function to wrap the time-dependent function, string, or array, and then multiply it with a \code{Qobj} as needed.

\inputpython{tutorial_snippets.tex}{3}{8}

String format representations should represent a valid Python function or expression that returns a complex number.  So for the above example, instead one could call \code{string\_form = QuTiP.QobjEvo([n, [a + ad, "cos(t)"]])}. Any NumPy or \code{scipy.special} functions can be referred to with \code{np} or \code{spe}.

Finally, if one is using data to represent the time dependence of a given object one may pass an array of data points, alongside a list of time steps at which those points are specified. Intermediate times will be interpolated with cubic splines.

\inputpython{tutorial_snippets.tex}{11}{16}

Once defined, \code{Qobevo} functions can be called to return the value of the \code{Qobj} at that time, as well as manipulated with the same mathematical rules as a \code{Qobj} and mixed with scalars, \code{Qobj} and other \code{QobjEvo} objects. 

Even more powerfully, \code{QobjEvo} functions can take arguments that can be quite complex, including objects or values that are derived from the solver they are used in. This allows one to directly implement both state and expectation-value based feedback. More details are provided in the tutorial, but, for example, an argument that will capture the state of the system during the evolution with \code{mesolve} can  be instantiated with \code{StateFeedback()}, and that state can then be used to inform the behavior of a function that is used in \code{mesolve} itself:

\inputpython{tutorial_snippets.tex}{18}{27}

\tocless\subsubsection{0002\_larmor-precession} \label{app:bloch_notebook}

\begin{figure}[t]
\includegraphics[width = 0.45\columnwidth]{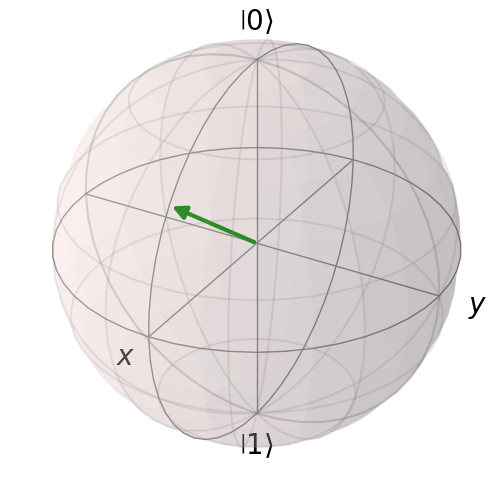}
\includegraphics[width = 0.45\columnwidth]{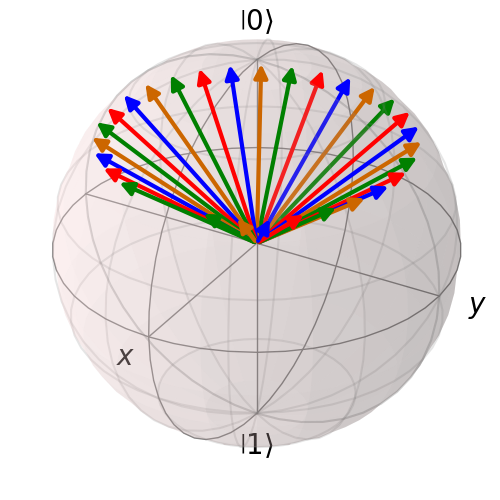}
\caption{Left shows the Bloch sphere representation of the state \code{psi = (2.0 * basis(2, 0) + basis(2, 1)).unit()}, while right shows the state at various times while precessing under a constant magnetic field. }\label{bloch}
\end{figure}

This tutorial demonstrates basic usage of the \code{sesolve()} function with an example of Larmor precession.  It demonstrates how to use in-built plotting functions to show the state of a spin on a Bloch sphere \fig{bloch}.  It then demonstrates how to solve the dynamics of such a spin in the presence of a magnetic field in the $Z$ direction, and obtain expectation value of observables, with \code{sesolve}, and plot the state on the Bloch sphere.

The tutorial then demonstrates how to use \code{Qobjevo} to represent time-dependent magnetic fields, and use those with \code{sesolve}.
\newline
\tocless\subsubsection{0003\_qubit-dynamics}

This tutorial shows how to model the dynamics of a qubit under the influence of noise using \code{mesolve}.  Like the previous example, it also demonstrates how to visualize these dynamics with a Bloch sphere.  This is shown in 
\fig{bloch2} where unitary, dephasing, and relaxation evolution are shown. For dephasing, the dynamics moves towards the center of the Bloch sphere, while for relaxation, it moves to a point within the volume.\newline
 
 \tocless\subsubsection{0004\_rabi-oscillations}

This tutorial shows how to simulate the dynamics of the Rabi model, one of the most fundamental models of quantum optics, which describes a single two-level atom interacting with a single cavity mode of light. The Hamiltonian for this system is,

\begin{equation}
H = \hbar \omega_c a^\dagger a + \frac{1}{2}\hbar\omega_a\sigma_z + \hbar g(a^\dagger + a)(\sigma_- + \sigma_+)
\end{equation}
which can be simplified, for weak coupling, with the rotating-wave approximation:
\begin{equation}
H_{\rm RWA} = \hbar \omega_c a^\dagger a + \frac{1}{2}\hbar\omega_a\sigma_z + \hbar g(a^\dagger\sigma_- + a\sigma_+) \, ,
\end{equation}
here $\omega_c$ and $\omega_a$ are the frequencies of the cavity and atom, while $g$ is the interaction strength between light and matter. This tutorial shows how to solve the dynamics of this model using \code{mesolve} when both the light and matter are in contact with a noisy environment.  The steps needed to do so are similar to other examples we have already described in this paper, so we will not linger on this example, but just mention that the example in this tutorial is only valid for weak-intermediate light-matter coupling, and in the ultra-strong light matter regime, when $g$ approaches $\omega_c$ and $\omega_a$, one should instead employ \code{brmesolve}, as described earlier for the interacting qubit example. This is explained in more detail in tutorial ``0009\_brmesolve\_cavity-QED''.\newline

\tocless\subsubsection{0005\_spin-chain}

This tutorial explains how to solve the dynamics of a Heisenberg spin-chain using \code{mesolve}. In construction it is similar to the example for the 1D Ising model we used as a bench-mark in the main text, but the Hamiltonian for the spin chain instead takes the form,

\begin{equation}
H = - \frac{1}{2}\sum_n^N h_n \sigma_z(n) - \frac{1}{2} \sum_n^{N-1} [ J_x^{(n)} \sigma_x(n) \sigma_x(n+1) + J_y^{(n)} \sigma_y(n) \sigma_y(n+1) +J_z^{(n)} \sigma_z(n) \sigma_z(n+1)] \, ,
\end{equation}
where $h_n$ is the magnetic field felt by spin $n$ while $J_k^{(n)}$ describes the nearest neighbour interaction strength in direction $k$ for spin $n$.  The example also includes dephasing on these spins, and demonstrates how to visualize dynamics of such a complex system.\newline

\tocless\subsubsection{0006\_photon\_birth\_death}

This tutorial is an example of how to use \code{mcsolve}, the Monte-Carlo solver, to simulate experimental results that appeared in \cite{Gleyzes2007}. The tutorial demonstrates how to reproduce figure 3 from this work, by simulating the creation and annihilation of photons inside an cavity, due to a thermal environment, when the cavity is initially prepared in the the single-photon Fock state.  

The model is very simple, consisting of the Hamiltonian  $H = a^\dagger a$, the aforementioned initial Fock state, and collapse operators describing the creation/annihilation of photons are appropriately weighted rates: $C_1 = \sqrt{\kappa (1 + \langle n \rangle)} \; a$ and $C_2 = \sqrt{\kappa \langle n \rangle} \; a^\dagger$, where $\kappa$ is the bare rate and $n$ is the thermal Bose-Einstein factor which depends on the temperature and the cavity frequency.  The tutorial then demonstrates how to solve this example using \code{mcsolve} and look at the dynamics for different numbers of trajectories.\newline

\tocless\subsubsection{0007\_brmesolve\_tls}

This tutorial demonstrates the basic usage of the Bloch-Redfield solver \code{brmesolve} for a single two-level system (TLS). This covers the same ground as our earlier explanation, but does mention one important feature we did not cover in the main paper; the calculation of steady states.  The function \code{R, ekets = bloch\_redfield\_tensor(H, [a\_op])} returns the Bloch-Redfield tensor, i.e., the right-hand-side of the equation of motion in superoperator form. This can then be used with the \code{steadystate(R)} function to find directly the steady-state solution of the Bloch-Redfield equation one is simulating.\newline

\tocless\subsubsection{0008\_brmesolve\_time\_dependence}

This tutorial explains how to use both time-dependent Hamiltonians and time-dependent dissipation with the Bloch-Redfield solver.  The functionality is equivalent to using time-dependence in other solvers.  One important point this tutorial covers is that with the Bloch-Redfield solver, the coupling to the environment must be a Hermitian operator.  Complex time-dependence can be included by splitting the operator into two parts, so that something like 

\begin{equation}
A = f(t)a + f(t)^* a^\dagger 
\end{equation}
can be implemented with \code{a\_ops = [([[a, 'exp(1j*t)'], [a.dag(), 'exp(-1j*t)']], f'{kappa} * (w >= 0)')]}, under the restriction that the the second function is the complex conjugate of the first one, and the second operator is the Hermitian conjugate of the first operator.\newline

\tocless\subsubsection{0009\_brmesolve\_cavity-QED}

In this tutorial, we present another example of when the Bloch-Redfield solver can be useful for constructing a more physically accurate master equation than relying on local dissipation alone.  The logic is similar to that in the main text of this work for the two-interacting qubit example, but instead, in this tutorial, concentrates on the Rabi model from quantum optics (described already in 0004\_rabi-oscillations).

The tutorial demonstrates that for weak coupling between light and matter, including cavity dissipation using \code{brmesolve}, or including local dissipation on the cavity alone via \code{mesolve}, produce similar results. However, for strong coupling, the local Lindblad model can fail to produce physical results (in essence, we can see continuous emission out of the cavity without any input;  a perpetuum mobile).\newline

\tocless\subsubsection{0010\_brmesolve\_phonon\_interaction}

This tutorial, contributed by K. A. Fischer from Stanford University,  is a complex example of how to use \code{brmesolve} to simulate the phonon-assisted initialization of a quantum dot, reproducing the results of an existing article \cite{PhysRevB.90.241404}.\newline

\tocless\subsubsection{0011\_floquet\_solver}

This tutorial describes the basic use of the Floquet Schrodinger and Master equation solvers, with an example akin to the description we used in the main text. \newline

\tocless\subsubsection{0012\_floquet\_formalism}

This tutorial describes some of the underpinnings of the Schrodinger and Master equation Floquet solvers in terms of the \code{FloquetBasis} and quasi-energies.  It is a very useful starting point for users who wish to use these objects for more niche applications.\newline

\tocless\subsubsection{0013\_nonmarkovian\_monte\_carlo}

This tutorial expands upon the use of the \code{nm\_mcsolve()} method for unravelling, in terms of Monte-Carlo trajectories, a master equation with non-Markovian rates (i.e., rates which are time-dependent and sometimes negative). 

As explained in the section on this solver, this is quite a nuanced method, and this tutorial provides several practical examples to help users get accustomed to it: (1) a  two-level atom in a photonic band-gap and (2) a two-qubit Redfield equation derived from when two qubits interact with a common bath.  It also describes how to employ the MPI feature in qutip to solve these trajectories in parallel on a super-computing cluster.\newline

\tocless\subsubsection{0015\_smesolve-heterodyne}

This example shows how to use \code{smesolve} to simulate heterodyne measurement of a cavity. In the earlier section on \code{smesolve} a similar example for homodyne measurement was given.  The key difference here is the measurement gives information on both quadratures of the system being measured.  The tutorial presents two approaches to defining and solving this type of problem.\newline

\tocless\subsubsection{0016\_smesolve-inefficient-detection}

This example demonstrates further utility of \code{smesolve} for modelling an important example from Wiseman and Milburn, Quantum measurement and control, section. 4.8.1.  In this example, a lossy cavity is monitored with an inefficient photodetector. This means the system undergoes evolution with two noise channels, one where photons are successfully detected, and one where they are not.  This is easily formulated as a Monte-Carlo master equation for use with \code{mcsolve}.  Furthermore, if this inefficient detection is used as part of a homodyne detection protocol, this can also be formulated as a stochastic master equation for use with \code{smesolve}.\newline

\tocless\subsubsection{0016\_smesolve-jc-photocurrent}
This example generalizes the previous so that the system being monitored includes an atom interacting with a cavity, as described by the Jaynes-Cummings model.  The output of the cavity is imperfectly detecting with  an inefficient photodetector, which is again simulated with \code{mcsolve}.\newline

\tocless\subsubsection{0018\_measures-trajectories-cats-kerr}

In this example, contributed by Fabrizio Minganti, reproduces published results in \cite{minganti1, minganti2, minganti3}. It is an extremely detailed example of modelling a nonlinear Kerr resonator undergoing parametric two-photon driving.  It demonstrates that how the cavity is measured, either via photon counting (with \code{mcsolve}) or homodyne detection (with \code{smesolve}), effects ones ability to understand the nature of the cavity state.\newline

\tocless\subsubsection{0019\_optomechanical-steadystate}

This tutorial serves to demonstrate the varying steady-state solvers in QuTiP using an archtypical example from optomechanics.  As well as showing the physics of optomechanics systems, and how the steady-state of an oscillator can be inspected for unphysical results, it demonstrates how the different methods and solvers in \code{steadystate()} can be invoked.\newline

\tocless\subsubsection{0020\_homodyned-Jaynes-Cummings-emission}

This example, contributed by K. A. Fisher and A. V. Domingues (reproducing results from \cite{PhysRevApplied.7.044002}), demonstrates how to obtain photonic correlation functions of a Jaynes-Cummings systems, with the purpose of probing the lowest-lying states as an effective two-level system. It also shows how to obtain second-order correlation functions from \code{mesolve()} manually (insteady of using utility functions).\newline

\tocless\subsubsection{0021\_quasi-steadystate-driven-system}

This notebook demonstrates the different ways steady-state information of periodically driven dissipative quantum systems can be calculated.  In particular, it demonstrates how properties of the periodically oscillating steady-state, averaged over one period, can be obtained with either \code{propagator\_steadystate()} or \code{steadystate\_floquet()}.

\subsection{Lectures}

The set of tutorials are a series of Lectures from Robert Johansson detailing use of QuTiP, from basics to complex applications. These were originally written for a taught invited course at Chalmers University in Sweden for an early version of QuTiP.  They have been updated and fixed to function with both QuTiP v4 and v5, and have proven to be a valuable for students and researchers everywhere.

\subsection{Quantum circuits and Pulse-level-circuit-simulation}

These two sets of tutorials detail use of QuTiP-QIP, from basic circuit simulations to complex pulse-level noise models.  They include basic examples, like a simple circuit implementing CNOT and Toffoli gates and a tutorial on how to export and import \code{openqasm} circuits, and more complex examples of quantum algorithms like the Quantum Fourier transform and the Deustch-Josza algorithm.

\subsection{Visualization}

QuTiP includes many visualization tools for common ways to present a quantum state, like the Bloch sphere and the Wigner Function. This set of tutorials demonstrates much of this functionality, alongside demonstrations of built-in animation functions, process tomography functions, and more.

\subsection{HEOM: Hierarchical Equations of Motion}

The HEOM tutorials demonstrate how to use QuTiP's hierarchical equations of motion solver for both fermionic and bosonic baths. These examples have been explained in detail in our recent publication \cite{lambert2020bofinheom}.

\subsection{Miscellaneous}

This repository is a place to include demonstrations of new QuTiP functionality, like the JAX backend, and more complex physical examples.

\subsection{QuTiP-notebooks}

This older repository \cite{qnotebooks} contains many more examples and tutorials that have not yet been ported to v5.  However, they remain an important resource, and overtime we hope to have all relevant notebooks updated and included in the official qutip-tutorials repository.

\bibliography{control,references}

\begin{thebibliography}{138}%
\makeatletter
\providecommand \@ifxundefined [1]{%
 \@ifx{#1\undefined}
}%
\providecommand \@ifnum [1]{%
 \ifnum #1\expandafter \@firstoftwo
 \else \expandafter \@secondoftwo
 \fi
}%
\providecommand \@ifx [1]{%
 \ifx #1\expandafter \@firstoftwo
 \else \expandafter \@secondoftwo
 \fi
}%
\providecommand \natexlab [1]{#1}%
\providecommand \enquote  [1]{``#1''}%
\providecommand \bibnamefont  [1]{#1}%
\providecommand \bibfnamefont [1]{#1}%
\providecommand \citenamefont [1]{#1}%
\providecommand \href@noop [0]{\@secondoftwo}%
\providecommand \href [0]{\begingroup \@sanitize@url \@href}%
\providecommand \@href[1]{\@@startlink{#1}\@@href}%
\providecommand \@@href[1]{\endgroup#1\@@endlink}%
\providecommand \@sanitize@url [0]{\catcode `\\12\catcode `\$12\catcode `\&12\catcode `\#12\catcode `\^12\catcode `\_12\catcode `\%12\relax}%
\providecommand \@@startlink[1]{}%
\providecommand \@@endlink[0]{}%
\providecommand \url  [0]{\begingroup\@sanitize@url \@url }%
\providecommand \@url [1]{\endgroup\@href {#1}{\urlprefix }}%
\providecommand \urlprefix  [0]{URL }%
\providecommand \Eprint [0]{\href }%
\providecommand \doibase [0]{https://doi.org/}%
\providecommand \selectlanguage [0]{\@gobble}%
\providecommand \bibinfo  [0]{\@secondoftwo}%
\providecommand \bibfield  [0]{\@secondoftwo}%
\providecommand \translation [1]{[#1]}%
\providecommand \BibitemOpen [0]{}%
\providecommand \bibitemStop [0]{}%
\providecommand \bibitemNoStop [0]{.\EOS\space}%
\providecommand \EOS [0]{\spacefactor3000\relax}%
\providecommand \BibitemShut  [1]{\csname bibitem#1\endcsname}%
\let\auto@bib@innerbib\@empty
\bibitem [{\citenamefont {Johansson}\ \emph {et~al.}(2012)\citenamefont {Johansson}, \citenamefont {Nation},\ and\ \citenamefont {Nori}}]{Johansson12}%
  \BibitemOpen
  \bibfield  {author} {\bibinfo {author} {\bibfnamefont {J.~R.}\ \bibnamefont {Johansson}}, \bibinfo {author} {\bibfnamefont {P.~D.}\ \bibnamefont {Nation}},\ and\ \bibinfo {author} {\bibfnamefont {F.}~\bibnamefont {Nori}},\ }\bibfield  {title} {\bibinfo {title} {{{QuTiP}}: {{An}} open-source {{Python}} framework for the dynamics of open quantum systems},\ }\href {https://doi.org/10.1016/j.cpc.2012.02.021} {\bibfield  {journal} {\bibinfo  {journal} {Comput. Phys. Commun.}\ }\textbf {\bibinfo {volume} {183}},\ \bibinfo {pages} {1760} (\bibinfo {year} {2012})}\BibitemShut {NoStop}%
\bibitem [{\citenamefont {Johansson}\ \emph {et~al.}(2013)\citenamefont {Johansson}, \citenamefont {Nation},\ and\ \citenamefont {Nori}}]{Johansson13}%
  \BibitemOpen
  \bibfield  {author} {\bibinfo {author} {\bibfnamefont {J.~R.}\ \bibnamefont {Johansson}}, \bibinfo {author} {\bibfnamefont {P.~D.}\ \bibnamefont {Nation}},\ and\ \bibinfo {author} {\bibfnamefont {F.}~\bibnamefont {Nori}},\ }\bibfield  {title} {\bibinfo {title} {{{QuTiP}} 2: {{A Python}} framework for the dynamics of open quantum systems},\ }\href {http://www.sciencedirect.com/science/article/pii/S0010465512003955} {\bibfield  {journal} {\bibinfo  {journal} {Comput. Phys. Commun.}\ }\textbf {\bibinfo {volume} {184}},\ \bibinfo {pages} {1234} (\bibinfo {year} {2013})}\BibitemShut {NoStop}%
\bibitem [{cod()}]{codeexamplesstatic}%
  \BibitemOpen
  \href {https://github.com/qutip/qutip-paper-v5-examples} {\bibinfo {title} {https://github.com/qutip/qutip-paper-v5-examples}}\BibitemShut {NoStop}%
\bibitem [{tut()}]{tutorials}%
  \BibitemOpen
  \href {https://github.com/qutip/qutip-tutorials} {\bibinfo {title} {https://github.com/qutip/qutip-tutorials}}\BibitemShut {NoStop}%
\bibitem [{\citenamefont {Harris}\ \emph {et~al.}(2020)\citenamefont {Harris}, \citenamefont {Millman}, \citenamefont {van~der Walt} \emph {et~al.}}]{harris2020array}%
  \BibitemOpen
  \bibfield  {author} {\bibinfo {author} {\bibfnamefont {C.~R.}\ \bibnamefont {Harris}}, \bibinfo {author} {\bibfnamefont {K.~J.}\ \bibnamefont {Millman}}, \bibinfo {author} {\bibfnamefont {S.~J.}\ \bibnamefont {van~der Walt}}, \emph {et~al.},\ }\bibfield  {title} {\bibinfo {title} {{Array programming with NumPy}},\ }\href {http://dx.doi.org/10.1038/s41586-020-2649-2} {\bibfield  {journal} {\bibinfo  {journal} {Nature}\ }\textbf {\bibinfo {volume} {585}},\ \bibinfo {pages} {357} (\bibinfo {year} {2020})}\BibitemShut {NoStop}%
\bibitem [{\citenamefont {Groth}\ \emph {et~al.}(2014)\citenamefont {Groth}, \citenamefont {Wimmer}, \citenamefont {Akhmerov},\ and\ \citenamefont {Waintal}}]{kwant}%
  \BibitemOpen
  \bibfield  {author} {\bibinfo {author} {\bibfnamefont {C.~W.}\ \bibnamefont {Groth}}, \bibinfo {author} {\bibfnamefont {M.}~\bibnamefont {Wimmer}}, \bibinfo {author} {\bibfnamefont {A.~R.}\ \bibnamefont {Akhmerov}},\ and\ \bibinfo {author} {\bibfnamefont {X.}~\bibnamefont {Waintal}},\ }\bibfield  {title} {\bibinfo {title} {Kwant: a software package for quantum transport},\ }\href {https://dx.doi.org/10.1088/1367-2630/16/6/063065} {\bibfield  {journal} {\bibinfo  {journal} {New Journal of Physics}\ }\textbf {\bibinfo {volume} {16}},\ \bibinfo {pages} {063065} (\bibinfo {year} {2014})}\BibitemShut {NoStop}%
\bibitem [{\citenamefont {Giannozzi}\ \emph {et~al.}(2009)\citenamefont {Giannozzi} \emph {et~al.}}]{espresso}%
  \BibitemOpen
  \bibfield  {author} {\bibinfo {author} {\bibfnamefont {P.}~\bibnamefont {Giannozzi}} \emph {et~al.},\ }\bibfield  {title} {\bibinfo {title} {{QUANTUM ESPRESSO}: a modular and open-source software project for quantum simulations of materials},\ }\href {https://dx.doi.org/10.1088/0953-8984/21/39/395502} {\bibfield  {journal} {\bibinfo  {journal} {Journal of Physics: Condensed Matter}\ }\textbf {\bibinfo {volume} {21}},\ \bibinfo {pages} {395502} (\bibinfo {year} {2009})}\BibitemShut {NoStop}%
\bibitem [{\citenamefont {McGibbon}\ \emph {et~al.}(2015)\citenamefont {McGibbon}, \citenamefont {Beauchamp}, \citenamefont {Harrigan}, \citenamefont {Klein}, \citenamefont {Swails}, \citenamefont {Hernández}, \citenamefont {Schwantes}, \citenamefont {Wang}, \citenamefont {Lane},\ and\ \citenamefont {Pande}}]{mdtraj}%
  \BibitemOpen
  \bibfield  {author} {\bibinfo {author} {\bibfnamefont {R.~T.}\ \bibnamefont {McGibbon}}, \bibinfo {author} {\bibfnamefont {K.~A.}\ \bibnamefont {Beauchamp}}, \bibinfo {author} {\bibfnamefont {M.~P.}\ \bibnamefont {Harrigan}}, \bibinfo {author} {\bibfnamefont {C.}~\bibnamefont {Klein}}, \bibinfo {author} {\bibfnamefont {J.~M.}\ \bibnamefont {Swails}}, \bibinfo {author} {\bibfnamefont {C.~X.}\ \bibnamefont {Hernández}}, \bibinfo {author} {\bibfnamefont {C.~R.}\ \bibnamefont {Schwantes}}, \bibinfo {author} {\bibfnamefont {L.-P.}\ \bibnamefont {Wang}}, \bibinfo {author} {\bibfnamefont {T.~J.}\ \bibnamefont {Lane}},\ and\ \bibinfo {author} {\bibfnamefont {V.~S.}\ \bibnamefont {Pande}},\ }\bibfield  {title} {\bibinfo {title} {{MDTraj}: A modern open library for the analysis of molecular dynamics trajectories},\ }\href {https://doi.org/10.1016/j.bpj.2015.08.015} {\bibfield  {journal} {\bibinfo  {journal} {Biophysical Journal}\ }\textbf {\bibinfo {volume} {109}},\ \bibinfo {pages} {1528} (\bibinfo {year}
  {2015})}\BibitemShut {NoStop}%
\bibitem [{\citenamefont {{The Astropy Collaboration}}\ \emph {et~al.}(2022)\citenamefont {{The Astropy Collaboration}} \emph {et~al.}}]{astropy}%
  \BibitemOpen
  \bibfield  {author} {\bibinfo {author} {\bibnamefont {{The Astropy Collaboration}}} \emph {et~al.},\ }\bibfield  {title} {\bibinfo {title} {{The Astropy Project}: Sustaining and growing a community-oriented open-source project and the latest major release (v5.0) of the core package},\ }\href {https://doi.org/10.3847/1538-4357/ac7c74} {\bibfield  {journal} {\bibinfo  {journal} {The Astrophysical Journal}\ }\textbf {\bibinfo {volume} {935}},\ \bibinfo {pages} {167} (\bibinfo {year} {2022})}\BibitemShut {NoStop}%
\bibitem [{\citenamefont {Zeng}\ \emph {et~al.}(2017)\citenamefont {Zeng}, \citenamefont {Johnson}, \citenamefont {Smith}, \citenamefont {Rubin}, \citenamefont {Reagor}, \citenamefont {Ryan},\ and\ \citenamefont {Rigetti}}]{Zeng_2017_Nature}%
  \BibitemOpen
  \bibfield  {author} {\bibinfo {author} {\bibfnamefont {W.}~\bibnamefont {Zeng}}, \bibinfo {author} {\bibfnamefont {B.}~\bibnamefont {Johnson}}, \bibinfo {author} {\bibfnamefont {R.}~\bibnamefont {Smith}}, \bibinfo {author} {\bibfnamefont {N.}~\bibnamefont {Rubin}}, \bibinfo {author} {\bibfnamefont {M.}~\bibnamefont {Reagor}}, \bibinfo {author} {\bibfnamefont {C.}~\bibnamefont {Ryan}},\ and\ \bibinfo {author} {\bibfnamefont {C.}~\bibnamefont {Rigetti}},\ }\bibfield  {title} {\bibinfo {title} {First quantum computers need smart software},\ }\href {https://doi.org/10.1038/549149a} {\bibfield  {journal} {\bibinfo  {journal} {Nature News}\ }\textbf {\bibinfo {volume} {549}},\ \bibinfo {pages} {149} (\bibinfo {year} {2017})}\BibitemShut {NoStop}%
\bibitem [{\citenamefont {Dargan}(2024)}]{quantuminsider}%
  \BibitemOpen
  \bibfield  {author} {\bibinfo {author} {\bibfnamefont {J.}~\bibnamefont {Dargan}},\ }\href {https://thequantuminsider.com/2022/05/27/quantum-computing-tools/} {\bibinfo {title} {The quantum insider: Top 35 open source quantum computing tools}} (\bibinfo {year} {2024})\BibitemShut {NoStop}%
\bibitem [{\citenamefont {Steiger}\ \emph {et~al.}(2018)\citenamefont {Steiger}, \citenamefont {H\"{a}ner},\ and\ \citenamefont {Troyer}}]{steiger2016projectq}%
  \BibitemOpen
  \bibfield  {author} {\bibinfo {author} {\bibfnamefont {D.~S.}\ \bibnamefont {Steiger}}, \bibinfo {author} {\bibfnamefont {T.}~\bibnamefont {H\"{a}ner}},\ and\ \bibinfo {author} {\bibfnamefont {M.}~\bibnamefont {Troyer}},\ }\bibfield  {title} {\bibinfo {title} {Project{Q}: an open source software framework for quantum computing},\ }\href {http://dx.doi.org/10.22331/q-2018-01-31-49} {\bibfield  {journal} {\bibinfo  {journal} {Quantum}\ }\textbf {\bibinfo {volume} {2}},\ \bibinfo {pages} {49} (\bibinfo {year} {2018})}\BibitemShut {NoStop}%
\bibitem [{\citenamefont {McClean}\ \emph {et~al.}(2020)\citenamefont {McClean} \emph {et~al.}}]{mcclean2017openfermion}%
  \BibitemOpen
  \bibfield  {author} {\bibinfo {author} {\bibfnamefont {J.~R.}\ \bibnamefont {McClean}} \emph {et~al.},\ }\bibfield  {title} {\bibinfo {title} {{OpenFermion}: the electronic structure package for quantum computers},\ }\href {http://dx.doi.org/10.1088/2058-9565/ab8ebc} {\bibfield  {journal} {\bibinfo  {journal} {Quantum Science and Technology}\ }\textbf {\bibinfo {volume} {5}},\ \bibinfo {pages} {034014} (\bibinfo {year} {2020})}\BibitemShut {NoStop}%
\bibitem [{\citenamefont {Javadi-Abhari}\ \emph {et~al.}(2024)\citenamefont {Javadi-Abhari}, \citenamefont {Treinish}, \citenamefont {Krsulich}, \citenamefont {Wood}, \citenamefont {Lishman}, \citenamefont {Gacon}, \citenamefont {Martiel}, \citenamefont {Nation}, \citenamefont {Bishop}, \citenamefont {Cross}, \citenamefont {Johnson},\ and\ \citenamefont {Gambetta}}]{2024qiskit}%
  \BibitemOpen
  \bibfield  {author} {\bibinfo {author} {\bibfnamefont {A.}~\bibnamefont {Javadi-Abhari}}, \bibinfo {author} {\bibfnamefont {M.}~\bibnamefont {Treinish}}, \bibinfo {author} {\bibfnamefont {K.}~\bibnamefont {Krsulich}}, \bibinfo {author} {\bibfnamefont {C.~J.}\ \bibnamefont {Wood}}, \bibinfo {author} {\bibfnamefont {J.}~\bibnamefont {Lishman}}, \bibinfo {author} {\bibfnamefont {J.}~\bibnamefont {Gacon}}, \bibinfo {author} {\bibfnamefont {S.}~\bibnamefont {Martiel}}, \bibinfo {author} {\bibfnamefont {P.~D.}\ \bibnamefont {Nation}}, \bibinfo {author} {\bibfnamefont {L.~S.}\ \bibnamefont {Bishop}}, \bibinfo {author} {\bibfnamefont {A.~W.}\ \bibnamefont {Cross}}, \bibinfo {author} {\bibfnamefont {B.~R.}\ \bibnamefont {Johnson}},\ and\ \bibinfo {author} {\bibfnamefont {J.~M.}\ \bibnamefont {Gambetta}},\ }\bibfield  {title} {\bibinfo {title} {Quantum computing with {Q}iskit},\ }\href {https://arxiv.org/abs/2405.08810} {\bibfield  {journal} {\bibinfo  {journal} {arXiv:2405.08810 [quant-ph]}\ } (\bibinfo {year}
  {2024})}\BibitemShut {NoStop}%
\bibitem [{\citenamefont {Tan}(2002)}]{qotbmtl}%
  \BibitemOpen
  \bibfield  {author} {\bibinfo {author} {\bibfnamefont {S.~M.}\ \bibnamefont {Tan}},\ }\href {https://qo.phy.auckland.ac.nz/toolbox/} {\bibinfo {title} {The quantum optics toolbox for {MATLAB}}} (\bibinfo {year} {2002})\BibitemShut {NoStop}%
\bibitem [{\citenamefont {Breuer}\ and\ \citenamefont {Petruccione}(2002)}]{breuer2002theory}%
  \BibitemOpen
  \bibfield  {author} {\bibinfo {author} {\bibfnamefont {H.-P.}\ \bibnamefont {Breuer}}\ and\ \bibinfo {author} {\bibfnamefont {F.}~\bibnamefont {Petruccione}},\ }\href {https://doi.org/10.1093/acprof:oso/9780199213900.001.0001} {\emph {\bibinfo {title} {The theory of open quantum systems}}}\ (\bibinfo  {publisher} {Oxford University Press},\ \bibinfo {year} {2002})\BibitemShut {NoStop}%
\bibitem [{\citenamefont {Lidar}(2019)}]{Lidar2019}%
  \BibitemOpen
  \bibfield  {author} {\bibinfo {author} {\bibfnamefont {D.~A.}\ \bibnamefont {Lidar}},\ }\bibfield  {title} {\bibinfo {title} {{Lecture Notes on the Theory of Open Quantum Systems}},\ }\href {http://arxiv.org/abs/1902.00967} {\bibfield  {journal} {\bibinfo  {journal} {arXiv:1902.00967 [quant-ph]}\ } (\bibinfo {year} {2019})}\BibitemShut {NoStop}%
\bibitem [{\citenamefont {Virtanen}\ \emph {et~al.}(2020)\citenamefont {Virtanen}, \citenamefont {Gommers}, \citenamefont {Oliphant}, \citenamefont {Haberland}, \citenamefont {Reddy}, \citenamefont {Cournapeau}, \citenamefont {Burovski}, \citenamefont {Peterson}, \citenamefont {Weckesser}, \citenamefont {Bright} \emph {et~al.}}]{virtanen2020scipy}%
  \BibitemOpen
  \bibfield  {author} {\bibinfo {author} {\bibfnamefont {P.}~\bibnamefont {Virtanen}}, \bibinfo {author} {\bibfnamefont {R.}~\bibnamefont {Gommers}}, \bibinfo {author} {\bibfnamefont {T.~E.}\ \bibnamefont {Oliphant}}, \bibinfo {author} {\bibfnamefont {M.}~\bibnamefont {Haberland}}, \bibinfo {author} {\bibfnamefont {T.}~\bibnamefont {Reddy}}, \bibinfo {author} {\bibfnamefont {D.}~\bibnamefont {Cournapeau}}, \bibinfo {author} {\bibfnamefont {E.}~\bibnamefont {Burovski}}, \bibinfo {author} {\bibfnamefont {P.}~\bibnamefont {Peterson}}, \bibinfo {author} {\bibfnamefont {W.}~\bibnamefont {Weckesser}}, \bibinfo {author} {\bibfnamefont {J.}~\bibnamefont {Bright}}, \emph {et~al.},\ }\bibfield  {title} {\bibinfo {title} {Sci{P}y 1.0: fundamental algorithms for scientific computing in {P}ython},\ }\href {https://doi.org/10.1038/s41592-019-0686-2} {\bibfield  {journal} {\bibinfo  {journal} {Nat. Methods}\ }\textbf {\bibinfo {volume} {17}},\ \bibinfo {pages} {261} (\bibinfo {year} {2020})}\BibitemShut {NoStop}%
\bibitem [{\citenamefont {Behnel}\ \emph {et~al.}(2011)\citenamefont {Behnel}, \citenamefont {Bradshaw}, \citenamefont {Citro}, \citenamefont {Dalcin}, \citenamefont {Seljebotn},\ and\ \citenamefont {Smith}}]{behnel2010cython}%
  \BibitemOpen
  \bibfield  {author} {\bibinfo {author} {\bibfnamefont {S.}~\bibnamefont {Behnel}}, \bibinfo {author} {\bibfnamefont {R.}~\bibnamefont {Bradshaw}}, \bibinfo {author} {\bibfnamefont {C.}~\bibnamefont {Citro}}, \bibinfo {author} {\bibfnamefont {L.}~\bibnamefont {Dalcin}}, \bibinfo {author} {\bibfnamefont {D.~S.}\ \bibnamefont {Seljebotn}},\ and\ \bibinfo {author} {\bibfnamefont {K.}~\bibnamefont {Smith}},\ }\bibfield  {title} {\bibinfo {title} {Cython: The best of both worlds},\ }\href {https://doi.org/10.1109/mcse.2010.118} {\bibfield  {journal} {\bibinfo  {journal} {Computing in Science {\&} Engineering}\ }\textbf {\bibinfo {volume} {13}},\ \bibinfo {pages} {31} (\bibinfo {year} {2011})}\BibitemShut {NoStop}%
\bibitem [{\citenamefont {Hunter}(2007)}]{hunter2007matplotlib}%
  \BibitemOpen
  \bibfield  {author} {\bibinfo {author} {\bibfnamefont {J.~D.}\ \bibnamefont {Hunter}},\ }\bibfield  {title} {\bibinfo {title} {{Matplotlib: A 2D graphics environment}},\ }\href {https://www.computer.org/csdl/magazine/cs/2007/03/c3090/13rRUwbJD0A} {\bibfield  {journal} {\bibinfo  {journal} {Comput. Sci. Eng.}\ }\textbf {\bibinfo {volume} {9}},\ \bibinfo {pages} {90} (\bibinfo {year} {2007})}\BibitemShut {NoStop}%
\bibitem [{\citenamefont {Li}\ \emph {et~al.}(2022)\citenamefont {Li}, \citenamefont {Ahmed}, \citenamefont {Saraogi}, \citenamefont {Lambert}, \citenamefont {Nori}, \citenamefont {Pitchford},\ and\ \citenamefont {Shammah}}]{li2021pulselevel}%
  \BibitemOpen
  \bibfield  {author} {\bibinfo {author} {\bibfnamefont {B.}~\bibnamefont {Li}}, \bibinfo {author} {\bibfnamefont {S.}~\bibnamefont {Ahmed}}, \bibinfo {author} {\bibfnamefont {S.}~\bibnamefont {Saraogi}}, \bibinfo {author} {\bibfnamefont {N.}~\bibnamefont {Lambert}}, \bibinfo {author} {\bibfnamefont {F.}~\bibnamefont {Nori}}, \bibinfo {author} {\bibfnamefont {A.}~\bibnamefont {Pitchford}},\ and\ \bibinfo {author} {\bibfnamefont {N.}~\bibnamefont {Shammah}},\ }\bibfield  {title} {\bibinfo {title} {Pulse-level noisy quantum circuits with qutip},\ }\href {http://dx.doi.org/10.22331/q-2022-01-24-630} {\bibfield  {journal} {\bibinfo  {journal} {Quantum}\ }\textbf {\bibinfo {volume} {6}},\ \bibinfo {pages} {630} (\bibinfo {year} {2022})}\BibitemShut {NoStop}%
\bibitem [{\citenamefont {Lambert}\ \emph {et~al.}(2023)\citenamefont {Lambert}, \citenamefont {Raheja}, \citenamefont {Cross}, \citenamefont {Menczel}, \citenamefont {Ahmed}, \citenamefont {Pitchford}, \citenamefont {Burgarth},\ and\ \citenamefont {Nori}}]{lambert2020bofinheom}%
  \BibitemOpen
  \bibfield  {author} {\bibinfo {author} {\bibfnamefont {N.}~\bibnamefont {Lambert}}, \bibinfo {author} {\bibfnamefont {T.}~\bibnamefont {Raheja}}, \bibinfo {author} {\bibfnamefont {S.}~\bibnamefont {Cross}}, \bibinfo {author} {\bibfnamefont {P.}~\bibnamefont {Menczel}}, \bibinfo {author} {\bibfnamefont {S.}~\bibnamefont {Ahmed}}, \bibinfo {author} {\bibfnamefont {A.}~\bibnamefont {Pitchford}}, \bibinfo {author} {\bibfnamefont {D.}~\bibnamefont {Burgarth}},\ and\ \bibinfo {author} {\bibfnamefont {F.}~\bibnamefont {Nori}},\ }\bibfield  {title} {\bibinfo {title} {{QuTiP-BoFiN}: A bosonic and fermionic numerical hierarchical-equations-of-motion library with applications in light-harvesting, quantum control, and single-molecule electronics},\ }\href {http://dx.doi.org/10.1103/PhysRevResearch.5.013181} {\bibfield  {journal} {\bibinfo  {journal} {Physical Review Research}\ }\textbf {\bibinfo {volume} {5}} (\bibinfo {year} {2023})}\BibitemShut {NoStop}%
\bibitem [{\citenamefont {Shammah}\ \emph {et~al.}(2018)\citenamefont {Shammah}, \citenamefont {Ahmed}, \citenamefont {Lambert}, \citenamefont {De~Liberato},\ and\ \citenamefont {Nori}}]{Shammah_2018}%
  \BibitemOpen
  \bibfield  {author} {\bibinfo {author} {\bibfnamefont {N.}~\bibnamefont {Shammah}}, \bibinfo {author} {\bibfnamefont {S.}~\bibnamefont {Ahmed}}, \bibinfo {author} {\bibfnamefont {N.}~\bibnamefont {Lambert}}, \bibinfo {author} {\bibfnamefont {S.}~\bibnamefont {De~Liberato}},\ and\ \bibinfo {author} {\bibfnamefont {F.}~\bibnamefont {Nori}},\ }\bibfield  {title} {\bibinfo {title} {Open quantum systems with local and collective incoherent processes: Efficient numerical simulations using permutational invariance},\ }\href {http://dx.doi.org/10.1103/PhysRevA.98.063815} {\bibfield  {journal} {\bibinfo  {journal} {Phys. Rev. A}\ }\textbf {\bibinfo {volume} {98}},\ \bibinfo {pages} {063815} (\bibinfo {year} {2018})}\BibitemShut {NoStop}%
\bibitem [{\citenamefont {{Google Summer of Code}}()}]{gsoc}%
  \BibitemOpen
  \bibfield  {author} {\bibinfo {author} {\bibnamefont {{Google Summer of Code}}},\ }\href {https://summerofcode.withgoogle.com/} {\bibinfo {title} {https://summerofcode.withgoogle.com/}}\BibitemShut {NoStop}%
\bibitem [{qgr()}]{qgrad}%
  \BibitemOpen
  \href {https://github.com/qgrad/qgrad} {\bibinfo {title} {https://github.com/qgrad/qgrad}}\BibitemShut {NoStop}%
\bibitem [{qja()}]{qjax}%
  \BibitemOpen
  \href {https://github.com/qutip/qutip-jax} {\bibinfo {title} {https://github.com/qutip/qutip-jax}}\BibitemShut {NoStop}%
\bibitem [{\citenamefont {Bradbury}\ \emph {et~al.}(2018)\citenamefont {Bradbury}, \citenamefont {Frostig}, \citenamefont {Hawkins}, \citenamefont {Johnson}, \citenamefont {Leary}, \citenamefont {Maclaurin}, \citenamefont {Necula}, \citenamefont {Paszke}, \citenamefont {Vander{P}las}, \citenamefont {Wanderman-{M}ilne},\ and\ \citenamefont {Zhang}}]{jax2018github}%
  \BibitemOpen
  \bibfield  {author} {\bibinfo {author} {\bibfnamefont {J.}~\bibnamefont {Bradbury}}, \bibinfo {author} {\bibfnamefont {R.}~\bibnamefont {Frostig}}, \bibinfo {author} {\bibfnamefont {P.}~\bibnamefont {Hawkins}}, \bibinfo {author} {\bibfnamefont {M.~J.}\ \bibnamefont {Johnson}}, \bibinfo {author} {\bibfnamefont {C.}~\bibnamefont {Leary}}, \bibinfo {author} {\bibfnamefont {D.}~\bibnamefont {Maclaurin}}, \bibinfo {author} {\bibfnamefont {G.}~\bibnamefont {Necula}}, \bibinfo {author} {\bibfnamefont {A.}~\bibnamefont {Paszke}}, \bibinfo {author} {\bibfnamefont {J.}~\bibnamefont {Vander{P}las}}, \bibinfo {author} {\bibfnamefont {S.}~\bibnamefont {Wanderman-{M}ilne}},\ and\ \bibinfo {author} {\bibfnamefont {Q.}~\bibnamefont {Zhang}},\ }\href {http://github.com/google/jax} {\bibinfo {title} {{JAX}: composable transformations of {P}ython+{N}um{P}y programs}} (\bibinfo {year} {2018})\BibitemShut {NoStop}%
\bibitem [{cup()}]{cupy}%
  \BibitemOpen
  \href {https://github.com/qutip/qutip-cupy} {\bibinfo {title} {https://github.com/qutip/qutip-cupy}}\BibitemShut {NoStop}%
\bibitem [{tfl()}]{tflow}%
  \BibitemOpen
  \href {https://github.com/qutip/qutip-tensorflow} {\bibinfo {title} {https://github.com/qutip/qutip-tensorflow}}\BibitemShut {NoStop}%
\bibitem [{tne()}]{tnetwork}%
  \BibitemOpen
  \href {https://github.com/qutip/qutip-tensornetwork} {\bibinfo {title} {https://github.com/qutip/qutip-tensornetwork}}\BibitemShut {NoStop}%
\bibitem [{arr()}]{arrayconsort}%
  \BibitemOpen
  \href {https://data-apis.org/array-api/latest} {\bibinfo {title} {https://data-apis.org/array-api/latest}}\BibitemShut {NoStop}%
\bibitem [{\citenamefont {Guilmin}\ \emph {et~al.}(2024)\citenamefont {Guilmin}, \citenamefont {Gautier}, \citenamefont {Bocquet},\ and\ \citenamefont {Genois}}]{guilmin2024dynamiqs}%
  \BibitemOpen
  \bibfield  {author} {\bibinfo {author} {\bibfnamefont {P.}~\bibnamefont {Guilmin}}, \bibinfo {author} {\bibfnamefont {R.}~\bibnamefont {Gautier}}, \bibinfo {author} {\bibfnamefont {A.}~\bibnamefont {Bocquet}},\ and\ \bibinfo {author} {\bibfnamefont {{\'{E}}.}~\bibnamefont {Genois}},\ }\bibfield  {title} {\bibinfo {title} {{Dynamiqs}: an open-source {P}ython library for {GPU}-accelerated and differentiable simulation of quantum systems}} (\bibinfo {year} {2024})\BibitemShut {NoStop}%
\bibitem [{\citenamefont {Luo}\ \emph {et~al.}(2020)\citenamefont {Luo}, \citenamefont {Liu}, \citenamefont {Zhang},\ and\ \citenamefont {Wang}}]{YaoFramework2019}%
  \BibitemOpen
  \bibfield  {author} {\bibinfo {author} {\bibfnamefont {X.-Z.}\ \bibnamefont {Luo}}, \bibinfo {author} {\bibfnamefont {J.-G.}\ \bibnamefont {Liu}}, \bibinfo {author} {\bibfnamefont {P.}~\bibnamefont {Zhang}},\ and\ \bibinfo {author} {\bibfnamefont {L.}~\bibnamefont {Wang}},\ }\bibfield  {title} {\bibinfo {title} {Yao.jl: {E}xtensible, {E}fficient {F}ramework for {Q}uantum {A}lgorithm {D}esign},\ }\href {https://doi.org/10.22331/q-2020-10-11-341} {\bibfield  {journal} {\bibinfo  {journal} {{Quantum}}\ }\textbf {\bibinfo {volume} {4}},\ \bibinfo {pages} {341} (\bibinfo {year} {2020})}\BibitemShut {NoStop}%
\bibitem [{\citenamefont {Nation}(2015)}]{nation2015steadystate}%
  \BibitemOpen
  \bibfield  {author} {\bibinfo {author} {\bibfnamefont {P.~D.}\ \bibnamefont {Nation}},\ }\bibfield  {title} {\bibinfo {title} {Steady-state solution methods for open quantum optical systems},\ }\href {https://arxiv.org/abs/1504.06768} {\bibfield  {journal} {\bibinfo  {journal} {arXiv:1504.06768 [quant-ph]}\ } (\bibinfo {year} {2015})}\BibitemShut {NoStop}%
\bibitem [{\citenamefont {Dalibard}\ \emph {et~al.}(1992)\citenamefont {Dalibard}, \citenamefont {Castin},\ and\ \citenamefont {M\o{}lmer}}]{mc1}%
  \BibitemOpen
  \bibfield  {author} {\bibinfo {author} {\bibfnamefont {J.}~\bibnamefont {Dalibard}}, \bibinfo {author} {\bibfnamefont {Y.}~\bibnamefont {Castin}},\ and\ \bibinfo {author} {\bibfnamefont {K.}~\bibnamefont {M\o{}lmer}},\ }\bibfield  {title} {\bibinfo {title} {Wave-function approach to dissipative processes in quantum optics},\ }\href {https://link.aps.org/doi/10.1103/PhysRevLett.68.580} {\bibfield  {journal} {\bibinfo  {journal} {Phys. Rev. Lett.}\ }\textbf {\bibinfo {volume} {68}},\ \bibinfo {pages} {580} (\bibinfo {year} {1992})}\BibitemShut {NoStop}%
\bibitem [{\citenamefont {Gneiting}\ \emph {et~al.}(2022)\citenamefont {Gneiting}, \citenamefont {Koottandavida}, \citenamefont {Rozhkov},\ and\ \citenamefont {Nori}}]{PhysRevResearch.4.023036}%
  \BibitemOpen
  \bibfield  {author} {\bibinfo {author} {\bibfnamefont {C.}~\bibnamefont {Gneiting}}, \bibinfo {author} {\bibfnamefont {A.}~\bibnamefont {Koottandavida}}, \bibinfo {author} {\bibfnamefont {A.~V.}\ \bibnamefont {Rozhkov}},\ and\ \bibinfo {author} {\bibfnamefont {F.}~\bibnamefont {Nori}},\ }\bibfield  {title} {\bibinfo {title} {Unraveling the topology of dissipative quantum systems},\ }\href {https://doi.org/10.1103/PhysRevResearch.4.023036} {\bibfield  {journal} {\bibinfo  {journal} {Phys. Rev. Res.}\ }\textbf {\bibinfo {volume} {4}},\ \bibinfo {pages} {023036} (\bibinfo {year} {2022})}\BibitemShut {NoStop}%
\bibitem [{\citenamefont {Gneiting}\ \emph {et~al.}(2021)\citenamefont {Gneiting}, \citenamefont {Rozhkov},\ and\ \citenamefont {Nori}}]{PhysRevA.104.062212}%
  \BibitemOpen
  \bibfield  {author} {\bibinfo {author} {\bibfnamefont {C.}~\bibnamefont {Gneiting}}, \bibinfo {author} {\bibfnamefont {A.~V.}\ \bibnamefont {Rozhkov}},\ and\ \bibinfo {author} {\bibfnamefont {F.}~\bibnamefont {Nori}},\ }\bibfield  {title} {\bibinfo {title} {Jump-time unraveling of {M}arkovian open quantum systems},\ }\href {https://doi.org/10.1103/PhysRevA.104.062212} {\bibfield  {journal} {\bibinfo  {journal} {Phys. Rev. A}\ }\textbf {\bibinfo {volume} {104}},\ \bibinfo {pages} {062212} (\bibinfo {year} {2021})}\BibitemShut {NoStop}%
\bibitem [{\citenamefont {Dum}\ \emph {et~al.}(1992)\citenamefont {Dum}, \citenamefont {Zoller},\ and\ \citenamefont {Ritsch}}]{mc2}%
  \BibitemOpen
  \bibfield  {author} {\bibinfo {author} {\bibfnamefont {R.}~\bibnamefont {Dum}}, \bibinfo {author} {\bibfnamefont {P.}~\bibnamefont {Zoller}},\ and\ \bibinfo {author} {\bibfnamefont {H.}~\bibnamefont {Ritsch}},\ }\bibfield  {title} {\bibinfo {title} {Monte carlo simulation of the atomic master equation for spontaneous emission},\ }\href {https://link.aps.org/doi/10.1103/PhysRevA.45.4879} {\bibfield  {journal} {\bibinfo  {journal} {Phys. Rev. A}\ }\textbf {\bibinfo {volume} {45}},\ \bibinfo {pages} {4879} (\bibinfo {year} {1992})}\BibitemShut {NoStop}%
\bibitem [{\citenamefont {M{\o}lmer}\ \emph {et~al.}(1993)\citenamefont {M{\o}lmer}, \citenamefont {Castin},\ and\ \citenamefont {Dalibard}}]{mc3}%
  \BibitemOpen
  \bibfield  {author} {\bibinfo {author} {\bibfnamefont {K.}~\bibnamefont {M{\o}lmer}}, \bibinfo {author} {\bibfnamefont {Y.}~\bibnamefont {Castin}},\ and\ \bibinfo {author} {\bibfnamefont {J.}~\bibnamefont {Dalibard}},\ }\bibfield  {title} {\bibinfo {title} {Monte carlo wave-function method in quantum optics},\ }\href {https://opg.optica.org/josab/abstract.cfm?URI=josab-10-3-524} {\bibfield  {journal} {\bibinfo  {journal} {J. Opt. Soc. Am. B}\ }\textbf {\bibinfo {volume} {10}},\ \bibinfo {pages} {524} (\bibinfo {year} {1993})}\BibitemShut {NoStop}%
\bibitem [{\citenamefont {Radaelli}\ \emph {et~al.}(2024)\citenamefont {Radaelli}, \citenamefont {Landi},\ and\ \citenamefont {Binder}}]{mcnew}%
  \BibitemOpen
  \bibfield  {author} {\bibinfo {author} {\bibfnamefont {M.}~\bibnamefont {Radaelli}}, \bibinfo {author} {\bibfnamefont {G.~T.}\ \bibnamefont {Landi}},\ and\ \bibinfo {author} {\bibfnamefont {F.~C.}\ \bibnamefont {Binder}},\ }\bibfield  {title} {\bibinfo {title} {Gillespie algorithm for quantum jump trajectories},\ }\href {https://doi.org/10.1103/PhysRevA.110.062212} {\bibfield  {journal} {\bibinfo  {journal} {Phys. Rev. A}\ }\textbf {\bibinfo {volume} {110}},\ \bibinfo {pages} {062212} (\bibinfo {year} {2024})}\BibitemShut {NoStop}%
\bibitem [{\citenamefont {Breuer}(2004)}]{BreuerPhysRevA2004}%
  \BibitemOpen
  \bibfield  {author} {\bibinfo {author} {\bibfnamefont {H.-P.}\ \bibnamefont {Breuer}},\ }\bibfield  {title} {\bibinfo {title} {Genuine quantum trajectories for non-{{Markovian}} processes},\ }\href {https://link.aps.org/doi/10.1103/PhysRevA.70.012106} {\bibfield  {journal} {\bibinfo  {journal} {Phys. Rev. A}\ }\textbf {\bibinfo {volume} {70}},\ \bibinfo {pages} {012106} (\bibinfo {year} {2004})}\BibitemShut {NoStop}%
\bibitem [{\citenamefont {Piilo}\ \emph {et~al.}(2008)\citenamefont {Piilo}, \citenamefont {Maniscalco}, \citenamefont {H{\"a}rk{\"o}nen},\ and\ \citenamefont {Suominen}}]{PiiloPhysRevLett2008}%
  \BibitemOpen
  \bibfield  {author} {\bibinfo {author} {\bibfnamefont {J.}~\bibnamefont {Piilo}}, \bibinfo {author} {\bibfnamefont {S.}~\bibnamefont {Maniscalco}}, \bibinfo {author} {\bibfnamefont {K.}~\bibnamefont {H{\"a}rk{\"o}nen}},\ and\ \bibinfo {author} {\bibfnamefont {K.-A.}\ \bibnamefont {Suominen}},\ }\bibfield  {title} {\bibinfo {title} {Non-{{Markovian Quantum Jumps}}},\ }\href {https://link.aps.org/doi/10.1103/PhysRevLett.100.180402} {\bibfield  {journal} {\bibinfo  {journal} {Phys. Rev. Lett.}\ }\textbf {\bibinfo {volume} {100}},\ \bibinfo {pages} {180402} (\bibinfo {year} {2008})}\BibitemShut {NoStop}%
\bibitem [{\citenamefont {Donvil}\ and\ \citenamefont {{Muratore-Ginanneschi}}(2023{\natexlab{a}})}]{DonvilOpenSystInfDyn2023}%
  \BibitemOpen
  \bibfield  {author} {\bibinfo {author} {\bibfnamefont {B.~I.~C.}\ \bibnamefont {Donvil}}\ and\ \bibinfo {author} {\bibfnamefont {P.}~\bibnamefont {{Muratore-Ginanneschi}}},\ }\bibfield  {title} {\bibinfo {title} {On the {{Unraveling}} of {{Open Quantum Dynamics}}},\ }\href {https://www.worldscientific.com/worldscinet/osid} {\bibfield  {journal} {\bibinfo  {journal} {Open Syst. Inf. Dyn.}\ } (\bibinfo {year} {2023}{\natexlab{a}})}\BibitemShut {NoStop}%
\bibitem [{\citenamefont {Breuer}\ \emph {et~al.}(1999)\citenamefont {Breuer}, \citenamefont {Kappler},\ and\ \citenamefont {Petruccione}}]{BreuerPhysRevA1999}%
  \BibitemOpen
  \bibfield  {author} {\bibinfo {author} {\bibfnamefont {H.-P.}\ \bibnamefont {Breuer}}, \bibinfo {author} {\bibfnamefont {B.}~\bibnamefont {Kappler}},\ and\ \bibinfo {author} {\bibfnamefont {F.}~\bibnamefont {Petruccione}},\ }\bibfield  {title} {\bibinfo {title} {Stochastic wave-function method for non-{{Markovian}} quantum master equations},\ }\href {https://link.aps.org/doi/10.1103/PhysRevA.59.1633} {\bibfield  {journal} {\bibinfo  {journal} {Phys. Rev. A}\ }\textbf {\bibinfo {volume} {59}},\ \bibinfo {pages} {1633} (\bibinfo {year} {1999})}\BibitemShut {NoStop}%
\bibitem [{\citenamefont {Hush}\ \emph {et~al.}(2015)\citenamefont {Hush}, \citenamefont {Lesanovsky},\ and\ \citenamefont {Garrahan}}]{HushPhysRevA2015}%
  \BibitemOpen
  \bibfield  {author} {\bibinfo {author} {\bibfnamefont {M.~R.}\ \bibnamefont {Hush}}, \bibinfo {author} {\bibfnamefont {I.}~\bibnamefont {Lesanovsky}},\ and\ \bibinfo {author} {\bibfnamefont {J.~P.}\ \bibnamefont {Garrahan}},\ }\bibfield  {title} {\bibinfo {title} {Generic map from non-{{Lindblad}} to {{Lindblad}} master equations},\ }\href {https://link.aps.org/doi/10.1103/PhysRevA.91.032113} {\bibfield  {journal} {\bibinfo  {journal} {Phys. Rev. A}\ }\textbf {\bibinfo {volume} {91}},\ \bibinfo {pages} {032113} (\bibinfo {year} {2015})}\BibitemShut {NoStop}%
\bibitem [{\citenamefont {Menczel}\ \emph {et~al.}(2024)\citenamefont {Menczel}, \citenamefont {Funo}, \citenamefont {Cirio}, \citenamefont {Lambert},\ and\ \citenamefont {Nori}}]{MenczelArXiv240111830Quant-Ph2024}%
  \BibitemOpen
  \bibfield  {author} {\bibinfo {author} {\bibfnamefont {P.}~\bibnamefont {Menczel}}, \bibinfo {author} {\bibfnamefont {K.}~\bibnamefont {Funo}}, \bibinfo {author} {\bibfnamefont {M.}~\bibnamefont {Cirio}}, \bibinfo {author} {\bibfnamefont {N.}~\bibnamefont {Lambert}},\ and\ \bibinfo {author} {\bibfnamefont {F.}~\bibnamefont {Nori}},\ }\bibfield  {title} {\bibinfo {title} {Non-hermitian pseudomodes for strongly coupled open quantum systems: Unravelings, correlations, and thermodynamics},\ }\href {https://doi.org/10.1103/PhysRevResearch.6.033237} {\bibfield  {journal} {\bibinfo  {journal} {Phys. Rev. Res.}\ }\textbf {\bibinfo {volume} {6}},\ \bibinfo {pages} {033237} (\bibinfo {year} {2024})}\BibitemShut {NoStop}%
\bibitem [{\citenamefont {Donvil}\ and\ \citenamefont {{Muratore-Ginanneschi}}(2022)}]{DonvilNatCommun2022}%
  \BibitemOpen
  \bibfield  {author} {\bibinfo {author} {\bibfnamefont {B.}~\bibnamefont {Donvil}}\ and\ \bibinfo {author} {\bibfnamefont {P.}~\bibnamefont {{Muratore-Ginanneschi}}},\ }\bibfield  {title} {\bibinfo {title} {Quantum trajectory framework for general time-local master equations},\ }\href {https://www.nature.com/articles/s41467-022-31533-8} {\bibfield  {journal} {\bibinfo  {journal} {Nat Commun}\ }\textbf {\bibinfo {volume} {13}},\ \bibinfo {pages} {4140} (\bibinfo {year} {2022})}\BibitemShut {NoStop}%
\bibitem [{\citenamefont {Donvil}\ and\ \citenamefont {{Muratore-Ginanneschi}}(2023{\natexlab{b}})}]{DonvilNewJPhys2023}%
  \BibitemOpen
  \bibfield  {author} {\bibinfo {author} {\bibfnamefont {B.}~\bibnamefont {Donvil}}\ and\ \bibinfo {author} {\bibfnamefont {P.}~\bibnamefont {{Muratore-Ginanneschi}}},\ }\bibfield  {title} {\bibinfo {title} {Unraveling-paired dynamical maps recover the input of quantum channels},\ }\href {https://dx.doi.org/10.1088/1367-2630/acd4dc} {\bibfield  {journal} {\bibinfo  {journal} {New J. Phys.}\ }\textbf {\bibinfo {volume} {25}},\ \bibinfo {pages} {053031} (\bibinfo {year} {2023}{\natexlab{b}})}\BibitemShut {NoStop}%
\bibitem [{\citenamefont {Davidovi{\'c}}(2020)}]{DavidovicQuantum2020}%
  \BibitemOpen
  \bibfield  {author} {\bibinfo {author} {\bibfnamefont {D.}~\bibnamefont {Davidovi{\'c}}},\ }\bibfield  {title} {\bibinfo {title} {Completely {{Positive}}, {{Simple}}, and {{Possibly Highly Accurate Approximation}} of the {{Redfield Equation}}},\ }\href {https://quantum-journal.org/papers/q-2020-09-21-326/} {\bibfield  {journal} {\bibinfo  {journal} {Quantum}\ }\textbf {\bibinfo {volume} {4}},\ \bibinfo {pages} {326} (\bibinfo {year} {2020})}\BibitemShut {NoStop}%
\bibitem [{\citenamefont {Gneiting}(2020)}]{GneitingPhysRevB2020}%
  \BibitemOpen
  \bibfield  {author} {\bibinfo {author} {\bibfnamefont {C.}~\bibnamefont {Gneiting}},\ }\bibfield  {title} {\bibinfo {title} {Disorder-dressed quantum evolution},\ }\href {https://link.aps.org/doi/10.1103/PhysRevB.101.214203} {\bibfield  {journal} {\bibinfo  {journal} {Phys. Rev. B}\ }\textbf {\bibinfo {volume} {101}},\ \bibinfo {pages} {214203} (\bibinfo {year} {2020})}\BibitemShut {NoStop}%
\bibitem [{\citenamefont {Groszkowski}\ \emph {et~al.}(2023)\citenamefont {Groszkowski}, \citenamefont {Seif}, \citenamefont {Koch},\ and\ \citenamefont {Clerk}}]{GroszkowskiQuantum2023}%
  \BibitemOpen
  \bibfield  {author} {\bibinfo {author} {\bibfnamefont {P.}~\bibnamefont {Groszkowski}}, \bibinfo {author} {\bibfnamefont {A.}~\bibnamefont {Seif}}, \bibinfo {author} {\bibfnamefont {J.}~\bibnamefont {Koch}},\ and\ \bibinfo {author} {\bibfnamefont {A.~A.}\ \bibnamefont {Clerk}},\ }\bibfield  {title} {\bibinfo {title} {Simple master equations for describing driven systems subject to classical non-{{Markovian}} noise},\ }\href {https://quantum-journal.org/papers/q-2023-04-06-972/} {\bibfield  {journal} {\bibinfo  {journal} {Quantum}\ }\textbf {\bibinfo {volume} {7}},\ \bibinfo {pages} {972} (\bibinfo {year} {2023})}\BibitemShut {NoStop}%
\bibitem [{\citenamefont {Floquet}(1883)}]{Floquet1883}%
  \BibitemOpen
  \bibfield  {author} {\bibinfo {author} {\bibfnamefont {G.}~\bibnamefont {Floquet}},\ }\bibfield  {title} {{\selectlanguage {french}\bibinfo {title} {Sur les \'equations diff\'erentielles lin\'eaires \`a coefficients p\'eriodiques}},\ }\href {http://www.numdam.org/articles/10.24033/asens.220/} {\bibfield  {journal} {\bibinfo  {journal} {Annales scientifiques de l'\'Ecole Normale Sup\'erieure}\ }\textbf {\bibinfo {volume} {2e s{\'e}rie, 12}},\ \bibinfo {pages} {47} (\bibinfo {year} {1883})}\BibitemShut {NoStop}%
\bibitem [{\citenamefont {Shirley}(1965)}]{Shirley1965}%
  \BibitemOpen
  \bibfield  {author} {\bibinfo {author} {\bibfnamefont {J.~H.}\ \bibnamefont {Shirley}},\ }\bibfield  {title} {\bibinfo {title} {Solution of the {S}chr\"odinger equation with a {H}amiltonian periodic in time},\ }\href {https://link.aps.org/doi/10.1103/PhysRev.138.B979} {\bibfield  {journal} {\bibinfo  {journal} {Phys. Rev.}\ }\textbf {\bibinfo {volume} {138}},\ \bibinfo {pages} {B979} (\bibinfo {year} {1965})}\BibitemShut {NoStop}%
\bibitem [{\citenamefont {Grifoni}\ and\ \citenamefont {Hänggi}(1998)}]{Grifoni1998}%
  \BibitemOpen
  \bibfield  {author} {\bibinfo {author} {\bibfnamefont {M.}~\bibnamefont {Grifoni}}\ and\ \bibinfo {author} {\bibfnamefont {P.}~\bibnamefont {Hänggi}},\ }\bibfield  {title} {\bibinfo {title} {Driven quantum tunneling},\ }\href {https://www.sciencedirect.com/science/article/pii/S0370157398000222} {\bibfield  {journal} {\bibinfo  {journal} {Physics Reports}\ }\textbf {\bibinfo {volume} {304}},\ \bibinfo {pages} {229} (\bibinfo {year} {1998})}\BibitemShut {NoStop}%
\bibitem [{\citenamefont {Creffield}(2003)}]{Creffield2003}%
  \BibitemOpen
  \bibfield  {author} {\bibinfo {author} {\bibfnamefont {C.}~\bibnamefont {Creffield}},\ }\bibfield  {title} {\bibinfo {title} {Location of crossings in the {F}loquet spectrum of a driven two-level system},\ }\href {https://link.aps.org/doi/10.1103/PhysRevB.67.165301} {\bibfield  {journal} {\bibinfo  {journal} {Phys. Rev. B}\ }\textbf {\bibinfo {volume} {67}},\ \bibinfo {pages} {165301} (\bibinfo {year} {2003})}\BibitemShut {NoStop}%
\bibitem [{\citenamefont {Clawson}\ and\ \citenamefont {Flagg}(2024)}]{fenton}%
  \BibitemOpen
  \bibfield  {author} {\bibinfo {author} {\bibfnamefont {F.}~\bibnamefont {Clawson}}\ and\ \bibinfo {author} {\bibfnamefont {E.~B.}\ \bibnamefont {Flagg}},\ }\bibfield  {title} {\bibinfo {title} {{Floquet-Lindblad} master equation approach to open quantum system dynamics},\ }\href {https://arxiv.org/abs/2410.18046} {\bibfield  {journal} {\bibinfo  {journal} {arXiv:2410.18046 [quant-ph]}\ } (\bibinfo {year} {2024})}\BibitemShut {NoStop}%
\bibitem [{\citenamefont {Tanimura}(2020)}]{tanimura_numerically_2020}%
  \BibitemOpen
  \bibfield  {author} {\bibinfo {author} {\bibfnamefont {Y.}~\bibnamefont {Tanimura}},\ }\bibfield  {title} {\bibinfo {title} {Numerically “exact” approach to open quantum dynamics: {The} hierarchical equations of motion ({HEOM})},\ }\href {https://doi.org/10.1063/5.0011599} {\bibfield  {journal} {\bibinfo  {journal} {The Journal of Chemical Physics}\ }\textbf {\bibinfo {volume} {153}},\ \bibinfo {pages} {020901} (\bibinfo {year} {2020})}\BibitemShut {NoStop}%
\bibitem [{\citenamefont {Tanimura}\ and\ \citenamefont {Kubo}(1989)}]{Tanimura1989}%
  \BibitemOpen
  \bibfield  {author} {\bibinfo {author} {\bibfnamefont {Y.}~\bibnamefont {Tanimura}}\ and\ \bibinfo {author} {\bibfnamefont {R.}~\bibnamefont {Kubo}},\ }\bibfield  {title} {\bibinfo {title} {Time evolution of a quantum system in contact with a nearly gaussian-markoffian noise bath},\ }\href {https://doi.org/10.1143/JPSJ.58.101} {\bibfield  {journal} {\bibinfo  {journal} {Journal of the Physical Society of Japan}\ }\textbf {\bibinfo {volume} {58}},\ \bibinfo {pages} {101} (\bibinfo {year} {1989})}\BibitemShut {NoStop}%
\bibitem [{\citenamefont {Cygorek}\ and\ \citenamefont {Gauger}(2024)}]{10.1063/5.0221182}%
  \BibitemOpen
  \bibfield  {author} {\bibinfo {author} {\bibfnamefont {M.}~\bibnamefont {Cygorek}}\ and\ \bibinfo {author} {\bibfnamefont {E.~M.}\ \bibnamefont {Gauger}},\ }\bibfield  {title} {\bibinfo {title} {{ACE}: A general-purpose non-{M}arkovian open quantum systems simulation toolkit based on process tensors},\ }\href {https://doi.org/10.1063/5.0221182} {\bibfield  {journal} {\bibinfo  {journal} {The Journal of Chemical Physics}\ }\textbf {\bibinfo {volume} {161}},\ \bibinfo {pages} {074111} (\bibinfo {year} {2024})}\BibitemShut {NoStop}%
\bibitem [{\citenamefont {Cirio}\ \emph {et~al.}(2023)\citenamefont {Cirio}, \citenamefont {Lambert}, \citenamefont {Liang}, \citenamefont {Kuo}, \citenamefont {Chen}, \citenamefont {Menczel}, \citenamefont {Funo},\ and\ \citenamefont {Nori}}]{PhysRevResearch.5.033011}%
  \BibitemOpen
  \bibfield  {author} {\bibinfo {author} {\bibfnamefont {M.}~\bibnamefont {Cirio}}, \bibinfo {author} {\bibfnamefont {N.}~\bibnamefont {Lambert}}, \bibinfo {author} {\bibfnamefont {P.}~\bibnamefont {Liang}}, \bibinfo {author} {\bibfnamefont {P.-C.}\ \bibnamefont {Kuo}}, \bibinfo {author} {\bibfnamefont {Y.-N.}\ \bibnamefont {Chen}}, \bibinfo {author} {\bibfnamefont {P.}~\bibnamefont {Menczel}}, \bibinfo {author} {\bibfnamefont {K.}~\bibnamefont {Funo}},\ and\ \bibinfo {author} {\bibfnamefont {F.}~\bibnamefont {Nori}},\ }\bibfield  {title} {\bibinfo {title} {Pseudofermion method for the exact description of fermionic environments: {F}rom single-molecule electronics to the {K}ondo resonance},\ }\href {https://doi.org/10.1103/PhysRevResearch.5.033011} {\bibfield  {journal} {\bibinfo  {journal} {Phys. Rev. Res.}\ }\textbf {\bibinfo {volume} {5}},\ \bibinfo {pages} {033011} (\bibinfo {year} {2023})}\BibitemShut {NoStop}%
\bibitem [{\citenamefont {Kuo}\ \emph {et~al.}(2023)\citenamefont {Kuo}, \citenamefont {Lambert}, \citenamefont {Cirio}, \citenamefont {Huang}, \citenamefont {Nori},\ and\ \citenamefont {Chen}}]{PhysRevResearch.5.043177}%
  \BibitemOpen
  \bibfield  {author} {\bibinfo {author} {\bibfnamefont {P.-C.}\ \bibnamefont {Kuo}}, \bibinfo {author} {\bibfnamefont {N.}~\bibnamefont {Lambert}}, \bibinfo {author} {\bibfnamefont {M.}~\bibnamefont {Cirio}}, \bibinfo {author} {\bibfnamefont {Y.-T.}\ \bibnamefont {Huang}}, \bibinfo {author} {\bibfnamefont {F.}~\bibnamefont {Nori}},\ and\ \bibinfo {author} {\bibfnamefont {Y.-N.}\ \bibnamefont {Chen}},\ }\bibfield  {title} {\bibinfo {title} {Kondo {QED}: The {K}ondo effect and photon trapping in a two-impurity {A}nderson model ultrastrongly coupled to light},\ }\href {https://doi.org/10.1103/PhysRevResearch.5.043177} {\bibfield  {journal} {\bibinfo  {journal} {Phys. Rev. Res.}\ }\textbf {\bibinfo {volume} {5}},\ \bibinfo {pages} {043177} (\bibinfo {year} {2023})}\BibitemShut {NoStop}%
\bibitem [{\citenamefont {Rahman}\ and\ \citenamefont {Kleinekathöfer}(2019)}]{Rahman_Kleinekathöfer_2019}%
  \BibitemOpen
  \bibfield  {author} {\bibinfo {author} {\bibfnamefont {H.}~\bibnamefont {Rahman}}\ and\ \bibinfo {author} {\bibfnamefont {U.}~\bibnamefont {Kleinekathöfer}},\ }\bibfield  {title} {\bibinfo {title} {Chebyshev hierarchical equations of motion for systems with arbitrary spectral densities and temperatures},\ }\href {https://doi.org/10.1063/1.5100102} {\bibfield  {journal} {\bibinfo  {journal} {The Journal of Chemical Physics}\ }\textbf {\bibinfo {volume} {150}},\ \bibinfo {pages} {244104} (\bibinfo {year} {2019})}\BibitemShut {NoStop}%
\bibitem [{\citenamefont {Ishizaki}\ and\ \citenamefont {Fleming}(2009)}]{IshizakiProcNatlAcadSci2009}%
  \BibitemOpen
  \bibfield  {author} {\bibinfo {author} {\bibfnamefont {A.}~\bibnamefont {Ishizaki}}\ and\ \bibinfo {author} {\bibfnamefont {G.~R.}\ \bibnamefont {Fleming}},\ }\bibfield  {title} {\bibinfo {title} {Theoretical examination of quantum coherence in a photosynthetic system at physiological temperature},\ }\href {https://doi.org/10.1073/pnas.0908989106} {\bibfield  {journal} {\bibinfo  {journal} {PNAS}\ }\textbf {\bibinfo {volume} {106}},\ \bibinfo {pages} {17255} (\bibinfo {year} {2009})}\BibitemShut {NoStop}%
\bibitem [{\citenamefont {H\"artle}\ \emph {et~al.}(2013)\citenamefont {H\"artle}, \citenamefont {Cohen}, \citenamefont {Reichman},\ and\ \citenamefont {Millis}}]{PhysRevB.88.235426}%
  \BibitemOpen
  \bibfield  {author} {\bibinfo {author} {\bibfnamefont {R.}~\bibnamefont {H\"artle}}, \bibinfo {author} {\bibfnamefont {G.}~\bibnamefont {Cohen}}, \bibinfo {author} {\bibfnamefont {D.~R.}\ \bibnamefont {Reichman}},\ and\ \bibinfo {author} {\bibfnamefont {A.~J.}\ \bibnamefont {Millis}},\ }\bibfield  {title} {\bibinfo {title} {Decoherence and lead-induced interdot coupling in nonequilibrium electron transport through interacting quantum dots: A hierarchical quantum master equation approach},\ }\href {https://doi.org/10.1103/PhysRevB.88.235426} {\bibfield  {journal} {\bibinfo  {journal} {Phys. Rev. B}\ }\textbf {\bibinfo {volume} {88}},\ \bibinfo {pages} {235426} (\bibinfo {year} {2013})}\BibitemShut {NoStop}%
\bibitem [{\citenamefont {Wenderoth}\ \emph {et~al.}(2016)\citenamefont {Wenderoth}, \citenamefont {B\"atge},\ and\ \citenamefont {H\"artle}}]{PhysRevB.94.121303}%
  \BibitemOpen
  \bibfield  {author} {\bibinfo {author} {\bibfnamefont {S.}~\bibnamefont {Wenderoth}}, \bibinfo {author} {\bibfnamefont {J.}~\bibnamefont {B\"atge}},\ and\ \bibinfo {author} {\bibfnamefont {R.}~\bibnamefont {H\"artle}},\ }\bibfield  {title} {\bibinfo {title} {Sharp peaks in the conductance of a double quantum dot and a quantum-dot spin valve at high temperatures: A hierarchical quantum master equation approach},\ }\href {https://doi.org/10.1103/PhysRevB.94.121303} {\bibfield  {journal} {\bibinfo  {journal} {Phys. Rev. B}\ }\textbf {\bibinfo {volume} {94}},\ \bibinfo {pages} {121303} (\bibinfo {year} {2016})}\BibitemShut {NoStop}%
\bibitem [{\citenamefont {Huang}\ \emph {et~al.}(2023)\citenamefont {Huang}, \citenamefont {Kuo}, \citenamefont {Lambert}, \citenamefont {Cirio}, \citenamefont {Cross}, \citenamefont {Yang}, \citenamefont {Nori},\ and\ \citenamefont {Chen}}]{Huang2023}%
  \BibitemOpen
  \bibfield  {author} {\bibinfo {author} {\bibfnamefont {Y.-T.}\ \bibnamefont {Huang}}, \bibinfo {author} {\bibfnamefont {P.-C.}\ \bibnamefont {Kuo}}, \bibinfo {author} {\bibfnamefont {N.}~\bibnamefont {Lambert}}, \bibinfo {author} {\bibfnamefont {M.}~\bibnamefont {Cirio}}, \bibinfo {author} {\bibfnamefont {S.}~\bibnamefont {Cross}}, \bibinfo {author} {\bibfnamefont {S.-L.}\ \bibnamefont {Yang}}, \bibinfo {author} {\bibfnamefont {F.}~\bibnamefont {Nori}},\ and\ \bibinfo {author} {\bibfnamefont {Y.-N.}\ \bibnamefont {Chen}},\ }\bibfield  {title} {\bibinfo {title} {An efficient {J}ulia framework for hierarchical equations of motion in open quantum systems},\ }\href {http://dx.doi.org/10.1038/s42005-023-01427-2} {\bibfield  {journal} {\bibinfo  {journal} {Communications Physics}\ }\textbf {\bibinfo {volume} {6}},\ \bibinfo {pages} {313} (\bibinfo {year} {2023})}\BibitemShut {NoStop}%
\bibitem [{\citenamefont {Lambert}\ \emph {et~al.}(2019)\citenamefont {Lambert}, \citenamefont {Ahmed}, \citenamefont {Cirio},\ and\ \citenamefont {Nori}}]{lambert2019modelling}%
  \BibitemOpen
  \bibfield  {author} {\bibinfo {author} {\bibfnamefont {N.}~\bibnamefont {Lambert}}, \bibinfo {author} {\bibfnamefont {S.}~\bibnamefont {Ahmed}}, \bibinfo {author} {\bibfnamefont {M.}~\bibnamefont {Cirio}},\ and\ \bibinfo {author} {\bibfnamefont {F.}~\bibnamefont {Nori}},\ }\bibfield  {title} {\bibinfo {title} {{Modelling the ultra-strongly coupled spin-boson model with unphysical modes}},\ }\href {http://www.nature.com/articles/s41467-019-11656-1} {\bibfield  {journal} {\bibinfo  {journal} {Nat. Commun.}\ }\textbf {\bibinfo {volume} {10}},\ \bibinfo {pages} {3721} (\bibinfo {year} {2019})}\BibitemShut {NoStop}%
\bibitem [{\citenamefont {Xu}\ \emph {et~al.}(2022)\citenamefont {Xu}, \citenamefont {Yan}, \citenamefont {Shi}, \citenamefont {Ankerhold},\ and\ \citenamefont {Stockburger}}]{xu_taming_2022}%
  \BibitemOpen
  \bibfield  {author} {\bibinfo {author} {\bibfnamefont {M.}~\bibnamefont {Xu}}, \bibinfo {author} {\bibfnamefont {Y.}~\bibnamefont {Yan}}, \bibinfo {author} {\bibfnamefont {Q.}~\bibnamefont {Shi}}, \bibinfo {author} {\bibfnamefont {J.}~\bibnamefont {Ankerhold}},\ and\ \bibinfo {author} {\bibfnamefont {J.~T.}\ \bibnamefont {Stockburger}},\ }\bibfield  {title} {{\selectlanguage {english}\bibinfo {title} {Taming {Quantum} {Noise} for {Efficient} {Low} {Temperature} {Simulations} of {Open} {Quantum} {Systems}}},\ }\href {https://link.aps.org/doi/10.1103/PhysRevLett.129.230601} {\bibfield  {journal} {\bibinfo  {journal} {Phys. Rev. Lett.}\ }\textbf {\bibinfo {volume} {129}},\ \bibinfo {pages} {230601} (\bibinfo {year} {2022})}\BibitemShut {NoStop}%
\bibitem [{\citenamefont {Wenderoth}\ \emph {et~al.}(2021)\citenamefont {Wenderoth}, \citenamefont {Breuer},\ and\ \citenamefont {Thoss}}]{Wenderoth_Breuer_Thoss_2021}%
  \BibitemOpen
  \bibfield  {author} {\bibinfo {author} {\bibfnamefont {S.}~\bibnamefont {Wenderoth}}, \bibinfo {author} {\bibfnamefont {H.-P.}\ \bibnamefont {Breuer}},\ and\ \bibinfo {author} {\bibfnamefont {M.}~\bibnamefont {Thoss}},\ }\bibfield  {title} {{\selectlanguage {english}\bibinfo {title} {Non-markovian effects in the spin-boson model at zero temperature}},\ }\href {https://link.aps.org/doi/10.1103/PhysRevA.104.012213} {\bibfield  {journal} {\bibinfo  {journal} {Physical Review A}\ }\textbf {\bibinfo {volume} {104}},\ \bibinfo {pages} {012213} (\bibinfo {year} {2021})}\BibitemShut {NoStop}%
\bibitem [{\citenamefont {Zhou}\ and\ \citenamefont {Shao}(2008)}]{Zhou_Shao_2008}%
  \BibitemOpen
  \bibfield  {author} {\bibinfo {author} {\bibfnamefont {Y.}~\bibnamefont {Zhou}}\ and\ \bibinfo {author} {\bibfnamefont {J.}~\bibnamefont {Shao}},\ }\bibfield  {title} {\bibinfo {title} {Solving the spin-boson model of strong dissipation with flexible random-deterministic scheme},\ }\href {https://doi.org/10.1063/1.2818095} {\bibfield  {journal} {\bibinfo  {journal} {The Journal of Chemical Physics}\ }\textbf {\bibinfo {volume} {128}},\ \bibinfo {pages} {034106} (\bibinfo {year} {2008})}\BibitemShut {NoStop}%
\bibitem [{\citenamefont {Wang}\ and\ \citenamefont {Thoss}(2008)}]{Wang_Thoss_2008}%
  \BibitemOpen
  \bibfield  {author} {\bibinfo {author} {\bibfnamefont {H.}~\bibnamefont {Wang}}\ and\ \bibinfo {author} {\bibfnamefont {M.}~\bibnamefont {Thoss}},\ }\bibfield  {title} {{\selectlanguage {english}\bibinfo {title} {From coherent motion to localization: dynamics of the spin-boson model at zero temperature}},\ }\href {https://iopscience.iop.org/article/10.1088/1367-2630/10/11/115005} {\bibfield  {journal} {\bibinfo  {journal} {New Journal of Physics}\ }\textbf {\bibinfo {volume} {10}},\ \bibinfo {pages} {115005} (\bibinfo {year} {2008})}\BibitemShut {NoStop}%
\bibitem [{\citenamefont {Takahashi}\ \emph {et~al.}(2024)\citenamefont {Takahashi}, \citenamefont {Rudge}, \citenamefont {Kaspar}, \citenamefont {Thoss},\ and\ \citenamefont {Borrelli}}]{Takahashi_Rudge_Kaspar_Thoss_Borrelli_2024}%
  \BibitemOpen
  \bibfield  {author} {\bibinfo {author} {\bibfnamefont {H.}~\bibnamefont {Takahashi}}, \bibinfo {author} {\bibfnamefont {S.}~\bibnamefont {Rudge}}, \bibinfo {author} {\bibfnamefont {C.}~\bibnamefont {Kaspar}}, \bibinfo {author} {\bibfnamefont {M.}~\bibnamefont {Thoss}},\ and\ \bibinfo {author} {\bibfnamefont {R.}~\bibnamefont {Borrelli}},\ }\bibfield  {title} {\bibinfo {title} {High accuracy exponential decomposition of bath correlation functions for arbitrary and structured spectral densities: Emerging methodologies and new approaches},\ }\href {https://doi.org/10.1063/5.0209348} {\bibfield  {journal} {\bibinfo  {journal} {The Journal of Chemical Physics}\ }\textbf {\bibinfo {volume} {160}},\ \bibinfo {pages} {204105} (\bibinfo {year} {2024})}\BibitemShut {NoStop}%
\bibitem [{\citenamefont {Suárez}\ and\ \citenamefont {Horodecki}(2025)}]{suárez2025makingnonmarkovianmasterequations}%
  \BibitemOpen
  \bibfield  {author} {\bibinfo {author} {\bibfnamefont {G.}~\bibnamefont {Suárez}}\ and\ \bibinfo {author} {\bibfnamefont {M.}~\bibnamefont {Horodecki}},\ }\bibfield  {title} {\bibinfo {title} {Making non-markovian master equations accessible with approximate environments},\ }\href {https://arxiv.org/abs/2506.22346} {\bibfield  {journal} {\bibinfo  {journal} {arXiv:2506.22346 [quant-ph]}\ } (\bibinfo {year} {2025})}\BibitemShut {NoStop}%
\bibitem [{ani()}]{animation_tutorial}%
  \BibitemOpen
  \href {https://github.com/qutip/qutip-tutorials/blob/main/tutorials-v5/visualization/animation-demo.md} {\bibinfo {title} {https://github.com/qutip/qutip-tutorials/blob/main/tutorials-v5/visualization/animation-demo.md}}\BibitemShut {NoStop}%
\bibitem [{enr()}]{enr_chain}%
  \BibitemOpen
  \href {https://github.com/qutip/qutip-tutorials/blob/main/tutorials-v5/miscellaneous/excitation-number-restricted-states-jc-chain.md} {\bibinfo {title} {https://github.com/qutip/qutip-tutorials/blob/main/tutorials-v5/miscellaneous/excitation-number-restricted-states-jc-chain.md}}\BibitemShut {NoStop}%
\bibitem [{\citenamefont {Lednev}\ \emph {et~al.}(2024)\citenamefont {Lednev}, \citenamefont {Garc\'{\i}a-Vidal},\ and\ \citenamefont {Feist}}]{feist}%
  \BibitemOpen
  \bibfield  {author} {\bibinfo {author} {\bibfnamefont {M.}~\bibnamefont {Lednev}}, \bibinfo {author} {\bibfnamefont {F.~J.}\ \bibnamefont {Garc\'{\i}a-Vidal}},\ and\ \bibinfo {author} {\bibfnamefont {J.}~\bibnamefont {Feist}},\ }\bibfield  {title} {\bibinfo {title} {Lindblad master equation capable of describing hybrid quantum systems in the ultrastrong coupling regime},\ }\href {https://doi.org/10.1103/PhysRevLett.132.106902} {\bibfield  {journal} {\bibinfo  {journal} {Phys. Rev. Lett.}\ }\textbf {\bibinfo {volume} {132}},\ \bibinfo {pages} {106902} (\bibinfo {year} {2024})}\BibitemShut {NoStop}%
\bibitem [{\citenamefont {Rom\'an-Roche}\ \emph {et~al.}(2020)\citenamefont {Rom\'an-Roche}, \citenamefont {S\'anchez-Burillo},\ and\ \citenamefont {Zueco}}]{zueco}%
  \BibitemOpen
  \bibfield  {author} {\bibinfo {author} {\bibfnamefont {J.}~\bibnamefont {Rom\'an-Roche}}, \bibinfo {author} {\bibfnamefont {E.}~\bibnamefont {S\'anchez-Burillo}},\ and\ \bibinfo {author} {\bibfnamefont {D.}~\bibnamefont {Zueco}},\ }\bibfield  {title} {\bibinfo {title} {Bound states in ultrastrong waveguide {QED}},\ }\href {https://link.aps.org/doi/10.1103/PhysRevA.102.023702} {\bibfield  {journal} {\bibinfo  {journal} {Phys. Rev. A}\ }\textbf {\bibinfo {volume} {102}},\ \bibinfo {pages} {023702} (\bibinfo {year} {2020})}\BibitemShut {NoStop}%
\bibitem [{\citenamefont {Zhang}\ \emph {et~al.}(2017)\citenamefont {Zhang}, \citenamefont {Liu}, \citenamefont {Wu}, \citenamefont {Jacobs},\ and\ \citenamefont {Nori}}]{Zhang2017}%
  \BibitemOpen
  \bibfield  {author} {\bibinfo {author} {\bibfnamefont {J.}~\bibnamefont {Zhang}}, \bibinfo {author} {\bibfnamefont {Y.-x.}\ \bibnamefont {Liu}}, \bibinfo {author} {\bibfnamefont {R.-B.}\ \bibnamefont {Wu}}, \bibinfo {author} {\bibfnamefont {K.}~\bibnamefont {Jacobs}},\ and\ \bibinfo {author} {\bibfnamefont {F.}~\bibnamefont {Nori}},\ }\bibfield  {title} {\bibinfo {title} {Quantum feedback: {T}heory, experiments, and applications},\ }\href {https://doi.org/10.1016/j.physrep.2017.02.003} {\bibfield  {journal} {\bibinfo  {journal} {Physics Reports}\ }\textbf {\bibinfo {volume} {679}},\ \bibinfo {pages} {1–60} (\bibinfo {year} {2017})}\BibitemShut {NoStop}%
\bibitem [{\citenamefont {Grimsmo}(2015)}]{arne}%
  \BibitemOpen
  \bibfield  {author} {\bibinfo {author} {\bibfnamefont {A.~L.}\ \bibnamefont {Grimsmo}},\ }\bibfield  {title} {\bibinfo {title} {Time-delayed quantum feedback control},\ }\href {https://link.aps.org/doi/10.1103/PhysRevLett.115.060402} {\bibfield  {journal} {\bibinfo  {journal} {Phys. Rev. Lett.}\ }\textbf {\bibinfo {volume} {115}},\ \bibinfo {pages} {060402} (\bibinfo {year} {2015})}\BibitemShut {NoStop}%
\bibitem [{\citenamefont {Arranz~Regidor}\ \emph {et~al.}(2021)\citenamefont {Arranz~Regidor}, \citenamefont {Crowder}, \citenamefont {Carmichael},\ and\ \citenamefont {Hughes}}]{stephenhughes1}%
  \BibitemOpen
  \bibfield  {author} {\bibinfo {author} {\bibfnamefont {S.}~\bibnamefont {Arranz~Regidor}}, \bibinfo {author} {\bibfnamefont {G.}~\bibnamefont {Crowder}}, \bibinfo {author} {\bibfnamefont {H.}~\bibnamefont {Carmichael}},\ and\ \bibinfo {author} {\bibfnamefont {S.}~\bibnamefont {Hughes}},\ }\bibfield  {title} {\bibinfo {title} {Modeling quantum light-matter interactions in waveguide {QED} with retardation, nonlinear interactions, and a time-delayed feedback: Matrix product states versus a space-discretized waveguide model},\ }\href {https://link.aps.org/doi/10.1103/PhysRevResearch.3.023030} {\bibfield  {journal} {\bibinfo  {journal} {Phys. Rev. Res.}\ }\textbf {\bibinfo {volume} {3}},\ \bibinfo {pages} {023030} (\bibinfo {year} {2021})}\BibitemShut {NoStop}%
\bibitem [{\citenamefont {Crowder}\ \emph {et~al.}(2022)\citenamefont {Crowder}, \citenamefont {Ramunno},\ and\ \citenamefont {Hughes}}]{PhysRevA.106.013714}%
  \BibitemOpen
  \bibfield  {author} {\bibinfo {author} {\bibfnamefont {G.}~\bibnamefont {Crowder}}, \bibinfo {author} {\bibfnamefont {L.}~\bibnamefont {Ramunno}},\ and\ \bibinfo {author} {\bibfnamefont {S.}~\bibnamefont {Hughes}},\ }\bibfield  {title} {\bibinfo {title} {Quantum trajectory theory and simulations of nonlinear spectra and multiphoton effects in waveguide-{QED} systems with a time-delayed coherent feedback},\ }\href {https://doi.org/10.1103/PhysRevA.106.013714} {\bibfield  {journal} {\bibinfo  {journal} {Phys. Rev. A}\ }\textbf {\bibinfo {volume} {106}},\ \bibinfo {pages} {013714} (\bibinfo {year} {2022})}\BibitemShut {NoStop}%
\bibitem [{\citenamefont {Dalc{\'i}n}\ \emph {et~al.}(2005)\citenamefont {Dalc{\'i}n}, \citenamefont {Paz},\ and\ \citenamefont {Storti}}]{DalcinJParallelDistribComput2005}%
  \BibitemOpen
  \bibfield  {author} {\bibinfo {author} {\bibfnamefont {L.}~\bibnamefont {Dalc{\'i}n}}, \bibinfo {author} {\bibfnamefont {R.}~\bibnamefont {Paz}},\ and\ \bibinfo {author} {\bibfnamefont {M.}~\bibnamefont {Storti}},\ }\bibfield  {title} {\bibinfo {title} {{{MPI}} for {{Python}}},\ }\href {https://www.sciencedirect.com/science/article/pii/S0743731505000560} {\bibfield  {journal} {\bibinfo  {journal} {J. Parallel Distrib. Comput.}\ }\textbf {\bibinfo {volume} {65}},\ \bibinfo {pages} {1108} (\bibinfo {year} {2005})}\BibitemShut {NoStop}%
\bibitem [{\citenamefont {Dalc{\'i}n}\ \emph {et~al.}(2008)\citenamefont {Dalc{\'i}n}, \citenamefont {Paz}, \citenamefont {Storti},\ and\ \citenamefont {D'El{\'i}a}}]{DalcinJParallelDistribComput2008}%
  \BibitemOpen
  \bibfield  {author} {\bibinfo {author} {\bibfnamefont {L.}~\bibnamefont {Dalc{\'i}n}}, \bibinfo {author} {\bibfnamefont {R.}~\bibnamefont {Paz}}, \bibinfo {author} {\bibfnamefont {M.}~\bibnamefont {Storti}},\ and\ \bibinfo {author} {\bibfnamefont {J.}~\bibnamefont {D'El{\'i}a}},\ }\bibfield  {title} {\bibinfo {title} {{{MPI}} for {{Python}}: {{Performance}} improvements and {{MPI-2}} extensions},\ }\href {https://www.sciencedirect.com/science/article/pii/S0743731507001712} {\bibfield  {journal} {\bibinfo  {journal} {J. Parallel Distrib. Comput.}\ }\textbf {\bibinfo {volume} {68}},\ \bibinfo {pages} {655} (\bibinfo {year} {2008})}\BibitemShut {NoStop}%
\bibitem [{\citenamefont {Dalcin}\ \emph {et~al.}(2011)\citenamefont {Dalcin}, \citenamefont {Paz}, \citenamefont {Kler},\ and\ \citenamefont {Cosimo}}]{DalcinAdvWaterResour2011}%
  \BibitemOpen
  \bibfield  {author} {\bibinfo {author} {\bibfnamefont {L.~D.}\ \bibnamefont {Dalcin}}, \bibinfo {author} {\bibfnamefont {R.~R.}\ \bibnamefont {Paz}}, \bibinfo {author} {\bibfnamefont {P.~A.}\ \bibnamefont {Kler}},\ and\ \bibinfo {author} {\bibfnamefont {A.}~\bibnamefont {Cosimo}},\ }\bibfield  {title} {\bibinfo {title} {Parallel distributed computing using {{Python}}},\ }\href {https://www.sciencedirect.com/science/article/pii/S0309170811000777} {\bibfield  {journal} {\bibinfo  {journal} {Adv. Water Resour.}\ }\bibinfo {series} {New {{Computational Methods}} and {{Software Tools}}},\ \textbf {\bibinfo {volume} {34}},\ \bibinfo {pages} {1124} (\bibinfo {year} {2011})}\BibitemShut {NoStop}%
\bibitem [{\citenamefont {Dalcin}\ and\ \citenamefont {Fang}(2021)}]{DalcinComputSciEng2021}%
  \BibitemOpen
  \bibfield  {author} {\bibinfo {author} {\bibfnamefont {L.}~\bibnamefont {Dalcin}}\ and\ \bibinfo {author} {\bibfnamefont {Y.-L.~L.}\ \bibnamefont {Fang}},\ }\bibfield  {title} {\bibinfo {title} {Mpi4py: {{Status Update After}} 12 {{Years}} of {{Development}}},\ }\href {https://ieeexplore.ieee.org/document/9439927} {\bibfield  {journal} {\bibinfo  {journal} {Comput. Sci. Eng.}\ }\textbf {\bibinfo {volume} {23}},\ \bibinfo {pages} {47} (\bibinfo {year} {2021})}\BibitemShut {NoStop}%
\bibitem [{\citenamefont {Caneva}\ \emph {et~al.}(2011)\citenamefont {Caneva}, \citenamefont {Calarco},\ and\ \citenamefont {Montangero}}]{crab}%
  \BibitemOpen
  \bibfield  {author} {\bibinfo {author} {\bibfnamefont {T.}~\bibnamefont {Caneva}}, \bibinfo {author} {\bibfnamefont {T.}~\bibnamefont {Calarco}},\ and\ \bibinfo {author} {\bibfnamefont {S.}~\bibnamefont {Montangero}},\ }\bibfield  {title} {\bibinfo {title} {Chopped random-basis quantum optimization},\ }\href {https://link.aps.org/doi/10.1103/PhysRevA.84.022326} {\bibfield  {journal} {\bibinfo  {journal} {Phys. Rev. A}\ }\textbf {\bibinfo {volume} {84}},\ \bibinfo {pages} {022326} (\bibinfo {year} {2011})}\BibitemShut {NoStop}%
\bibitem [{\citenamefont {Khaneja}\ \emph {et~al.}(2005)\citenamefont {Khaneja}, \citenamefont {Reiss}, \citenamefont {Kehlet}, \citenamefont {Schulte-Herbrüggen},\ and\ \citenamefont {Glaser}}]{grape}%
  \BibitemOpen
  \bibfield  {author} {\bibinfo {author} {\bibfnamefont {N.}~\bibnamefont {Khaneja}}, \bibinfo {author} {\bibfnamefont {T.}~\bibnamefont {Reiss}}, \bibinfo {author} {\bibfnamefont {C.}~\bibnamefont {Kehlet}}, \bibinfo {author} {\bibfnamefont {T.}~\bibnamefont {Schulte-Herbrüggen}},\ and\ \bibinfo {author} {\bibfnamefont {S.~J.}\ \bibnamefont {Glaser}},\ }\bibfield  {title} {\bibinfo {title} {Optimal control of coupled spin dynamics: design of {NMR} pulse sequences by gradient ascent algorithms},\ }\href {https://www.sciencedirect.com/science/article/pii/S1090780704003696} {\bibfield  {journal} {\bibinfo  {journal} {Journal of Magnetic Resonance}\ }\textbf {\bibinfo {volume} {172}},\ \bibinfo {pages} {296} (\bibinfo {year} {2005})}\BibitemShut {NoStop}%
\bibitem [{qqt()}]{qqtrl}%
  \BibitemOpen
  \href {https://github.com/qutip/qutip-qtrl} {\bibinfo {title} {https://github.com/qutip/qutip-qtrl}}\BibitemShut {NoStop}%
\bibitem [{qqo()}]{qqoc}%
  \BibitemOpen
  \href {https://github.com/qutip/qutip-qoc} {\bibinfo {title} {https://github.com/qutip/qutip-qoc}}\BibitemShut {NoStop}%
\bibitem [{\citenamefont {Machnes}\ \emph {et~al.}(2018)\citenamefont {Machnes}, \citenamefont {Ass\'emat}, \citenamefont {Tannor},\ and\ \citenamefont {Wilhelm}}]{MachnesGOAT}%
  \BibitemOpen
  \bibfield  {author} {\bibinfo {author} {\bibfnamefont {S.}~\bibnamefont {Machnes}}, \bibinfo {author} {\bibfnamefont {E.}~\bibnamefont {Ass\'emat}}, \bibinfo {author} {\bibfnamefont {D.}~\bibnamefont {Tannor}},\ and\ \bibinfo {author} {\bibfnamefont {F.~K.}\ \bibnamefont {Wilhelm}},\ }\bibfield  {title} {\bibinfo {title} {Tunable, flexible, and efficient optimization of control pulses for practical qubits},\ }\href {https://link.aps.org/doi/10.1103/PhysRevLett.120.150401} {\bibfield  {journal} {\bibinfo  {journal} {Phys. Rev. Lett.}\ }\textbf {\bibinfo {volume} {120}},\ \bibinfo {pages} {150401} (\bibinfo {year} {2018})}\BibitemShut {NoStop}%
\bibitem [{\citenamefont {Goerz}\ \emph {et~al.}(2022)\citenamefont {Goerz}, \citenamefont {Carrasco},\ and\ \citenamefont {Malinovsky}}]{Goerz2022}%
  \BibitemOpen
  \bibfield  {author} {\bibinfo {author} {\bibfnamefont {M.~H.}\ \bibnamefont {Goerz}}, \bibinfo {author} {\bibfnamefont {S.~C.}\ \bibnamefont {Carrasco}},\ and\ \bibinfo {author} {\bibfnamefont {V.~S.}\ \bibnamefont {Malinovsky}},\ }\bibfield  {title} {\bibinfo {title} {Quantum optimal control via semi-automatic differentiation},\ }\href {https://doi.org/10.22331/q-2022-12-07-871} {\bibfield  {journal} {\bibinfo  {journal} {Quantum}\ }\textbf {\bibinfo {volume} {6}},\ \bibinfo {pages} {871} (\bibinfo {year} {2022})}\BibitemShut {NoStop}%
\bibitem [{\citenamefont {Zong}\ \emph {et~al.}(2021)\citenamefont {Zong}, \citenamefont {Sun}, \citenamefont {Dong}, \citenamefont {Run}, \citenamefont {Xiang}, \citenamefont {Zhan}, \citenamefont {Wang}, \citenamefont {Fei}, \citenamefont {Wu}, \citenamefont {Jin}, \citenamefont {Xiao}, \citenamefont {Jia}, \citenamefont {Duan}, \citenamefont {Wu}, \citenamefont {Yin},\ and\ \citenamefont {Guo}}]{grape_sc}%
  \BibitemOpen
  \bibfield  {author} {\bibinfo {author} {\bibfnamefont {Z.}~\bibnamefont {Zong}}, \bibinfo {author} {\bibfnamefont {Z.}~\bibnamefont {Sun}}, \bibinfo {author} {\bibfnamefont {Z.}~\bibnamefont {Dong}}, \bibinfo {author} {\bibfnamefont {C.}~\bibnamefont {Run}}, \bibinfo {author} {\bibfnamefont {L.}~\bibnamefont {Xiang}}, \bibinfo {author} {\bibfnamefont {Z.}~\bibnamefont {Zhan}}, \bibinfo {author} {\bibfnamefont {Q.}~\bibnamefont {Wang}}, \bibinfo {author} {\bibfnamefont {Y.}~\bibnamefont {Fei}}, \bibinfo {author} {\bibfnamefont {Y.}~\bibnamefont {Wu}}, \bibinfo {author} {\bibfnamefont {W.}~\bibnamefont {Jin}}, \bibinfo {author} {\bibfnamefont {C.}~\bibnamefont {Xiao}}, \bibinfo {author} {\bibfnamefont {Z.}~\bibnamefont {Jia}}, \bibinfo {author} {\bibfnamefont {P.}~\bibnamefont {Duan}}, \bibinfo {author} {\bibfnamefont {J.}~\bibnamefont {Wu}}, \bibinfo {author} {\bibfnamefont {Y.}~\bibnamefont {Yin}},\ and\ \bibinfo {author} {\bibfnamefont {G.}~\bibnamefont {Guo}},\ }\bibfield  {title} {\bibinfo {title}
  {Optimization of a controlled-$z$ gate with data-driven gradient-ascent pulse engineering in a superconducting-qubit system},\ }\href {https://link.aps.org/doi/10.1103/PhysRevApplied.15.064005} {\bibfield  {journal} {\bibinfo  {journal} {Phys. Rev. Appl.}\ }\textbf {\bibinfo {volume} {15}},\ \bibinfo {pages} {064005} (\bibinfo {year} {2021})}\BibitemShut {NoStop}%
\bibitem [{\citenamefont {Goerz}\ \emph {et~al.}(2019)\citenamefont {Goerz}, \citenamefont {Basilewitsch}, \citenamefont {Gago-Encinas}, \citenamefont {Krauss}, \citenamefont {Horn}, \citenamefont {Reich},\ and\ \citenamefont {Koch}}]{Goerz_2019_SciPost}%
  \BibitemOpen
  \bibfield  {author} {\bibinfo {author} {\bibfnamefont {M.~H.}\ \bibnamefont {Goerz}}, \bibinfo {author} {\bibfnamefont {D.}~\bibnamefont {Basilewitsch}}, \bibinfo {author} {\bibfnamefont {F.}~\bibnamefont {Gago-Encinas}}, \bibinfo {author} {\bibfnamefont {M.~G.}\ \bibnamefont {Krauss}}, \bibinfo {author} {\bibfnamefont {K.~P.}\ \bibnamefont {Horn}}, \bibinfo {author} {\bibfnamefont {D.~M.}\ \bibnamefont {Reich}},\ and\ \bibinfo {author} {\bibfnamefont {C.~P.}\ \bibnamefont {Koch}},\ }\bibfield  {title} {\bibinfo {title} {{Krotov: A Python implementation of {K}rotov's method for quantum optimal control}},\ }\href {https://scipost.org/10.21468/SciPostPhys.7.6.080} {\bibfield  {journal} {\bibinfo  {journal} {SciPost Phys.}\ }\textbf {\bibinfo {volume} {7}},\ \bibinfo {pages} {80} (\bibinfo {year} {2019})}\BibitemShut {NoStop}%
\bibitem [{\citenamefont {Araki}\ \emph {et~al.}(2023)\citenamefont {Araki}, \citenamefont {Nori},\ and\ \citenamefont {Gneiting}}]{PhysRevA.107.032609}%
  \BibitemOpen
  \bibfield  {author} {\bibinfo {author} {\bibfnamefont {T.}~\bibnamefont {Araki}}, \bibinfo {author} {\bibfnamefont {F.}~\bibnamefont {Nori}},\ and\ \bibinfo {author} {\bibfnamefont {C.}~\bibnamefont {Gneiting}},\ }\bibfield  {title} {\bibinfo {title} {Robust quantum control with disorder-dressed evolution},\ }\href {https://doi.org/10.1103/PhysRevA.107.032609} {\bibfield  {journal} {\bibinfo  {journal} {Phys. Rev. A}\ }\textbf {\bibinfo {volume} {107}},\ \bibinfo {pages} {032609} (\bibinfo {year} {2023})}\BibitemShut {NoStop}%
\bibitem [{\citenamefont {Müller}\ \emph {et~al.}(2022)\citenamefont {Müller}, \citenamefont {Said}, \citenamefont {Jelezko}, \citenamefont {Calarco},\ and\ \citenamefont {Montangero}}]{M_ller_2022}%
  \BibitemOpen
  \bibfield  {author} {\bibinfo {author} {\bibfnamefont {M.~M.}\ \bibnamefont {Müller}}, \bibinfo {author} {\bibfnamefont {R.~S.}\ \bibnamefont {Said}}, \bibinfo {author} {\bibfnamefont {F.}~\bibnamefont {Jelezko}}, \bibinfo {author} {\bibfnamefont {T.}~\bibnamefont {Calarco}},\ and\ \bibinfo {author} {\bibfnamefont {S.}~\bibnamefont {Montangero}},\ }\bibfield  {title} {\bibinfo {title} {One decade of quantum optimal control in the chopped random basis},\ }\href {http://dx.doi.org/10.1088/1361-6633/ac723c} {\bibfield  {journal} {\bibinfo  {journal} {Reports on Progress in Physics}\ }\textbf {\bibinfo {volume} {85}},\ \bibinfo {pages} {076001} (\bibinfo {year} {2022})}\BibitemShut {NoStop}%
\bibitem [{\citenamefont {Corti}\ \emph {et~al.}(2022)\citenamefont {Corti}, \citenamefont {Banchi},\ and\ \citenamefont {Cidronali}}]{corti_robustness_2022}%
  \BibitemOpen
  \bibfield  {author} {\bibinfo {author} {\bibfnamefont {H.~A.}\ \bibnamefont {Corti}}, \bibinfo {author} {\bibfnamefont {L.}~\bibnamefont {Banchi}},\ and\ \bibinfo {author} {\bibfnamefont {A.}~\bibnamefont {Cidronali}},\ }\bibfield  {title} {\bibinfo {title} {Robustness of a universal gate set implementation in transmon systems via {Chopped} {Random} {Basis} optimal control},\ }\href {https://www.sciencedirect.com/science/article/pii/S0375960122002018} {\bibfield  {journal} {\bibinfo  {journal} {Physics Letters A}\ }\textbf {\bibinfo {volume} {438}},\ \bibinfo {pages} {128119} (\bibinfo {year} {2022})}\BibitemShut {NoStop}%
\bibitem [{\citenamefont {Riaz}\ \emph {et~al.}(2019)\citenamefont {Riaz}, \citenamefont {Shuang},\ and\ \citenamefont {Qamar}}]{riaz_optimal_2019}%
  \BibitemOpen
  \bibfield  {author} {\bibinfo {author} {\bibfnamefont {B.}~\bibnamefont {Riaz}}, \bibinfo {author} {\bibfnamefont {C.}~\bibnamefont {Shuang}},\ and\ \bibinfo {author} {\bibfnamefont {S.}~\bibnamefont {Qamar}},\ }\bibfield  {title} {\bibinfo {title} {Optimal control methods for quantum gate preparation: a comparative study},\ }\href {https://doi.org/10.1007/s11128-019-2190-0} {\bibfield  {journal} {\bibinfo  {journal} {Quantum Information Processing}\ }\textbf {\bibinfo {volume} {18}},\ \bibinfo {pages} {100} (\bibinfo {year} {2019})}\BibitemShut {NoStop}%
\bibitem [{\citenamefont {Baydin}\ \emph {et~al.}(2018)\citenamefont {Baydin}, \citenamefont {Pearlmutter}, \citenamefont {Radul},\ and\ \citenamefont {Siskind}}]{BaydinPR15}%
  \BibitemOpen
  \bibfield  {author} {\bibinfo {author} {\bibfnamefont {A.~G.}\ \bibnamefont {Baydin}}, \bibinfo {author} {\bibfnamefont {B.~A.}\ \bibnamefont {Pearlmutter}}, \bibinfo {author} {\bibfnamefont {A.~A.}\ \bibnamefont {Radul}},\ and\ \bibinfo {author} {\bibfnamefont {J.~M.}\ \bibnamefont {Siskind}},\ }\bibfield  {title} {\bibinfo {title} {Automatic differentiation in machine learning: a survey},\ }\href {http://jmlr.org/papers/v18/17-468.html} {\bibfield  {journal} {\bibinfo  {journal} {Journal of Machine Learning Research}\ }\textbf {\bibinfo {volume} {18}},\ \bibinfo {pages} {1} (\bibinfo {year} {2018})}\BibitemShut {NoStop}%
\bibitem [{\citenamefont {Song}\ \emph {et~al.}(2022)\citenamefont {Song}, \citenamefont {Li}, \citenamefont {Hai}, \citenamefont {Guo},\ and\ \citenamefont {Deng}}]{Song22}%
  \BibitemOpen
  \bibfield  {author} {\bibinfo {author} {\bibfnamefont {Y.}~\bibnamefont {Song}}, \bibinfo {author} {\bibfnamefont {J.}~\bibnamefont {Li}}, \bibinfo {author} {\bibfnamefont {Y.-J.}\ \bibnamefont {Hai}}, \bibinfo {author} {\bibfnamefont {Q.}~\bibnamefont {Guo}},\ and\ \bibinfo {author} {\bibfnamefont {X.-H.}\ \bibnamefont {Deng}},\ }\bibfield  {title} {\bibinfo {title} {Optimizing quantum control pulses with complex constraints and few variables through autodifferentiation},\ }\href {https://link.aps.org/doi/10.1103/PhysRevA.105.012616} {\bibfield  {journal} {\bibinfo  {journal} {Phys. Rev. A}\ }\textbf {\bibinfo {volume} {105}},\ \bibinfo {pages} {012616} (\bibinfo {year} {2022})}\BibitemShut {NoStop}%
\bibitem [{qqi()}]{qqip}%
  \BibitemOpen
  \href {https://github.com/qutip/qutip-qip} {\bibinfo {title} {https://github.com/qutip/qutip-qip}}\BibitemShut {NoStop}%
\bibitem [{\citenamefont {Buluta}\ and\ \citenamefont {Nori}(2009)}]{Buluta2009}%
  \BibitemOpen
  \bibfield  {author} {\bibinfo {author} {\bibfnamefont {I.}~\bibnamefont {Buluta}}\ and\ \bibinfo {author} {\bibfnamefont {F.}~\bibnamefont {Nori}},\ }\bibfield  {title} {\bibinfo {title} {Quantum simulators},\ }\href {https://doi.org/10.1126/science.1177838} {\bibfield  {journal} {\bibinfo  {journal} {Science}\ }\textbf {\bibinfo {volume} {326}},\ \bibinfo {pages} {108–111} (\bibinfo {year} {2009})}\BibitemShut {NoStop}%
\bibitem [{\citenamefont {Georgescu}\ \emph {et~al.}(2014)\citenamefont {Georgescu}, \citenamefont {Ashhab},\ and\ \citenamefont {Nori}}]{Georgescu2014}%
  \BibitemOpen
  \bibfield  {author} {\bibinfo {author} {\bibfnamefont {I.~M.}\ \bibnamefont {Georgescu}}, \bibinfo {author} {\bibfnamefont {S.}~\bibnamefont {Ashhab}},\ and\ \bibinfo {author} {\bibfnamefont {F.}~\bibnamefont {Nori}},\ }\bibfield  {title} {\bibinfo {title} {Quantum simulation},\ }\href {https://doi.org/10.1103/revmodphys.86.153} {\bibfield  {journal} {\bibinfo  {journal} {Reviews of Modern Physics}\ }\textbf {\bibinfo {volume} {86}},\ \bibinfo {pages} {153–185} (\bibinfo {year} {2014})}\BibitemShut {NoStop}%
\bibitem [{\citenamefont {Nation}\ \emph {et~al.}(2012)\citenamefont {Nation}, \citenamefont {Johansson}, \citenamefont {Blencowe},\ and\ \citenamefont {Nori}}]{RevModPhys.84.1}%
  \BibitemOpen
  \bibfield  {author} {\bibinfo {author} {\bibfnamefont {P.~D.}\ \bibnamefont {Nation}}, \bibinfo {author} {\bibfnamefont {J.~R.}\ \bibnamefont {Johansson}}, \bibinfo {author} {\bibfnamefont {M.~P.}\ \bibnamefont {Blencowe}},\ and\ \bibinfo {author} {\bibfnamefont {F.}~\bibnamefont {Nori}},\ }\bibfield  {title} {\bibinfo {title} {Colloquium: Stimulating uncertainty: {A}mplifying the quantum vacuum with superconducting circuits},\ }\href {https://doi.org/10.1103/RevModPhys.84.1} {\bibfield  {journal} {\bibinfo  {journal} {Rev. Mod. Phys.}\ }\textbf {\bibinfo {volume} {84}},\ \bibinfo {pages} {1} (\bibinfo {year} {2012})}\BibitemShut {NoStop}%
\bibitem [{\citenamefont {Nielsen}\ and\ \citenamefont {Chuang}(2000)}]{nielsen2002quantum}%
  \BibitemOpen
  \bibfield  {author} {\bibinfo {author} {\bibfnamefont {M.~A.}\ \bibnamefont {Nielsen}}\ and\ \bibinfo {author} {\bibfnamefont {I.~L.}\ \bibnamefont {Chuang}},\ }\href {https://doi.org/10.1017/CBO9780511976667} {\emph {\bibinfo {title} {Quantum Computation and Quantum Information}}}\ (\bibinfo  {publisher} {Cambridge University Press},\ \bibinfo {year} {2000})\BibitemShut {NoStop}%
\bibitem [{\citenamefont {Lloyd}(1996)}]{lloyd96}%
  \BibitemOpen
  \bibfield  {author} {\bibinfo {author} {\bibfnamefont {S.}~\bibnamefont {Lloyd}},\ }\bibfield  {title} {\bibinfo {title} {Universal quantum simulators},\ }\href {https://www.science.org/doi/abs/10.1126/science.273.5278.1073} {\bibfield  {journal} {\bibinfo  {journal} {Science}\ }\textbf {\bibinfo {volume} {273}},\ \bibinfo {pages} {1073} (\bibinfo {year} {1996})}\BibitemShut {NoStop}%
\bibitem [{\citenamefont {Aharonov}\ and\ \citenamefont {Ta-Shma}(2003)}]{10.1145/780542.780546}%
  \BibitemOpen
  \bibfield  {author} {\bibinfo {author} {\bibfnamefont {D.}~\bibnamefont {Aharonov}}\ and\ \bibinfo {author} {\bibfnamefont {A.}~\bibnamefont {Ta-Shma}},\ }\bibfield  {title} {\bibinfo {title} {Adiabatic quantum state generation and statistical zero knowledge},\ }in\ \href {https://doi.org/10.1145/780542.780546} {\emph {\bibinfo {booktitle} {Proceedings of the Thirty-Fifth Annual ACM Symposium on Theory of Computing}}},\ \bibinfo {series and number} {STOC '03}\ (\bibinfo  {publisher} {Association for Computing Machinery},\ \bibinfo {address} {New York, NY, USA},\ \bibinfo {year} {2003})\ p.\ \bibinfo {pages} {20–29}\BibitemShut {NoStop}%
\bibitem [{\citenamefont {Childs}\ and\ \citenamefont {Wiebe}(2012)}]{10.5555/2481569.2481570}%
  \BibitemOpen
  \bibfield  {author} {\bibinfo {author} {\bibfnamefont {A.~M.}\ \bibnamefont {Childs}}\ and\ \bibinfo {author} {\bibfnamefont {N.}~\bibnamefont {Wiebe}},\ }\bibfield  {title} {\bibinfo {title} {Hamiltonian simulation using linear combinations of unitary operations},\ }\href@noop {} {\bibfield  {journal} {\bibinfo  {journal} {Quantum Info. Comput.}\ }\textbf {\bibinfo {volume} {12}},\ \bibinfo {pages} {901–924} (\bibinfo {year} {2012})}\BibitemShut {NoStop}%
\bibitem [{\citenamefont {Berry}\ \emph {et~al.}(2015)\citenamefont {Berry}, \citenamefont {Childs},\ and\ \citenamefont {Kothari}}]{7354428}%
  \BibitemOpen
  \bibfield  {author} {\bibinfo {author} {\bibfnamefont {D.~W.}\ \bibnamefont {Berry}}, \bibinfo {author} {\bibfnamefont {A.~M.}\ \bibnamefont {Childs}},\ and\ \bibinfo {author} {\bibfnamefont {R.}~\bibnamefont {Kothari}},\ }\bibfield  {title} {\bibinfo {title} {Hamiltonian simulation with nearly optimal dependence on all parameters},\ }in\ \href {https://doi.org/10.1109/FOCS.2015.54} {\emph {\bibinfo {booktitle} {2015 IEEE 56th Annual Symposium on Foundations of Computer Science}}}\ (\bibinfo {year} {2015})\ pp.\ \bibinfo {pages} {792--809}\BibitemShut {NoStop}%
\bibitem [{\citenamefont {Low}\ and\ \citenamefont {Chuang}(2019)}]{Low_2019}%
  \BibitemOpen
  \bibfield  {author} {\bibinfo {author} {\bibfnamefont {G.~H.}\ \bibnamefont {Low}}\ and\ \bibinfo {author} {\bibfnamefont {I.~L.}\ \bibnamefont {Chuang}},\ }\bibfield  {title} {\bibinfo {title} {Hamiltonian simulation by qubitization},\ }\href {https://doi.org/10.22331/q-2019-07-12-163} {\bibfield  {journal} {\bibinfo  {journal} {Quantum}\ }\textbf {\bibinfo {volume} {3}},\ \bibinfo {pages} {163} (\bibinfo {year} {2019})}\BibitemShut {NoStop}%
\bibitem [{\citenamefont {Lambert}\ \emph {et~al.}(2024)\citenamefont {Lambert}, \citenamefont {Cirio}, \citenamefont {Lin}, \citenamefont {Menczel}, \citenamefont {Liang},\ and\ \citenamefont {Nori}}]{dse1}%
  \BibitemOpen
  \bibfield  {author} {\bibinfo {author} {\bibfnamefont {N.}~\bibnamefont {Lambert}}, \bibinfo {author} {\bibfnamefont {M.}~\bibnamefont {Cirio}}, \bibinfo {author} {\bibfnamefont {J.-D.}\ \bibnamefont {Lin}}, \bibinfo {author} {\bibfnamefont {P.}~\bibnamefont {Menczel}}, \bibinfo {author} {\bibfnamefont {P.}~\bibnamefont {Liang}},\ and\ \bibinfo {author} {\bibfnamefont {F.}~\bibnamefont {Nori}},\ }\bibfield  {title} {\bibinfo {title} {Fixing detailed balance in ancilla-based dissipative state engineering},\ }\href {https://doi.org/10.1103/PhysRevResearch.6.043229} {\bibfield  {journal} {\bibinfo  {journal} {Phys. Rev. Res.}\ }\textbf {\bibinfo {volume} {6}},\ \bibinfo {pages} {043229} (\bibinfo {year} {2024})}\BibitemShut {NoStop}%
\bibitem [{\citenamefont {{\em et al.}}(2024)}]{dse2}%
  \BibitemOpen
  \bibfield  {author} {\bibinfo {author} {\bibfnamefont {X.~M.}\ \bibnamefont {{\em et al.}}},\ }\bibfield  {title} {\bibinfo {title} {Stable quantum-correlated many-body states through engineered dissipation},\ }\href {http://dx.doi.org/10.1126/science.adh9932} {\bibfield  {journal} {\bibinfo  {journal} {Science}\ }\textbf {\bibinfo {volume} {383}},\ \bibinfo {pages} {1332–1337} (\bibinfo {year} {2024})}\BibitemShut {NoStop}%
\bibitem [{\citenamefont {Cubitt}(2023)}]{dse3}%
  \BibitemOpen
  \bibfield  {author} {\bibinfo {author} {\bibfnamefont {T.~S.}\ \bibnamefont {Cubitt}},\ }\bibfield  {title} {\bibinfo {title} {Dissipative ground state preparation and the dissipative quantum eigensolver},\ }\href {https://arxiv.org/abs/2303.11962} {\bibfield  {journal} {\bibinfo  {journal} {arXiv:2303.11962 [quant-ph]}\ } (\bibinfo {year} {2023})}\BibitemShut {NoStop}%
\bibitem [{\citenamefont {Raghunandan}\ \emph {et~al.}(2020)\citenamefont {Raghunandan}, \citenamefont {Wolf}, \citenamefont {Ospelkaus}, \citenamefont {Schmidt},\ and\ \citenamefont {Weimer}}]{dse4}%
  \BibitemOpen
  \bibfield  {author} {\bibinfo {author} {\bibfnamefont {M.}~\bibnamefont {Raghunandan}}, \bibinfo {author} {\bibfnamefont {F.}~\bibnamefont {Wolf}}, \bibinfo {author} {\bibfnamefont {C.}~\bibnamefont {Ospelkaus}}, \bibinfo {author} {\bibfnamefont {P.~O.}\ \bibnamefont {Schmidt}},\ and\ \bibinfo {author} {\bibfnamefont {H.}~\bibnamefont {Weimer}},\ }\bibfield  {title} {\bibinfo {title} {Initialization of quantum simulators by sympathetic cooling},\ }\href {https://doi.org/10.1126/sciadv.aaw9268} {\bibfield  {journal} {\bibinfo  {journal} {Science Advances}\ }\textbf {\bibinfo {volume} {6}},\ \bibinfo {pages} {eaaw9268} (\bibinfo {year} {2020})}\BibitemShut {NoStop}%
\bibitem [{\citenamefont {Polla}\ \emph {et~al.}(2021)\citenamefont {Polla}, \citenamefont {Herasymenko},\ and\ \citenamefont {O'Brien}}]{dse5}%
  \BibitemOpen
  \bibfield  {author} {\bibinfo {author} {\bibfnamefont {S.}~\bibnamefont {Polla}}, \bibinfo {author} {\bibfnamefont {Y.}~\bibnamefont {Herasymenko}},\ and\ \bibinfo {author} {\bibfnamefont {T.~E.}\ \bibnamefont {O'Brien}},\ }\bibfield  {title} {\bibinfo {title} {Quantum digital cooling},\ }\href {https://link.aps.org/doi/10.1103/PhysRevA.104.012414} {\bibfield  {journal} {\bibinfo  {journal} {Phys. Rev. A}\ }\textbf {\bibinfo {volume} {104}},\ \bibinfo {pages} {012414} (\bibinfo {year} {2021})}\BibitemShut {NoStop}%
\bibitem [{\citenamefont {Zeng}\ \emph {et~al.}(2023)\citenamefont {Zeng}, \citenamefont {Zhou}, \citenamefont {Rinaldi}, \citenamefont {Gneiting},\ and\ \citenamefont {Nori}}]{PhysRevLett.131.050601}%
  \BibitemOpen
  \bibfield  {author} {\bibinfo {author} {\bibfnamefont {Y.}~\bibnamefont {Zeng}}, \bibinfo {author} {\bibfnamefont {Z.-Y.}\ \bibnamefont {Zhou}}, \bibinfo {author} {\bibfnamefont {E.}~\bibnamefont {Rinaldi}}, \bibinfo {author} {\bibfnamefont {C.}~\bibnamefont {Gneiting}},\ and\ \bibinfo {author} {\bibfnamefont {F.}~\bibnamefont {Nori}},\ }\bibfield  {title} {\bibinfo {title} {Approximate autonomous quantum error correction with reinforcement learning},\ }\href {https://link.aps.org/doi/10.1103/PhysRevLett.131.050601} {\bibfield  {journal} {\bibinfo  {journal} {Phys. Rev. Lett.}\ }\textbf {\bibinfo {volume} {131}},\ \bibinfo {pages} {050601} (\bibinfo {year} {2023})}\BibitemShut {NoStop}%
\bibitem [{\citenamefont {Ding}\ \emph {et~al.}(2024)\citenamefont {Ding}, \citenamefont {Li},\ and\ \citenamefont {Lin}}]{lindbladsim1}%
  \BibitemOpen
  \bibfield  {author} {\bibinfo {author} {\bibfnamefont {Z.}~\bibnamefont {Ding}}, \bibinfo {author} {\bibfnamefont {X.}~\bibnamefont {Li}},\ and\ \bibinfo {author} {\bibfnamefont {L.}~\bibnamefont {Lin}},\ }\bibfield  {title} {\bibinfo {title} {Simulating open quantum systems using {H}amiltonian simulations},\ }\href {https://doi.org/10.1103/PRXQuantum.5.020332} {\bibfield  {journal} {\bibinfo  {journal} {PRX Quantum}\ }\textbf {\bibinfo {volume} {5}} (\bibinfo {year} {2024})},\ \Eprint {https://arxiv.org/abs/2311.15533} {arXiv:2311.15533 [quant-ph]} \BibitemShut {NoStop}%
\bibitem [{\citenamefont {Cleve}\ and\ \citenamefont {Wang}(2017)}]{lindbladsim2}%
  \BibitemOpen
  \bibfield  {author} {\bibinfo {author} {\bibfnamefont {R.}~\bibnamefont {Cleve}}\ and\ \bibinfo {author} {\bibfnamefont {C.}~\bibnamefont {Wang}},\ }\bibfield  {title} {\bibinfo {title} {Efficient quantum algorithms for simulating {L}indblad evolution},\ }\href {https://doi.org/10.48550/arXiv.1612.09512} {\bibfield  {journal} {\bibinfo  {journal} {arXiv:1612.09512 [quant-ph]}\ } (\bibinfo {year} {2017})}\BibitemShut {NoStop}%
\bibitem [{\citenamefont {Groszkowski}\ and\ \citenamefont {Koch}(2021)}]{Groszkowski2021}%
  \BibitemOpen
  \bibfield  {author} {\bibinfo {author} {\bibfnamefont {P.}~\bibnamefont {Groszkowski}}\ and\ \bibinfo {author} {\bibfnamefont {J.}~\bibnamefont {Koch}},\ }\bibfield  {title} {\bibinfo {title} {{S}qubits: a python package for superconducting qubits},\ }\href {http://dx.doi.org/10.22331/q-2021-11-17-583} {\bibfield  {journal} {\bibinfo  {journal} {Quantum}\ }\textbf {\bibinfo {volume} {5}},\ \bibinfo {pages} {583} (\bibinfo {year} {2021})}\BibitemShut {NoStop}%
\bibitem [{\citenamefont {Rajabzadeh}\ \emph {et~al.}(2023)\citenamefont {Rajabzadeh}, \citenamefont {Wang}, \citenamefont {Lee}, \citenamefont {Makihara}, \citenamefont {Guo},\ and\ \citenamefont {Safavi-Naeini}}]{Rajabzadeh2023}%
  \BibitemOpen
  \bibfield  {author} {\bibinfo {author} {\bibfnamefont {T.}~\bibnamefont {Rajabzadeh}}, \bibinfo {author} {\bibfnamefont {Z.}~\bibnamefont {Wang}}, \bibinfo {author} {\bibfnamefont {N.}~\bibnamefont {Lee}}, \bibinfo {author} {\bibfnamefont {T.}~\bibnamefont {Makihara}}, \bibinfo {author} {\bibfnamefont {Y.}~\bibnamefont {Guo}},\ and\ \bibinfo {author} {\bibfnamefont {A.~H.}\ \bibnamefont {Safavi-Naeini}},\ }\bibfield  {title} {\bibinfo {title} {Analysis of arbitrary superconducting quantum circuits accompanied by a python package: {SQ}circuit},\ }\href {http://dx.doi.org/10.22331/q-2023-09-25-1118} {\bibfield  {journal} {\bibinfo  {journal} {Quantum}\ }\textbf {\bibinfo {volume} {7}},\ \bibinfo {pages} {1118} (\bibinfo {year} {2023})}\BibitemShut {NoStop}%
\bibitem [{\citenamefont {Efthymiou}\ \emph {et~al.}(2021)\citenamefont {Efthymiou}, \citenamefont {Ramos-Calderer}, \citenamefont {Bravo-Prieto}, \citenamefont {P{\'{e}}rez-Salinas}, \citenamefont {Garc{\'{\i}}a-Mart{\'{\i}}n}, \citenamefont {Garcia-Saez}, \citenamefont {Latorre},\ and\ \citenamefont {Carrazza}}]{qibo}%
  \BibitemOpen
  \bibfield  {author} {\bibinfo {author} {\bibfnamefont {S.}~\bibnamefont {Efthymiou}}, \bibinfo {author} {\bibfnamefont {S.}~\bibnamefont {Ramos-Calderer}}, \bibinfo {author} {\bibfnamefont {C.}~\bibnamefont {Bravo-Prieto}}, \bibinfo {author} {\bibfnamefont {A.}~\bibnamefont {P{\'{e}}rez-Salinas}}, \bibinfo {author} {\bibfnamefont {D.}~\bibnamefont {Garc{\'{\i}}a-Mart{\'{\i}}n}}, \bibinfo {author} {\bibfnamefont {A.}~\bibnamefont {Garcia-Saez}}, \bibinfo {author} {\bibfnamefont {J.~I.}\ \bibnamefont {Latorre}},\ and\ \bibinfo {author} {\bibfnamefont {S.}~\bibnamefont {Carrazza}},\ }\bibfield  {title} {\bibinfo {title} {Qibo: a framework for quantum simulation with hardware acceleration},\ }\href {https://doi.org/10.1088/2058-9565/ac39f5} {\bibfield  {journal} {\bibinfo  {journal} {Quantum Science and Technology}\ }\textbf {\bibinfo {volume} {7}},\ \bibinfo {pages} {015018} (\bibinfo {year} {2021})}\BibitemShut {NoStop}%
\bibitem [{cud()}]{cudaq}%
  \BibitemOpen
  \href {https://developer.nvidia.com/cuda-q} {\bibinfo {title} {https://developer.nvidia.com/cuda-q}}\BibitemShut {NoStop}%
\bibitem [{\citenamefont {Silvério}\ \emph {et~al.}(2024)\citenamefont {Silvério}, \citenamefont {Grijalva}, \citenamefont {Cornillot}, \citenamefont {Henriet}, \citenamefont {Ajdnik}, \citenamefont {Karalekas}, \citenamefont {Leclerc}, \citenamefont {de~Terrasson}, \citenamefont {Vignoli}, \citenamefont {Jbrem}, \citenamefont {D'Arcangelo}, \citenamefont {Dalyac}, \citenamefont {Louis-PaulHenry}, \citenamefont {Wennersteen}, \citenamefont {Gessa}, \citenamefont {Emmanuel}, \citenamefont {Dutta}, \citenamefont {Shammah}, \citenamefont {Codoscope}, \citenamefont {MatthieuMoreau}, \citenamefont {Tsai}, \citenamefont {Gondhalekar}, \citenamefont {Rava}, \citenamefont {Panigrahi}, \citenamefont {Harold}, \citenamefont {Julius}, \citenamefont {Tallot}, \citenamefont {Oliver}, \citenamefont {Slimane33},\ and\ \citenamefont {WingCode}}]{pulser}%
  \BibitemOpen
  \bibfield  {author} {\bibinfo {author} {\bibfnamefont {H.}~\bibnamefont {Silvério}}, \bibinfo {author} {\bibfnamefont {S.}~\bibnamefont {Grijalva}}, \bibinfo {author} {\bibfnamefont {A.}~\bibnamefont {Cornillot}}, \bibinfo {author} {\bibfnamefont {L.}~\bibnamefont {Henriet}}, \bibinfo {author} {\bibfnamefont {L.}~\bibnamefont {Ajdnik}}, \bibinfo {author} {\bibfnamefont {P.}~\bibnamefont {Karalekas}}, \bibinfo {author} {\bibfnamefont {L.}~\bibnamefont {Leclerc}}, \bibinfo {author} {\bibfnamefont {C.}~\bibnamefont {de~Terrasson}}, \bibinfo {author} {\bibfnamefont {L.}~\bibnamefont {Vignoli}}, \bibinfo {author} {\bibnamefont {Jbrem}}, \bibinfo {author} {\bibfnamefont {M.}~\bibnamefont {D'Arcangelo}}, \bibinfo {author} {\bibfnamefont {C.}~\bibnamefont {Dalyac}}, \bibinfo {author} {\bibnamefont {Louis-PaulHenry}}, \bibinfo {author} {\bibfnamefont {A.}~\bibnamefont {Wennersteen}}, \bibinfo {author} {\bibfnamefont {D.}~\bibnamefont {Gessa}}, \bibinfo {author} {\bibfnamefont {L.}~\bibnamefont {Emmanuel}}, \bibinfo
  {author} {\bibfnamefont {R.}~\bibnamefont {Dutta}}, \bibinfo {author} {\bibfnamefont {N.}~\bibnamefont {Shammah}}, \bibinfo {author} {\bibnamefont {Codoscope}}, \bibinfo {author} {\bibnamefont {MatthieuMoreau}}, \bibinfo {author} {\bibfnamefont {R.}~\bibnamefont {Tsai}}, \bibinfo {author} {\bibfnamefont {Y.}~\bibnamefont {Gondhalekar}}, \bibinfo {author} {\bibfnamefont {A.~B.}\ \bibnamefont {Rava}}, \bibinfo {author} {\bibfnamefont {A.}~\bibnamefont {Panigrahi}}, \bibinfo {author} {\bibnamefont {Harold}}, \bibinfo {author} {\bibnamefont {Julius}}, \bibinfo {author} {\bibfnamefont {L.-J.}\ \bibnamefont {Tallot}}, \bibinfo {author} {\bibnamefont {Oliver}}, \bibinfo {author} {\bibnamefont {Slimane33}},\ and\ \bibinfo {author} {\bibnamefont {WingCode}},\ }\href {https://doi.org/10.5281/zenodo.13950061} {\bibinfo {title} {pasqal-io/pulser: v1.1.0}} (\bibinfo {year} {2024})\BibitemShut {NoStop}%
\bibitem [{\citenamefont {Buluta}\ \emph {et~al.}(2011)\citenamefont {Buluta}, \citenamefont {Ashhab},\ and\ \citenamefont {Nori}}]{buluta2011natural}%
  \BibitemOpen
  \bibfield  {author} {\bibinfo {author} {\bibfnamefont {I.}~\bibnamefont {Buluta}}, \bibinfo {author} {\bibfnamefont {S.}~\bibnamefont {Ashhab}},\ and\ \bibinfo {author} {\bibfnamefont {F.}~\bibnamefont {Nori}},\ }\bibfield  {title} {\bibinfo {title} {Natural and artificial atoms for quantum computation},\ }\href {http://dx.doi.org/10.1088/0034-4885/74/10/104401} {\bibfield  {journal} {\bibinfo  {journal} {Rep. Prog. Phys.}\ }\textbf {\bibinfo {volume} {74}},\ \bibinfo {pages} {104401} (\bibinfo {year} {2011})}\BibitemShut {NoStop}%
\bibitem [{\citenamefont {Gu}\ \emph {et~al.}(2017)\citenamefont {Gu}, \citenamefont {Kockum}, \citenamefont {Miranowicz}, \citenamefont {Liu},\ and\ \citenamefont {Nori}}]{Gu2017}%
  \BibitemOpen
  \bibfield  {author} {\bibinfo {author} {\bibfnamefont {X.}~\bibnamefont {Gu}}, \bibinfo {author} {\bibfnamefont {A.~F.}\ \bibnamefont {Kockum}}, \bibinfo {author} {\bibfnamefont {A.}~\bibnamefont {Miranowicz}}, \bibinfo {author} {\bibfnamefont {Y.-x.}\ \bibnamefont {Liu}},\ and\ \bibinfo {author} {\bibfnamefont {F.}~\bibnamefont {Nori}},\ }\bibfield  {title} {\bibinfo {title} {Microwave photonics with superconducting quantum circuits},\ }\href {https://doi.org/10.1016/j.physrep.2017.10.002} {\bibfield  {journal} {\bibinfo  {journal} {Physics Reports}\ }\textbf {\bibinfo {volume} {718–719}},\ \bibinfo {pages} {1–102} (\bibinfo {year} {2017})}\BibitemShut {NoStop}%
\bibitem [{\citenamefont {Kockum}\ and\ \citenamefont {Nori}(2019)}]{Kockum2019}%
  \BibitemOpen
  \bibfield  {author} {\bibinfo {author} {\bibfnamefont {A.~F.}\ \bibnamefont {Kockum}}\ and\ \bibinfo {author} {\bibfnamefont {F.}~\bibnamefont {Nori}},\ }\bibinfo {title} {Quantum bits with {Josephson Junctions}},\ in\ \href {https://doi.org/10.1007/978-3-030-20726-7_17} {\emph {\bibinfo {booktitle} {Fundamentals and Frontiers of the Josephson Effect}}}\ (\bibinfo  {publisher} {Springer International Publishing},\ \bibinfo {year} {2019})\ p.\ \bibinfo {pages} {703–741}\BibitemShut {NoStop}%
\bibitem [{\citenamefont {Cheng}\ \emph {et~al.}(2023)\citenamefont {Cheng}, \citenamefont {Deng}, \citenamefont {Gu}, \citenamefont {He}, \citenamefont {Hu}, \citenamefont {Huang}, \citenamefont {Li}, \citenamefont {Lin}, \citenamefont {Lu}, \citenamefont {Lu}, \citenamefont {Qiu}, \citenamefont {Wang}, \citenamefont {Xin}, \citenamefont {Yu}, \citenamefont {Yung}, \citenamefont {Zeng}, \citenamefont {Zhang}, \citenamefont {Zhong}, \citenamefont {Peng}, \citenamefont {Nori},\ and\ \citenamefont {Yu}}]{Cheng2023}%
  \BibitemOpen
  \bibfield  {author} {\bibinfo {author} {\bibfnamefont {B.}~\bibnamefont {Cheng}}, \bibinfo {author} {\bibfnamefont {X.-H.}\ \bibnamefont {Deng}}, \bibinfo {author} {\bibfnamefont {X.}~\bibnamefont {Gu}}, \bibinfo {author} {\bibfnamefont {Y.}~\bibnamefont {He}}, \bibinfo {author} {\bibfnamefont {G.}~\bibnamefont {Hu}}, \bibinfo {author} {\bibfnamefont {P.}~\bibnamefont {Huang}}, \bibinfo {author} {\bibfnamefont {J.}~\bibnamefont {Li}}, \bibinfo {author} {\bibfnamefont {B.-C.}\ \bibnamefont {Lin}}, \bibinfo {author} {\bibfnamefont {D.}~\bibnamefont {Lu}}, \bibinfo {author} {\bibfnamefont {Y.}~\bibnamefont {Lu}}, \bibinfo {author} {\bibfnamefont {C.}~\bibnamefont {Qiu}}, \bibinfo {author} {\bibfnamefont {H.}~\bibnamefont {Wang}}, \bibinfo {author} {\bibfnamefont {T.}~\bibnamefont {Xin}}, \bibinfo {author} {\bibfnamefont {S.}~\bibnamefont {Yu}}, \bibinfo {author} {\bibfnamefont {M.-H.}\ \bibnamefont {Yung}}, \bibinfo {author} {\bibfnamefont {J.}~\bibnamefont {Zeng}}, \bibinfo {author} {\bibfnamefont
  {S.}~\bibnamefont {Zhang}}, \bibinfo {author} {\bibfnamefont {Y.}~\bibnamefont {Zhong}}, \bibinfo {author} {\bibfnamefont {X.}~\bibnamefont {Peng}}, \bibinfo {author} {\bibfnamefont {F.}~\bibnamefont {Nori}},\ and\ \bibinfo {author} {\bibfnamefont {D.}~\bibnamefont {Yu}},\ }\bibfield  {title} {\bibinfo {title} {Noisy intermediate-scale quantum computers},\ }\href {http://dx.doi.org/10.1007/s11467-022-1249-z} {\bibfield  {journal} {\bibinfo  {journal} {Frontiers of Physics}\ }\textbf {\bibinfo {volume} {18}},\ \bibinfo {pages} {21308} (\bibinfo {year} {2023})}\BibitemShut {NoStop}%
\bibitem [{\citenamefont {Shillito}\ \emph {et~al.}(2021)\citenamefont {Shillito}, \citenamefont {Gross}, \citenamefont {Di~Paolo}, \citenamefont {Genois},\ and\ \citenamefont {Blais}}]{dysolve}%
  \BibitemOpen
  \bibfield  {author} {\bibinfo {author} {\bibfnamefont {R.}~\bibnamefont {Shillito}}, \bibinfo {author} {\bibfnamefont {J.~A.}\ \bibnamefont {Gross}}, \bibinfo {author} {\bibfnamefont {A.}~\bibnamefont {Di~Paolo}}, \bibinfo {author} {\bibfnamefont {E.}~\bibnamefont {Genois}},\ and\ \bibinfo {author} {\bibfnamefont {A.}~\bibnamefont {Blais}},\ }\bibfield  {title} {\bibinfo {title} {Fast and differentiable simulation of driven quantum systems},\ }\href {https://doi.org/10.1103/PhysRevResearch.3.033266} {\bibfield  {journal} {\bibinfo  {journal} {Phys. Rev. Res.}\ }\textbf {\bibinfo {volume} {3}},\ \bibinfo {pages} {033266} (\bibinfo {year} {2021})}\BibitemShut {NoStop}%
\bibitem [{qut()}]{qutip-cuquantum}%
  \BibitemOpen
  \href {https://github.com/qutip/qutip-cuquantum} {\bibinfo {title} {https://github.com/qutip/qutip-cuquantum}}\BibitemShut {NoStop}%
\bibitem [{\citenamefont {Bayraktar}\ \emph {et~al.}(2023)\citenamefont {Bayraktar}, \citenamefont {Charara}, \citenamefont {Clark}, \citenamefont {Cohen}, \citenamefont {Costa}, \citenamefont {Fang}, \citenamefont {Gao}, \citenamefont {Guan}, \citenamefont {Gunnels}, \citenamefont {Haidar}, \citenamefont {Hehn}, \citenamefont {Hohnerbach}, \citenamefont {Jones}, \citenamefont {Lubowe}, \citenamefont {Lyakh}, \citenamefont {Morino}, \citenamefont {Springer}, \citenamefont {Stanwyck}, \citenamefont {Terentyev}, \citenamefont {Varadhan}, \citenamefont {Wong},\ and\ \citenamefont {Yamaguchi}}]{cuquantum}%
  \BibitemOpen
  \bibfield  {author} {\bibinfo {author} {\bibfnamefont {H.}~\bibnamefont {Bayraktar}}, \bibinfo {author} {\bibfnamefont {A.}~\bibnamefont {Charara}}, \bibinfo {author} {\bibfnamefont {D.}~\bibnamefont {Clark}}, \bibinfo {author} {\bibfnamefont {S.}~\bibnamefont {Cohen}}, \bibinfo {author} {\bibfnamefont {T.}~\bibnamefont {Costa}}, \bibinfo {author} {\bibfnamefont {Y.-L.~L.}\ \bibnamefont {Fang}}, \bibinfo {author} {\bibfnamefont {Y.}~\bibnamefont {Gao}}, \bibinfo {author} {\bibfnamefont {J.}~\bibnamefont {Guan}}, \bibinfo {author} {\bibfnamefont {J.}~\bibnamefont {Gunnels}}, \bibinfo {author} {\bibfnamefont {A.}~\bibnamefont {Haidar}}, \bibinfo {author} {\bibfnamefont {A.}~\bibnamefont {Hehn}}, \bibinfo {author} {\bibfnamefont {M.}~\bibnamefont {Hohnerbach}}, \bibinfo {author} {\bibfnamefont {M.}~\bibnamefont {Jones}}, \bibinfo {author} {\bibfnamefont {T.}~\bibnamefont {Lubowe}}, \bibinfo {author} {\bibfnamefont {D.}~\bibnamefont {Lyakh}}, \bibinfo {author} {\bibfnamefont {S.}~\bibnamefont {Morino}},
  \bibinfo {author} {\bibfnamefont {P.}~\bibnamefont {Springer}}, \bibinfo {author} {\bibfnamefont {S.}~\bibnamefont {Stanwyck}}, \bibinfo {author} {\bibfnamefont {I.}~\bibnamefont {Terentyev}}, \bibinfo {author} {\bibfnamefont {S.}~\bibnamefont {Varadhan}}, \bibinfo {author} {\bibfnamefont {J.}~\bibnamefont {Wong}},\ and\ \bibinfo {author} {\bibfnamefont {T.}~\bibnamefont {Yamaguchi}},\ }\bibfield  {title} {\bibinfo {title} {cuquantum sdk: A high-performance library for accelerating quantum science},\ }in\ \href {https://doi.org/10.1109/QCE57702.2023.00119} {\emph {\bibinfo {booktitle} {2023 IEEE International Conference on Quantum Computing and Engineering (QCE)}}},\ Vol.~\bibinfo {volume} {01}\ (\bibinfo {year} {2023})\ pp.\ \bibinfo {pages} {1050--1061}\BibitemShut {NoStop}%
\bibitem [{\citenamefont {Kr{\"a}mer}\ \emph {et~al.}(2018)\citenamefont {Kr{\"a}mer}, \citenamefont {Plankensteiner}, \citenamefont {Ostermann},\ and\ \citenamefont {Ritsch}}]{kramer2018quantumoptics}%
  \BibitemOpen
  \bibfield  {author} {\bibinfo {author} {\bibfnamefont {S.}~\bibnamefont {Kr{\"a}mer}}, \bibinfo {author} {\bibfnamefont {D.}~\bibnamefont {Plankensteiner}}, \bibinfo {author} {\bibfnamefont {L.}~\bibnamefont {Ostermann}},\ and\ \bibinfo {author} {\bibfnamefont {H.}~\bibnamefont {Ritsch}},\ }\bibfield  {title} {\bibinfo {title} {{QuantumOptics. jl}: A {J}ulia framework for simulating open quantum systems},\ }\href {https://doi.org/10.1016/j.cpc.2018.02.004} {\bibfield  {journal} {\bibinfo  {journal} {Computer Physics Communications}\ }\textbf {\bibinfo {volume} {227}},\ \bibinfo {pages} {109} (\bibinfo {year} {2018})}\BibitemShut {NoStop}%
\bibitem [{\citenamefont {Gilchrist}\ \emph {et~al.}(2005)\citenamefont {Gilchrist}, \citenamefont {Langford},\ and\ \citenamefont {Nielsen}}]{PhysRevA.71.062310}%
  \BibitemOpen
  \bibfield  {author} {\bibinfo {author} {\bibfnamefont {A.}~\bibnamefont {Gilchrist}}, \bibinfo {author} {\bibfnamefont {N.~K.}\ \bibnamefont {Langford}},\ and\ \bibinfo {author} {\bibfnamefont {M.~A.}\ \bibnamefont {Nielsen}},\ }\bibfield  {title} {\bibinfo {title} {Distance measures to compare real and ideal quantum processes},\ }\href {https://doi.org/10.1103/PhysRevA.71.062310} {\bibfield  {journal} {\bibinfo  {journal} {Phys. Rev. A}\ }\textbf {\bibinfo {volume} {71}},\ \bibinfo {pages} {062310} (\bibinfo {year} {2005})}\BibitemShut {NoStop}%
\bibitem [{qno()}]{qnotebooks}%
  \BibitemOpen
  \href {https://github.com/qutip/qutip-notebooks} {\bibinfo {title} {https://github.com/qutip/qutip-notebooks}}\BibitemShut {NoStop}%
\bibitem [{\citenamefont {Gleyzes}\ \emph {et~al.}(2007)\citenamefont {Gleyzes}, \citenamefont {Kuhr}, \citenamefont {Guerlin}, \citenamefont {Bernu}, \citenamefont {Deléglise}, \citenamefont {Busk~Hoff}, \citenamefont {Brune}, \citenamefont {Raimond},\ and\ \citenamefont {Haroche}}]{Gleyzes2007}%
  \BibitemOpen
  \bibfield  {author} {\bibinfo {author} {\bibfnamefont {S.}~\bibnamefont {Gleyzes}}, \bibinfo {author} {\bibfnamefont {S.}~\bibnamefont {Kuhr}}, \bibinfo {author} {\bibfnamefont {C.}~\bibnamefont {Guerlin}}, \bibinfo {author} {\bibfnamefont {J.}~\bibnamefont {Bernu}}, \bibinfo {author} {\bibfnamefont {S.}~\bibnamefont {Deléglise}}, \bibinfo {author} {\bibfnamefont {U.}~\bibnamefont {Busk~Hoff}}, \bibinfo {author} {\bibfnamefont {M.}~\bibnamefont {Brune}}, \bibinfo {author} {\bibfnamefont {J.-M.}\ \bibnamefont {Raimond}},\ and\ \bibinfo {author} {\bibfnamefont {S.}~\bibnamefont {Haroche}},\ }\bibfield  {title} {\bibinfo {title} {Quantum jumps of light recording the birth and death of a photon in a cavity},\ }\href {http://dx.doi.org/10.1038/nature05589} {\bibfield  {journal} {\bibinfo  {journal} {Nature}\ }\textbf {\bibinfo {volume} {446}},\ \bibinfo {pages} {297–300} (\bibinfo {year} {2007})}\BibitemShut {NoStop}%
\bibitem [{\citenamefont {Ardelt}\ \emph {et~al.}(2014)\citenamefont {Ardelt}, \citenamefont {Hanschke}, \citenamefont {Fischer}, \citenamefont {M\"uller}, \citenamefont {Kleinkauf}, \citenamefont {Koller}, \citenamefont {Bechtold}, \citenamefont {Simmet}, \citenamefont {Wierzbowski}, \citenamefont {Riedl}, \citenamefont {Abstreiter},\ and\ \citenamefont {Finley}}]{PhysRevB.90.241404}%
  \BibitemOpen
  \bibfield  {author} {\bibinfo {author} {\bibfnamefont {P.-L.}\ \bibnamefont {Ardelt}}, \bibinfo {author} {\bibfnamefont {L.}~\bibnamefont {Hanschke}}, \bibinfo {author} {\bibfnamefont {K.~A.}\ \bibnamefont {Fischer}}, \bibinfo {author} {\bibfnamefont {K.}~\bibnamefont {M\"uller}}, \bibinfo {author} {\bibfnamefont {A.}~\bibnamefont {Kleinkauf}}, \bibinfo {author} {\bibfnamefont {M.}~\bibnamefont {Koller}}, \bibinfo {author} {\bibfnamefont {A.}~\bibnamefont {Bechtold}}, \bibinfo {author} {\bibfnamefont {T.}~\bibnamefont {Simmet}}, \bibinfo {author} {\bibfnamefont {J.}~\bibnamefont {Wierzbowski}}, \bibinfo {author} {\bibfnamefont {H.}~\bibnamefont {Riedl}}, \bibinfo {author} {\bibfnamefont {G.}~\bibnamefont {Abstreiter}},\ and\ \bibinfo {author} {\bibfnamefont {J.~J.}\ \bibnamefont {Finley}},\ }\bibfield  {title} {\bibinfo {title} {Dissipative preparation of the exciton and biexciton in self-assembled quantum dots on picosecond time scales},\ }\href {https://link.aps.org/doi/10.1103/PhysRevB.90.241404}
  {\bibfield  {journal} {\bibinfo  {journal} {Phys. Rev. B}\ }\textbf {\bibinfo {volume} {90}},\ \bibinfo {pages} {241404} (\bibinfo {year} {2014})}\BibitemShut {NoStop}%
\bibitem [{\citenamefont {Bartolo}\ \emph {et~al.}(2017)\citenamefont {Bartolo}, \citenamefont {Minganti}, \citenamefont {Lolli},\ and\ \citenamefont {Ciuti}}]{minganti1}%
  \BibitemOpen
  \bibfield  {author} {\bibinfo {author} {\bibfnamefont {N.}~\bibnamefont {Bartolo}}, \bibinfo {author} {\bibfnamefont {F.}~\bibnamefont {Minganti}}, \bibinfo {author} {\bibfnamefont {J.}~\bibnamefont {Lolli}},\ and\ \bibinfo {author} {\bibfnamefont {C.}~\bibnamefont {Ciuti}},\ }\bibfield  {title} {\bibinfo {title} {Homodyne versus photon-counting quantum trajectories for dissipative {K}err resonators with two-photon driving},\ }\href {http://dx.doi.org/10.1140/epjst/e2016-60385-8} {\bibfield  {journal} {\bibinfo  {journal} {The European Physical Journal Special Topics}\ }\textbf {\bibinfo {volume} {226}},\ \bibinfo {pages} {2705–2713} (\bibinfo {year} {2017})}\BibitemShut {NoStop}%
\bibitem [{\citenamefont {Minganti}\ \emph {et~al.}(2016)\citenamefont {Minganti}, \citenamefont {Bartolo}, \citenamefont {Lolli}, \citenamefont {Casteels},\ and\ \citenamefont {Ciuti}}]{minganti2}%
  \BibitemOpen
  \bibfield  {author} {\bibinfo {author} {\bibfnamefont {F.}~\bibnamefont {Minganti}}, \bibinfo {author} {\bibfnamefont {N.}~\bibnamefont {Bartolo}}, \bibinfo {author} {\bibfnamefont {J.}~\bibnamefont {Lolli}}, \bibinfo {author} {\bibfnamefont {W.}~\bibnamefont {Casteels}},\ and\ \bibinfo {author} {\bibfnamefont {C.}~\bibnamefont {Ciuti}},\ }\bibfield  {title} {\bibinfo {title} {Exact results for schr\"{o}dinger cats in driven-dissipative systems and their feedback control},\ }\href {http://dx.doi.org/10.1038/srep26987} {\bibfield  {journal} {\bibinfo  {journal} {Scientific Reports}\ }\textbf {\bibinfo {volume} {6}} (\bibinfo {year} {2016})}\BibitemShut {NoStop}%
\bibitem [{\citenamefont {Bartolo}\ \emph {et~al.}(2016)\citenamefont {Bartolo}, \citenamefont {Minganti}, \citenamefont {Casteels},\ and\ \citenamefont {Ciuti}}]{minganti3}%
  \BibitemOpen
  \bibfield  {author} {\bibinfo {author} {\bibfnamefont {N.}~\bibnamefont {Bartolo}}, \bibinfo {author} {\bibfnamefont {F.}~\bibnamefont {Minganti}}, \bibinfo {author} {\bibfnamefont {W.}~\bibnamefont {Casteels}},\ and\ \bibinfo {author} {\bibfnamefont {C.}~\bibnamefont {Ciuti}},\ }\bibfield  {title} {\bibinfo {title} {Exact steady state of a kerr resonator with one- and two-photon driving and dissipation: Controllable wigner-function multimodality and dissipative phase transitions},\ }\href {https://link.aps.org/doi/10.1103/PhysRevA.94.033841} {\bibfield  {journal} {\bibinfo  {journal} {Phys. Rev. A}\ }\textbf {\bibinfo {volume} {94}},\ \bibinfo {pages} {033841} (\bibinfo {year} {2016})}\BibitemShut {NoStop}%
\bibitem [{\citenamefont {Fischer}\ \emph {et~al.}(2017)\citenamefont {Fischer}, \citenamefont {Kelaita}, \citenamefont {Sapra}, \citenamefont {Dory}, \citenamefont {Lagoudakis}, \citenamefont {M\"uller},\ and\ \citenamefont {Vu\ifmmode \check{c}\else \v{c}\fi{}kovi\ifmmode~\acute{c}\else \'{c}\fi{}}}]{PhysRevApplied.7.044002}%
  \BibitemOpen
  \bibfield  {author} {\bibinfo {author} {\bibfnamefont {K.~A.}\ \bibnamefont {Fischer}}, \bibinfo {author} {\bibfnamefont {Y.~A.}\ \bibnamefont {Kelaita}}, \bibinfo {author} {\bibfnamefont {N.~V.}\ \bibnamefont {Sapra}}, \bibinfo {author} {\bibfnamefont {C.}~\bibnamefont {Dory}}, \bibinfo {author} {\bibfnamefont {K.~G.}\ \bibnamefont {Lagoudakis}}, \bibinfo {author} {\bibfnamefont {K.}~\bibnamefont {M\"uller}},\ and\ \bibinfo {author} {\bibfnamefont {J.}~\bibnamefont {Vu\ifmmode \check{c}\else \v{c}\fi{}kovi\ifmmode~\acute{c}\else \'{c}\fi{}}},\ }\bibfield  {title} {\bibinfo {title} {On-chip architecture for self-homodyned nonclassical light},\ }\href {https://link.aps.org/doi/10.1103/PhysRevApplied.7.044002} {\bibfield  {journal} {\bibinfo  {journal} {Phys. Rev. Appl.}\ }\textbf {\bibinfo {volume} {7}},\ \bibinfo {pages} {044002} (\bibinfo {year} {2017})}\BibitemShut {NoStop}%
\end{thebibliography}%


%
\end{document}